\documentclass[11pt]{article}
\usepackage[utf8]{inputenc}
\usepackage[T1]{fontenc}
\usepackage{lmodern}
\usepackage{amsmath, amssymb, amsthm, mathtools, bm, mathrsfs}
\usepackage{geometry}
\usepackage{graphicx}
\usepackage{xcolor}
\usepackage{booktabs}
\usepackage{longtable}
\usepackage{array}
\usepackage{enumitem}
\usepackage[labelfont=bf,labelsep=colon]{caption}
\usepackage{microtype}
\usepackage{etoolbox}
\usepackage[colorlinks=true, linkcolor=blue!70!black, citecolor=blue!70!black, urlcolor=blue!70!black, pdfauthor={Patrick Hayden}, pdftitle={Robust Passive Quantum Memory in Three Dimensions, Part I: Thermal and Zero-Temperature Perturbative Stability}]{hyperref}
\newif\ifstandardproofs
\standardproofstrue

\definecolor{toclink}{rgb}{0,0,0}

\newcommand{\leanHostType}{github} 
\newcommand{\leanGitHubRoot}{https://github.com/pmhayden/passive-quantum-memory}
\newcommand{\leanDocGenRoot}{https://pmhayden.github.io/passive-quantum-memory/docs}
\newcommand{\leanRepoRev}{main}
\definecolor{leanheadcolor}{rgb}{0.08, 0.22, 0.42}
\definecolor{leancodecolor}{rgb}{0.10, 0.24, 0.45}
\newif\ifshowleanlinks
\showleanlinkstrue

\begingroup\catcode`\#=12 \gdef\leanHash{#}\endgroup
\begingroup\catcode`\_=12 \gdef\leanUnderscore{_}\endgroup

\makeatletter

\gdef\currentleankey{}
\newcommand{\leanlink}[1]{\gdef\currentleankey{#1}\ignorespaces}

\newcommand{\lean@setrepourl}{%
  \ifdefstring{\leanHostType}{github}{%
    \edef\lean@resolvedurl{\leanGitHubRoot/tree/\leanRepoRev}%
  }{%
    \edef\lean@resolvedurl{\leanDocGenRoot/}%
  }%
}

\newcommand{\lean@setfileurl}[1]{%
  \edef\lean@resolvedurl{\leanGitHubRoot/blob/\leanRepoRev/#1}%
}

\newcommand{\lean@setdeclurl}[1]{%
  \ifcsdef{lean@file@#1}{}{%
    \PackageError{lean_links}{Unknown Lean registry key '#1'}{Run scripts/generate_lean_links.py}%
  }%
  \ifdefstring{\leanHostType}{github}{%
    \edef\lean@resolvedurl{\leanGitHubRoot/blob/\leanRepoRev/\csuse{lean@file@#1}\leanHash L\csuse{lean@start@#1}-L\csuse{lean@end@#1}}%
  }{%
    \edef\lean@resolvedurl{\leanDocGenRoot/\csuse{lean@modpath@#1}.html\leanHash\csuse{lean@fqname@#1}}%
  }%
}

\newcommand{\leanthmhead}[3]{%
  \edef\lean@tmpkey{\currentleankey}%
  \gdef\currentleankey{}%
  \ifshowleanlinks
    \ifdefempty{\lean@tmpkey}{%
      \thmname{#1}\thmnumber{\@ifnotempty{#1}{ }\@upn{#2}}%
    }{%
      \lean@setdeclurl{\lean@tmpkey}%
      {\hypersetup{urlcolor=leanheadcolor}\expandafter\href\expandafter{\lean@resolvedurl}{\thmname{#1}\thmnumber{\@ifnotempty{#1}{ }\@upn{#2}}}}%
    }%
  \else
    \thmname{#1}\thmnumber{\@ifnotempty{#1}{ }\@upn{#2}}%
  \fi
  \thmnote{ {\mdseries\upshape(#3)}}%
}

\def\lean@breakablepath#1{\lean@breakablepath@aux#1/\lean@stop}
\def\lean@breakablepath@aux#1/#2\lean@stop{%
  #1\ifx\relax#2\relax\else/\allowbreak\lean@breakablepath@aux#2\lean@stop\fi}

\DeclareRobustCommand{\leanrepolink}{%
  \ifshowleanlinks
    \begingroup
      \lean@setrepourl
      \hypersetup{urlcolor=leancodecolor}%
      \expandafter\href\expandafter{\lean@resolvedurl}{\texttt{passive-quantum-memory}}%
    \endgroup
  \else
    \texttt{passive-quantum-memory}%
  \fi
}

\DeclareRobustCommand{\leanfile}[2][]{%
  \ifshowleanlinks
    \begingroup
      \lean@setfileurl{#2}%
      \hypersetup{urlcolor=leancodecolor}%
      \ifstrempty{#1}{%
        \expandafter\href\expandafter{\lean@resolvedurl}{\texttt{\lean@breakablepath{#2}}}%
      }{%
        \expandafter\href\expandafter{\lean@resolvedurl}{\texttt{\lean@breakablepath{#1}}}%
      }%
    \endgroup
  \else
    \ifstrempty{#1}{%
      \texttt{\lean@breakablepath{#2}}%
    }{%
      \texttt{\lean@breakablepath{#1}}%
    }%
  \fi
}

\DeclareRobustCommand{\leandecl}[1]{%
  \ifcsdef{lean@disp@#1}{}{%
    \PackageError{lean_links}{Unknown Lean declaration key '#1'}{Run scripts/generate_lean_links.py}%
  }%
  \ifshowleanlinks
    \begingroup
      \lean@setdeclurl{#1}%
      \hypersetup{urlcolor=leancodecolor}%
      \expandafter\href\expandafter{\lean@resolvedurl}{\texttt{\csuse{lean@disp@#1}}}%
    \endgroup
  \else
    \texttt{\csuse{lean@disp@#1}}%
  \fi
}

\makeatother

\newtheoremstyle{leanplain}{}{}{\itshape}{}{\bfseries}{.}{5pt plus 1pt minus 1pt}{\leanthmhead{#1}{#2}{#3}}
\newtheoremstyle{leandef}{}{}{\normalfont}{}{\bfseries}{.}{5pt plus 1pt minus 1pt}{\leanthmhead{#1}{#2}{#3}}

\csdef{lean@file@def:3d-css-complex}{Defs/F2ChainAndCSS.lean}
\csdef{lean@start@def:3d-css-complex}{140}
\csdef{lean@end@def:3d-css-complex}{146}
\csdef{lean@module@def:3d-css-complex}{Defs.F2ChainAndCSS}
\csdef{lean@modpath@def:3d-css-complex}{Defs/F2ChainAndCSS}
\csdef{lean@fqname@def:3d-css-complex}{CSSCode3D}
\csdef{lean@disp@def:3d-css-complex}{CSSCode3\allowbreak{}D}
\csdef{lean@file@def:well-separated-base}{Defs/Step1PeriodicLLL.lean}
\csdef{lean@start@def:well-separated-base}{96}
\csdef{lean@end@def:well-separated-base}{121}
\csdef{lean@module@def:well-separated-base}{Defs.Step1PeriodicLLL}
\csdef{lean@modpath@def:well-separated-base}{Defs/Step1PeriodicLLL}
\csdef{lean@fqname@def:well-separated-base}{ValidLLLSphericalCapPerturbation}
\csdef{lean@disp@def:well-separated-base}{Valid\allowbreak{}LLLSpherical\allowbreak{}Cap\allowbreak{}Perturbation}
\csdef{lean@file@lem:seed-code}{Defs/BDLParamsAndSeed.lean}
\csdef{lean@start@lem:seed-code}{190}
\csdef{lean@end@lem:seed-code}{196}
\csdef{lean@module@lem:seed-code}{Defs.BDLParamsAndSeed}
\csdef{lean@modpath@lem:seed-code}{Defs/BDLParamsAndSeed}
\csdef{lean@fqname@lem:seed-code}{bdlSurfaceSeedCode}
\csdef{lean@disp@lem:seed-code}{bdl\allowbreak{}Surface\allowbreak{}Seed\allowbreak{}Code}
\csdef{lean@file@prop:step1-lll}{Proofs/Step1LLLDensity/SphericalCapAndLLL.lean}
\csdef{lean@start@prop:step1-lll}{125}
\csdef{lean@end@prop:step1-lll}{130}
\csdef{lean@module@prop:step1-lll}{Proofs.Step1LLLDensity.SphericalCapAndLLL}
\csdef{lean@modpath@prop:step1-lll}{Proofs/Step1LLLDensity/SphericalCapAndLLL}
\csdef{lean@fqname@prop:step1-lll}{bdl\leanUnderscore step1\leanUnderscore lll\leanUnderscore perturbation\leanUnderscore exists}
\csdef{lean@disp@prop:step1-lll}{bdl\_\allowbreak{}step1\_\allowbreak{}lll\_\allowbreak{}perturbation\_\allowbreak{}exists}
\csdef{lean@file@def:three-layer-mapping-cone}{Defs/Step2CheckMatrices.lean}
\csdef{lean@start@def:three-layer-mapping-cone}{657}
\csdef{lean@end@def:three-layer-mapping-cone}{661}
\csdef{lean@module@def:three-layer-mapping-cone}{Defs.Step2CheckMatrices}
\csdef{lean@modpath@def:three-layer-mapping-cone}{Defs/Step2CheckMatrices}
\csdef{lean@fqname@def:three-layer-mapping-cone}{bdlStepX}
\csdef{lean@disp@def:three-layer-mapping-cone}{bdl\allowbreak{}Step\allowbreak{}X}
\csdef{lean@file@prop:mapping-cone-invariants}{Proofs/HierarchicalClusterDecoder/ClusterDecoderSynthesis.lean}
\csdef{lean@start@prop:mapping-cone-invariants}{334}
\csdef{lean@end@prop:mapping-cone-invariants}{337}
\csdef{lean@module@prop:mapping-cone-invariants}{Proofs.HierarchicalClusterDecoder.ClusterDecoderSynthesis}
\csdef{lean@modpath@prop:mapping-cone-invariants}{Proofs/HierarchicalClusterDecoder/ClusterDecoderSynthesis}
\csdef{lean@fqname@prop:mapping-cone-invariants}{WeldedHierarchy.BDLDecoder.hasBDLConeInvariants\leanUnderscore bdlIteratedBundled}
\csdef{lean@disp@prop:mapping-cone-invariants}{has\allowbreak{}BDLCone\allowbreak{}Invariants\_\allowbreak{}bdl\allowbreak{}Iterated\allowbreak{}Bundled}
\csdef{lean@file@def:rca}{Proofs/ConeAcyclicityRCA/LocalCocycleDualityAndBalls.lean}
\csdef{lean@start@def:rca}{58}
\csdef{lean@end@def:rca}{61}
\csdef{lean@module@def:rca}{Proofs.ConeAcyclicityRCA.LocalCocycleDualityAndBalls}
\csdef{lean@modpath@def:rca}{Proofs/ConeAcyclicityRCA/LocalCocycleDualityAndBalls}
\csdef{lean@fqname@def:rca}{WeldedHierarchy.locallyGeneratedInBall}
\csdef{lean@disp@def:rca}{locally\allowbreak{}Generated\allowbreak{}In\allowbreak{}Ball}
\csdef{lean@file@def:tqo}{Defs/LocalHamiltonian.lean}
\csdef{lean@start@def:tqo}{173}
\csdef{lean@end@def:tqo}{182}
\csdef{lean@module@def:tqo}{Defs.LocalHamiltonian}
\csdef{lean@modpath@def:tqo}{Defs/LocalHamiltonian}
\csdef{lean@fqname@def:tqo}{SatisfiesTQO1}
\csdef{lean@disp@def:tqo}{Satisfies\allowbreak{}TQO1}
\csdef{lean@file@prop:trinity-equivalence}{Proofs/InhomogeneousPeierls/OperatorTQOFromF2TQO.lean}
\csdef{lean@start@prop:trinity-equivalence}{315}
\csdef{lean@end@prop:trinity-equivalence}{422}
\csdef{lean@module@prop:trinity-equivalence}{Proofs.InhomogeneousPeierls.OperatorTQOFromF2TQO}
\csdef{lean@modpath@prop:trinity-equivalence}{Proofs/InhomogeneousPeierls/OperatorTQOFromF2TQO}
\csdef{lean@fqname@prop:trinity-equivalence}{staticCSSSystem\leanUnderscore satisfiesOperatorTQO}
\csdef{lean@disp@prop:trinity-equivalence}{static\allowbreak{}CSSSystem\_\allowbreak{}satisfies\allowbreak{}Operator\allowbreak{}TQO}
\csdef{lean@file@prop:classical-hpl}{Proofs/HierarchicalClusterDecoder/IteratedLogicalOperatorLifts.lean}
\csdef{lean@start@prop:classical-hpl}{201}
\csdef{lean@end@prop:classical-hpl}{220}
\csdef{lean@module@prop:classical-hpl}{Proofs.HierarchicalClusterDecoder.IteratedLogicalOperatorLifts}
\csdef{lean@modpath@prop:classical-hpl}{Proofs/HierarchicalClusterDecoder/IteratedLogicalOperatorLifts}
\csdef{lean@fqname@prop:classical-hpl}{WeldedHierarchy.hasUniqueLogicalHomology\leanUnderscore bdlIterated}
\csdef{lean@disp@prop:classical-hpl}{has\allowbreak{}Unique\allowbreak{}Logical\allowbreak{}Homology\_\allowbreak{}bdl\allowbreak{}Iterated}
\csdef{lean@file@lem:clean-shell-subcylinder}{Proofs/ConeAcyclicityRCA/RescaledRCASynthesis.lean}
\csdef{lean@start@lem:clean-shell-subcylinder}{309}
\csdef{lean@end@lem:clean-shell-subcylinder}{356}
\csdef{lean@module@lem:clean-shell-subcylinder}{Proofs.ConeAcyclicityRCA.RescaledRCASynthesis}
\csdef{lean@modpath@lem:clean-shell-subcylinder}{Proofs/ConeAcyclicityRCA/RescaledRCASynthesis}
\csdef{lean@fqname@lem:clean-shell-subcylinder}{WeldedHierarchy.bdlHalfStep\leanUnderscore subcylinder\leanUnderscore filling}
\csdef{lean@disp@lem:clean-shell-subcylinder}{bdl\allowbreak{}Half\allowbreak{}Step\_\allowbreak{}subcylinder\_\allowbreak{}filling}
\csdef{lean@file@thm:rca-welded-hierarchy}{Proofs/ConeAcyclicityRCA/RescaledRCASynthesis.lean}
\csdef{lean@start@thm:rca-welded-hierarchy}{452}
\csdef{lean@end@thm:rca-welded-hierarchy}{511}
\csdef{lean@module@thm:rca-welded-hierarchy}{Proofs.ConeAcyclicityRCA.RescaledRCASynthesis}
\csdef{lean@modpath@thm:rca-welded-hierarchy}{Proofs/ConeAcyclicityRCA/RescaledRCASynthesis}
\csdef{lean@fqname@thm:rca-welded-hierarchy}{WeldedHierarchy.bdlWeldedCode\leanUnderscore inf\leanUnderscore cone\leanUnderscore acyclicity}
\csdef{lean@disp@thm:rca-welded-hierarchy}{bdl\allowbreak{}Welded\allowbreak{}Code\_\allowbreak{}inf\_\allowbreak{}cone\_\allowbreak{}acyclicity}
\csdef{lean@file@lem:stage1-cylinder-cleaner}{Proofs/HierarchicalClusterDecoder/TwoStageCleanerComposition.lean}
\csdef{lean@start@lem:stage1-cylinder-cleaner}{122}
\csdef{lean@end@lem:stage1-cylinder-cleaner}{164}
\csdef{lean@module@lem:stage1-cylinder-cleaner}{Proofs.HierarchicalClusterDecoder.TwoStageCleanerComposition}
\csdef{lean@modpath@lem:stage1-cylinder-cleaner}{Proofs/HierarchicalClusterDecoder/TwoStageCleanerComposition}
\csdef{lean@fqname@lem:stage1-cylinder-cleaner}{WeldedHierarchy.BDLDecoder.cleanCylinderX\leanUnderscore spec}
\csdef{lean@disp@lem:stage1-cylinder-cleaner}{clean\allowbreak{}Cylinder\allowbreak{}X\_\allowbreak{}spec}
\csdef{lean@file@lem:stage2-star-cleaner}{Proofs/HierarchicalClusterDecoder/TwoStageCleanerComposition.lean}
\csdef{lean@start@lem:stage2-star-cleaner}{416}
\csdef{lean@end@lem:stage2-star-cleaner}{467}
\csdef{lean@module@lem:stage2-star-cleaner}{Proofs.HierarchicalClusterDecoder.TwoStageCleanerComposition}
\csdef{lean@modpath@lem:stage2-star-cleaner}{Proofs/HierarchicalClusterDecoder/TwoStageCleanerComposition}
\csdef{lean@fqname@lem:stage2-star-cleaner}{WeldedHierarchy.BDLDecoder.cleanStarQ\leanUnderscore spec}
\csdef{lean@disp@lem:stage2-star-cleaner}{clean\allowbreak{}Star\allowbreak{}Q\_\allowbreak{}spec}
\csdef{lean@file@thm:16-27-contraction}{Proofs/HierarchicalClusterDecoder/TwoHopTannerContraction.lean}
\csdef{lean@start@thm:16-27-contraction}{349}
\csdef{lean@end@thm:16-27-contraction}{379}
\csdef{lean@module@thm:16-27-contraction}{Proofs.HierarchicalClusterDecoder.TwoHopTannerContraction}
\csdef{lean@modpath@thm:16-27-contraction}{Proofs/HierarchicalClusterDecoder/TwoHopTannerContraction}
\csdef{lean@fqname@thm:16-27-contraction}{WeldedHierarchy.BDLDecoder.rxCleanedCoarseSyndrome\leanUnderscore twoHop\leanUnderscore contraction}
\csdef{lean@disp@thm:16-27-contraction}{rx\allowbreak{}Cleaned\allowbreak{}Coarse\allowbreak{}Syndrome\_\allowbreak{}two\allowbreak{}Hop\_\allowbreak{}contraction}
\csdef{lean@file@def:multiscale-cluster-decoder}{Proofs/HierarchicalClusterDecoder/DecodedLogicalCommutation.lean}
\csdef{lean@start@def:multiscale-cluster-decoder}{41}
\csdef{lean@end@def:multiscale-cluster-decoder}{54}
\csdef{lean@module@def:multiscale-cluster-decoder}{Proofs.HierarchicalClusterDecoder.DecodedLogicalCommutation}
\csdef{lean@modpath@def:multiscale-cluster-decoder}{Proofs/HierarchicalClusterDecoder/DecodedLogicalCommutation}
\csdef{lean@fqname@def:multiscale-cluster-decoder}{WeldedHierarchy.BDLDecoder.HasBallParityWitnessFamily}
\csdef{lean@disp@def:multiscale-cluster-decoder}{Has\allowbreak{}Ball\allowbreak{}Parity\allowbreak{}Witness\allowbreak{}Family}
\csdef{lean@file@prop:cluster-decoder-witness}{Proofs/HierarchicalClusterDecoder/ClusterDecoderSynthesis.lean}
\csdef{lean@start@prop:cluster-decoder-witness}{378}
\csdef{lean@end@prop:cluster-decoder-witness}{428}
\csdef{lean@module@prop:cluster-decoder-witness}{Proofs.HierarchicalClusterDecoder.ClusterDecoderSynthesis}
\csdef{lean@modpath@prop:cluster-decoder-witness}{Proofs/HierarchicalClusterDecoder/ClusterDecoderSynthesis}
\csdef{lean@fqname@prop:cluster-decoder-witness}{WeldedHierarchy.BDLDecoder.hasBallParityWitnessFamily\leanUnderscore bdlIterated\leanUnderscore succ}
\csdef{lean@disp@prop:cluster-decoder-witness}{has\allowbreak{}Ball\allowbreak{}Parity\allowbreak{}Witness\allowbreak{}Family\_\allowbreak{}bdl\allowbreak{}Iterated\_\allowbreak{}succ}
\csdef{lean@file@lem:percolating-cluster-bounds}{Proofs/TwoAxisMemorySynthesis/MacroscopicLambdaPowImmunity.lean}
\csdef{lean@start@lem:percolating-cluster-bounds}{41}
\csdef{lean@end@lem:percolating-cluster-bounds}{62}
\csdef{lean@module@lem:percolating-cluster-bounds}{Proofs.TwoAxisMemorySynthesis.MacroscopicLambdaPowImmunity}
\csdef{lean@modpath@lem:percolating-cluster-bounds}{Proofs/TwoAxisMemorySynthesis/MacroscopicLambdaPowImmunity}
\csdef{lean@fqname@lem:percolating-cluster-bounds}{hasBallParityWitnessFamily\leanUnderscore bdlWelded\leanUnderscore lambdaPow}
\csdef{lean@disp@lem:percolating-cluster-bounds}{has\allowbreak{}Ball\allowbreak{}Parity\allowbreak{}Witness\allowbreak{}Family\_\allowbreak{}bdl\allowbreak{}Welded\_\allowbreak{}lambda\allowbreak{}Pow}
\csdef{lean@file@thm:walsh-fourier-peierls}{Proofs/HierarchicalClusterDecoder/WalshFourierCosetAndHomology.lean}
\csdef{lean@start@thm:walsh-fourier-peierls}{215}
\csdef{lean@end@thm:walsh-fourier-peierls}{258}
\csdef{lean@module@thm:walsh-fourier-peierls}{Proofs.HierarchicalClusterDecoder.WalshFourierCosetAndHomology}
\csdef{lean@modpath@thm:walsh-fourier-peierls}{Proofs/HierarchicalClusterDecoder/WalshFourierCosetAndHomology}
\csdef{lean@fqname@thm:walsh-fourier-peierls}{WeldedHierarchy.coset\leanUnderscore syndromeWeight\leanUnderscore sum\leanUnderscore le}
\csdef{lean@disp@thm:walsh-fourier-peierls}{coset\_\allowbreak{}syndrome\allowbreak{}Weight\_\allowbreak{}sum\_\allowbreak{}le}
\csdef{lean@file@def:safe-decoded-observables}{Challenge.lean}
\csdef{lean@start@def:safe-decoded-observables}{79}
\csdef{lean@end@def:safe-decoded-observables}{112}
\csdef{lean@module@def:safe-decoded-observables}{Challenge}
\csdef{lean@modpath@def:safe-decoded-observables}{Challenge}
\csdef{lean@fqname@def:safe-decoded-observables}{PassiveQuantumMemory.HasSafeDecodedLogicalObservables}
\csdef{lean@disp@def:safe-decoded-observables}{Has\allowbreak{}Safe\allowbreak{}Decoded\allowbreak{}Logical\allowbreak{}Observables}
\csdef{lean@file@prop:safe-subspace-immunity}{Proofs/TwoAxisMemorySynthesis/Part1TwoAxisMemorySynthesis.lean}
\csdef{lean@start@prop:safe-subspace-immunity}{551}
\csdef{lean@end@prop:safe-subspace-immunity}{618}
\csdef{lean@module@prop:safe-subspace-immunity}{Proofs.TwoAxisMemorySynthesis.Part1TwoAxisMemorySynthesis}
\csdef{lean@modpath@prop:safe-subspace-immunity}{Proofs/TwoAxisMemorySynthesis/Part1TwoAxisMemorySynthesis}
\csdef{lean@fqname@prop:safe-subspace-immunity}{Part1TwoAxisMemorySynthesis.bdl\leanUnderscore inhomogeneous\leanUnderscore codeSlot\leanUnderscore memory\leanUnderscore package}
\csdef{lean@disp@prop:safe-subspace-immunity}{bdl\_\allowbreak{}inhomogeneous\_\allowbreak{}code\allowbreak{}Slot\_\allowbreak{}memory\_\allowbreak{}package}
\csdef{lean@file@thm:unperturbed-kms-lifetime}{Challenge.lean}
\csdef{lean@start@thm:unperturbed-kms-lifetime}{146}
\csdef{lean@end@thm:unperturbed-kms-lifetime}{194}
\csdef{lean@module@thm:unperturbed-kms-lifetime}{Challenge}
\csdef{lean@modpath@thm:unperturbed-kms-lifetime}{Challenge}
\csdef{lean@fqname@thm:unperturbed-kms-lifetime}{PassiveQuantumMemory.SatisfiesUnperturbedThermalAndKMSStability}
\csdef{lean@disp@thm:unperturbed-kms-lifetime}{Satisfies\allowbreak{}Unperturbed\allowbreak{}Thermal\allowbreak{}And\allowbreak{}KMSStability}
\csdef{lean@file@def:banach-quasilocal}{Defs/SuperpolyDecay.lean}
\csdef{lean@start@def:banach-quasilocal}{41}
\csdef{lean@end@def:banach-quasilocal}{42}
\csdef{lean@module@def:banach-quasilocal}{Defs.SuperpolyDecay}
\csdef{lean@modpath@def:banach-quasilocal}{Defs/SuperpolyDecay}
\csdef{lean@fqname@def:banach-quasilocal}{IsSuperpolynomialDecay}
\csdef{lean@disp@def:banach-quasilocal}{Is\allowbreak{}Superpolynomial\allowbreak{}Decay}
\csdef{lean@file@lem:dsuper-closure}{Proofs/LiebRobinson/SuperpolyDecayClosure.lean}
\csdef{lean@start@lem:dsuper-closure}{85}
\csdef{lean@end@lem:dsuper-closure}{93}
\csdef{lean@module@lem:dsuper-closure}{Proofs.LiebRobinson.SuperpolyDecayClosure}
\csdef{lean@modpath@lem:dsuper-closure}{Proofs/LiebRobinson/SuperpolyDecayClosure}
\csdef{lean@fqname@lem:dsuper-closure}{isSuperpolynomialDecay\leanUnderscore mul\leanUnderscore pow}
\csdef{lean@disp@lem:dsuper-closure}{is\allowbreak{}Superpolynomial\allowbreak{}Decay\_\allowbreak{}mul\_\allowbreak{}pow}
\csdef{lean@file@prop:bochner-time-convolution}{Proofs/KatoODE/SchwartzCollarLightCone.lean}
\csdef{lean@start@prop:bochner-time-convolution}{290}
\csdef{lean@end@prop:bochner-time-convolution}{448}
\csdef{lean@module@prop:bochner-time-convolution}{Proofs.KatoODE.SchwartzCollarLightCone}
\csdef{lean@modpath@prop:bochner-time-convolution}{Proofs/KatoODE/SchwartzCollarLightCone}
\csdef{lean@fqname@prop:bochner-time-convolution}{norm\leanUnderscore bochnerFilteredOp\leanUnderscore interpolated\leanUnderscore sub\leanUnderscore condExp\leanUnderscore schwartz\leanUnderscore le}
\csdef{lean@disp@prop:bochner-time-convolution}{norm\_\allowbreak{}bochner\allowbreak{}Filtered\allowbreak{}Op\_\allowbreak{}interpolated\_\allowbreak{}sub\_\allowbreak{}cond\allowbreak{}Exp\_\allowbreak{}schwartz\_\allowbreak{}le}
\csdef{lean@file@prop:qac-block-diagonalization}{Proofs/HastingsKomaLightCone/SpectralLowPassFilterClosure.lean}
\csdef{lean@start@prop:qac-block-diagonalization}{278}
\csdef{lean@end@prop:qac-block-diagonalization}{379}
\csdef{lean@module@prop:qac-block-diagonalization}{Proofs.HastingsKomaLightCone.SpectralLowPassFilterClosure}
\csdef{lean@modpath@prop:qac-block-diagonalization}{Proofs/HastingsKomaLightCone/SpectralLowPassFilterClosure}
\csdef{lean@fqname@prop:qac-block-diagonalization}{kato\leanUnderscore hastings\leanUnderscore band\leanUnderscore limited\leanUnderscore qac\leanUnderscore site\leanUnderscore family\leanUnderscore of\leanUnderscore gap\leanUnderscore bootstrap}
\csdef{lean@disp@prop:qac-block-diagonalization}{kato\_\allowbreak{}hastings\_\allowbreak{}band\_\allowbreak{}limited\_\allowbreak{}qac\_\allowbreak{}site\_\allowbreak{}family\_\allowbreak{}of\_\allowbreak{}gap\_\allowbreak{}bootstrap}
\csdef{lean@file@lem:norm-transfer}{Proofs/TwoAxisMemorySynthesis/BootstrappedQACMemoryReduction.lean}
\csdef{lean@start@lem:norm-transfer}{97}
\csdef{lean@end@lem:norm-transfer}{127}
\csdef{lean@module@lem:norm-transfer}{Proofs.TwoAxisMemorySynthesis.BootstrappedQACMemoryReduction}
\csdef{lean@modpath@lem:norm-transfer}{Proofs/TwoAxisMemorySynthesis/BootstrappedQACMemoryReduction}
\csdef{lean@fqname@lem:norm-transfer}{bdlInhomogeneousSystem\leanUnderscore cor1\leanUnderscore loc}
\csdef{lean@disp@lem:norm-transfer}{bdl\allowbreak{}Inhomogeneous\allowbreak{}System\_\allowbreak{}cor1\_\allowbreak{}loc}
\csdef{lean@file@prop:single-cube-telescoping}{Proofs/CenteredBallBHM/CenteredBallFormAndGapBootstrap.lean}
\csdef{lean@start@prop:single-cube-telescoping}{315}
\csdef{lean@end@prop:single-cube-telescoping}{370}
\csdef{lean@module@prop:single-cube-telescoping}{Proofs.CenteredBallBHM.CenteredBallFormAndGapBootstrap}
\csdef{lean@modpath@prop:single-cube-telescoping}{Proofs/CenteredBallBHM/CenteredBallFormAndGapBootstrap}
\csdef{lean@fqname@prop:single-cube-telescoping}{centeredSiteOp\leanUnderscore diffD\leanUnderscore and\leanUnderscore rem\leanUnderscore bounds\leanUnderscore of\leanUnderscore multiScale\leanUnderscore boxBall\leanUnderscore condExp\leanUnderscore and\leanUnderscore cor1\leanUnderscore loc\leanUnderscore of\leanUnderscore offDiag}
\csdef{lean@disp@prop:single-cube-telescoping}{centered\allowbreak{}Site\allowbreak{}Op\_\allowbreak{}diff\allowbreak{}D\_\allowbreak{}and\_\allowbreak{}rem\_\allowbreak{}bounds\_\allowbreak{}of\_\allowbreak{}multi\allowbreak{}Scale\_\allowbreak{}box\allowbreak{}Ball\_\allowbreak{}cond\allowbreak{}Exp\_\allowbreak{}and\_\allowbreak{}cor1\_\allowbreak{}loc\_\allowbreak{}of\_\allowbreak{}off\allowbreak{}Diag}
\csdef{lean@file@thm:gap-bootstrap}{Proofs/TwoAxisMemorySynthesis/ZeroOffDiagQACMasterReduction.lean}
\csdef{lean@start@thm:gap-bootstrap}{45}
\csdef{lean@end@thm:gap-bootstrap}{132}
\csdef{lean@module@thm:gap-bootstrap}{Proofs.TwoAxisMemorySynthesis.ZeroOffDiagQACMasterReduction}
\csdef{lean@modpath@thm:gap-bootstrap}{Proofs/TwoAxisMemorySynthesis/ZeroOffDiagQACMasterReduction}
\csdef{lean@fqname@thm:gap-bootstrap}{indexedSpectralGap\leanUnderscore bootstrap\leanUnderscore on\leanUnderscore Icc\leanUnderscore of\leanUnderscore siteFamily\leanUnderscore boxBall\leanUnderscore norm\leanUnderscore of\leanUnderscore cor1\leanUnderscore loc}
\csdef{lean@disp@thm:gap-bootstrap}{indexed\allowbreak{}Spectral\allowbreak{}Gap\_\allowbreak{}bootstrap\_\allowbreak{}on\_\allowbreak{}Icc\_\allowbreak{}of\_\allowbreak{}site\allowbreak{}Family\_\allowbreak{}box\allowbreak{}Ball\_\allowbreak{}norm\_\allowbreak{}of\_\allowbreak{}cor1\_\allowbreak{}loc}
\csdef{lean@file@thm:zero-temp-dressed-memory}{Challenge.lean}
\csdef{lean@start@thm:zero-temp-dressed-memory}{198}
\csdef{lean@end@thm:zero-temp-dressed-memory}{228}
\csdef{lean@module@thm:zero-temp-dressed-memory}{Challenge}
\csdef{lean@modpath@thm:zero-temp-dressed-memory}{Challenge}
\csdef{lean@fqname@thm:zero-temp-dressed-memory}{PassiveQuantumMemory.SatisfiesZeroTempPerturbativeMemoryStability}
\csdef{lean@disp@thm:zero-temp-dressed-memory}{Satisfies\allowbreak{}Zero\allowbreak{}Temp\allowbreak{}Perturbative\allowbreak{}Memory\allowbreak{}Stability}
\csdef{lean@file@thm:main}{Challenge.lean}
\csdef{lean@start@thm:main}{335}
\csdef{lean@end@thm:main}{370}
\csdef{lean@module@thm:main}{Challenge}
\csdef{lean@modpath@thm:main}{Challenge}
\csdef{lean@fqname@thm:main}{PassiveQuantumMemory.part1\leanUnderscore thermally\leanUnderscore and\leanUnderscore perturbatively\leanUnderscore stable\leanUnderscore memory}
\csdef{lean@disp@thm:main}{part1\_\allowbreak{}thermally\_\allowbreak{}and\_\allowbreak{}perturbatively\_\allowbreak{}stable\_\allowbreak{}memory}

\csdef{lean@file@decl:fq-part1-main}{Challenge.lean}
\csdef{lean@start@decl:fq-part1-main}{335}
\csdef{lean@end@decl:fq-part1-main}{370}
\csdef{lean@module@decl:fq-part1-main}{Challenge}
\csdef{lean@modpath@decl:fq-part1-main}{Challenge}
\csdef{lean@fqname@decl:fq-part1-main}{PassiveQuantumMemory.part1\leanUnderscore thermally\leanUnderscore and\leanUnderscore perturbatively\leanUnderscore stable\leanUnderscore memory}
\csdef{lean@disp@decl:fq-part1-main}{Passive\allowbreak{}Quantum\allowbreak{}Memory.\allowbreak{}part1\_\allowbreak{}thermally\_\allowbreak{}and\_\allowbreak{}perturbatively\_\allowbreak{}stable\_\allowbreak{}memory}
\csdef{lean@file@decl:fq-part2-main}{Challenge.lean}
\csdef{lean@start@decl:fq-part2-main}{376}
\csdef{lean@end@decl:fq-part2-main}{413}
\csdef{lean@module@decl:fq-part2-main}{Challenge}
\csdef{lean@modpath@decl:fq-part2-main}{Challenge}
\csdef{lean@fqname@decl:fq-part2-main}{PassiveQuantumMemory.part2\leanUnderscore fixed\leanUnderscore temp\leanUnderscore perturbative\leanUnderscore thermal\leanUnderscore and\leanUnderscore kms\leanUnderscore stability}
\csdef{lean@disp@decl:fq-part2-main}{Passive\allowbreak{}Quantum\allowbreak{}Memory.\allowbreak{}part2\_\allowbreak{}fixed\_\allowbreak{}temp\_\allowbreak{}perturbative\_\allowbreak{}thermal\_\allowbreak{}and\_\allowbreak{}kms\_\allowbreak{}stability}
\csdef{lean@file@decl:snapPointToOpenBox}{Defs/BDLSubstitutionCode.lean}
\csdef{lean@start@decl:snapPointToOpenBox}{38}
\csdef{lean@end@decl:snapPointToOpenBox}{41}
\csdef{lean@module@decl:snapPointToOpenBox}{Defs.BDLSubstitutionCode}
\csdef{lean@modpath@decl:snapPointToOpenBox}{Defs/BDLSubstitutionCode}
\csdef{lean@fqname@decl:snapPointToOpenBox}{snapPointToOpenBox}
\csdef{lean@disp@decl:snapPointToOpenBox}{snap\allowbreak{}Point\allowbreak{}To\allowbreak{}Open\allowbreak{}Box}
\csdef{lean@file@decl:part1-main}{Challenge.lean}
\csdef{lean@start@decl:part1-main}{335}
\csdef{lean@end@decl:part1-main}{370}
\csdef{lean@module@decl:part1-main}{Challenge}
\csdef{lean@modpath@decl:part1-main}{Challenge}
\csdef{lean@fqname@decl:part1-main}{PassiveQuantumMemory.part1\leanUnderscore thermally\leanUnderscore and\leanUnderscore perturbatively\leanUnderscore stable\leanUnderscore memory}
\csdef{lean@disp@decl:part1-main}{part1\_\allowbreak{}thermally\_\allowbreak{}and\_\allowbreak{}perturbatively\_\allowbreak{}stable\_\allowbreak{}memory}
\csdef{lean@file@decl:part1-solution}{Solution.lean}
\csdef{lean@start@decl:part1-solution}{279}
\csdef{lean@end@decl:part1-solution}{312}
\csdef{lean@module@decl:part1-solution}{Solution}
\csdef{lean@modpath@decl:part1-solution}{Solution}
\csdef{lean@fqname@decl:part1-solution}{PassiveQuantumMemory.part1\leanUnderscore thermally\leanUnderscore and\leanUnderscore perturbatively\leanUnderscore stable\leanUnderscore memory}
\csdef{lean@disp@decl:part1-solution}{Solution.\allowbreak{}part1\_\allowbreak{}thermally\_\allowbreak{}and\_\allowbreak{}perturbatively\_\allowbreak{}stable\_\allowbreak{}memory}
\csdef{lean@file@decl:part1-main-proof}{Proofs/TwoAxisMemorySynthesis/Part1TwoAxisMemorySynthesis.lean}
\csdef{lean@start@decl:part1-main-proof}{771}
\csdef{lean@end@decl:part1-main-proof}{882}
\csdef{lean@module@decl:part1-main-proof}{Proofs.TwoAxisMemorySynthesis.Part1TwoAxisMemorySynthesis}
\csdef{lean@modpath@decl:part1-main-proof}{Proofs/TwoAxisMemorySynthesis/Part1TwoAxisMemorySynthesis}
\csdef{lean@fqname@decl:part1-main-proof}{Part1TwoAxisMemorySynthesis.part1\leanUnderscore thermally\leanUnderscore and\leanUnderscore perturbatively\leanUnderscore stable\leanUnderscore memory\leanUnderscore main}
\csdef{lean@disp@decl:part1-main-proof}{part1\_\allowbreak{}thermally\_\allowbreak{}and\_\allowbreak{}perturbatively\_\allowbreak{}stable\_\allowbreak{}memory\_\allowbreak{}main}
\csdef{lean@file@decl:part2-main}{Challenge.lean}
\csdef{lean@start@decl:part2-main}{376}
\csdef{lean@end@decl:part2-main}{413}
\csdef{lean@module@decl:part2-main}{Challenge}
\csdef{lean@modpath@decl:part2-main}{Challenge}
\csdef{lean@fqname@decl:part2-main}{PassiveQuantumMemory.part2\leanUnderscore fixed\leanUnderscore temp\leanUnderscore perturbative\leanUnderscore thermal\leanUnderscore and\leanUnderscore kms\leanUnderscore stability}
\csdef{lean@disp@decl:part2-main}{part2\_\allowbreak{}fixed\_\allowbreak{}temp\_\allowbreak{}perturbative\_\allowbreak{}thermal\_\allowbreak{}and\_\allowbreak{}kms\_\allowbreak{}stability}
\csdef{lean@file@decl:part2-solution}{Solution.lean}
\csdef{lean@start@decl:part2-solution}{314}
\csdef{lean@end@decl:part2-solution}{349}
\csdef{lean@module@decl:part2-solution}{Solution}
\csdef{lean@modpath@decl:part2-solution}{Solution}
\csdef{lean@fqname@decl:part2-solution}{PassiveQuantumMemory.part2\leanUnderscore fixed\leanUnderscore temp\leanUnderscore perturbative\leanUnderscore thermal\leanUnderscore and\leanUnderscore kms\leanUnderscore stability}
\csdef{lean@disp@decl:part2-solution}{Solution.\allowbreak{}part2\_\allowbreak{}fixed\_\allowbreak{}temp\_\allowbreak{}perturbative\_\allowbreak{}thermal\_\allowbreak{}and\_\allowbreak{}kms\_\allowbreak{}stability}
\csdef{lean@file@decl:part2-main-proof}{Proofs/PerturbativeThermalSynthesis/Part2PerturbativeThermalSynthesis.lean}
\csdef{lean@start@decl:part2-main-proof}{535}
\csdef{lean@end@decl:part2-main-proof}{692}
\csdef{lean@module@decl:part2-main-proof}{Proofs.PerturbativeThermalSynthesis.Part2PerturbativeThermalSynthesis}
\csdef{lean@modpath@decl:part2-main-proof}{Proofs/PerturbativeThermalSynthesis/Part2PerturbativeThermalSynthesis}
\csdef{lean@fqname@decl:part2-main-proof}{Part2PerturbativeThermalSynthesis.part2\leanUnderscore fixed\leanUnderscore temp\leanUnderscore perturbative\leanUnderscore thermal\leanUnderscore and\leanUnderscore kms\leanUnderscore stability\leanUnderscore main}
\csdef{lean@disp@decl:part2-main-proof}{part2\_\allowbreak{}fixed\_\allowbreak{}temp\_\allowbreak{}perturbative\_\allowbreak{}thermal\_\allowbreak{}and\_\allowbreak{}kms\_\allowbreak{}stability\_\allowbreak{}main}
\csdef{lean@file@decl:SatisfiesFixedTempPerturbativeThermalAndKMSStability}{Challenge.lean}
\csdef{lean@start@decl:SatisfiesFixedTempPerturbativeThermalAndKMSStability}{233}
\csdef{lean@end@decl:SatisfiesFixedTempPerturbativeThermalAndKMSStability}{326}
\csdef{lean@module@decl:SatisfiesFixedTempPerturbativeThermalAndKMSStability}{Challenge}
\csdef{lean@modpath@decl:SatisfiesFixedTempPerturbativeThermalAndKMSStability}{Challenge}
\csdef{lean@fqname@decl:SatisfiesFixedTempPerturbativeThermalAndKMSStability}{PassiveQuantumMemory.SatisfiesFixedTempPerturbativeThermalAndKMSStability}
\csdef{lean@disp@decl:SatisfiesFixedTempPerturbativeThermalAndKMSStability}{Satisfies\allowbreak{}Fixed\allowbreak{}Temp\allowbreak{}Perturbative\allowbreak{}Thermal\allowbreak{}And\allowbreak{}KMSStability}
\csdef{lean@file@decl:F2ChainComplex}{Defs/F2ChainAndCSS.lean}
\csdef{lean@start@decl:F2ChainComplex}{74}
\csdef{lean@end@decl:F2ChainComplex}{80}
\csdef{lean@module@decl:F2ChainComplex}{Defs.F2ChainAndCSS}
\csdef{lean@modpath@decl:F2ChainComplex}{Defs/F2ChainAndCSS}
\csdef{lean@fqname@decl:F2ChainComplex}{F2ChainComplex}
\csdef{lean@disp@decl:F2ChainComplex}{F2\allowbreak{}Chain\allowbreak{}Complex}
\csdef{lean@file@decl:CSSCode3D}{Defs/F2ChainAndCSS.lean}
\csdef{lean@start@decl:CSSCode3D}{140}
\csdef{lean@end@decl:CSSCode3D}{146}
\csdef{lean@module@decl:CSSCode3D}{Defs.F2ChainAndCSS}
\csdef{lean@modpath@decl:CSSCode3D}{Defs/F2ChainAndCSS}
\csdef{lean@fqname@decl:CSSCode3D}{CSSCode3D}
\csdef{lean@disp@decl:CSSCode3D}{CSSCode3\allowbreak{}D}
\csdef{lean@file@decl:BundledCSSCode3D}{Defs/BDLParamsAndSeed.lean}
\csdef{lean@start@decl:BundledCSSCode3D}{36}
\csdef{lean@end@decl:BundledCSSCode3D}{51}
\csdef{lean@module@decl:BundledCSSCode3D}{Defs.BDLParamsAndSeed}
\csdef{lean@modpath@decl:BundledCSSCode3D}{Defs/BDLParamsAndSeed}
\csdef{lean@fqname@decl:BundledCSSCode3D}{BundledCSSCode3D}
\csdef{lean@disp@decl:BundledCSSCode3D}{Bundled\allowbreak{}CSSCode3\allowbreak{}D}
\csdef{lean@file@decl:LocalCSSDescriptor}{Defs/LocalCSSDescriptor.lean}
\csdef{lean@start@decl:LocalCSSDescriptor}{45}
\csdef{lean@end@decl:LocalCSSDescriptor}{50}
\csdef{lean@module@decl:LocalCSSDescriptor}{Defs.LocalCSSDescriptor}
\csdef{lean@modpath@decl:LocalCSSDescriptor}{Defs/LocalCSSDescriptor}
\csdef{lean@fqname@decl:LocalCSSDescriptor}{LocalCSSDescriptor}
\csdef{lean@disp@decl:LocalCSSDescriptor}{Local\allowbreak{}CSSDescriptor}
\csdef{lean@file@decl:ValidLLLSphericalCapPerturbation}{Defs/Step1PeriodicLLL.lean}
\csdef{lean@start@decl:ValidLLLSphericalCapPerturbation}{96}
\csdef{lean@end@decl:ValidLLLSphericalCapPerturbation}{121}
\csdef{lean@module@decl:ValidLLLSphericalCapPerturbation}{Defs.Step1PeriodicLLL}
\csdef{lean@modpath@decl:ValidLLLSphericalCapPerturbation}{Defs/Step1PeriodicLLL}
\csdef{lean@fqname@decl:ValidLLLSphericalCapPerturbation}{ValidLLLSphericalCapPerturbation}
\csdef{lean@disp@decl:ValidLLLSphericalCapPerturbation}{Valid\allowbreak{}LLLSpherical\allowbreak{}Cap\allowbreak{}Perturbation}
\csdef{lean@file@decl:ActiveSimplex2}{Defs/Step1PeriodicLLL.lean}
\csdef{lean@start@decl:ActiveSimplex2}{58}
\csdef{lean@end@decl:ActiveSimplex2}{61}
\csdef{lean@module@decl:ActiveSimplex2}{Defs.Step1PeriodicLLL}
\csdef{lean@modpath@decl:ActiveSimplex2}{Defs/Step1PeriodicLLL}
\csdef{lean@fqname@decl:ActiveSimplex2}{ActiveSimplex2}
\csdef{lean@disp@decl:ActiveSimplex2}{Active\allowbreak{}Simplex2}
\csdef{lean@file@decl:bdlPerturbStep}{Defs/Step1PeriodicLLL.lean}
\csdef{lean@start@decl:bdlPerturbStep}{45}
\csdef{lean@end@decl:bdlPerturbStep}{55}
\csdef{lean@module@decl:bdlPerturbStep}{Defs.Step1PeriodicLLL}
\csdef{lean@modpath@decl:bdlPerturbStep}{Defs/Step1PeriodicLLL}
\csdef{lean@fqname@decl:bdlPerturbStep}{bdlPerturbStep}
\csdef{lean@disp@decl:bdlPerturbStep}{bdl\allowbreak{}Perturb\allowbreak{}Step}
\csdef{lean@file@decl:bdlLLLPerturbation}{Defs/Step1PeriodicLLL.lean}
\csdef{lean@start@decl:bdlLLLPerturbation}{224}
\csdef{lean@end@decl:bdlLLLPerturbation}{229}
\csdef{lean@module@decl:bdlLLLPerturbation}{Defs.Step1PeriodicLLL}
\csdef{lean@modpath@decl:bdlLLLPerturbation}{Defs/Step1PeriodicLLL}
\csdef{lean@fqname@decl:bdlLLLPerturbation}{bdlLLLPerturbation}
\csdef{lean@disp@decl:bdlLLLPerturbation}{bdl\allowbreak{}LLLPerturbation}
\csdef{lean@file@decl:bdlSurfaceSeedCode}{Defs/BDLParamsAndSeed.lean}
\csdef{lean@start@decl:bdlSurfaceSeedCode}{190}
\csdef{lean@end@decl:bdlSurfaceSeedCode}{196}
\csdef{lean@module@decl:bdlSurfaceSeedCode}{Defs.BDLParamsAndSeed}
\csdef{lean@modpath@decl:bdlSurfaceSeedCode}{Defs/BDLParamsAndSeed}
\csdef{lean@fqname@decl:bdlSurfaceSeedCode}{bdlSurfaceSeedCode}
\csdef{lean@disp@decl:bdlSurfaceSeedCode}{bdl\allowbreak{}Surface\allowbreak{}Seed\allowbreak{}Code}
\csdef{lean@file@decl:bdlSurfaceSeedBundled}{Defs/BDLParamsAndSeed.lean}
\csdef{lean@start@decl:bdlSurfaceSeedBundled}{199}
\csdef{lean@end@decl:bdlSurfaceSeedBundled}{204}
\csdef{lean@module@decl:bdlSurfaceSeedBundled}{Defs.BDLParamsAndSeed}
\csdef{lean@modpath@decl:bdlSurfaceSeedBundled}{Defs/BDLParamsAndSeed}
\csdef{lean@fqname@decl:bdlSurfaceSeedBundled}{bdlSurfaceSeedBundled}
\csdef{lean@disp@decl:bdlSurfaceSeedBundled}{bdl\allowbreak{}Surface\allowbreak{}Seed\allowbreak{}Bundled}
\csdef{lean@file@decl:bdl-step1-lll-unit-ball-multiplicity}{Proofs/Step1LLLDensity/Step1LLLMainSynthesis.lean}
\csdef{lean@start@decl:bdl-step1-lll-unit-ball-multiplicity}{32}
\csdef{lean@end@decl:bdl-step1-lll-unit-ball-multiplicity}{488}
\csdef{lean@module@decl:bdl-step1-lll-unit-ball-multiplicity}{Proofs.Step1LLLDensity.Step1LLLMainSynthesis}
\csdef{lean@modpath@decl:bdl-step1-lll-unit-ball-multiplicity}{Proofs/Step1LLLDensity/Step1LLLMainSynthesis}
\csdef{lean@fqname@decl:bdl-step1-lll-unit-ball-multiplicity}{WeldedHierarchy.bdl\leanUnderscore step1\leanUnderscore lll\leanUnderscore unit\leanUnderscore ball\leanUnderscore multiplicity}
\csdef{lean@disp@decl:bdl-step1-lll-unit-ball-multiplicity}{bdl\_\allowbreak{}step1\_\allowbreak{}lll\_\allowbreak{}unit\_\allowbreak{}ball\_\allowbreak{}multiplicity}
\csdef{lean@file@decl:exists-good-assignment-conditional-lll}{Proofs/Step1LLLDensity/ConditionalBadVertexLLL.lean}
\csdef{lean@start@decl:exists-good-assignment-conditional-lll}{254}
\csdef{lean@end@decl:exists-good-assignment-conditional-lll}{334}
\csdef{lean@module@decl:exists-good-assignment-conditional-lll}{Proofs.Step1LLLDensity.ConditionalBadVertexLLL}
\csdef{lean@modpath@decl:exists-good-assignment-conditional-lll}{Proofs/Step1LLLDensity/ConditionalBadVertexLLL}
\csdef{lean@fqname@decl:exists-good-assignment-conditional-lll}{WeldedHierarchy.exists\leanUnderscore good\leanUnderscore assignment\leanUnderscore of\leanUnderscore conditional\leanUnderscore bad\leanUnderscore vertices\leanUnderscore lll}
\csdef{lean@disp@decl:exists-good-assignment-conditional-lll}{exists\_\allowbreak{}good\_\allowbreak{}assignment\_\allowbreak{}of\_\allowbreak{}conditional\_\allowbreak{}bad\_\allowbreak{}vertices\_\allowbreak{}lll}
\csdef{lean@file@decl:four-planes-cramer-residual}{Proofs/Step1LLLDensity/RationalSphericalCapGeometry.lean}
\csdef{lean@start@decl:four-planes-cramer-residual}{96}
\csdef{lean@end@decl:four-planes-cramer-residual}{126}
\csdef{lean@module@decl:four-planes-cramer-residual}{Proofs.Step1LLLDensity.RationalSphericalCapGeometry}
\csdef{lean@modpath@decl:four-planes-cramer-residual}{Proofs/Step1LLLDensity/RationalSphericalCapGeometry}
\csdef{lean@fqname@decl:four-planes-cramer-residual}{WeldedHierarchy.four\leanUnderscore planes\leanUnderscore cramer\leanUnderscore residual\leanUnderscore bound}
\csdef{lean@disp@decl:four-planes-cramer-residual}{four\_\allowbreak{}planes\_\allowbreak{}cramer\_\allowbreak{}residual\_\allowbreak{}bound}
\csdef{lean@file@decl:card-activeSimplex-posQ-filter}{Proofs/Step1LLLDensity/DiscreteCapSlabPointBounds.lean}
\csdef{lean@start@decl:card-activeSimplex-posQ-filter}{57}
\csdef{lean@end@decl:card-activeSimplex-posQ-filter}{77}
\csdef{lean@module@decl:card-activeSimplex-posQ-filter}{Proofs.Step1LLLDensity.DiscreteCapSlabPointBounds}
\csdef{lean@modpath@decl:card-activeSimplex-posQ-filter}{Proofs/Step1LLLDensity/DiscreteCapSlabPointBounds}
\csdef{lean@fqname@decl:card-activeSimplex-posQ-filter}{WeldedHierarchy.card\leanUnderscore activeSimplex\leanUnderscore posQ\leanUnderscore filter\leanUnderscore le}
\csdef{lean@disp@decl:card-activeSimplex-posQ-filter}{card\_\allowbreak{}active\allowbreak{}Simplex\_\allowbreak{}pos\allowbreak{}Q\_\allowbreak{}filter\_\allowbreak{}le}
\csdef{lean@file@decl:dominant-cramer-cascade}{Proofs/Step1LLLDensity/DominantCramerCascade.lean}
\csdef{lean@start@decl:dominant-cramer-cascade}{162}
\csdef{lean@end@decl:dominant-cramer-cascade}{372}
\csdef{lean@module@decl:dominant-cramer-cascade}{Proofs.Step1LLLDensity.DominantCramerCascade}
\csdef{lean@modpath@decl:dominant-cramer-cascade}{Proofs/Step1LLLDensity/DominantCramerCascade}
\csdef{lean@fqname@decl:dominant-cramer-cascade}{WeldedHierarchy.card\leanUnderscore le\leanUnderscore bdlDensityTPrime\leanUnderscore of\leanUnderscore three\leanUnderscore stage\leanUnderscore dominant\leanUnderscore cramer\leanUnderscore cascade}
\csdef{lean@disp@decl:dominant-cramer-cascade}{card\_\allowbreak{}le\_\allowbreak{}bdl\allowbreak{}Density\allowbreak{}TPrime\_\allowbreak{}of\_\allowbreak{}three\_\allowbreak{}stage\_\allowbreak{}dominant\_\allowbreak{}cramer\_\allowbreak{}cascade}
\csdef{lean@file@decl:bdl-random-lll-unit-ball-density}{Proofs/Step2DegreeAndWeld/DegreeOverlapAndDensityPreservation.lean}
\csdef{lean@start@decl:bdl-random-lll-unit-ball-density}{283}
\csdef{lean@end@decl:bdl-random-lll-unit-ball-density}{287}
\csdef{lean@module@decl:bdl-random-lll-unit-ball-density}{Proofs.Step2DegreeAndWeld.DegreeOverlapAndDensityPreservation}
\csdef{lean@modpath@decl:bdl-random-lll-unit-ball-density}{Proofs/Step2DegreeAndWeld/DegreeOverlapAndDensityPreservation}
\csdef{lean@fqname@decl:bdl-random-lll-unit-ball-density}{WeldedHierarchy.bdl\leanUnderscore random\leanUnderscore lll\leanUnderscore unit\leanUnderscore ball\leanUnderscore density}
\csdef{lean@disp@decl:bdl-random-lll-unit-ball-density}{bdl\_\allowbreak{}random\_\allowbreak{}lll\_\allowbreak{}unit\_\allowbreak{}ball\_\allowbreak{}density}
\csdef{lean@file@decl:BDLStepXQ}{Defs/Step2CylinderComplex.lean}
\csdef{lean@start@decl:BDLStepXQ}{419}
\csdef{lean@end@decl:BDLStepXQ}{422}
\csdef{lean@module@decl:BDLStepXQ}{Defs.Step2CylinderComplex}
\csdef{lean@modpath@decl:BDLStepXQ}{Defs/Step2CylinderComplex}
\csdef{lean@fqname@decl:BDLStepXQ}{BDLStepXQ}
\csdef{lean@disp@decl:BDLStepXQ}{BDLStep\allowbreak{}XQ}
\csdef{lean@file@decl:bdlStepX-commutativity}{Defs/Step2CheckMatrices.lean}
\csdef{lean@start@decl:bdlStepX-commutativity}{617}
\csdef{lean@end@decl:bdlStepX-commutativity}{631}
\csdef{lean@module@decl:bdlStepX-commutativity}{Defs.Step2CheckMatrices}
\csdef{lean@modpath@decl:bdlStepX-commutativity}{Defs/Step2CheckMatrices}
\csdef{lean@fqname@decl:bdlStepX-commutativity}{bdlStepX\leanUnderscore commutativity}
\csdef{lean@disp@decl:bdlStepX-commutativity}{bdl\allowbreak{}Step\allowbreak{}X\_\allowbreak{}commutativity}
\csdef{lean@file@decl:bdlStepX}{Defs/Step2CheckMatrices.lean}
\csdef{lean@start@decl:bdlStepX}{657}
\csdef{lean@end@decl:bdlStepX}{661}
\csdef{lean@module@decl:bdlStepX}{Defs.Step2CheckMatrices}
\csdef{lean@modpath@decl:bdlStepX}{Defs/Step2CheckMatrices}
\csdef{lean@fqname@decl:bdlStepX}{bdlStepX}
\csdef{lean@disp@decl:bdlStepX}{bdl\allowbreak{}Step\allowbreak{}X}
\csdef{lean@file@decl:bdlStepZ}{Defs/Step2CheckMatrices.lean}
\csdef{lean@start@decl:bdlStepZ}{664}
\csdef{lean@end@decl:bdlStepZ}{665}
\csdef{lean@module@decl:bdlStepZ}{Defs.Step2CheckMatrices}
\csdef{lean@modpath@decl:bdlStepZ}{Defs/Step2CheckMatrices}
\csdef{lean@fqname@decl:bdlStepZ}{bdlStepZ}
\csdef{lean@disp@decl:bdlStepZ}{bdl\allowbreak{}Step\allowbreak{}Z}
\csdef{lean@file@decl:bdlWeldedCode}{Defs/Step2CheckMatrices.lean}
\csdef{lean@start@decl:bdlWeldedCode}{676}
\csdef{lean@end@decl:bdlWeldedCode}{678}
\csdef{lean@module@decl:bdlWeldedCode}{Defs.Step2CheckMatrices}
\csdef{lean@modpath@decl:bdlWeldedCode}{Defs/Step2CheckMatrices}
\csdef{lean@fqname@decl:bdlWeldedCode}{bdlWeldedCode}
\csdef{lean@disp@decl:bdlWeldedCode}{bdl\allowbreak{}Welded\allowbreak{}Code}
\csdef{lean@file@decl:hasBoundedCheckDegreeOverlap-bdlIterated}{Proofs/Step2DegreeAndWeld/DegreeOverlapAndDensityPreservation.lean}
\csdef{lean@start@decl:hasBoundedCheckDegreeOverlap-bdlIterated}{201}
\csdef{lean@end@decl:hasBoundedCheckDegreeOverlap-bdlIterated}{209}
\csdef{lean@module@decl:hasBoundedCheckDegreeOverlap-bdlIterated}{Proofs.Step2DegreeAndWeld.DegreeOverlapAndDensityPreservation}
\csdef{lean@modpath@decl:hasBoundedCheckDegreeOverlap-bdlIterated}{Proofs/Step2DegreeAndWeld/DegreeOverlapAndDensityPreservation}
\csdef{lean@fqname@decl:hasBoundedCheckDegreeOverlap-bdlIterated}{WeldedHierarchy.hasBoundedCheckDegreeOverlap\leanUnderscore bdlIterated}
\csdef{lean@disp@decl:hasBoundedCheckDegreeOverlap-bdlIterated}{has\allowbreak{}Bounded\allowbreak{}Check\allowbreak{}Degree\allowbreak{}Overlap\_\allowbreak{}bdl\allowbreak{}Iterated}
\csdef{lean@file@decl:hasBoundedSimplexDegreeD-32}{Proofs/Step2DegreeAndWeld/CheckToSimplexDegreeReduction.lean}
\csdef{lean@start@decl:hasBoundedSimplexDegreeD-32}{293}
\csdef{lean@end@decl:hasBoundedSimplexDegreeD-32}{296}
\csdef{lean@module@decl:hasBoundedSimplexDegreeD-32}{Proofs.Step2DegreeAndWeld.CheckToSimplexDegreeReduction}
\csdef{lean@modpath@decl:hasBoundedSimplexDegreeD-32}{Proofs/Step2DegreeAndWeld/CheckToSimplexDegreeReduction}
\csdef{lean@fqname@decl:hasBoundedSimplexDegreeD-32}{WeldedHierarchy.hasBoundedSimplexDegreeD\leanUnderscore 32\leanUnderscore of\leanUnderscore hasBoundedCheckDegreeOverlap}
\csdef{lean@disp@decl:hasBoundedSimplexDegreeD-32}{has\allowbreak{}Bounded\allowbreak{}Simplex\allowbreak{}Degree\allowbreak{}D\_\allowbreak{}32\_\allowbreak{}of\_\allowbreak{}has\allowbreak{}Bounded\allowbreak{}Check\allowbreak{}Degree\allowbreak{}Overlap}
\csdef{lean@file@decl:hasEveryQubitChecked-bdlIterated}{Proofs/ConeAcyclicityRCA/IteratedCylinderLinkInvariants.lean}
\csdef{lean@start@decl:hasEveryQubitChecked-bdlIterated}{606}
\csdef{lean@end@decl:hasEveryQubitChecked-bdlIterated}{613}
\csdef{lean@module@decl:hasEveryQubitChecked-bdlIterated}{Proofs.ConeAcyclicityRCA.IteratedCylinderLinkInvariants}
\csdef{lean@modpath@decl:hasEveryQubitChecked-bdlIterated}{Proofs/ConeAcyclicityRCA/IteratedCylinderLinkInvariants}
\csdef{lean@fqname@decl:hasEveryQubitChecked-bdlIterated}{WeldedHierarchy.hasEveryQubitChecked\leanUnderscore bdlIterated}
\csdef{lean@disp@decl:hasEveryQubitChecked-bdlIterated}{has\allowbreak{}Every\allowbreak{}Qubit\allowbreak{}Checked\_\allowbreak{}bdl\allowbreak{}Iterated}
\csdef{lean@file@decl:hasConnectedXLinks-bdlIterated}{Proofs/ConeAcyclicityRCA/IteratedCylinderLinkInvariants.lean}
\csdef{lean@start@decl:hasConnectedXLinks-bdlIterated}{622}
\csdef{lean@end@decl:hasConnectedXLinks-bdlIterated}{632}
\csdef{lean@module@decl:hasConnectedXLinks-bdlIterated}{Proofs.ConeAcyclicityRCA.IteratedCylinderLinkInvariants}
\csdef{lean@modpath@decl:hasConnectedXLinks-bdlIterated}{Proofs/ConeAcyclicityRCA/IteratedCylinderLinkInvariants}
\csdef{lean@fqname@decl:hasConnectedXLinks-bdlIterated}{WeldedHierarchy.hasConnectedXLinks\leanUnderscore bdlIterated}
\csdef{lean@disp@decl:hasConnectedXLinks-bdlIterated}{has\allowbreak{}Connected\allowbreak{}XLinks\_\allowbreak{}bdl\allowbreak{}Iterated}
\csdef{lean@file@decl:bdl-random-interaction-range8}{Proofs/Step1LLLDensity/CylinderInteractionLocality.lean}
\csdef{lean@start@decl:bdl-random-interaction-range8}{628}
\csdef{lean@end@decl:bdl-random-interaction-range8}{630}
\csdef{lean@module@decl:bdl-random-interaction-range8}{Proofs.Step1LLLDensity.CylinderInteractionLocality}
\csdef{lean@modpath@decl:bdl-random-interaction-range8}{Proofs/Step1LLLDensity/CylinderInteractionLocality}
\csdef{lean@fqname@decl:bdl-random-interaction-range8}{WeldedHierarchy.bdl\leanUnderscore random\leanUnderscore interaction\leanUnderscore range8}
\csdef{lean@disp@decl:bdl-random-interaction-range8}{bdl\_\allowbreak{}random\_\allowbreak{}interaction\_\allowbreak{}range8}
\csdef{lean@file@decl:locallyGeneratedInBall}{Proofs/ConeAcyclicityRCA/LocalCocycleDualityAndBalls.lean}
\csdef{lean@start@decl:locallyGeneratedInBall}{58}
\csdef{lean@end@decl:locallyGeneratedInBall}{61}
\csdef{lean@module@decl:locallyGeneratedInBall}{Proofs.ConeAcyclicityRCA.LocalCocycleDualityAndBalls}
\csdef{lean@modpath@decl:locallyGeneratedInBall}{Proofs/ConeAcyclicityRCA/LocalCocycleDualityAndBalls}
\csdef{lean@fqname@decl:locallyGeneratedInBall}{WeldedHierarchy.locallyGeneratedInBall}
\csdef{lean@disp@decl:locallyGeneratedInBall}{locally\allowbreak{}Generated\allowbreak{}In\allowbreak{}Ball}
\csdef{lean@file@decl:scaleDownRadius-bound}{Proofs/ConeAcyclicityRCA/DualConeAcyclicityTransfer.lean}
\csdef{lean@start@decl:scaleDownRadius-bound}{211}
\csdef{lean@end@decl:scaleDownRadius-bound}{233}
\csdef{lean@module@decl:scaleDownRadius-bound}{Proofs.ConeAcyclicityRCA.DualConeAcyclicityTransfer}
\csdef{lean@modpath@decl:scaleDownRadius-bound}{Proofs/ConeAcyclicityRCA/DualConeAcyclicityTransfer}
\csdef{lean@fqname@decl:scaleDownRadius-bound}{WeldedHierarchy.scaleDownRadius\leanUnderscore bound}
\csdef{lean@disp@decl:scaleDownRadius-bound}{scale\allowbreak{}Down\allowbreak{}Radius\_\allowbreak{}bound}
\csdef{lean@file@decl:SatisfiesTQO1}{Defs/LocalHamiltonian.lean}
\csdef{lean@start@decl:SatisfiesTQO1}{173}
\csdef{lean@end@decl:SatisfiesTQO1}{182}
\csdef{lean@module@decl:SatisfiesTQO1}{Defs.LocalHamiltonian}
\csdef{lean@modpath@decl:SatisfiesTQO1}{Defs/LocalHamiltonian}
\csdef{lean@fqname@decl:SatisfiesTQO1}{SatisfiesTQO1}
\csdef{lean@disp@decl:SatisfiesTQO1}{Satisfies\allowbreak{}TQO1}
\csdef{lean@file@decl:SatisfiesLinearTQO2}{Defs/LocalHamiltonian.lean}
\csdef{lean@start@decl:SatisfiesLinearTQO2}{199}
\csdef{lean@end@decl:SatisfiesLinearTQO2}{203}
\csdef{lean@module@decl:SatisfiesLinearTQO2}{Defs.LocalHamiltonian}
\csdef{lean@modpath@decl:SatisfiesLinearTQO2}{Defs/LocalHamiltonian}
\csdef{lean@fqname@decl:SatisfiesLinearTQO2}{SatisfiesLinearTQO2}
\csdef{lean@disp@decl:SatisfiesLinearTQO2}{Satisfies\allowbreak{}Linear\allowbreak{}TQO2}
\csdef{lean@file@decl:SatisfiesBoxF2LinearTQO}{Defs/RescaledRCA.lean}
\csdef{lean@start@decl:SatisfiesBoxF2LinearTQO}{141}
\csdef{lean@end@decl:SatisfiesBoxF2LinearTQO}{163}
\csdef{lean@module@decl:SatisfiesBoxF2LinearTQO}{Defs.RescaledRCA}
\csdef{lean@modpath@decl:SatisfiesBoxF2LinearTQO}{Defs/RescaledRCA}
\csdef{lean@fqname@decl:SatisfiesBoxF2LinearTQO}{SatisfiesBoxF2LinearTQO}
\csdef{lean@disp@decl:SatisfiesBoxF2LinearTQO}{Satisfies\allowbreak{}Box\allowbreak{}F2\allowbreak{}Linear\allowbreak{}TQO}
\csdef{lean@file@decl:local-gen-iff-localCocycleExtends}{Proofs/ConeAcyclicityRCA/LocalCocycleDualityAndBalls.lean}
\csdef{lean@start@decl:local-gen-iff-localCocycleExtends}{263}
\csdef{lean@end@decl:local-gen-iff-localCocycleExtends}{268}
\csdef{lean@module@decl:local-gen-iff-localCocycleExtends}{Proofs.ConeAcyclicityRCA.LocalCocycleDualityAndBalls}
\csdef{lean@modpath@decl:local-gen-iff-localCocycleExtends}{Proofs/ConeAcyclicityRCA/LocalCocycleDualityAndBalls}
\csdef{lean@fqname@decl:local-gen-iff-localCocycleExtends}{WeldedHierarchy.local\leanUnderscore gen\leanUnderscore iff\leanUnderscore localCocycleExtends}
\csdef{lean@disp@decl:local-gen-iff-localCocycleExtends}{local\_\allowbreak{}gen\_\allowbreak{}iff\_\allowbreak{}local\allowbreak{}Cocycle\allowbreak{}Extends}
\csdef{lean@file@decl:boxF2LinearTQO-stabilizer-decomp}{Proofs/F2TQOToOperatorTQO.lean}
\csdef{lean@start@decl:boxF2LinearTQO-stabilizer-decomp}{77}
\csdef{lean@end@decl:boxF2LinearTQO-stabilizer-decomp}{102}
\csdef{lean@module@decl:boxF2LinearTQO-stabilizer-decomp}{Proofs.F2TQOToOperatorTQO}
\csdef{lean@modpath@decl:boxF2LinearTQO-stabilizer-decomp}{Proofs/F2TQOToOperatorTQO}
\csdef{lean@fqname@decl:boxF2LinearTQO-stabilizer-decomp}{boxF2LinearTQO\leanUnderscore stabilizer\leanUnderscore decomp}
\csdef{lean@disp@decl:boxF2LinearTQO-stabilizer-decomp}{box\allowbreak{}F2\allowbreak{}Linear\allowbreak{}TQO\_\allowbreak{}stabilizer\_\allowbreak{}decomp}
\csdef{lean@file@decl:staticCSSSystem-satisfiesOperatorTQO}{Proofs/InhomogeneousPeierls/OperatorTQOFromF2TQO.lean}
\csdef{lean@start@decl:staticCSSSystem-satisfiesOperatorTQO}{315}
\csdef{lean@end@decl:staticCSSSystem-satisfiesOperatorTQO}{422}
\csdef{lean@module@decl:staticCSSSystem-satisfiesOperatorTQO}{Proofs.InhomogeneousPeierls.OperatorTQOFromF2TQO}
\csdef{lean@modpath@decl:staticCSSSystem-satisfiesOperatorTQO}{Proofs/InhomogeneousPeierls/OperatorTQOFromF2TQO}
\csdef{lean@fqname@decl:staticCSSSystem-satisfiesOperatorTQO}{staticCSSSystem\leanUnderscore satisfiesOperatorTQO}
\csdef{lean@disp@decl:staticCSSSystem-satisfiesOperatorTQO}{static\allowbreak{}CSSSystem\_\allowbreak{}satisfies\allowbreak{}Operator\allowbreak{}TQO}
\csdef{lean@file@decl:hasUniqueLogicalHomology-bdlIterated}{Proofs/HierarchicalClusterDecoder/IteratedLogicalOperatorLifts.lean}
\csdef{lean@start@decl:hasUniqueLogicalHomology-bdlIterated}{201}
\csdef{lean@end@decl:hasUniqueLogicalHomology-bdlIterated}{220}
\csdef{lean@module@decl:hasUniqueLogicalHomology-bdlIterated}{Proofs.HierarchicalClusterDecoder.IteratedLogicalOperatorLifts}
\csdef{lean@modpath@decl:hasUniqueLogicalHomology-bdlIterated}{Proofs/HierarchicalClusterDecoder/IteratedLogicalOperatorLifts}
\csdef{lean@fqname@decl:hasUniqueLogicalHomology-bdlIterated}{WeldedHierarchy.hasUniqueLogicalHomology\leanUnderscore bdlIterated}
\csdef{lean@disp@decl:hasUniqueLogicalHomology-bdlIterated}{has\allowbreak{}Unique\allowbreak{}Logical\allowbreak{}Homology\_\allowbreak{}bdl\allowbreak{}Iterated}
\csdef{lean@file@decl:cylinderResidualCocycle}{Proofs/ConeAcyclicityRCA/Layer1ConeCocycle.lean}
\csdef{lean@start@decl:cylinderResidualCocycle}{604}
\csdef{lean@end@decl:cylinderResidualCocycle}{609}
\csdef{lean@module@decl:cylinderResidualCocycle}{Proofs.ConeAcyclicityRCA.Layer1ConeCocycle}
\csdef{lean@modpath@decl:cylinderResidualCocycle}{Proofs/ConeAcyclicityRCA/Layer1ConeCocycle}
\csdef{lean@fqname@decl:cylinderResidualCocycle}{WeldedHierarchy.cylinderResidualCocycle}
\csdef{lean@disp@decl:cylinderResidualCocycle}{cylinder\allowbreak{}Residual\allowbreak{}Cocycle}
\csdef{lean@file@decl:coarseProjCocycle}{Proofs/ConeAcyclicityRCA/CoarseProjCocycleAndSeedLinks.lean}
\csdef{lean@start@decl:coarseProjCocycle}{185}
\csdef{lean@end@decl:coarseProjCocycle}{188}
\csdef{lean@module@decl:coarseProjCocycle}{Proofs.ConeAcyclicityRCA.CoarseProjCocycleAndSeedLinks}
\csdef{lean@modpath@decl:coarseProjCocycle}{Proofs/ConeAcyclicityRCA/CoarseProjCocycleAndSeedLinks}
\csdef{lean@fqname@decl:coarseProjCocycle}{WeldedHierarchy.coarseProjCocycle}
\csdef{lean@disp@decl:coarseProjCocycle}{coarse\allowbreak{}Proj\allowbreak{}Cocycle}
\csdef{lean@file@decl:liftCylinderLZ}{Proofs/Step2DegreeAndWeld/LogicalChainLiftsAndPreservation.lean}
\csdef{lean@start@decl:liftCylinderLZ}{103}
\csdef{lean@end@decl:liftCylinderLZ}{109}
\csdef{lean@module@decl:liftCylinderLZ}{Proofs.Step2DegreeAndWeld.LogicalChainLiftsAndPreservation}
\csdef{lean@modpath@decl:liftCylinderLZ}{Proofs/Step2DegreeAndWeld/LogicalChainLiftsAndPreservation}
\csdef{lean@fqname@decl:liftCylinderLZ}{liftCylinderLZ}
\csdef{lean@disp@decl:liftCylinderLZ}{lift\allowbreak{}Cylinder\allowbreak{}LZ}
\csdef{lean@file@decl:liftCylinderLX}{Proofs/Step2DegreeAndWeld/LogicalChainLiftsAndPreservation.lean}
\csdef{lean@start@decl:liftCylinderLX}{112}
\csdef{lean@end@decl:liftCylinderLX}{117}
\csdef{lean@module@decl:liftCylinderLX}{Proofs.Step2DegreeAndWeld.LogicalChainLiftsAndPreservation}
\csdef{lean@modpath@decl:liftCylinderLX}{Proofs/Step2DegreeAndWeld/LogicalChainLiftsAndPreservation}
\csdef{lean@fqname@decl:liftCylinderLX}{liftCylinderLX}
\csdef{lean@disp@decl:liftCylinderLX}{lift\allowbreak{}Cylinder\allowbreak{}LX}
\csdef{lean@file@decl:subcylinderCandidateJ}{Proofs/ConeAcyclicityRCA/SubcylinderRayConstancyAndCheckpoints.lean}
\csdef{lean@start@decl:subcylinderCandidateJ}{315}
\csdef{lean@end@decl:subcylinderCandidateJ}{319}
\csdef{lean@module@decl:subcylinderCandidateJ}{Proofs.ConeAcyclicityRCA.SubcylinderRayConstancyAndCheckpoints}
\csdef{lean@modpath@decl:subcylinderCandidateJ}{Proofs/ConeAcyclicityRCA/SubcylinderRayConstancyAndCheckpoints}
\csdef{lean@fqname@decl:subcylinderCandidateJ}{WeldedHierarchy.subcylinderCandidateJ}
\csdef{lean@disp@decl:subcylinderCandidateJ}{subcylinder\allowbreak{}Candidate\allowbreak{}J}
\csdef{lean@file@decl:not-both-mem-ball3D-subcylinderCandidateJ}{Proofs/ConeAcyclicityRCA/SubcylinderRayConstancyAndCheckpoints.lean}
\csdef{lean@start@decl:not-both-mem-ball3D-subcylinderCandidateJ}{327}
\csdef{lean@end@decl:not-both-mem-ball3D-subcylinderCandidateJ}{408}
\csdef{lean@module@decl:not-both-mem-ball3D-subcylinderCandidateJ}{Proofs.ConeAcyclicityRCA.SubcylinderRayConstancyAndCheckpoints}
\csdef{lean@modpath@decl:not-both-mem-ball3D-subcylinderCandidateJ}{Proofs/ConeAcyclicityRCA/SubcylinderRayConstancyAndCheckpoints}
\csdef{lean@fqname@decl:not-both-mem-ball3D-subcylinderCandidateJ}{WeldedHierarchy.not\leanUnderscore both\leanUnderscore mem\leanUnderscore ball3D\leanUnderscore of\leanUnderscore subcylinderCandidateJ}
\csdef{lean@disp@decl:not-both-mem-ball3D-subcylinderCandidateJ}{not\_\allowbreak{}both\_\allowbreak{}mem\_\allowbreak{}ball3\allowbreak{}D\_\allowbreak{}of\_\allowbreak{}subcylinder\allowbreak{}Candidate\allowbreak{}J}
\csdef{lean@file@decl:bdlHalfStep-subcylinder-filling}{Proofs/ConeAcyclicityRCA/RescaledRCASynthesis.lean}
\csdef{lean@start@decl:bdlHalfStep-subcylinder-filling}{309}
\csdef{lean@end@decl:bdlHalfStep-subcylinder-filling}{356}
\csdef{lean@module@decl:bdlHalfStep-subcylinder-filling}{Proofs.ConeAcyclicityRCA.RescaledRCASynthesis}
\csdef{lean@modpath@decl:bdlHalfStep-subcylinder-filling}{Proofs/ConeAcyclicityRCA/RescaledRCASynthesis}
\csdef{lean@fqname@decl:bdlHalfStep-subcylinder-filling}{WeldedHierarchy.bdlHalfStep\leanUnderscore subcylinder\leanUnderscore filling}
\csdef{lean@disp@decl:bdlHalfStep-subcylinder-filling}{bdl\allowbreak{}Half\allowbreak{}Step\_\allowbreak{}subcylinder\_\allowbreak{}filling}
\csdef{lean@file@decl:bdlWeldedCode-inf-cone-acyclicity}{Proofs/ConeAcyclicityRCA/RescaledRCASynthesis.lean}
\csdef{lean@start@decl:bdlWeldedCode-inf-cone-acyclicity}{452}
\csdef{lean@end@decl:bdlWeldedCode-inf-cone-acyclicity}{511}
\csdef{lean@module@decl:bdlWeldedCode-inf-cone-acyclicity}{Proofs.ConeAcyclicityRCA.RescaledRCASynthesis}
\csdef{lean@modpath@decl:bdlWeldedCode-inf-cone-acyclicity}{Proofs/ConeAcyclicityRCA/RescaledRCASynthesis}
\csdef{lean@fqname@decl:bdlWeldedCode-inf-cone-acyclicity}{WeldedHierarchy.bdlWeldedCode\leanUnderscore inf\leanUnderscore cone\leanUnderscore acyclicity}
\csdef{lean@disp@decl:bdlWeldedCode-inf-cone-acyclicity}{bdl\allowbreak{}Welded\allowbreak{}Code\_\allowbreak{}inf\_\allowbreak{}cone\_\allowbreak{}acyclicity}
\csdef{lean@file@decl:bdlLocalDescriptor-satisfiesBoxF2LinearTQO}{Proofs/ConeAcyclicityRCA/SixCheckpointCleanShell.lean}
\csdef{lean@start@decl:bdlLocalDescriptor-satisfiesBoxF2LinearTQO}{341}
\csdef{lean@end@decl:bdlLocalDescriptor-satisfiesBoxF2LinearTQO}{544}
\csdef{lean@module@decl:bdlLocalDescriptor-satisfiesBoxF2LinearTQO}{Proofs.ConeAcyclicityRCA.SixCheckpointCleanShell}
\csdef{lean@modpath@decl:bdlLocalDescriptor-satisfiesBoxF2LinearTQO}{Proofs/ConeAcyclicityRCA/SixCheckpointCleanShell}
\csdef{lean@fqname@decl:bdlLocalDescriptor-satisfiesBoxF2LinearTQO}{bdlLocalDescriptor\leanUnderscore satisfiesBoxF2LinearTQO}
\csdef{lean@disp@decl:bdlLocalDescriptor-satisfiesBoxF2LinearTQO}{bdl\allowbreak{}Local\allowbreak{}Descriptor\_\allowbreak{}satisfies\allowbreak{}Box\allowbreak{}F2\allowbreak{}Linear\allowbreak{}TQO}
\csdef{lean@file@decl:cyl-sector-pointwise-bound}{Proofs/HierarchicalClusterDecoder/Stage1CylinderSubadditivity.lean}
\csdef{lean@start@decl:cyl-sector-pointwise-bound}{227}
\csdef{lean@end@decl:cyl-sector-pointwise-bound}{231}
\csdef{lean@module@decl:cyl-sector-pointwise-bound}{Proofs.HierarchicalClusterDecoder.Stage1CylinderSubadditivity}
\csdef{lean@modpath@decl:cyl-sector-pointwise-bound}{Proofs/HierarchicalClusterDecoder/Stage1CylinderSubadditivity}
\csdef{lean@fqname@decl:cyl-sector-pointwise-bound}{WeldedHierarchy.BDLDecoder.cyl\leanUnderscore sector\leanUnderscore pointwise\leanUnderscore bound}
\csdef{lean@disp@decl:cyl-sector-pointwise-bound}{cyl\_\allowbreak{}sector\_\allowbreak{}pointwise\_\allowbreak{}bound}
\csdef{lean@file@decl:cleanCylinderX-spec}{Proofs/HierarchicalClusterDecoder/TwoStageCleanerComposition.lean}
\csdef{lean@start@decl:cleanCylinderX-spec}{122}
\csdef{lean@end@decl:cleanCylinderX-spec}{164}
\csdef{lean@module@decl:cleanCylinderX-spec}{Proofs.HierarchicalClusterDecoder.TwoStageCleanerComposition}
\csdef{lean@modpath@decl:cleanCylinderX-spec}{Proofs/HierarchicalClusterDecoder/TwoStageCleanerComposition}
\csdef{lean@fqname@decl:cleanCylinderX-spec}{WeldedHierarchy.BDLDecoder.cleanCylinderX\leanUnderscore spec}
\csdef{lean@disp@decl:cleanCylinderX-spec}{clean\allowbreak{}Cylinder\allowbreak{}X\_\allowbreak{}spec}
\csdef{lean@file@decl:exists-local-cylinder-cleaner}{Proofs/HierarchicalClusterDecoder/Stage1CylinderCleaner.lean}
\csdef{lean@start@decl:exists-local-cylinder-cleaner}{590}
\csdef{lean@end@decl:exists-local-cylinder-cleaner}{644}
\csdef{lean@module@decl:exists-local-cylinder-cleaner}{Proofs.HierarchicalClusterDecoder.Stage1CylinderCleaner}
\csdef{lean@modpath@decl:exists-local-cylinder-cleaner}{Proofs/HierarchicalClusterDecoder/Stage1CylinderCleaner}
\csdef{lean@fqname@decl:exists-local-cylinder-cleaner}{WeldedHierarchy.BDLDecoder.exists\leanUnderscore local\leanUnderscore cylinder\leanUnderscore cleaner}
\csdef{lean@disp@decl:exists-local-cylinder-cleaner}{exists\_\allowbreak{}local\_\allowbreak{}cylinder\_\allowbreak{}cleaner}
\csdef{lean@file@decl:three-mul-min-le}{Proofs/HierarchicalClusterDecoder/Stage2StarComplementSymmetry.lean}
\csdef{lean@start@decl:three-mul-min-le}{176}
\csdef{lean@end@decl:three-mul-min-le}{178}
\csdef{lean@module@decl:three-mul-min-le}{Proofs.HierarchicalClusterDecoder.Stage2StarComplementSymmetry}
\csdef{lean@modpath@decl:three-mul-min-le}{Proofs/HierarchicalClusterDecoder/Stage2StarComplementSymmetry}
\csdef{lean@fqname@decl:three-mul-min-le}{WeldedHierarchy.BDLDecoder.three\leanUnderscore mul\leanUnderscore min\leanUnderscore le}
\csdef{lean@disp@decl:three-mul-min-le}{three\_\allowbreak{}mul\_\allowbreak{}min\_\allowbreak{}le}
\csdef{lean@file@decl:cleanStarQ-spec}{Proofs/HierarchicalClusterDecoder/TwoStageCleanerComposition.lean}
\csdef{lean@start@decl:cleanStarQ-spec}{416}
\csdef{lean@end@decl:cleanStarQ-spec}{467}
\csdef{lean@module@decl:cleanStarQ-spec}{Proofs.HierarchicalClusterDecoder.TwoStageCleanerComposition}
\csdef{lean@modpath@decl:cleanStarQ-spec}{Proofs/HierarchicalClusterDecoder/TwoStageCleanerComposition}
\csdef{lean@fqname@decl:cleanStarQ-spec}{WeldedHierarchy.BDLDecoder.cleanStarQ\leanUnderscore spec}
\csdef{lean@disp@decl:cleanStarQ-spec}{clean\allowbreak{}Star\allowbreak{}Q\_\allowbreak{}spec}
\csdef{lean@file@decl:exists-local-star-cleaner}{Proofs/HierarchicalClusterDecoder/Stage2StarCleaner.lean}
\csdef{lean@start@decl:exists-local-star-cleaner}{521}
\csdef{lean@end@decl:exists-local-star-cleaner}{648}
\csdef{lean@module@decl:exists-local-star-cleaner}{Proofs.HierarchicalClusterDecoder.Stage2StarCleaner}
\csdef{lean@modpath@decl:exists-local-star-cleaner}{Proofs/HierarchicalClusterDecoder/Stage2StarCleaner}
\csdef{lean@fqname@decl:exists-local-star-cleaner}{WeldedHierarchy.BDLDecoder.exists\leanUnderscore local\leanUnderscore star\leanUnderscore cleaner}
\csdef{lean@disp@decl:exists-local-star-cleaner}{exists\_\allowbreak{}local\_\allowbreak{}star\_\allowbreak{}cleaner}
\csdef{lean@file@decl:cleanTwoStageRX}{Proofs/HierarchicalClusterDecoder/TwoStageCleanerComposition.lean}
\csdef{lean@start@decl:cleanTwoStageRX}{263}
\csdef{lean@end@decl:cleanTwoStageRX}{269}
\csdef{lean@module@decl:cleanTwoStageRX}{Proofs.HierarchicalClusterDecoder.TwoStageCleanerComposition}
\csdef{lean@modpath@decl:cleanTwoStageRX}{Proofs/HierarchicalClusterDecoder/TwoStageCleanerComposition}
\csdef{lean@fqname@decl:cleanTwoStageRX}{WeldedHierarchy.BDLDecoder.cleanTwoStageRX}
\csdef{lean@disp@decl:cleanTwoStageRX}{clean\allowbreak{}Two\allowbreak{}Stage\allowbreak{}RX}
\csdef{lean@file@decl:rxCleanedCoarseSyndrome}{Proofs/HierarchicalClusterDecoder/TwoStageCleanerComposition.lean}
\csdef{lean@start@decl:rxCleanedCoarseSyndrome}{280}
\csdef{lean@end@decl:rxCleanedCoarseSyndrome}{285}
\csdef{lean@module@decl:rxCleanedCoarseSyndrome}{Proofs.HierarchicalClusterDecoder.TwoStageCleanerComposition}
\csdef{lean@modpath@decl:rxCleanedCoarseSyndrome}{Proofs/HierarchicalClusterDecoder/TwoStageCleanerComposition}
\csdef{lean@fqname@decl:rxCleanedCoarseSyndrome}{WeldedHierarchy.BDLDecoder.rxCleanedCoarseSyndrome}
\csdef{lean@disp@decl:rxCleanedCoarseSyndrome}{rx\allowbreak{}Cleaned\allowbreak{}Coarse\allowbreak{}Syndrome}
\csdef{lean@file@decl:weightNetExportXQ-cylCleanerCand-eq}{Proofs/HierarchicalClusterDecoder/Stage1CylinderSubadditivity.lean}
\csdef{lean@start@decl:weightNetExportXQ-cylCleanerCand-eq}{321}
\csdef{lean@end@decl:weightNetExportXQ-cylCleanerCand-eq}{334}
\csdef{lean@module@decl:weightNetExportXQ-cylCleanerCand-eq}{Proofs.HierarchicalClusterDecoder.Stage1CylinderSubadditivity}
\csdef{lean@modpath@decl:weightNetExportXQ-cylCleanerCand-eq}{Proofs/HierarchicalClusterDecoder/Stage1CylinderSubadditivity}
\csdef{lean@fqname@decl:weightNetExportXQ-cylCleanerCand-eq}{WeldedHierarchy.BDLDecoder.weightNetExportXQ\leanUnderscore cylCleanerCand\leanUnderscore eq}
\csdef{lean@disp@decl:weightNetExportXQ-cylCleanerCand-eq}{weight\allowbreak{}Net\allowbreak{}Export\allowbreak{}XQ\_\allowbreak{}cyl\allowbreak{}Cleaner\allowbreak{}Cand\_\allowbreak{}eq}
\csdef{lean@file@decl:rxCleanedCoarseSyndrome-twoHop-contraction}{Proofs/HierarchicalClusterDecoder/TwoHopTannerContraction.lean}
\csdef{lean@start@decl:rxCleanedCoarseSyndrome-twoHop-contraction}{349}
\csdef{lean@end@decl:rxCleanedCoarseSyndrome-twoHop-contraction}{379}
\csdef{lean@module@decl:rxCleanedCoarseSyndrome-twoHop-contraction}{Proofs.HierarchicalClusterDecoder.TwoHopTannerContraction}
\csdef{lean@modpath@decl:rxCleanedCoarseSyndrome-twoHop-contraction}{Proofs/HierarchicalClusterDecoder/TwoHopTannerContraction}
\csdef{lean@fqname@decl:rxCleanedCoarseSyndrome-twoHop-contraction}{WeldedHierarchy.BDLDecoder.rxCleanedCoarseSyndrome\leanUnderscore twoHop\leanUnderscore contraction}
\csdef{lean@disp@decl:rxCleanedCoarseSyndrome-twoHop-contraction}{rx\allowbreak{}Cleaned\allowbreak{}Coarse\allowbreak{}Syndrome\_\allowbreak{}two\allowbreak{}Hop\_\allowbreak{}contraction}
\csdef{lean@file@decl:liftWitnessFamily-rx-twoHop-step}{Proofs/HierarchicalClusterDecoder/WitnessLiftingAndParityImmunity.lean}
\csdef{lean@start@decl:liftWitnessFamily-rx-twoHop-step}{253}
\csdef{lean@end@decl:liftWitnessFamily-rx-twoHop-step}{294}
\csdef{lean@module@decl:liftWitnessFamily-rx-twoHop-step}{Proofs.HierarchicalClusterDecoder.WitnessLiftingAndParityImmunity}
\csdef{lean@modpath@decl:liftWitnessFamily-rx-twoHop-step}{Proofs/HierarchicalClusterDecoder/WitnessLiftingAndParityImmunity}
\csdef{lean@fqname@decl:liftWitnessFamily-rx-twoHop-step}{WeldedHierarchy.BDLDecoder.liftWitnessFamily\leanUnderscore rx\leanUnderscore twoHop\leanUnderscore step}
\csdef{lean@disp@decl:liftWitnessFamily-rx-twoHop-step}{lift\allowbreak{}Witness\allowbreak{}Family\_\allowbreak{}rx\_\allowbreak{}two\allowbreak{}Hop\_\allowbreak{}step}
\csdef{lean@file@decl:HasBallParityWitnessFamily}{Proofs/HierarchicalClusterDecoder/DecodedLogicalCommutation.lean}
\csdef{lean@start@decl:HasBallParityWitnessFamily}{41}
\csdef{lean@end@decl:HasBallParityWitnessFamily}{54}
\csdef{lean@module@decl:HasBallParityWitnessFamily}{Proofs.HierarchicalClusterDecoder.DecodedLogicalCommutation}
\csdef{lean@modpath@decl:HasBallParityWitnessFamily}{Proofs/HierarchicalClusterDecoder/DecodedLogicalCommutation}
\csdef{lean@fqname@decl:HasBallParityWitnessFamily}{WeldedHierarchy.BDLDecoder.HasBallParityWitnessFamily}
\csdef{lean@disp@decl:HasBallParityWitnessFamily}{Has\allowbreak{}Ball\allowbreak{}Parity\allowbreak{}Witness\allowbreak{}Family}
\csdef{lean@file@decl:hasBallParityWitnessFamily-bdlIterated-succ}{Proofs/HierarchicalClusterDecoder/ClusterDecoderSynthesis.lean}
\csdef{lean@start@decl:hasBallParityWitnessFamily-bdlIterated-succ}{378}
\csdef{lean@end@decl:hasBallParityWitnessFamily-bdlIterated-succ}{428}
\csdef{lean@module@decl:hasBallParityWitnessFamily-bdlIterated-succ}{Proofs.HierarchicalClusterDecoder.ClusterDecoderSynthesis}
\csdef{lean@modpath@decl:hasBallParityWitnessFamily-bdlIterated-succ}{Proofs/HierarchicalClusterDecoder/ClusterDecoderSynthesis}
\csdef{lean@fqname@decl:hasBallParityWitnessFamily-bdlIterated-succ}{WeldedHierarchy.BDLDecoder.hasBallParityWitnessFamily\leanUnderscore bdlIterated\leanUnderscore succ}
\csdef{lean@disp@decl:hasBallParityWitnessFamily-bdlIterated-succ}{has\allowbreak{}Ball\allowbreak{}Parity\allowbreak{}Witness\allowbreak{}Family\_\allowbreak{}bdl\allowbreak{}Iterated\_\allowbreak{}succ}
\csdef{lean@file@decl:dualPairing-bdlWeldedLZ-bdlWeldedLX}{Proofs/HierarchicalClusterDecoder/ClusterDecoderSynthesis.lean}
\csdef{lean@start@decl:dualPairing-bdlWeldedLZ-bdlWeldedLX}{480}
\csdef{lean@end@decl:dualPairing-bdlWeldedLZ-bdlWeldedLX}{482}
\csdef{lean@module@decl:dualPairing-bdlWeldedLZ-bdlWeldedLX}{Proofs.HierarchicalClusterDecoder.ClusterDecoderSynthesis}
\csdef{lean@modpath@decl:dualPairing-bdlWeldedLZ-bdlWeldedLX}{Proofs/HierarchicalClusterDecoder/ClusterDecoderSynthesis}
\csdef{lean@fqname@decl:dualPairing-bdlWeldedLZ-bdlWeldedLX}{WeldedHierarchy.BDLDecoder.dualPairing\leanUnderscore bdlWeldedLZ\leanUnderscore bdlWeldedLX}
\csdef{lean@disp@decl:dualPairing-bdlWeldedLZ-bdlWeldedLX}{dual\allowbreak{}Pairing\_\allowbreak{}bdl\allowbreak{}Welded\allowbreak{}LZ\_\allowbreak{}bdl\allowbreak{}Welded\allowbreak{}LX}
\csdef{lean@file@decl:coset-syndromeWeight-sum-le}{Proofs/HierarchicalClusterDecoder/WalshFourierCosetAndHomology.lean}
\csdef{lean@start@decl:coset-syndromeWeight-sum-le}{215}
\csdef{lean@end@decl:coset-syndromeWeight-sum-le}{258}
\csdef{lean@module@decl:coset-syndromeWeight-sum-le}{Proofs.HierarchicalClusterDecoder.WalshFourierCosetAndHomology}
\csdef{lean@modpath@decl:coset-syndromeWeight-sum-le}{Proofs/HierarchicalClusterDecoder/WalshFourierCosetAndHomology}
\csdef{lean@fqname@decl:coset-syndromeWeight-sum-le}{WeldedHierarchy.coset\leanUnderscore syndromeWeight\leanUnderscore sum\leanUnderscore le}
\csdef{lean@disp@decl:coset-syndromeWeight-sum-le}{coset\_\allowbreak{}syndrome\allowbreak{}Weight\_\allowbreak{}sum\_\allowbreak{}le}
\csdef{lean@file@decl:three-div-two-pow-le-clusterRawWeight}{Proofs/HierarchicalClusterDecoder/ClusterDecoderSynthesis.lean}
\csdef{lean@start@decl:three-div-two-pow-le-clusterRawWeight}{505}
\csdef{lean@end@decl:three-div-two-pow-le-clusterRawWeight}{512}
\csdef{lean@module@decl:three-div-two-pow-le-clusterRawWeight}{Proofs.HierarchicalClusterDecoder.ClusterDecoderSynthesis}
\csdef{lean@modpath@decl:three-div-two-pow-le-clusterRawWeight}{Proofs/HierarchicalClusterDecoder/ClusterDecoderSynthesis}
\csdef{lean@fqname@decl:three-div-two-pow-le-clusterRawWeight}{WeldedHierarchy.BDLDecoder.three\leanUnderscore div\leanUnderscore two\leanUnderscore pow\leanUnderscore le\leanUnderscore clusterRawWeight}
\csdef{lean@disp@decl:three-div-two-pow-le-clusterRawWeight}{three\_\allowbreak{}div\_\allowbreak{}two\_\allowbreak{}pow\_\allowbreak{}le\_\allowbreak{}cluster\allowbreak{}Raw\allowbreak{}Weight}
\csdef{lean@file@decl:hasBallParityWitnessFamily-bdlWelded-lambdaPow}{Proofs/TwoAxisMemorySynthesis/MacroscopicLambdaPowImmunity.lean}
\csdef{lean@start@decl:hasBallParityWitnessFamily-bdlWelded-lambdaPow}{41}
\csdef{lean@end@decl:hasBallParityWitnessFamily-bdlWelded-lambdaPow}{62}
\csdef{lean@module@decl:hasBallParityWitnessFamily-bdlWelded-lambdaPow}{Proofs.TwoAxisMemorySynthesis.MacroscopicLambdaPowImmunity}
\csdef{lean@modpath@decl:hasBallParityWitnessFamily-bdlWelded-lambdaPow}{Proofs/TwoAxisMemorySynthesis/MacroscopicLambdaPowImmunity}
\csdef{lean@fqname@decl:hasBallParityWitnessFamily-bdlWelded-lambdaPow}{hasBallParityWitnessFamily\leanUnderscore bdlWelded\leanUnderscore lambdaPow}
\csdef{lean@disp@decl:hasBallParityWitnessFamily-bdlWelded-lambdaPow}{has\allowbreak{}Ball\allowbreak{}Parity\allowbreak{}Witness\allowbreak{}Family\_\allowbreak{}bdl\allowbreak{}Welded\_\allowbreak{}lambda\allowbreak{}Pow}
\csdef{lean@file@decl:gibbsFailureProb-bdlSafeSyndromeSet-le}{Proofs/InhomogeneousPeierls/InhomogeneousPeierlsMain.lean}
\csdef{lean@start@decl:gibbsFailureProb-bdlSafeSyndromeSet-le}{292}
\csdef{lean@end@decl:gibbsFailureProb-bdlSafeSyndromeSet-le}{315}
\csdef{lean@module@decl:gibbsFailureProb-bdlSafeSyndromeSet-le}{Proofs.InhomogeneousPeierls.InhomogeneousPeierlsMain}
\csdef{lean@modpath@decl:gibbsFailureProb-bdlSafeSyndromeSet-le}{Proofs/InhomogeneousPeierls/InhomogeneousPeierlsMain}
\csdef{lean@fqname@decl:gibbsFailureProb-bdlSafeSyndromeSet-le}{gibbsFailureProb\leanUnderscore bdlSafeSyndromeSet\leanUnderscore le}
\csdef{lean@disp@decl:gibbsFailureProb-bdlSafeSyndromeSet-le}{gibbs\allowbreak{}Failure\allowbreak{}Prob\_\allowbreak{}bdl\allowbreak{}Safe\allowbreak{}Syndrome\allowbreak{}Set\_\allowbreak{}le}
\csdef{lean@file@decl:HasSafeDecodedLogicalObservables}{Challenge.lean}
\csdef{lean@start@decl:HasSafeDecodedLogicalObservables}{79}
\csdef{lean@end@decl:HasSafeDecodedLogicalObservables}{112}
\csdef{lean@module@decl:HasSafeDecodedLogicalObservables}{Challenge}
\csdef{lean@modpath@decl:HasSafeDecodedLogicalObservables}{Challenge}
\csdef{lean@fqname@decl:HasSafeDecodedLogicalObservables}{PassiveQuantumMemory.HasSafeDecodedLogicalObservables}
\csdef{lean@disp@decl:HasSafeDecodedLogicalObservables}{Has\allowbreak{}Safe\allowbreak{}Decoded\allowbreak{}Logical\allowbreak{}Observables}
\csdef{lean@file@decl:bdlSafeSyndromeSet-cube-safe-immunity-lambdaPow}{Proofs/TwoAxisMemorySynthesis/MacroscopicLambdaPowImmunity.lean}
\csdef{lean@start@decl:bdlSafeSyndromeSet-cube-safe-immunity-lambdaPow}{296}
\csdef{lean@end@decl:bdlSafeSyndromeSet-cube-safe-immunity-lambdaPow}{330}
\csdef{lean@module@decl:bdlSafeSyndromeSet-cube-safe-immunity-lambdaPow}{Proofs.TwoAxisMemorySynthesis.MacroscopicLambdaPowImmunity}
\csdef{lean@modpath@decl:bdlSafeSyndromeSet-cube-safe-immunity-lambdaPow}{Proofs/TwoAxisMemorySynthesis/MacroscopicLambdaPowImmunity}
\csdef{lean@fqname@decl:bdlSafeSyndromeSet-cube-safe-immunity-lambdaPow}{bdlSafeSyndromeSet\leanUnderscore cube\leanUnderscore op\leanUnderscore safe\leanUnderscore immunity\leanUnderscore lambdaPow}
\csdef{lean@disp@decl:bdlSafeSyndromeSet-cube-safe-immunity-lambdaPow}{bdl\allowbreak{}Safe\allowbreak{}Syndrome\allowbreak{}Set\_\allowbreak{}cube\_\allowbreak{}op\_\allowbreak{}safe\_\allowbreak{}immunity\_\allowbreak{}lambda\allowbreak{}Pow}
\csdef{lean@file@decl:bdl-macroscopic-safe-projector-exists}{Proofs/TwoAxisMemorySynthesis/Part1TwoAxisMemorySynthesis.lean}
\csdef{lean@start@decl:bdl-macroscopic-safe-projector-exists}{551}
\csdef{lean@end@decl:bdl-macroscopic-safe-projector-exists}{618}
\csdef{lean@module@decl:bdl-macroscopic-safe-projector-exists}{Proofs.TwoAxisMemorySynthesis.Part1TwoAxisMemorySynthesis}
\csdef{lean@modpath@decl:bdl-macroscopic-safe-projector-exists}{Proofs/TwoAxisMemorySynthesis/Part1TwoAxisMemorySynthesis}
\csdef{lean@fqname@decl:bdl-macroscopic-safe-projector-exists}{Part1TwoAxisMemorySynthesis.bdl\leanUnderscore inhomogeneous\leanUnderscore codeSlot\leanUnderscore memory\leanUnderscore package}
\csdef{lean@disp@decl:bdl-macroscopic-safe-projector-exists}{bdl\_\allowbreak{}inhomogeneous\_\allowbreak{}code\allowbreak{}Slot\_\allowbreak{}memory\_\allowbreak{}package}
\csdef{lean@file@decl:opKMSLindbladGenerator}{Defs/KMSAndGibbs.lean}
\csdef{lean@start@decl:opKMSLindbladGenerator}{288}
\csdef{lean@end@decl:opKMSLindbladGenerator}{300}
\csdef{lean@module@decl:opKMSLindbladGenerator}{Defs.KMSAndGibbs}
\csdef{lean@modpath@decl:opKMSLindbladGenerator}{Defs/KMSAndGibbs}
\csdef{lean@fqname@decl:opKMSLindbladGenerator}{opKMSLindbladGenerator}
\csdef{lean@disp@decl:opKMSLindbladGenerator}{op\allowbreak{}KMSLindblad\allowbreak{}Generator}
\csdef{lean@file@decl:SatisfiesKMSLindbladODE}{Defs/KMSAndGibbs.lean}
\csdef{lean@start@decl:SatisfiesKMSLindbladODE}{355}
\csdef{lean@end@decl:SatisfiesKMSLindbladODE}{373}
\csdef{lean@module@decl:SatisfiesKMSLindbladODE}{Defs.KMSAndGibbs}
\csdef{lean@modpath@decl:SatisfiesKMSLindbladODE}{Defs/KMSAndGibbs}
\csdef{lean@fqname@decl:SatisfiesKMSLindbladODE}{SatisfiesKMSLindbladODE}
\csdef{lean@disp@decl:SatisfiesKMSLindbladODE}{Satisfies\allowbreak{}KMSLindblad\allowbreak{}ODE}
\csdef{lean@file@decl:SatisfiesUnperturbedThermalAndKMSStability}{Challenge.lean}
\csdef{lean@start@decl:SatisfiesUnperturbedThermalAndKMSStability}{146}
\csdef{lean@end@decl:SatisfiesUnperturbedThermalAndKMSStability}{194}
\csdef{lean@module@decl:SatisfiesUnperturbedThermalAndKMSStability}{Challenge}
\csdef{lean@modpath@decl:SatisfiesUnperturbedThermalAndKMSStability}{Challenge}
\csdef{lean@fqname@decl:SatisfiesUnperturbedThermalAndKMSStability}{PassiveQuantumMemory.SatisfiesUnperturbedThermalAndKMSStability}
\csdef{lean@disp@decl:SatisfiesUnperturbedThermalAndKMSStability}{Satisfies\allowbreak{}Unperturbed\allowbreak{}Thermal\allowbreak{}And\allowbreak{}KMSStability}
\csdef{lean@file@decl:kms-fourth-root-projector-bound}{Proofs/KMSSafeSubspaceODE/KMSFourthRootAndPinchedGibbs.lean}
\csdef{lean@start@decl:kms-fourth-root-projector-bound}{97}
\csdef{lean@end@decl:kms-fourth-root-projector-bound}{146}
\csdef{lean@module@decl:kms-fourth-root-projector-bound}{Proofs.KMSSafeSubspaceODE.KMSFourthRootAndPinchedGibbs}
\csdef{lean@modpath@decl:kms-fourth-root-projector-bound}{Proofs/KMSSafeSubspaceODE/KMSFourthRootAndPinchedGibbs}
\csdef{lean@fqname@decl:kms-fourth-root-projector-bound}{kms\leanUnderscore fourth\leanUnderscore root\leanUnderscore projector\leanUnderscore bound}
\csdef{lean@disp@decl:kms-fourth-root-projector-bound}{kms\_\allowbreak{}fourth\_\allowbreak{}root\_\allowbreak{}projector\_\allowbreak{}bound}
\csdef{lean@file@decl:isCompletelyPositiveMap-davies}{Proofs/TwoAxisMemorySynthesis/DaviesGKLSCompletePositivity.lean}
\csdef{lean@start@decl:isCompletelyPositiveMap-davies}{658}
\csdef{lean@end@decl:isCompletelyPositiveMap-davies}{677}
\csdef{lean@module@decl:isCompletelyPositiveMap-davies}{Proofs.TwoAxisMemorySynthesis.DaviesGKLSCompletePositivity}
\csdef{lean@modpath@decl:isCompletelyPositiveMap-davies}{Proofs/TwoAxisMemorySynthesis/DaviesGKLSCompletePositivity}
\csdef{lean@fqname@decl:isCompletelyPositiveMap-davies}{isCompletelyPositiveMap\leanUnderscore opDaviesThermalSemigroup}
\csdef{lean@disp@decl:isCompletelyPositiveMap-davies}{is\allowbreak{}Completely\allowbreak{}Positive\allowbreak{}Map\_\allowbreak{}op\allowbreak{}Davies\allowbreak{}Thermal\allowbreak{}Semigroup}
\csdef{lean@file@decl:IsCubeRangeBath}{Defs/KMSAndGibbs.lean}
\csdef{lean@start@decl:IsCubeRangeBath}{241}
\csdef{lean@end@decl:IsCubeRangeBath}{246}
\csdef{lean@module@decl:IsCubeRangeBath}{Defs.KMSAndGibbs}
\csdef{lean@modpath@decl:IsCubeRangeBath}{Defs/KMSAndGibbs}
\csdef{lean@fqname@decl:IsCubeRangeBath}{IsCubeRangeBath}
\csdef{lean@disp@decl:IsCubeRangeBath}{Is\allowbreak{}Cube\allowbreak{}Range\allowbreak{}Bath}
\csdef{lean@file@decl:opDaviesThermalGenerator}{Defs/KMSAndGibbs.lean}
\csdef{lean@start@decl:opDaviesThermalGenerator}{530}
\csdef{lean@end@decl:opDaviesThermalGenerator}{542}
\csdef{lean@module@decl:opDaviesThermalGenerator}{Defs.KMSAndGibbs}
\csdef{lean@modpath@decl:opDaviesThermalGenerator}{Defs/KMSAndGibbs}
\csdef{lean@fqname@decl:opDaviesThermalGenerator}{opDaviesThermalGenerator}
\csdef{lean@disp@decl:opDaviesThermalGenerator}{op\allowbreak{}Davies\allowbreak{}Thermal\allowbreak{}Generator}
\csdef{lean@file@decl:opDaviesThermalSemigroup-cptp}{Proofs/TwoAxisMemorySynthesis/DaviesGeneratorBridge.lean}
\csdef{lean@start@decl:opDaviesThermalSemigroup-cptp}{1339}
\csdef{lean@end@decl:opDaviesThermalSemigroup-cptp}{1491}
\csdef{lean@module@decl:opDaviesThermalSemigroup-cptp}{Proofs.TwoAxisMemorySynthesis.DaviesGeneratorBridge}
\csdef{lean@modpath@decl:opDaviesThermalSemigroup-cptp}{Proofs/TwoAxisMemorySynthesis/DaviesGeneratorBridge}
\csdef{lean@fqname@decl:opDaviesThermalSemigroup-cptp}{opDaviesThermalSemigroup\leanUnderscore cptp\leanUnderscore and\leanUnderscore ode}
\csdef{lean@disp@decl:opDaviesThermalSemigroup-cptp}{op\allowbreak{}Davies\allowbreak{}Thermal\allowbreak{}Semigroup\_\allowbreak{}cptp\_\allowbreak{}and\_\allowbreak{}ode}
\csdef{lean@file@decl:unperturbed-cube-kms-davies}{Proofs/TwoAxisMemorySynthesis/DaviesGeneratorBridge.lean}
\csdef{lean@start@decl:unperturbed-cube-kms-davies}{1498}
\csdef{lean@end@decl:unperturbed-cube-kms-davies}{1680}
\csdef{lean@module@decl:unperturbed-cube-kms-davies}{Proofs.TwoAxisMemorySynthesis.DaviesGeneratorBridge}
\csdef{lean@modpath@decl:unperturbed-cube-kms-davies}{Proofs/TwoAxisMemorySynthesis/DaviesGeneratorBridge}
\csdef{lean@fqname@decl:unperturbed-cube-kms-davies}{unperturbed\leanUnderscore cube\leanUnderscore kms\leanUnderscore and\leanUnderscore davies\leanUnderscore lifetime\leanUnderscore bounds}
\csdef{lean@disp@decl:unperturbed-cube-kms-davies}{unperturbed\_\allowbreak{}cube\_\allowbreak{}kms\_\allowbreak{}and\_\allowbreak{}davies\_\allowbreak{}lifetime\_\allowbreak{}bounds}
\csdef{lean@file@decl:dressed-cube-telescoping}{Proofs/KMSSafeSubspaceODE/CubeDressedCommutatorTelescoping.lean}
\csdef{lean@start@decl:dressed-cube-telescoping}{537}
\csdef{lean@end@decl:dressed-cube-telescoping}{711}
\csdef{lean@module@decl:dressed-cube-telescoping}{Proofs.KMSSafeSubspaceODE.CubeDressedCommutatorTelescoping}
\csdef{lean@modpath@decl:dressed-cube-telescoping}{Proofs/KMSSafeSubspaceODE/CubeDressedCommutatorTelescoping}
\csdef{lean@fqname@decl:dressed-cube-telescoping}{dressed\leanUnderscore cube\leanUnderscore safe\leanUnderscore commutator\leanUnderscore telescoping}
\csdef{lean@disp@decl:dressed-cube-telescoping}{dressed\_\allowbreak{}cube\_\allowbreak{}safe\_\allowbreak{}commutator\_\allowbreak{}telescoping}
\csdef{lean@file@decl:IsSuperpolynomialDecay}{Defs/SuperpolyDecay.lean}
\csdef{lean@start@decl:IsSuperpolynomialDecay}{41}
\csdef{lean@end@decl:IsSuperpolynomialDecay}{42}
\csdef{lean@module@decl:IsSuperpolynomialDecay}{Defs.SuperpolyDecay}
\csdef{lean@modpath@decl:IsSuperpolynomialDecay}{Defs/SuperpolyDecay}
\csdef{lean@fqname@decl:IsSuperpolynomialDecay}{IsSuperpolynomialDecay}
\csdef{lean@disp@decl:IsSuperpolynomialDecay}{Is\allowbreak{}Superpolynomial\allowbreak{}Decay}
\csdef{lean@file@decl:QuasiLocalPerturbation}{Defs/LocalHamiltonian.lean}
\csdef{lean@start@decl:QuasiLocalPerturbation}{261}
\csdef{lean@end@decl:QuasiLocalPerturbation}{268}
\csdef{lean@module@decl:QuasiLocalPerturbation}{Defs.LocalHamiltonian}
\csdef{lean@modpath@decl:QuasiLocalPerturbation}{Defs/LocalHamiltonian}
\csdef{lean@fqname@decl:QuasiLocalPerturbation}{QuasiLocalPerturbation}
\csdef{lean@disp@decl:QuasiLocalPerturbation}{Quasi\allowbreak{}Local\allowbreak{}Perturbation}
\csdef{lean@file@decl:isSuperpolynomialDecay-mul-pow}{Proofs/LiebRobinson/SuperpolyDecayClosure.lean}
\csdef{lean@start@decl:isSuperpolynomialDecay-mul-pow}{85}
\csdef{lean@end@decl:isSuperpolynomialDecay-mul-pow}{93}
\csdef{lean@module@decl:isSuperpolynomialDecay-mul-pow}{Proofs.LiebRobinson.SuperpolyDecayClosure}
\csdef{lean@modpath@decl:isSuperpolynomialDecay-mul-pow}{Proofs/LiebRobinson/SuperpolyDecayClosure}
\csdef{lean@fqname@decl:isSuperpolynomialDecay-mul-pow}{isSuperpolynomialDecay\leanUnderscore mul\leanUnderscore pow}
\csdef{lean@disp@decl:isSuperpolynomialDecay-mul-pow}{is\allowbreak{}Superpolynomial\allowbreak{}Decay\_\allowbreak{}mul\_\allowbreak{}pow}
\csdef{lean@file@decl:norm-bochnerFilteredOp-schwartz-le}{Proofs/KatoODE/SchwartzCollarLightCone.lean}
\csdef{lean@start@decl:norm-bochnerFilteredOp-schwartz-le}{290}
\csdef{lean@end@decl:norm-bochnerFilteredOp-schwartz-le}{448}
\csdef{lean@module@decl:norm-bochnerFilteredOp-schwartz-le}{Proofs.KatoODE.SchwartzCollarLightCone}
\csdef{lean@modpath@decl:norm-bochnerFilteredOp-schwartz-le}{Proofs/KatoODE/SchwartzCollarLightCone}
\csdef{lean@fqname@decl:norm-bochnerFilteredOp-schwartz-le}{norm\leanUnderscore bochnerFilteredOp\leanUnderscore interpolated\leanUnderscore sub\leanUnderscore condExp\leanUnderscore schwartz\leanUnderscore le}
\csdef{lean@disp@decl:norm-bochnerFilteredOp-schwartz-le}{norm\_\allowbreak{}bochner\allowbreak{}Filtered\allowbreak{}Op\_\allowbreak{}interpolated\_\allowbreak{}sub\_\allowbreak{}cond\allowbreak{}Exp\_\allowbreak{}schwartz\_\allowbreak{}le}
\csdef{lean@file@decl:kato-hastings-subspace-intertwining}{Proofs/HastingsKomaLightCone/KatoHastingsSubspaceIntertwining.lean}
\csdef{lean@start@decl:kato-hastings-subspace-intertwining}{417}
\csdef{lean@end@decl:kato-hastings-subspace-intertwining}{427}
\csdef{lean@module@decl:kato-hastings-subspace-intertwining}{Proofs.HastingsKomaLightCone.KatoHastingsSubspaceIntertwining}
\csdef{lean@modpath@decl:kato-hastings-subspace-intertwining}{Proofs/HastingsKomaLightCone/KatoHastingsSubspaceIntertwining}
\csdef{lean@fqname@decl:kato-hastings-subspace-intertwining}{kato\leanUnderscore hastings\leanUnderscore subspace\leanUnderscore intertwining\leanUnderscore w\leanUnderscore BP\leanUnderscore of\leanUnderscore gap\leanUnderscore bootstrap}
\csdef{lean@disp@decl:kato-hastings-subspace-intertwining}{kato\_\allowbreak{}hastings\_\allowbreak{}subspace\_\allowbreak{}intertwining\_\allowbreak{}w\_\allowbreak{}BP\_\allowbreak{}of\_\allowbreak{}gap\_\allowbreak{}bootstrap}
\csdef{lean@file@decl:kato-hastings-band-limited-qac}{Proofs/HastingsKomaLightCone/SpectralLowPassFilterClosure.lean}
\csdef{lean@start@decl:kato-hastings-band-limited-qac}{278}
\csdef{lean@end@decl:kato-hastings-band-limited-qac}{379}
\csdef{lean@module@decl:kato-hastings-band-limited-qac}{Proofs.HastingsKomaLightCone.SpectralLowPassFilterClosure}
\csdef{lean@modpath@decl:kato-hastings-band-limited-qac}{Proofs/HastingsKomaLightCone/SpectralLowPassFilterClosure}
\csdef{lean@fqname@decl:kato-hastings-band-limited-qac}{kato\leanUnderscore hastings\leanUnderscore band\leanUnderscore limited\leanUnderscore qac\leanUnderscore site\leanUnderscore family\leanUnderscore of\leanUnderscore gap\leanUnderscore bootstrap}
\csdef{lean@disp@decl:kato-hastings-band-limited-qac}{kato\_\allowbreak{}hastings\_\allowbreak{}band\_\allowbreak{}limited\_\allowbreak{}qac\_\allowbreak{}site\_\allowbreak{}family\_\allowbreak{}of\_\allowbreak{}gap\_\allowbreak{}bootstrap}
\csdef{lean@file@decl:twoBlockOffDiagProj-bochnerFilteredSiteOp-eq-zero}{Proofs/HastingsKomaLightCone/KatoHastingsSubspaceIntertwining.lean}
\csdef{lean@start@decl:twoBlockOffDiagProj-bochnerFilteredSiteOp-eq-zero}{104}
\csdef{lean@end@decl:twoBlockOffDiagProj-bochnerFilteredSiteOp-eq-zero}{133}
\csdef{lean@module@decl:twoBlockOffDiagProj-bochnerFilteredSiteOp-eq-zero}{Proofs.HastingsKomaLightCone.KatoHastingsSubspaceIntertwining}
\csdef{lean@modpath@decl:twoBlockOffDiagProj-bochnerFilteredSiteOp-eq-zero}{Proofs/HastingsKomaLightCone/KatoHastingsSubspaceIntertwining}
\csdef{lean@fqname@decl:twoBlockOffDiagProj-bochnerFilteredSiteOp-eq-zero}{twoBlockOffDiagProj\leanUnderscore and\leanUnderscore norm\leanUnderscore bochnerFilteredSiteOp\leanUnderscore katoFlowUnitary\leanUnderscore eq\leanUnderscore zero}
\csdef{lean@disp@decl:twoBlockOffDiagProj-bochnerFilteredSiteOp-eq-zero}{two\allowbreak{}Block\allowbreak{}Off\allowbreak{}Diag\allowbreak{}Proj\_\allowbreak{}and\_\allowbreak{}norm\_\allowbreak{}bochner\allowbreak{}Filtered\allowbreak{}Site\allowbreak{}Op\_\allowbreak{}kato\allowbreak{}Flow\allowbreak{}Unitary\_\allowbreak{}eq\_\allowbreak{}zero}
\csdef{lean@file@decl:norm-kato-flow-unitary-single-site-le}{Proofs/KatoODE/NonAutonomousUnitaryLightCone.lean}
\csdef{lean@start@decl:norm-kato-flow-unitary-single-site-le}{183}
\csdef{lean@end@decl:norm-kato-flow-unitary-single-site-le}{228}
\csdef{lean@module@decl:norm-kato-flow-unitary-single-site-le}{Proofs.KatoODE.NonAutonomousUnitaryLightCone}
\csdef{lean@modpath@decl:norm-kato-flow-unitary-single-site-le}{Proofs/KatoODE/NonAutonomousUnitaryLightCone}
\csdef{lean@fqname@decl:norm-kato-flow-unitary-single-site-le}{norm\leanUnderscore kato\leanUnderscore flow\leanUnderscore unitary\leanUnderscore single\leanUnderscore site\leanUnderscore box\leanUnderscore ball\leanUnderscore sub\leanUnderscore cond\leanUnderscore exp\leanUnderscore le\leanUnderscore J\leanUnderscore mul}
\csdef{lean@disp@decl:norm-kato-flow-unitary-single-site-le}{norm\_\allowbreak{}kato\_\allowbreak{}flow\_\allowbreak{}unitary\_\allowbreak{}single\_\allowbreak{}site\_\allowbreak{}box\_\allowbreak{}ball\_\allowbreak{}sub\_\allowbreak{}cond\_\allowbreak{}exp\_\allowbreak{}le\_\allowbreak{}J\_\allowbreak{}mul}
\csdef{lean@file@decl:bdlInhomogeneousSystem-cor1-loc}{Proofs/TwoAxisMemorySynthesis/BootstrappedQACMemoryReduction.lean}
\csdef{lean@start@decl:bdlInhomogeneousSystem-cor1-loc}{97}
\csdef{lean@end@decl:bdlInhomogeneousSystem-cor1-loc}{127}
\csdef{lean@module@decl:bdlInhomogeneousSystem-cor1-loc}{Proofs.TwoAxisMemorySynthesis.BootstrappedQACMemoryReduction}
\csdef{lean@modpath@decl:bdlInhomogeneousSystem-cor1-loc}{Proofs/TwoAxisMemorySynthesis/BootstrappedQACMemoryReduction}
\csdef{lean@fqname@decl:bdlInhomogeneousSystem-cor1-loc}{bdlInhomogeneousSystem\leanUnderscore cor1\leanUnderscore loc}
\csdef{lean@disp@decl:bdlInhomogeneousSystem-cor1-loc}{bdl\allowbreak{}Inhomogeneous\allowbreak{}System\_\allowbreak{}cor1\_\allowbreak{}loc}
\csdef{lean@file@decl:telescoping-norm-bounds-of-approx-P0}{Proofs/TwoBlockDecomposition/GroundBlockScalarShift.lean}
\csdef{lean@start@decl:telescoping-norm-bounds-of-approx-P0}{172}
\csdef{lean@end@decl:telescoping-norm-bounds-of-approx-P0}{230}
\csdef{lean@module@decl:telescoping-norm-bounds-of-approx-P0}{Proofs.TwoBlockDecomposition.GroundBlockScalarShift}
\csdef{lean@modpath@decl:telescoping-norm-bounds-of-approx-P0}{Proofs/TwoBlockDecomposition/GroundBlockScalarShift}
\csdef{lean@fqname@decl:telescoping-norm-bounds-of-approx-P0}{telescoping\leanUnderscore norm\leanUnderscore bounds\leanUnderscore of\leanUnderscore approx\leanUnderscore P0}
\csdef{lean@disp@decl:telescoping-norm-bounds-of-approx-P0}{telescoping\_\allowbreak{}norm\_\allowbreak{}bounds\_\allowbreak{}of\_\allowbreak{}approx\_\allowbreak{}P0}
\csdef{lean@file@decl:conj-sub-H0-centered-telescoping}{Proofs/CenteredBallBHM/CenteredBallOverlapAndTQO1Diff.lean}
\csdef{lean@start@decl:conj-sub-H0-centered-telescoping}{507}
\csdef{lean@end@decl:conj-sub-H0-centered-telescoping}{528}
\csdef{lean@module@decl:conj-sub-H0-centered-telescoping}{Proofs.CenteredBallBHM.CenteredBallOverlapAndTQO1Diff}
\csdef{lean@modpath@decl:conj-sub-H0-centered-telescoping}{Proofs/CenteredBallBHM/CenteredBallOverlapAndTQO1Diff}
\csdef{lean@fqname@decl:conj-sub-H0-centered-telescoping}{conj\leanUnderscore sub\leanUnderscore H0\leanUnderscore sub\leanUnderscore shift\leanUnderscore eq\leanUnderscore centered\leanUnderscore telescoping\leanUnderscore of\leanUnderscore siteFamily\leanUnderscore boxBall}
\csdef{lean@disp@decl:conj-sub-H0-centered-telescoping}{conj\_\allowbreak{}sub\_\allowbreak{}H0\_\allowbreak{}sub\_\allowbreak{}shift\_\allowbreak{}eq\_\allowbreak{}centered\_\allowbreak{}telescoping\_\allowbreak{}of\_\allowbreak{}site\allowbreak{}Family\_\allowbreak{}box\allowbreak{}Ball}
\csdef{lean@file@decl:centeredSiteOp-diffD-and-rem-bounds}{Proofs/CenteredBallBHM/CenteredBallFormAndGapBootstrap.lean}
\csdef{lean@start@decl:centeredSiteOp-diffD-and-rem-bounds}{315}
\csdef{lean@end@decl:centeredSiteOp-diffD-and-rem-bounds}{370}
\csdef{lean@module@decl:centeredSiteOp-diffD-and-rem-bounds}{Proofs.CenteredBallBHM.CenteredBallFormAndGapBootstrap}
\csdef{lean@modpath@decl:centeredSiteOp-diffD-and-rem-bounds}{Proofs/CenteredBallBHM/CenteredBallFormAndGapBootstrap}
\csdef{lean@fqname@decl:centeredSiteOp-diffD-and-rem-bounds}{centeredSiteOp\leanUnderscore diffD\leanUnderscore and\leanUnderscore rem\leanUnderscore bounds\leanUnderscore of\leanUnderscore multiScale\leanUnderscore boxBall\leanUnderscore condExp\leanUnderscore and\leanUnderscore cor1\leanUnderscore loc\leanUnderscore of\leanUnderscore offDiag}
\csdef{lean@disp@decl:centeredSiteOp-diffD-and-rem-bounds}{centered\allowbreak{}Site\allowbreak{}Op\_\allowbreak{}diff\allowbreak{}D\_\allowbreak{}and\_\allowbreak{}rem\_\allowbreak{}bounds\_\allowbreak{}of\_\allowbreak{}multi\allowbreak{}Scale\_\allowbreak{}box\allowbreak{}Ball\_\allowbreak{}cond\allowbreak{}Exp\_\allowbreak{}and\_\allowbreak{}cor1\_\allowbreak{}loc\_\allowbreak{}of\_\allowbreak{}off\allowbreak{}Diag}
\csdef{lean@file@decl:relative-form-bound-of-siteFamily-boxBall}{Proofs/CenteredBallBHM/CenteredBallFormAndGapBootstrap.lean}
\csdef{lean@start@decl:relative-form-bound-of-siteFamily-boxBall}{107}
\csdef{lean@end@decl:relative-form-bound-of-siteFamily-boxBall}{179}
\csdef{lean@module@decl:relative-form-bound-of-siteFamily-boxBall}{Proofs.CenteredBallBHM.CenteredBallFormAndGapBootstrap}
\csdef{lean@modpath@decl:relative-form-bound-of-siteFamily-boxBall}{Proofs/CenteredBallBHM/CenteredBallFormAndGapBootstrap}
\csdef{lean@fqname@decl:relative-form-bound-of-siteFamily-boxBall}{relative\leanUnderscore form\leanUnderscore bound\leanUnderscore of\leanUnderscore siteFamily\leanUnderscore boxBall}
\csdef{lean@disp@decl:relative-form-bound-of-siteFamily-boxBall}{relative\_\allowbreak{}form\_\allowbreak{}bound\_\allowbreak{}of\_\allowbreak{}site\allowbreak{}Family\_\allowbreak{}box\allowbreak{}Ball}
\csdef{lean@file@decl:indexedSpectralGap-bootstrap}{Proofs/TwoAxisMemorySynthesis/ZeroOffDiagQACMasterReduction.lean}
\csdef{lean@start@decl:indexedSpectralGap-bootstrap}{45}
\csdef{lean@end@decl:indexedSpectralGap-bootstrap}{132}
\csdef{lean@module@decl:indexedSpectralGap-bootstrap}{Proofs.TwoAxisMemorySynthesis.ZeroOffDiagQACMasterReduction}
\csdef{lean@modpath@decl:indexedSpectralGap-bootstrap}{Proofs/TwoAxisMemorySynthesis/ZeroOffDiagQACMasterReduction}
\csdef{lean@fqname@decl:indexedSpectralGap-bootstrap}{indexedSpectralGap\leanUnderscore bootstrap\leanUnderscore on\leanUnderscore Icc\leanUnderscore of\leanUnderscore siteFamily\leanUnderscore boxBall\leanUnderscore norm\leanUnderscore of\leanUnderscore cor1\leanUnderscore loc}
\csdef{lean@disp@decl:indexedSpectralGap-bootstrap}{indexed\allowbreak{}Spectral\allowbreak{}Gap\_\allowbreak{}bootstrap\_\allowbreak{}on\_\allowbreak{}Icc\_\allowbreak{}of\_\allowbreak{}site\allowbreak{}Family\_\allowbreak{}box\allowbreak{}Ball\_\allowbreak{}norm\_\allowbreak{}of\_\allowbreak{}cor1\_\allowbreak{}loc}
\csdef{lean@file@decl:bdl-zero-temp-dressing-bundle-exists}{Proofs/TwoAxisMemorySynthesis/ZeroOffDiagQACMasterReduction.lean}
\csdef{lean@start@decl:bdl-zero-temp-dressing-bundle-exists}{485}
\csdef{lean@end@decl:bdl-zero-temp-dressing-bundle-exists}{540}
\csdef{lean@module@decl:bdl-zero-temp-dressing-bundle-exists}{Proofs.TwoAxisMemorySynthesis.ZeroOffDiagQACMasterReduction}
\csdef{lean@modpath@decl:bdl-zero-temp-dressing-bundle-exists}{Proofs/TwoAxisMemorySynthesis/ZeroOffDiagQACMasterReduction}
\csdef{lean@fqname@decl:bdl-zero-temp-dressing-bundle-exists}{bdl\leanUnderscore inhomogeneous\leanUnderscore zero\leanUnderscore temp\leanUnderscore dressing\leanUnderscore bundle\leanUnderscore exists}
\csdef{lean@disp@decl:bdl-zero-temp-dressing-bundle-exists}{bdl\_\allowbreak{}inhomogeneous\_\allowbreak{}zero\_\allowbreak{}temp\_\allowbreak{}dressing\_\allowbreak{}bundle\_\allowbreak{}exists}
\csdef{lean@file@decl:SatisfiesZeroTempPerturbativeMemoryStability}{Challenge.lean}
\csdef{lean@start@decl:SatisfiesZeroTempPerturbativeMemoryStability}{198}
\csdef{lean@end@decl:SatisfiesZeroTempPerturbativeMemoryStability}{228}
\csdef{lean@module@decl:SatisfiesZeroTempPerturbativeMemoryStability}{Challenge}
\csdef{lean@modpath@decl:SatisfiesZeroTempPerturbativeMemoryStability}{Challenge}
\csdef{lean@fqname@decl:SatisfiesZeroTempPerturbativeMemoryStability}{PassiveQuantumMemory.SatisfiesZeroTempPerturbativeMemoryStability}
\csdef{lean@disp@decl:SatisfiesZeroTempPerturbativeMemoryStability}{Satisfies\allowbreak{}Zero\allowbreak{}Temp\allowbreak{}Perturbative\allowbreak{}Memory\allowbreak{}Stability}
\csdef{lean@file@decl:dressed-ground-excited-form-bounds}{Proofs/CenteredBallBHM/ExcitedBandGapFromRelativeForm.lean}
\csdef{lean@start@decl:dressed-ground-excited-form-bounds}{277}
\csdef{lean@end@decl:dressed-ground-excited-form-bounds}{335}
\csdef{lean@module@decl:dressed-ground-excited-form-bounds}{Proofs.CenteredBallBHM.ExcitedBandGapFromRelativeForm}
\csdef{lean@modpath@decl:dressed-ground-excited-form-bounds}{Proofs/CenteredBallBHM/ExcitedBandGapFromRelativeForm}
\csdef{lean@fqname@decl:dressed-ground-excited-form-bounds}{dressed\leanUnderscore ground\leanUnderscore and\leanUnderscore excited\leanUnderscore band\leanUnderscore quadratic\leanUnderscore form\leanUnderscore bounds}
\csdef{lean@disp@decl:dressed-ground-excited-form-bounds}{dressed\_\allowbreak{}ground\_\allowbreak{}and\_\allowbreak{}excited\_\allowbreak{}band\_\allowbreak{}quadratic\_\allowbreak{}form\_\allowbreak{}bounds}
\csdef{lean@file@decl:coherent-dressed-pinched-commutator}{Proofs/CenteredBallBHM/CenteredBallFormAndGapBootstrap.lean}
\csdef{lean@start@decl:coherent-dressed-pinched-commutator}{40}
\csdef{lean@end@decl:coherent-dressed-pinched-commutator}{105}
\csdef{lean@module@decl:coherent-dressed-pinched-commutator}{Proofs.CenteredBallBHM.CenteredBallFormAndGapBootstrap}
\csdef{lean@modpath@decl:coherent-dressed-pinched-commutator}{Proofs/CenteredBallBHM/CenteredBallFormAndGapBootstrap}
\csdef{lean@fqname@decl:coherent-dressed-pinched-commutator}{coherent\leanUnderscore dressed\leanUnderscore pinched\leanUnderscore commutator\leanUnderscore le\leanUnderscore of\leanUnderscore siteFamily\leanUnderscore boxBall}
\csdef{lean@disp@decl:coherent-dressed-pinched-commutator}{coherent\_\allowbreak{}dressed\_\allowbreak{}pinched\_\allowbreak{}commutator\_\allowbreak{}le\_\allowbreak{}of\_\allowbreak{}site\allowbreak{}Family\_\allowbreak{}box\allowbreak{}Ball}
\csdef{lean@file@decl:norm-dressed-pinched-commutator-qac-tail}{Proofs/CenteredBallBHM/QACTailPinchedCommutators.lean}
\csdef{lean@start@decl:norm-dressed-pinched-commutator-qac-tail}{76}
\csdef{lean@end@decl:norm-dressed-pinched-commutator-qac-tail}{110}
\csdef{lean@module@decl:norm-dressed-pinched-commutator-qac-tail}{Proofs.CenteredBallBHM.QACTailPinchedCommutators}
\csdef{lean@modpath@decl:norm-dressed-pinched-commutator-qac-tail}{Proofs/CenteredBallBHM/QACTailPinchedCommutators}
\csdef{lean@fqname@decl:norm-dressed-pinched-commutator-qac-tail}{norm\leanUnderscore dressed\leanUnderscore pinched\leanUnderscore commutator\leanUnderscore le\leanUnderscore two\leanUnderscore mul\leanUnderscore qac\leanUnderscore tail\leanUnderscore clm}
\csdef{lean@disp@decl:norm-dressed-pinched-commutator-qac-tail}{norm\_\allowbreak{}dressed\_\allowbreak{}pinched\_\allowbreak{}commutator\_\allowbreak{}le\_\allowbreak{}two\_\allowbreak{}mul\_\allowbreak{}qac\_\allowbreak{}tail\_\allowbreak{}clm}
\csdef{lean@file@decl:zero-temp-heisenberg-ground-drift}{Proofs/TwoAxisMemorySynthesis/AutonomousHeisenbergGroundDrift.lean}
\csdef{lean@start@decl:zero-temp-heisenberg-ground-drift}{364}
\csdef{lean@end@decl:zero-temp-heisenberg-ground-drift}{395}
\csdef{lean@module@decl:zero-temp-heisenberg-ground-drift}{Proofs.TwoAxisMemorySynthesis.AutonomousHeisenbergGroundDrift}
\csdef{lean@modpath@decl:zero-temp-heisenberg-ground-drift}{Proofs/TwoAxisMemorySynthesis/AutonomousHeisenbergGroundDrift}
\csdef{lean@fqname@decl:zero-temp-heisenberg-ground-drift}{zero\leanUnderscore temp\leanUnderscore heisenberg\leanUnderscore ground\leanUnderscore drift\leanUnderscore of\leanUnderscore relative\leanUnderscore form}
\csdef{lean@disp@decl:zero-temp-heisenberg-ground-drift}{zero\_\allowbreak{}temp\_\allowbreak{}heisenberg\_\allowbreak{}ground\_\allowbreak{}drift\_\allowbreak{}of\_\allowbreak{}relative\_\allowbreak{}form}
\csdef{lean@file@decl:satisfiesZeroTemp-of-dressing-bundle}{Proofs/TwoAxisMemorySynthesis/Part1TwoAxisMemorySynthesis.lean}
\csdef{lean@start@decl:satisfiesZeroTemp-of-dressing-bundle}{126}
\csdef{lean@end@decl:satisfiesZeroTemp-of-dressing-bundle}{214}
\csdef{lean@module@decl:satisfiesZeroTemp-of-dressing-bundle}{Proofs.TwoAxisMemorySynthesis.Part1TwoAxisMemorySynthesis}
\csdef{lean@modpath@decl:satisfiesZeroTemp-of-dressing-bundle}{Proofs/TwoAxisMemorySynthesis/Part1TwoAxisMemorySynthesis}
\csdef{lean@fqname@decl:satisfiesZeroTemp-of-dressing-bundle}{Part1TwoAxisMemorySynthesis.satisfiesZeroTempPerturbativeMemoryStability\leanUnderscore of\leanUnderscore dressing\leanUnderscore bundle}
\csdef{lean@disp@decl:satisfiesZeroTemp-of-dressing-bundle}{satisfies\allowbreak{}Zero\allowbreak{}Temp\allowbreak{}Perturbative\allowbreak{}Memory\allowbreak{}Stability\_\allowbreak{}of\_\allowbreak{}dressing\_\allowbreak{}bundle}

\DeclarePairedDelimiter\ceil{\lceil}{\rceil}
\DeclarePairedDelimiter\floor{\lfloor}{\rfloor}
\newcommand{\norm}[1]{\lVert #1 \rVert}
\newcommand{\abs}[1]{\lvert #1 \rvert}
\newcommand{\ip}[2]{\langle #1, #2 \rangle}
\newcommand{\ket}[1]{\lvert #1 \rangle}

\newcommand{\N}{\mathbb{N}}
\newcommand{\Z}{\mathbb{Z}}
\newcommand{\R}{\mathbb{R}}
\newcommand{\C}{\mathbb{C}}
\newcommand{\F}{\mathbb{F}_2}
\newcommand{\bbE}{\mathbb{E}}

\newcommand{\cA}{\mathcal{A}}
\newcommand{\cB}{\mathcal{B}}
\newcommand{\cC}{\mathcal{C}}
\newcommand{\cD}{\mathcal{D}}
\newcommand{\cE}{\mathcal{E}}
\newcommand{\cF}{\mathcal{F}}
\newcommand{\cG}{\mathcal{G}}
\newcommand{\cH}{\mathcal{H}}
\newcommand{\cI}{\mathcal{I}}
\newcommand{\cK}{\mathcal{K}}
\newcommand{\cL}{\mathcal{L}}

\newcommand{\cO}{\mathcal{O}}
\newcommand{\cP}{\mathcal{P}}

\newcommand{\cR}{\mathcal{R}}
\newcommand{\cS}{\mathcal{S}}
\newcommand{\cT}{\mathcal{T}}

\newcommand{\cW}{\mathcal{W}}

\newcommand{\id}{\mathrm{id}}
\newcommand{\loc}{\mathrm{loc}}

\newcommand{\splitE}{\mathrm{split}}
\newcommand{\diam}{\operatorname{diam}}
\newcommand{\dist}{\operatorname{dist}}
\newcommand{\supp}{\operatorname{supp}}
\newcommand{\im}{\operatorname{im}}

\newcommand{\Tr}{\operatorname{Tr}}

\newcommand{\wt}{\operatorname{wt}}
\newcommand{\RCA}{\mathrm{RCA}}
\newcommand{\alt}{\operatorname{alt}}
\newcommand{\aff}{\operatorname{aff}}
\newcommand{\syn}{\mathrm{syn}}

\newcommand{\rank}{\operatorname{rank}}
\newcommand{\Cone}{\operatorname{Cone}}
\DeclareMathOperator*{\argmin}{arg\,min}

\theoremstyle{leanplain}
\newtheorem{theorem}{Theorem}[section]
\newtheorem*{informaltheorem}{Theorem}
\newtheorem{lemma}[theorem]{Lemma}
\newtheorem{proposition}[theorem]{Proposition}

\theoremstyle{leandef}
\newtheorem{definition}[theorem]{Definition}

\newtheorem{remark}[theorem]{Remark}

\title{\textbf{Robust Passive Quantum Memory in Three Dimensions, Part~I:\\ Thermal and Zero-Temperature Perturbative Stability}}

\author{Patrick Hayden\\[6pt]
{\small\textit{Leinweber Institute for Theoretical Physics, Stanford University, Stanford, CA 94305}}\\[2pt]
{\small\textit{Google DeepMind, Mountain View, CA 94043}}}

\date{November 2026}

\begin{document}

\maketitle

\begin{abstract}
Whether a three-dimensional quantum many-body system with geometrically local, bounded-degree interactions can passively protect encoded quantum information has remained a central open question since Kitaev introduced topological quantum computation. Passive protection requires stability against both thermal noise at nonzero temperature $T > 0$ and generic static Hamiltonian perturbations $V$. Balasubramanian, Davydova, and Lin (BDL) recently established unperturbed ($V = 0$) three-dimensional thermal self-correction using a recursively defined family of stabilizer Hamiltonians. However, they left open whether their construction remains stable against quantum perturbations, even at zero temperature. In this paper (Part~I), we show that a slight variant of the BDL code family achieves stability along both boundary axes of the perturbation--temperature phase diagram as the system size grows. Following BDL, at $V = 0$ it is thermally stable below a critical temperature $T_c > 0$, achieving a stretched-exponentially long memory lifetime when coupled to any local Markovian (Davies) thermal bath generated by single-site system--bath interactions satisfying detailed balance with bounded coupling rates. At zero temperature ($T = 0$) under arbitrary weak, superpolynomially decaying quantum perturbations, the family maintains a uniform $\Omega(1)$ bulk spectral gap, superpolynomially small ground-space splitting, and a superpolynomially long closed-system unitary memory lifetime for quasi-locally dressed logical observables. To prove this zero-temperature stability, we establish local topological quantum order across the multiscale hierarchy using a linearly dilated relaxation of the usual Bravyi--Hastings--Michalakis criteria. In the companion paper (Part~II), we combine the homological and operator-algebraic tools developed here with finite-temperature worldline cluster expansions to establish passive self-correction throughout the positive-temperature perturbed interior $(T > 0, V \neq 0)$.
\end{abstract}

\noindent\rule{\textwidth}{0.4pt}
\vspace{-18pt}
\setcounter{tocdepth}{2}
{
  \hypersetup{linkcolor=toclink}
  \tableofcontents
}
\noindent\rule{\textwidth}{0.4pt}
\vspace{8pt}

\section{Introduction}
\label{sec:intro}

\subsection{Passive Quantum Memory}

\begin{figure}[t]
\centering
\includegraphics[width=0.78\textwidth]{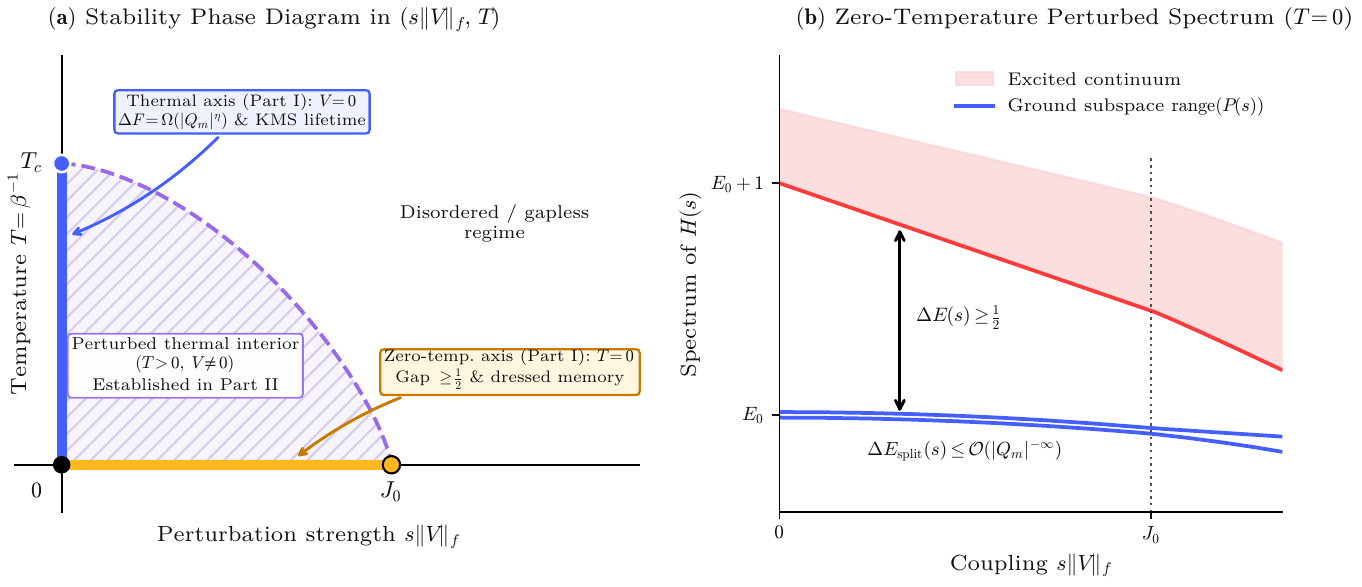}
\caption{Two-axis stability framework and zero-temperature perturbed low-energy spectrum for the multiscale welded memory $\cC_m$ on $\abs{Q_m}$ physical qubits. \textbf{(a)} Stability regimes in perturbation strength $s\norm{V}_f$ and temperature $T = \beta^{-1}$: Part~I establishes unperturbed positive-temperature Gibbs and Markovian Davies (Kubo--Martin--Schwinger (KMS) symmetric) self-correction along the vertical axis ($V = 0,\, 0 < T < T_c$, blue segment) with free-energy barrier $\Delta F = \Omega(\abs{Q_m}^\eta)$, and zero-temperature superpolynomial perturbative stability along the horizontal axis ($T = 0,\, 0 \le s\norm{V}_f \le J_0$, amber segment); the combined positive-temperature perturbed interior ($T > 0,\, V \neq 0$, shaded violet region) is established in Part~II~\cite{HaydenPartII}. \textbf{(b)} Zero-temperature low-energy spectrum of $H(s) = H_0 + s V$ versus $s\norm{V}_f$, showing the dressed ground space $\mathrm{range}(P(s))$ (projector $P(s)$) around ground energy $E_0$ (blue band, whose energy splitting $\Delta E_{\splitE}(s)$ decays faster than any inverse power of $\abs{Q_m}$, denoted $\cO(\abs{Q_m}^{-\infty})$ in the plot) separated from the excited continuum (shaded red region) by a uniform spectral gap $\Delta E(s) \ge 1/2$ up to $J_0$ (vertical dotted line).}
\label{fig:two-axes-stability}
\end{figure}

A passive self-correcting quantum memory is a quantum many-body system of $N$ physical spins governed by a geometrically local Hamiltonian in $\R^D$ whose degenerate ground-state manifold encodes quantum information and preserves it over macroscopic timescales without active syndrome measurement or feedback control~\cite{kitaev2003fault, dennis2002topological, alicki2010thermal, chesi2010thermodynamic, terhal2015quantum, brown2016quantum}. Just as a classical two-dimensional Ising ferromagnet below its Curie temperature separates its two phases by a domain-wall free-energy barrier~\cite{peierls1936ising, griffiths1964peierls} (protecting a classical bit over exponentially long timescales), a quantum spin system can store a qubit passively only if its low-energy topological sector withstands two physically distinct mechanisms of decoherence (Figure~\ref{fig:two-axes-stability}):
\begin{enumerate}[label=\textbf{(\roman*)}, leftmargin=2.2em, itemsep=3pt]
    \item \textbf{Thermodynamic and Markovian stability at positive temperature ($T > 0$)}: Under weak coupling to a Markovian thermal bath~\cite{davies1974markovian, lindblad1976generators, GKS1976} at inverse temperature $\beta = T^{-1} > \beta_c = T_c^{-1} > 0$, thermal fluctuations must encounter a macroscopically growing dimensionless equilibrium free-energy barrier $\Delta F = \Omega(N^\eta)$ ($\eta > 0$)~\cite{chesi2010thermodynamic, siva2017topological, baspin2025free} separating distinct logical homology sectors in both bit-flip ($X$) and phase-flip ($Z$) Pauli sectors, yielding a stretched-exponential memory relaxation time $\tau_{\mathrm{mem}} = \exp(\Omega(N^\eta))$~\cite{dennis2002topological, alicki2010thermal, chesi2010thermodynamic} (cf.~\cite{bergamaschi2026rapid} for rapid within-sector mixing).
    \item \textbf{Perturbative and coherent dynamical stability ($V \neq 0$)}: When the idealized commuting stabilizer Hamiltonian $H_0 = \sum_i Q_i$ (a sum of frustration-free check projectors $Q_i$) is deformed to $H(s) = H_0 + s V$ ($s \in [0, 1]$) by generic weak, superpolynomially decaying quantum interactions $V$ of local strength $J$ below a constant threshold $J_0 > 0$, the bulk spectral gap $\Delta E(s)$ above the ground-state manifold must remain bounded below by an $\Omega(1)$ constant ($\Delta E(s) \ge 1/2$), while the ground-space energy splitting $\Delta E_{\splitE}(s)$ decays faster than any inverse power of the system size~\cite{kitaev2003fault, BHM10, BH11}, protecting quasi-adiabatically dressed logical observables in the ground space over superpolynomially long times~\cite{hastings2005quasi, BHM10, BMNS12} (whereas undecoded bare strings fail in $O(1)$ time, and undressed 2D states fail in $O(\log N)$ time even after decoding~\cite{pastawski2010limitations}).
\end{enumerate}
Neither requirement is routine in three spatial dimensions. A Hamiltonian with a robust spectral gap against static perturbations $V$ but only an $O(1)$ thermal barrier, such as Kitaev's two-dimensional toric code~\cite{kitaev2003fault, alicki2009thermalization} or the three-dimensional $(1,2)$ toric code~\cite{castelnovo2008topological, yoshida2011feasibility}, rapidly loses its encoded state at any positive temperature as point-like anyonic defects diffuse across the lattice at constant energy cost. Conversely, an architecture engineered for finite-temperature syndrome confinement by breaking translation invariance across spatial scales must still maintain uniform local topological quantum order so that static stray fields or two-body spin couplings do not lift the ground-state degeneracy or close the excitation gap.

Building on the multiscale welded code architecture introduced by Balasubramanian, Davydova, and Lin~\cite{BDL24}, in this two-part series we study a three-layer variant $\cC_m$ across spatial scales $m \ge 0$ on an active qubit set $Q_m$, realized at even scales $m = 2k + 2$ ($k \ge 0$) as bounded-degree commuting-projector Hamiltonians $H_0 = H_{0,k}$ on open cubic boxes of side $L_k$ ($N = \Theta(L_k^3)$ lattice spins, with inactive $\ket{0}$-pinned slots). We establish the $(J, T)$ stability phase diagram (with a $T_{\min}$-dependent coupling threshold at fixed positive temperature) under deformations $H(s) = H_0 + s V$ ($s \in [0, 1]$) of weighted local perturbation norm $\norm{V}_f \le J$ for a superpolynomial spatial decay profile $f \in \cD_{\mathrm{super}}$ (defined in Section~\ref{subsec:dsuper-banach}). \textbf{Part~I} (the present paper) introduces the three-layer gadget modification and a syndrome-only multiscale cluster decoder with safe-syndrome observables, and proves stability independently along both boundary axes of Figure~\ref{fig:two-axes-stability}(a): positive-temperature Gibbs and Markovian Davies memory stability at $V = 0$ ($0 < T < T_c$), and zero-temperature superpolynomial perturbative and closed-system dynamical memory stability for $0 \le J \le J_0$ ($T = 0$). Establishing stability along these two boundary axes does not by itself settle the interior where $T > 0$ and $V \neq 0$ act together: at positive temperature the bulk carries a finite density of thermal defects, and quantum perturbations can move and mix those excitations in ways absent along either axis alone. The companion paper \textbf{Part~II}~\cite{HaydenPartII} (previewed in Section~\ref{subsec:part-ii-preview}) resolves this interplay, establishing passive self-correction across every fixed positive-temperature window $[T_{\min}, T_c)$ ($T_{\min} > 0$) for weak finite-range quantum perturbations and across the cryogenic scaling regime $0 \le T \le (C_{\mathrm{scale}}\log L_k)^{-1}$ down to $T = 0$ for all $f \in \cD_{\mathrm{super}}$.

\begin{figure}[t]
\centering
\includegraphics[width=0.90\textwidth]{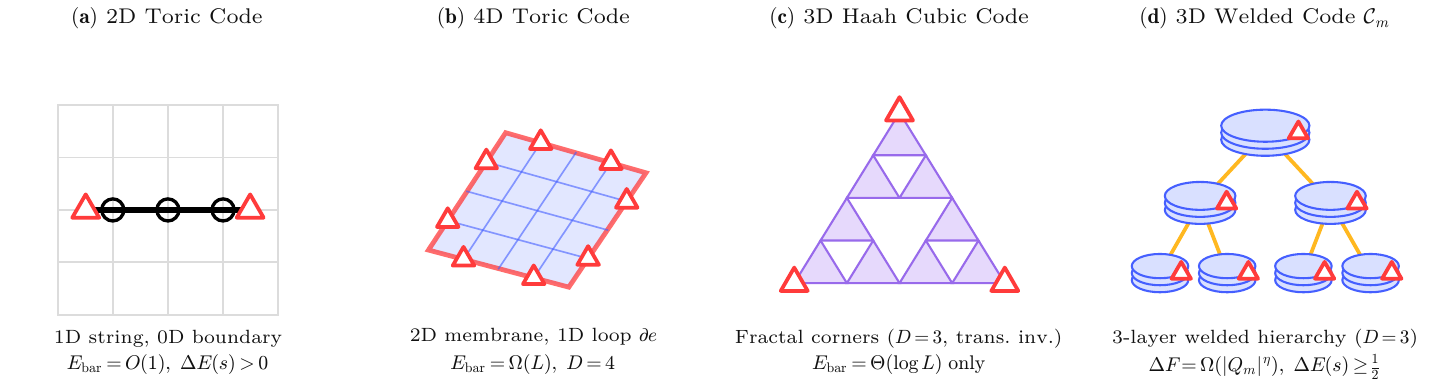}
\caption{Excitation geometry, zero-temperature energy barrier $E_{\mathrm{bar}}$, and spectral gap $\Delta E(s)$ across dimensions; open black rings ($\circ$) mark flipped physical qubits and open red triangles ($\triangle$) mark violated stabilizer checks. \textbf{(a)} In the 2D toric code, a string-like logical path (thick black segment with open black rings ($\circ$) along the chain and open red triangles ($\triangle$) at its two endpoints) carries only $O(1)$ syndrome weight. \textbf{(b)} In the 4D $(2,2)$ toric code, a 2D membrane error $e$ (shaded blue plaquettes) is bounded by a closed 1D syndrome loop $\partial e$ (red perimeter with $\triangle$ markers) of length $\Theta(L)$. \textbf{(c)} In 3D translation-invariant fracton models (Haah's cubic code), Sierpi\'nski-prism errors (purple triangles) create corner syndrome defects ($\triangle$) with an $O(\log L)$ barrier that saturates at finite temperature. \textbf{(d)} In the 3D multiscale welded code $\cC_m$ on $\abs{Q_m}$ physical qubits, stacked light-blue ellipses represent three-layer cylinder gadgets connected across consecutive hierarchy scales by amber welding links with open red triangles ($\triangle$) on their rims, amplifying connected syndrome clusters by $27/16 > 1$ per two-step cycle.}
\label{fig:dimensional-landscape}
\end{figure}

Over the past two decades, work on regular $D$-dimensional lattices of linear size $L = \Theta(N^{1/D})$ has clarified how sharply the prospects for passive self-correction depend on the spatial dimension (Figure~\ref{fig:dimensional-landscape}). In four and two dimensions, the picture is clear-cut:
\begin{itemize}[leftmargin=1.8em, itemsep=3pt]
    \item \textbf{Four dimensions ($D = 4$) and membrane loop tension}: In Kitaev's four-dimensional $(2,2)$ toric code~\cite{dennis2002topological}, both $X$-type and $Z$-type errors are supported on two-dimensional surfaces whose syndromes form closed one-dimensional boundary loops with linear energy tension. Because any sequence of single-qubit flips connecting distinct logical sectors must sweep out a surface of diameter $\Theta(L)$ with boundary perimeter $\Theta(L)$, Peierls contour arguments~\cite{peierls1936ising, griffiths1964peierls} establish a polynomial free-energy barrier $\Delta F = \Theta(N^{1/4})$ and thermal self-correction below $T_c > 0$~\cite{dennis2002topological, alicki2010thermal, chesi2010thermodynamic}, while local loop contractibility on balls of radius $r$ ensures perturbative stability with a constant additive collar $R(r) = r + 2$~\cite{BHM10}.
    \item \textbf{Two dimensions ($D = 2$) and string-like no-go theorems}: In two dimensions, Bravyi and Terhal~\cite{bravyi2009no} proved that any geometrically local stabilizer code supports string-like logical operators with $O(1)$ energy barrier (extended in~\cite{BPT10} to $k_{\mathrm{log}} d^2 = O(N)$ tradeoffs between logical qubits $k_{\mathrm{log}}$ and distance $d$, and to commuting frustration-free models with local topological order in~\cite{landon2013local}), while Alicki, Fannes, and Horodecki~\cite{alicki2009thermalization} proved $O(1)$ Markovian thermalization for the 2D toric code.
\end{itemize}

The three-dimensional case ($D = 3$) sits on the boundary between these two regimes. In the three-dimensional $(1,2)$ toric code, one logical operator is membrane-like with loop syndromes, but its conjugate partner is string-like with point-like endpoint defects, leaving an $O(1)$ energy barrier and $O(1)$-time thermal decoherence~\cite{castelnovo2008topological, alicki2009thermalization, chesi2010thermodynamic}---a limitation shared by all translation-invariant 3D stabilizer codes with constant ground-state degeneracy~\cite{yoshida2011feasibility}. Building on Chamon's fracton model~\cite{chamon2005quantum}, Haah's cubic code~\cite{haah2011local} eliminated string-like logical operators altogether in favor of immobile fractal excitations with a logarithmic energy barrier $\Theta(\log L)$~\cite{haah2011local, bravyi2011energy}, and Haah subsequently proved via Laurent-polynomial algebra~\cite{haah2013commuting} that no exact, degenerate, translation-invariant commuting qubit Pauli Hamiltonian in $\R^3$ can exceed an $O(\log L)$ barrier. At positive temperature, this logarithmic barrier yields ``partial self-correction'': Bravyi and Haah~\cite{bravyi2013quantum} proved that the cubic code's memory lifetime grows as $L^{\Omega(\beta)}$ up to a thermal crossover scale $L_{\mathrm{th}}^*(\beta) = e^{c\beta}$ ($c > 0$), above which thermally nucleated defects overwhelm the barrier so the lifetime saturates at $e^{O(\beta^2)}$~\cite{bravyi2013quantum, siva2017topological} (with rapid thermal mixing recently established at all $T > 0$~\cite{stengele2026haah}).

Surpassing the $\Theta(\log L)$ ceiling requires relaxing at least one hypothesis of Haah's no-go theorem---either moving beyond independent commuting Pauli checks via non-Pauli symmetry-protected domain walls~\cite{li2024domain} or locally redundant stabilizers~\cite{lin2024proposals}, or breaking spatial translation invariance. Following the latter route, Michnicki~\cite{michnicki20143d} welded together three-dimensional toric code blocks of varying sizes to achieve the first power-law zero-temperature energy barrier in $\R^3$, and subsequent constructions achieved or conjectured polynomial energy barriers through fractal and prefractal geometries~\cite{brell2016proposal, lin2024proposals} (cf.\ Ising models on fractal lattices~\cite{vezzani2003spontaneous}), three-dimensional embeddings of asymptotically good quantum low-density parity-check (LDPC) codes~\cite{panteleev2022asymptotically, leverrier2022quantum, portnoy2023local, lin2023geometrically, li2024transform}, and topological layer codes~\cite{williamson2023layer, gu2025layer, williamson2025partial}.

Yet a macroscopic \emph{energy} barrier $\Delta E$ at $T = 0$ does not guarantee a macroscopic \emph{free-energy} barrier $\Delta F = \Delta E - T \Delta S$ at fixed positive temperature $T > 0$. In both Michnicki's welded code~\cite{siva2017topological} and three-dimensional layer codes~\cite{baspin2025free}, the configurational entropy $\Delta S$ of defect pathways grows fast enough at any fixed $T > 0$ to collapse the free-energy barrier, leaving only partial self-correction up to $L_{\mathrm{th}}^*(\beta) = e^{\Omega(\beta)}$~\cite{siva2017topological, gu2025layer, williamson2025partial}. Balasubramanian, Davydova, and Lin~\cite{BDL24} overcame this entropic barrier collapse by iterating cellular code welding across $\Theta(\log L)$ spatial scales, establishing genuine thermal self-correction for an unperturbed 3D stabilizer Hamiltonian ($V = 0$) below a positive critical temperature $T_c > 0$ (and, with Thorngren, proving positive-temperature classical metastability for perturbed 2D welded spin models~\cite{BDLT26}). Whether such a multiscale three-dimensional quantum code remains perturbatively stable once generic quantum interactions $V \neq 0$ are turned on---either at $T = 0$ or at positive temperature $T > 0$---remained open.

Stated informally as an existence result on open three-dimensional cubic lattices---deferring code-specific definitions and the precise statement to Theorem~\ref{thm:main} in Section~\ref{subsec:main-theorem-synthesis}---the main theorem of Part~I establishes both boundary-axis properties of Figure~\ref{fig:two-axes-stability}(a) on a common code family:

\begin{informaltheorem}[Informal; see Theorem~\ref{thm:main} for the precise statement]
There exist positive constants $T_c$ and $\eta$, together with a family of frustration-free, commuting Pauli stabilizer Hamiltonians $H_0$ on open three-dimensional cubic lattices of arbitrarily large linear size $L$, each encoding a single logical qubit. Every stabilizer check acts on $O(1)$ qubits within $O(1)$ distance, every qubit participates in $O(1)$ checks, and the following two stability properties hold uniformly as $L \to \infty$:
\begin{enumerate}[label=\textbf{(\arabic*)}, leftmargin=2.2em, itemsep=3pt]
    \item \textbf{Thermal self-correction below $T_c$ ($V = 0$, $0 < T < T_c$)}: At any temperature $T < T_c$, there is a pair of decoded logical observables for which thermal excitations encounter a macroscopic equilibrium free-energy barrier $\Delta F = \Omega(N^\eta)$. When $H_0$ is coupled to any local Markovian (Davies) thermal bath at temperature $T$ with bounded coupling rates, the encoded qubit survives in the thermal state for stretched-exponentially long times, $\tau_{\mathrm{mem}} = \exp\bigl(\Omega(N^\eta)\bigr)$.
    \item \textbf{Zero-temperature perturbative stability ($T = 0$, $V \neq 0$)}: For any fixed superpolynomial spatial decay rate, there exists a coupling threshold $J_0 > 0$ such that for every static Hermitian perturbation $V$ on the lattice of local strength at most $J_0$ decaying at that rate, the perturbed Hamiltonian $H = H_0 + V$ retains an $\Omega(1)$ spectral gap above a quasi-degenerate two-dimensional ground space whose energy splitting decays faster than any inverse power of $N$. Moreover, a quasi-local unitary transformation dresses the decoded logical observables so that they remain immune to local errors and preserve the encoded qubit under closed-system evolution $e^{-itH}$ for times growing faster than any polynomial in $N$.
\end{enumerate}
\end{informaltheorem}

Modulo the minor distinction between a two-layer and a three-layer welded code (and the accompanying decoder adaptation discussed below), Part~\textbf{(1)} is the thermal self-correction theorem of Balasubramanian, Davydova, and Lin~\cite{BDL24}, which we reprove here (and certify in Lean~4) on the same three-layer code family that supports our zero-temperature perturbative stability theorem in Part~\textbf{(2)}.

\subsection{Multiscale Welded Codes}

To circumvent the translation-invariant point-charge and fractal-generator obstructions in three-dimensional qubit Pauli stabilizer codes~\cite{yoshida2011feasibility, haah2013commuting}, Balasubramanian, Davydova, and Lin~\cite{BDL24} recently introduced a multiscale welded code architecture in $\R^3$ that builds a hierarchy of codes $\cC_0 \to \cC_1 \to \cdots \to \cC_m$ across $m$ spatial scales (Sections~2--3 of~\cite{BDL24}). We adopt their foundational multiscale hierarchy and two-step inductive structure: one starts from a constant-size planar surface code seed $\cC_0$ encoding a single logical qubit, where $X$-type checks and $Z$-type checks each act on a constant number of nearby physical qubits. At each odd scale, a primal cellular replacement step substitutes every $X$-check with a subdivided hollow cylinder of physical qubits and $Z$-checks, welded together along star-shaped interfaces at the coarse qubits via a height-2 cochain complex diagram (Section~2.3 and Section~3.2 of~\cite{BDL24}, following~\cite{yuan2026unified}); this replacement leaves the encoded logical qubit unchanged while amplifying the $Z$-check syndromes produced by $X$-type bit-flip errors. At each even scale, a dual replacement step applies the exact same substitution with the roles of $X$ and $Z$ interchanged, amplifying the $X$-check syndromes produced by $Z$-type phase-flip errors. Between replacement steps, the complex is spatially rescaled by a constant factor $\Lambda > 1$ and its vertex coordinates in $\R^3$ are slightly perturbed using the Lov\'asz Local Lemma (LLL)~\cite{erdos1975problems, moser2009constructive, moser2010constructive, chandrasekaran2013deterministic} (Section~3.1 of~\cite{BDL24}, building on subdivided manifold embeddings~\cite{gromov2012generalizations, freedman2021building, portnoy2023local}) so that physical cells never crowd together: across all scales $m$, every unit ball in $\R^3$ contains at most $O(1)$ qubits and checks, and every check acts on at most $O(1)$ qubits within Euclidean distance $O(1)$.

Our code construction departs locally from~\cite{BDL24} in using three horizontal layers rather than two in each cylinder and star replacement gadget. This modification grew out of difficulties we encountered when attempting to formalize that work in Lean~4 while verifying the correctness of our perturbative stability argument: at the time, we thought those difficulties pointed to a flaw in the original paper and moved to the three-layer code in order to get around them. More careful study after completing the formalization indicated that the fault lay with our formalization effort rather than with~\cite{BDL24}; nonetheless, moving to the three-layer code mildly simplifies the proof, so we retain it here. In the two-layer gadget, each coarse dual check is replicated across two parallel horizontal sheets, and sweeping internal defects outward onto those copies requires combinatorial string and star-arm pairing to resolve $1$-versus-$1$ ties between the sheets. Adding a third horizontal layer turns those ties into a strict $2$-versus-$1$ majority on each interface arm individually.

Establishing that this three-layer welded code family is also perturbatively stable against generic static quantum interactions $H_0 + s V$ at zero temperature (a question left to future work in~\cite{BDL24}) and protects dressed logical observables over superpolynomially long times requires overcoming three challenges across decoding, local homology, and many-body operator theory:
\begin{enumerate}[label=\textbf{(\arabic*)}, leftmargin=2.2em, itemsep=3pt]
    \item \textbf{Single-qubit recovery chains versus ball-protected parity decoding}: In the unperturbed setting ($V = 0$), a local thermal bath flips one spin at a time, so it suffices that the decoder be immune to single-qubit errors on typical thermal syndromes~\cite{BDL24}. Once static quantum perturbations $H_0 + s V$ are turned on at $T = 0$, however, quasi-adiabatic continuation spreads both local errors and the block-diagonalized Hamiltonian across macroscopic spatial collars. Protecting dressed logical observables against coherent drift and local noise therefore requires a bare safe subspace on which the decoded logical operators commute with arbitrary multi-qubit operators supported on macroscopic balls whose radius grows as a power of the system size (a protection radius reused at $T > 0, V \neq 0$ in Part~II~\cite{HaydenPartII}). Rather than lifting local cleaners across scales into a physical recovery chain, our \emph{multiscale cluster decoder} passes only coarse-grained syndromes and a single logical parity bit down to a constant-size base welded code, where local homological acyclicity anchors an inductive family of ball-restricted parity witnesses with macroscopic protection radius. Working purely with syndromes and a parity bit also streamlines formalization: it avoids lifted recovery chains---whose physical support can jump by macroscopic stabilizers under a single spin flip---and replaces the multi-level decoding graph and depth-first cluster counting of~\cite{BDL24} at $V = 0$ with an elementary subset witness pool and a Fourier coset Peierls bound.
    \item \textbf{Linearly dilated local homology collars and scale-matched renormalization stopping}: In the perturbative stability theory of Bravyi, Hastings, and Michalakis~\cite{BHM10} and Bravyi and Hastings~\cite{BH11}, conventional topological codes such as the 2D and 4D toric codes satisfy local topological quantum order with a constant additive collar ($\mathrm{TQO\text{-}2}$): any local operator supported in a Euclidean ball $B(x_0, r) := \{y \in \R^3 : \norm{y - x_0}_2 \le r\}$ of radius $r$ centered at $x_0 \in \R^3$ that annihilates the global ground space already annihilates the local ground space on $B(x_0, r + O(1))$. In a multiscale welded code, however, a syndrome-free stabilizer supported in $B(x_0, r)$ that straddles a welding interface at scale $\Theta(r)$ can only be generated by parity checks that sweep across the adjacent cylinders at that scale, requiring a \emph{linearly dilated collar} $B(x_0, \ceil{C r})$ with dilation factor $C \ge 1$. We call this property $\mathrm{Linear\ TQO\text{-}2}(C)$ and unify it with local ground-state indistinguishability $\mathrm{TQO\text{-}1}(L^*)$ on balls of radius up to $L^* = \Theta(\Lambda^m)$ into a single homological condition called \emph{Relative Cone Acyclicity} $\RCA(L^*, C)$ (Definitions~\ref{def:rca}--\ref{def:tqo}). Furthermore, if one attempts to prove local acyclicity by pulling a chain supported in $B(x_0, r)$ all the way down to $\cC_0$ via the Homological Perturbation Lemma (HPL)~\cite{brown1965twisted, gugenheim1972chain, crainic2004perturbation}, each coarse-graining step divides the radius by $\Lambda$ while adding an $O(1)$ gadget collar ($r \mapsto r/\Lambda + O(1)$), stalling at an $O(1)$ fixed-point radius that inflates to the full system size $\Theta(\Lambda^m)$ when lifted back to scale $m$. We resolve this by stopping the coarse-graining at the scale-matched depth $\ceil{\log_\Lambda r}$ where the rescaled radius first reaches $O(1)$, and contracting the chain locally across two consecutive replacement steps via a six-checkpoint single-simplex altitude ray selection before lifting back to scale $m$.
    \item \textbf{Band-limited quasi-adiabatic continuation and centered-ball telescoping}: At each even hierarchy depth $m = 2k + 2$, snapping the cells of $\cC_m$ onto an open cubic lattice $\Gamma_k = \{0, \dots, L_k - 1\}^3$ of side length $L_k = \Lambda^{2k+4}$ yields a commuting-projector Hamiltonian $H_{0,k}$. Under the deformation $H(s) = H_{0,k} + s V$, quasi-adiabatic continuation (QAC)~\cite{hastings2005quasi, BHM10, BH11, BMNS12, MZ13} (building on Kato's adiabatic parallel transport~\cite{kato1950adiabatic} and the exact band-limited spectral flow of Bachmann, Michalakis, Nachtergaele, and Sims~\cite{BMNS12}) furnishes a quasi-local unitary $U(s)$ that rotates the perturbed ground space back to the unperturbed code space, while a smooth compact-frequency low-pass filter decomposes $U(s)^\dagger H(s) U(s) - H_{0,k}$ into site-anchored block-diagonal operators $X_u(s)$ ($u \in \Gamma_k$) whose spatial tails belong to the class $\cD_{\mathrm{super}}$ of functions decaying faster than any inverse polynomial (Definition~\ref{def:banach-quasilocal}). Subtracting each site operator's ground-space expectation value centers $X_u(s)$ so that it annihilates the unperturbed ground space. Because $\cD_{\mathrm{super}}$ is closed under polynomial volume weighting, tail summation, and affine argument rescaling $r \mapsto (r - a)/C$, truncating at an intermediate radius $\Lambda^{k-4}$ (well inside both the local indistinguishability cutoff and the decoder protection radius) allows us to telescope each centered site operator across concentric linearly dilated balls around $u$ on the entire Hilbert space. This yields a two-sided relative operator-form sandwich, a persistent bulk spectral gap $\Delta E(s) > 1/2$, dressed safe-subspace single-site immunity, and a superpolynomial zero-temperature unitary memory lifetime. (Because the operator exponential $X \mapsto e^{-\beta X}$ is not operator-monotone, this full-spectrum operator sandwich does not by itself bound the physical perturbed Gibbs logical failure probability at fixed positive temperature $T > 0$; overcoming spatially fluctuating bulk vacuum shifts is the subject of Part~II~\cite{HaydenPartII}.)
\end{enumerate}

\subsection{Outline of the Paper}

\begin{figure}[t]
\centering
\includegraphics[width=0.92\textwidth]{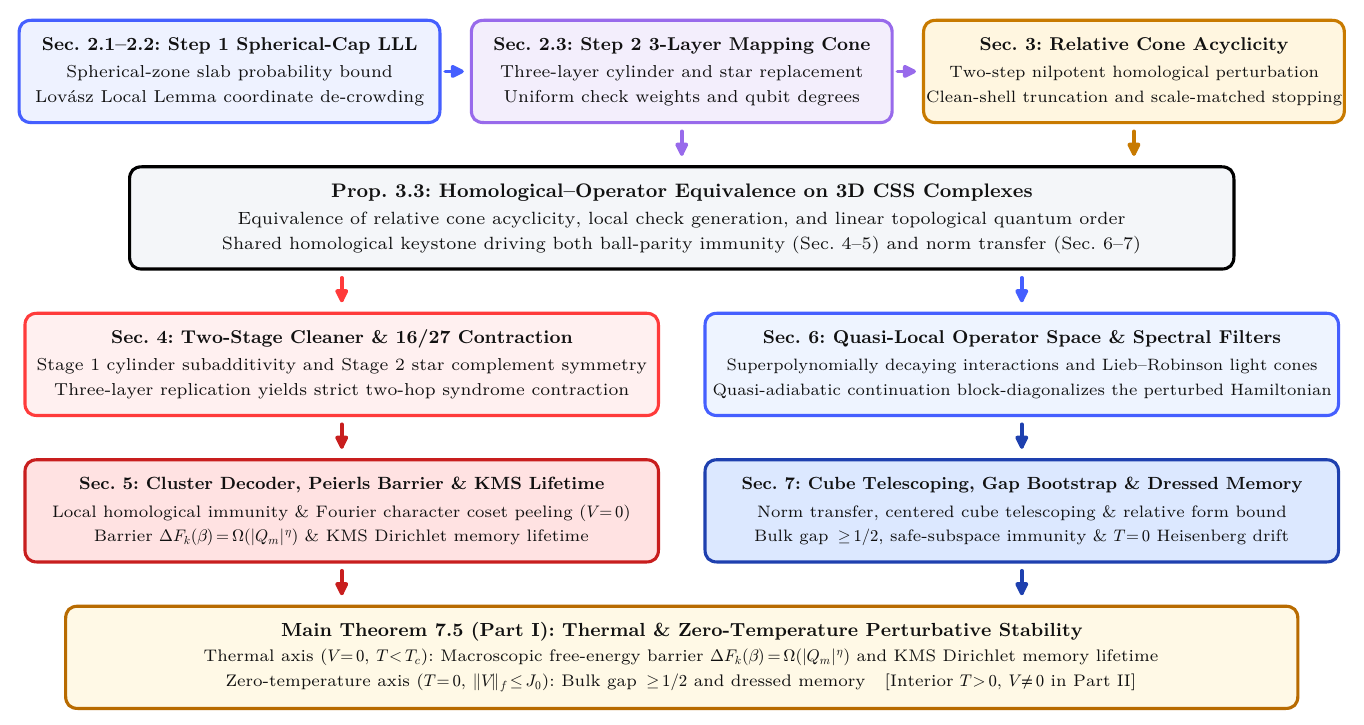}
\caption{Logical architecture of Part~I. The top row (blue, violet, and amber boxes) depicts the shared geometric and homological foundation in Sections~\ref{sec:bdl-construction}--\ref{sec:linear-tqo2-proof}, establishing uniform bounded geometry (check weights $d_X, d_Z \le 8$, range $r_0 = 8$, density $\rho_0 = O(1)$) and Relative Cone Acyclicity $\RCA(L^*, C)$. By the homological--operator equivalence (Proposition~\ref{prop:trinity-equivalence}, central gray banner), $\RCA(L^*, C)$ branches into the red left column ($V = 0$, $T < T_c$, Sections~\ref{sec:two-stage-cleaner}--\ref{sec:cluster-decoder-peierls}: two-stage $16/27$ syndrome cleaning, the multiscale cluster decoder, the coset Peierls bound, safe decoded observables, and Markovian Davies / KMS thermal memory lifetime) and the blue right column ($T = 0$, $J \le J_0$, Sections~\ref{sec:qac-banach}--\ref{sec:telescoping-main}: band-limited QAC block diagonalization, centered-ball single-cube telescoping, gap bootstrap, and dressed zero-temperature memory stability), reuniting in Theorem~\ref{thm:main}.}
\label{fig:proof-roadmap}
\end{figure}

Figure~\ref{fig:proof-roadmap} illustrates how the sections of the paper fit together:
\begin{itemize}[leftmargin=1.8em, itemsep=2.5pt]
    \item \textbf{Section~\ref{sec:bdl-construction}} defines bounded-geometry 3D Calderbank--Shor--Steane (CSS) complexes and active $2$-simplex altitude and density regularity, proves the Step~1 spherical-cap coordinate de-crowding theorem via spherical-zone measure bounds and vertex-disjoint simplex tuple amplification~\cite{BDL24}, and adapts the cellular mapping-cone replacement of~\cite{BDL24} to the Step~2 three-layer functor $\cR_X^{(3)}$ across the directed acyclic graph of cylinder, star, and replicated-check sectors.
    \item \textbf{Section~\ref{sec:linear-tqo2-proof}} introduces Relative Cone Acyclicity $\RCA(R_0, C)$ at cutoff radius $R_0$ and collar factor $C$, proves its equivalence to local check generation and $\mathrm{TQO\text{-}1}(R_0) \wedge \mathrm{Linear\ TQO\text{-}2}(C)$, establishes first-order termination of the Homological Perturbation Lemma series via the three-sector grading, and combines six-checkpoint single-simplex altitude ray selection, two-step sub-cylinder filling, and alternating primal/dual scale-matched stopping to prove $\RCA(L^*, C)$ for $\cC_m$.
    \item \textbf{Section~\ref{sec:two-stage-cleaner}} constructs the two-stage local syndrome cleaner and coarse syndrome map, proving the $\frac{4}{3}$ cylinder export bound via $\Z_{\ge 0}$-subadditivity across the three reference layers, the $\frac{4}{3}$ star majority bound and exact three-layer replication via layer-complement symmetry, and the combined $\frac{16}{27}$ contraction over two-hop fine gadget neighborhoods of coarse check subsets.
    \item \textbf{Section~\ref{sec:cluster-decoder-peierls}} introduces the multiscale cluster decoder $\cD_{\mathrm{cluster}}$ and inductive ball-restricted parity witness families of minimum weight $(3/2)^k$, proves the classical coset Peierls bound, constructs the open-box spin Hamiltonian $H_{0,k}$, safe-syndrome projector $\Pi_{\mathrm{safe},k}$, and decoded logical observables $O_{\mathrm{dec},k}^Z, O_{\mathrm{dec},k}^X$ (Section~\ref{subsec:safe-decoded-observables}), and establishes the stretched-exponential unperturbed memory lifetime under bounded-range KMS-symmetric double-commutator and Markovian Davies thermal dynamics (Section~\ref{subsec:unperturbed-kms-lifetime}).
    \item \textbf{Section~\ref{sec:qac-banach}} introduces the superpolynomial decay class $\cD_{\mathrm{super}}$ and site-anchored quasi-local operator space $\cB_f(\Gamma_k)$ ($f \in \cD_{\mathrm{super}}$), establishes algebraic Lieb--Robinson and band-limited spectral-filter light-cone bounds, and derives direct site-filtered block diagonalization of the quasi-adiabatically rotated Hamiltonian into site-anchored operators commuting with the code projector.
    \item \textbf{Section~\ref{sec:telescoping-main}} proves local-to-global ground-projector norm transfer on linearly dilated balls, centered single-cube telescoping up to the intermediate cutoff radius $\Lambda^{k-4}$, the full-space relative form bound and continuous open--closed spectral gap bootstrap, and zero-temperature dressed safe-subspace immunity and closed-system dynamical memory stability (Section~\ref{subsec:zero-temp-dressed-memory}), culminating in the precise statement and proof of Theorem~\ref{thm:main} (Section~\ref{subsec:main-theorem-synthesis}).
    \item \textbf{Section~\ref{sec:discussion}} distills the main lessons from the proof, explains why naive unitary QAC dressing does not settle positive-temperature perturbed self-correction and previews the phase-diagram proof of Part~II~\cite{HaydenPartII} (Section~\ref{subsec:part-ii-preview}), and discusses open problems.
\end{itemize}

\subsection*{AI Use and Formal Verification Statement}
\label{subsec:ai-use-statement}

The author views this project as an experiment in human--AI research collaboration and publication. While the research direction, problem formulation, and overall proof strategy were set by the author, Gemini 4 Argon performed most of the detailed mathematical reasoning and proof construction. Both the Argon and Carbon versions of Gemini 4 assisted in drafting the manuscript.

To verify our main claims, standalone existential formulations of the core physical conclusions of Theorem~\ref{thm:main} and of the finite-range positive-temperature stability theorem of Part~II~\cite{HaydenPartII} are stated in \texttt{Challenge.lean} (over minimal spin-lattice definitions) and machine-checked in Lean~4 (\texttt{v4.35.0})~\cite{moura2021lean4} on Mathlib~\cite{mathlib2020} via \texttt{lake comparator}. Rather than formalizing the manuscript statement-by-statement, the Lean development is organized solely around discharging those two challenge theorems: classical background, converse implications, and auxiliary corollaries not needed on that path remain paper-only, and in several places Lean adopts more direct proof routes, specialized parameter regimes, or approximate operator bounds that suffice for the challenge conclusions. Where a numbered heading has a Lean~4 counterpart---often a specialization, sub-lemma, or predicate---it is hyperlinked as a navigational aid; Appendix~\ref{app:formal-correspondence} summarizes the trust boundary and scope of what is certified.

\section{The Three-Layer Welded Hierarchy}
\label{sec:bdl-construction}

\subsection{3D CSS Complexes}
\label{subsec:css-complexes}

Following Sections~2.1--2.2 of~\cite{BDL24}, we begin by formalizing three-dimensional CSS codes~\cite{calderbank1996good, steane1996multiple} as homological chain complexes~\cite{freedman2001projective, kitaev2003fault, bravyi2014homological} equipped with spatial embeddings in $\R^3$ and uniform bounded-geometry invariants. In a CSS code, $X$-type checks correspond to $0$-cells $M_X$, physical qubits correspond to $1$-cells $Q$, and $Z$-type checks correspond to $2$-cells $M_Z$. Because we study both primal ($X$-error / $Z$-syndrome) and dual ($Z$-error / $X$-syndrome) excitations, we track the coboundary operators $\delta^0, \delta^1$ alongside their transpose boundary operators $\partial_1 = (\delta^0)^\top$ and $\partial_2 = (\delta^1)^\top$, as well as the spatial positions of all cells in $\R^3$.

\leanlink{def:3d-css-complex}
\begin{definition}[{3D CSS Chain Complex and Bounded-Geometry Invariants~\cite[Sections~2.1--2.2]{BDL24}}]
\label{def:3d-css-complex}
A \emph{3D CSS complex} $(\cC, \Phi)$ consists of a three-term chain complex of finite-dimensional $\F$-vector spaces equipped with distinguished standard bases $(M_X, Q, M_Z)$ and a spatial embedding $\Phi : M_X \sqcup Q \sqcup M_Z \to \R^3$:
\begin{equation}
    \label{eq:css-chain-cochain}
    \F^{M_X} \underset{\partial_1}{\overset{\delta^0}{\rightleftarrows}} \F^Q \underset{\partial_2}{\overset{\delta^1}{\rightleftarrows}} \F^{M_Z}, \qquad \delta^1 \circ \delta^0 = 0, \quad \partial_1 \circ \partial_2 = 0, \quad \partial_k = (\delta^{k-1})^\top,
\end{equation}
where $M_X$ indexes the $X$-type checks ($0$-cells), $Q$ indexes the physical qubits ($1$-cells), and $M_Z$ indexes the $Z$-type checks ($2$-cells); we write $C^0(\cC) := \F^{M_X}$, $C^1(\cC) := \F^Q$, and $C^2(\cC) := \F^{M_Z}$ (with first cohomology $H^1(\cC) := \ker \delta^1 / \im \delta^0$, first homology $H_1(\cC) := \ker \partial_1 / \im \partial_2$, standard basis indicators $\mathbf{1}_u$, set cardinality $\abs{S}$, Hamming weight $\abs{v} = \wt_n(v) := \abs{\supp(v)}$ and all-ones vector $\mathbf{1}_n$ ($v \in \F^n$), $\F$-pairing $\ip{u}{v} := \sum_i u_i v_i \in \F$, and minimum logical distances $d_{\mathrm{log}}^X(\cC) := \min_{c \in \ker \delta^1 \setminus \im \delta^0}\abs{c}$, $d_{\mathrm{log}}^Z(\cC) := \min_{c \in \ker \partial_1 \setminus \im \partial_2}\abs{c}$, $d_{\mathrm{log}}(\cC) := \min(d_{\mathrm{log}}^X(\cC), d_{\mathrm{log}}^Z(\cC))$). For each $x \in M_X$ and $z \in M_Z$, we denote by $\partial_X(x) := \supp(\delta^0 \mathbf{1}_x) \subseteq Q$ and $\partial_Z(z) := \supp(\partial_2 \mathbf{1}_z) \subseteq Q$ the sets of qubits on which $x$ and $z$ act, and by $\partial_X^\top(q) := \{x \in M_X : q \in \partial_X(x)\}$ and $\partial_Z^\top(q) := \{z \in M_Z : q \in \partial_Z(z)\}$ the sets of checks incident to a qubit $q \in Q$. Its \emph{CSS dual complex} is $(\cC^\vee, \Phi^\vee) := (\F^{M_Z} \rightleftarrows \F^Q \rightleftarrows \F^{M_X}, \Phi)$ with $\delta_{\cC^\vee}^0 = (\delta_{\cC}^1)^\top = \partial_2$ and $\delta_{\cC^\vee}^1 = (\delta_{\cC}^0)^\top = \partial_1$. For integers $d_0 \ge 1$ and reals $r_0, \rho_0 > 0$, we say that $(\cC, \Phi)$ has \emph{$(d_0, r_0, \rho_0)$-bounded geometry} in $\R^3$ if the following four conditions hold:
\begin{enumerate}[label=\textbf{(\alph*)}, leftmargin=2.2em, itemsep=2.5pt]
    \item \textbf{Uniform LDPC check and qubit degrees}: Every check satisfies $1 \le \abs{\partial_X(x)} \le d_X \le d_0$ and $1 \le \abs{\partial_Z(z)} \le d_Z \le d_0$ (where $d_X := \max_{x \in M_X}\abs{\partial_X(x)}$ and $d_Z := \max_{z \in M_Z}\abs{\partial_Z(z)}$), and every qubit $q \in Q$ participates in $1 \le \abs{\partial_X^\top(q)} \le d_{QX} \le d_0$ $X$-checks and $1 \le \abs{\partial_Z^\top(q)} \le d_{QZ} \le d_0$ $Z$-checks (where $d_{QX} := \max_{q \in Q}\abs{\partial_X^\top(q)}$ and $d_{QZ} := \max_{q \in Q}\abs{\partial_Z^\top(q)}$). In the three-layer welded hierarchy $\cC_m$, we maintain the sharp bounds $d_X, d_Z \le 8$ and $1 \le \abs{\partial_X^\top(q)}, \abs{\partial_Z^\top(q)} \le 4$ ($d_{QX}, d_{QZ} \le 4$) across all scales $m \ge 0$.
    \item \textbf{Check orthogonality, two-qubit overlap, and connected links}: For every pair $(x, z) \in M_X \times M_Z$, the shared support satisfies $\abs{\partial_X(x) \cap \partial_Z(z)} \in \{0, 2\}$, so the Pauli stabilizers $A_x = \prod_{q \in \partial_X(x)} X_q$ and $B_z = \prod_{q \in \partial_Z(z)} Z_q$ commute ($[A_x, B_z] = 0$). Moreover, for each $x \in M_X$, the bipartite incidence link graph $G_x$ between $\partial_X(x)$ and its co-incident $Z$-checks $\{z \in M_Z : \partial_X(x) \cap \partial_Z(z) \neq \emptyset\}$ (joined whenever $q \in \partial_X(x) \cap \partial_Z(z)$; resp.\ $G_z$ for $z \in M_Z$) is connected, with check-vertex degree $2$ and qubit-vertex degree at most $d_{QZ} \le d_0$ (resp.\ $d_{QX} \le d_0$; in $\cC_0$, $d_{QX}, d_{QZ} \le 2$ gives simple paths, whereas $d_{QX}, d_{QZ} \le 4$ for $m \ge 1$).
    \item \textbf{Geometric locality range $r_0$}: For every incident pair $x \sim q$ or $q \sim z$, the Euclidean distance in $\R^3$ satisfies $\norm{\Phi(x) - \Phi(q)}_2 \le r_0$ and $\norm{\Phi(q) - \Phi(z)}_2 \le r_0$.
    \item \textbf{Uniform spatial density $\rho_0$}: Every Euclidean unit ball $B(p, 1) \subset \R^3$ contains at most $\rho_0$ cell centers from each sector of $\Phi(M_X \sqcup Q \sqcup M_Z)$.
\end{enumerate}
\end{definition}

\begin{figure}[t]
\centering
\includegraphics[width=0.90\textwidth]{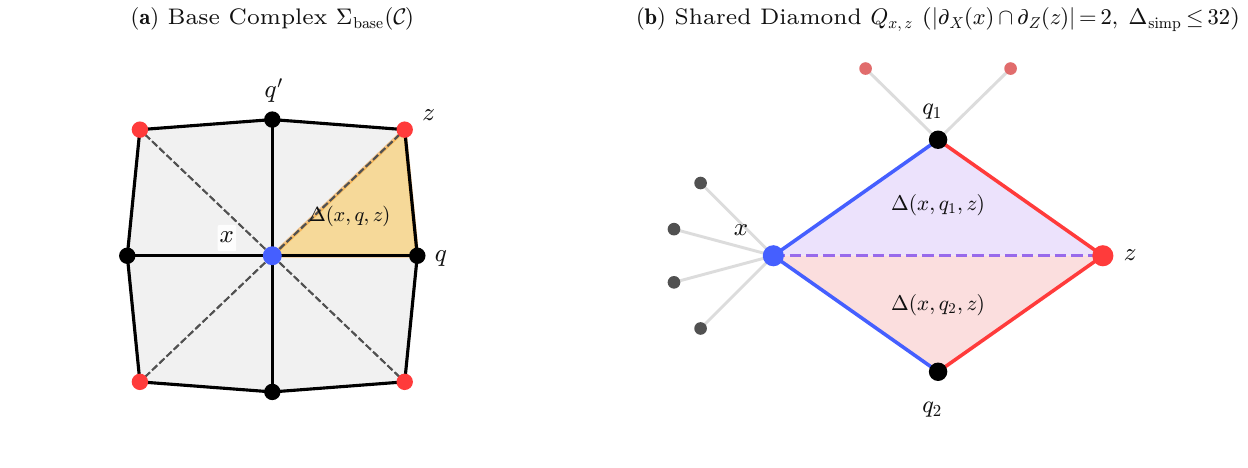}
\caption{From the bipartite CSS Tanner graph of $(\cC, \Phi)$ to the two-dimensional incidence base complex $\Sigma_{\mathrm{base}}(\cC)$ in $\R^3$; blue circles denote $X$-checks ($0$-cells $M_X$), black circles denote physical qubits ($1$-cells $Q$), and red circles denote $Z$-checks ($2$-cells $M_Z$). \textbf{(a)} Every incident triple $(x, q, z)$ spans a shaded yellow $2$-simplex $\Delta(x, q, z)$ in $\Sigma_{\mathrm{base}}(\cC)$. \textbf{(b)} Because every co-incident check pair $(x,z) \in M_X \times M_Z$ shares exactly two qubits $q_1, q_2 \in \partial_X(x) \cap \partial_Z(z)$, the two incident triangles $\Delta(x, q_1, z)$ (purple shading) and $\Delta(x, q_2, z)$ (pink shading) glue along the dashed diagonal $(x, z)$ to form a four-sided diamond $Q_{x,z}$ (each cell belongs to at most $\Delta_{\mathrm{simp}} := \max(d_X d_{QZ}, d_{QX} d_{QZ}, d_Z d_{QX}) \le 32$ triangles; Proposition~\ref{prop:step1-lll}).}
\label{fig:tanner-base-simplices}
\end{figure}

Following Section~2.2 and Section~3.1 of~\cite{BDL24}, to control how replacement gadgets sit in $\R^3$ after spatial dilation and perturbation, we associate to $(\cC, \Phi)$ its two-dimensional incidence base complex $\Sigma_{\mathrm{base}}(\cC)$ (the Tanner square/simplicial complex of~\cite[Section~2.2]{BDL24}, following~\cite{li2024transform}), whose vertices are $M_X \sqcup Q \sqcup M_Z$, whose active edges join incident pairs $(x, q)$, $(q, z)$, and co-incident check pairs $(x, z)$ ($\partial_X(x) \cap \partial_Z(z) \neq \emptyset$), and whose active $2$-simplices (triangles) index incident triples $(x, q, z)$ (Figure~\ref{fig:tanner-base-simplices}). By check orthogonality $\abs{\partial_X(x) \cap \partial_Z(z)} \in \{0, 2\}$, every co-incident pair $(x, z)$ shares exactly two qubits $q_1, q_2 \in \partial_X(x) \cap \partial_Z(z)$, so the two triangles $\Delta(x, q_1, z)$ and $\Delta(x, q_2, z)$ glue along $(x, z)$ to form an incidence diamond $Q_{x,z}$ encoding check commutativity~\cite{li2024transform}. Neither our bounded-density recursion nor our local cone-acyclicity proof requires vertex-disjoint simplices of $\Sigma_{\mathrm{base}}(\cC)$ to remain separated in $\R^3$ or adjacent simplices to maintain a lower-bounded dihedral angle; it suffices to track active edge-length bounds, single-triangle altitudes, and unit-ball $2$-simplex multiplicity.

\leanlink{def:well-separated-base}
\begin{definition}[{Incidence Base Complex $\Sigma_{\mathrm{base}}(\cC)$ and Regular Simplex Geometry~\cite[Section~2.2 and Section~3.1]{BDL24}, \cite{li2024transform}}]
\label{def:well-separated-base}
Given a 3D CSS complex $(\cC, \Phi)$, let $\Sigma_{\mathrm{base}}(\cC)$ be its two-dimensional incidence simplicial complex, where each active $2$-simplex $\tau = (x, q, z)$ ($q \in \partial_X(x) \cap \partial_Z(z)$) is mapped to the convex hull $\Delta(\Phi(x), \Phi(q), \Phi(z)) \subset \R^3$ of its vertex positions. For parameters $0 < L_{\min} \le L_{\max}$, $h_{\min} > 0$, and $\rho_1 \ge 1$, the embedding $\Phi$ has \emph{$(L_{\min}, L_{\max}, h_{\min}, \rho_1)$-regular simplex geometry} if:
\begin{enumerate}[label=\textbf{(\alph*)}, leftmargin=2.2em, itemsep=2.5pt]
    \item \textbf{Check-qubit edge-length bounds}: Every check-qubit active edge $e \in \Sigma_{\mathrm{base}}(\cC)$ joining $(x, q)$ or $(q, z)$ has Euclidean length in $[L_{\min}, L_{\max}]$ (so every co-incident check diagonal $(x, z)$ has length at most $2 L_{\max}$).
    \item \textbf{Single-simplex altitude regularity}: Every active $2$-simplex $\tau = (x, q, z) \in \Sigma_{\mathrm{base}}(\cC)$ is non-degenerate with all three perpendicular altitudes at least $h_{\min}$.
    \item \textbf{Bounded unit-ball $2$-simplex multiplicity}: Every Euclidean unit ball $B(p, 1) \subset \R^3$ intersects the convex hulls of at most $\rho_1$ active $2$-simplices of $\Sigma_{\mathrm{base}}(\cC)$.
\end{enumerate}
\end{definition}

\begin{figure}[t]
\centering
\includegraphics[width=0.84\textwidth]{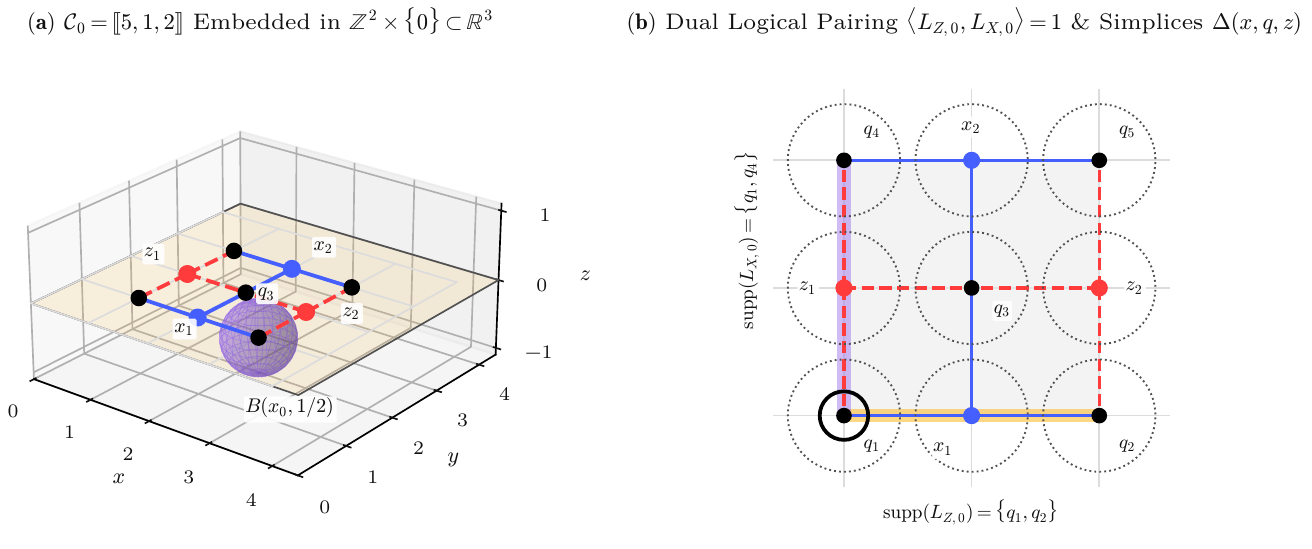}
\caption{The planar $[\![5, 1, 2]\!]$ surface code seed $(\cC_0, \Phi_0)$ embedded in $\Z^2 \times \{0\} \subset \R^3$, with five physical qubits $q_1, \dots, q_5$ (black circles), two weight-3 $X$-checks $x_1, x_2$ (blue circles, solid blue incidence lines), and two weight-3 $Z$-checks $z_1, z_2$ (red circles, dashed red incidence lines). \textbf{(a)} Three-dimensional perspective in the $z = 0$ plane (amber shading), showing an example clearance ball $B(x_0, 1/2)$ (violet sphere) around $x_0 = \Phi_0(q_2)$. \textbf{(b)} Top-down view showing the four gray incidence diamonds formed by the eight right-isosceles triangles $\Delta(x,q,z)$, the interior-disjoint dotted clearance circles of radius $1/2$ around each cell center, and the minimum-weight logical representatives $L_{Z,0}$ supported on $\{q_1, q_2\}$ (thick amber horizontal bar) and $L_{X,0}$ on $\{q_1, q_4\}$ (thick violet vertical bar) with unit mod-$2$ pairing $\ip{L_{Z,0}}{L_{X,0}} = 1$ at the ringed corner qubit $q_1$.}
\label{fig:seed-code-embedding}
\end{figure}

As in~\cite[Section~3.3]{BDL24}, the renormalization hierarchy starts from a constant-size planar surface code~\cite{bravyi1998quantum, freedman2001projective} $(\cC_0, \Phi_0)$ embedded in $\R^2 \times \{0\} \subset \R^3$ (Figure~\ref{fig:seed-code-embedding}). Specifically, we use the same minimal $[\![5, 1, 2]\!]$ planar surface code seed as~\cite[Section~3.3]{BDL24}, which has 5 qubits, two weight-3 $X$-checks, two weight-3 $Z$-checks, encodes a single logical qubit ($\dim_{\F} H^1(\cC_0) = \dim_{\F} H_1(\cC_0) = 1$), and whose eight incident right-isosceles triangles $\Delta(x, q, z)$ form four incidence diamonds $Q_{x,z}$ tiling a $2 \times 2$ planar square without self-intersection; Lemma~\ref{lem:seed-code} records its explicit bounded-geometry and simplex-regularity constants to initialize the multiscale induction.

\leanlink{lem:seed-code}
\begin{lemma}[{Base Case: The $[\![5, 1, 2]\!]$ Planar Surface Code Seed $(\cC_0, \Phi_0)$~\cite[Section~3.3]{BDL24}}]
\label{lem:seed-code}
Let $\cC_0$ be the $[\![5, 1, 2]\!]$ planar surface code with 5 qubits $Q_0 = \{q_{\mathrm{SW}}, q_{\mathrm{SE}}, q_{\mathrm{NW}}, q_{\mathrm{NE}}, q_{\mathrm{C}}\}$ at $(-1,-1,0)$, $(1,-1,0)$, $(-1,1,0)$, $(1,1,0)$, $(0,0,0)$, two weight-3 $X$-checks $M_{X,0} = \{x_{\mathrm{L}}, x_{\mathrm{R}}\}$ at $(-1,0,0), (1,0,0)$ with supports $\{q_{\mathrm{NW}}, q_{\mathrm{SW}}, q_{\mathrm{C}}\}$ and $\{q_{\mathrm{NE}}, q_{\mathrm{SE}}, q_{\mathrm{C}}\}$, and two weight-3 $Z$-checks $M_{Z,0} = \{z_{\mathrm{T}}, z_{\mathrm{B}}\}$ at $(0,1,0), (0,-1,0)$ with supports $\{q_{\mathrm{NW}}, q_{\mathrm{NE}}, q_{\mathrm{C}}\}$ and $\{q_{\mathrm{SW}}, q_{\mathrm{SE}}, q_{\mathrm{C}}\}$ (shifted and rotated by $\pi/2$ onto $\{0,1,2\}^2 \times \{0\}$ in our formalization and $[1,3]^2 \times \{0\}$ in Figure~\ref{fig:seed-code-embedding}). Then $(\cC_0, \Phi_0)$ is self-dual ($\cC_0^\vee \cong \cC_0$ under a $\pi/2$ planar rotation), satisfies $\dim_{\F} H^1(\cC_0) = \dim_{\F} H_1(\cC_0) = 1$ (spanned by dual representatives $(L_{X,0}, L_{Z,0}) \in \ker\delta_{\cC_0}^1 \times \ker\partial_1^{\cC_0}$ with $\ip{L_{Z,0}}{L_{X,0}} = 1$) and distances $d_{\mathrm{log}}^X(\cC_0) = d_{\mathrm{log}}^Z(\cC_0) = 2$, has $(3, \sqrt{2}, 9)$-bounded geometry, and its eight incident triangles form a $(1, 1, 1/\sqrt{2}, 8)$-regular simplicial base complex (with diagonal length $\sqrt{2}$).
\end{lemma}

\ifstandardproofs
\begin{proof}
Each of $x_{\mathrm{L}}, x_{\mathrm{R}}, z_{\mathrm{T}}, z_{\mathrm{B}}$ has weight $3$; each outer corner qubit has $d_{QX} = d_{QZ} = 1$ and the central qubit $q_{\mathrm{C}}$ has $d_{QX} = d_{QZ} = 2$. Every pair $(x, z) \in M_{X,0} \times M_{Z,0}$ shares exactly two qubits (the central qubit $q_{\mathrm{C}}$ and one corner qubit), so $\abs{\partial_X(x) \cap \partial_Z(z)} = 2$, $\delta_{\cC_0}^1 \circ \delta_{\cC_0}^0 = 0$, and each check incidence link (such as $q_{\mathrm{NW}} - z_{\mathrm{T}} - q_{\mathrm{C}} - z_{\mathrm{B}} - q_{\mathrm{SW}}$ for $x_{\mathrm{L}}$) is a connected $4$-edge path. Since $\rank(\delta_{\cC_0}^0) = \rank(\delta_{\cC_0}^1) = 2$, we have $\dim_{\F} H^1(\cC_0) = 5 - 2 - 2 = 1$, with minimum-weight logical operators $X_{q_{\mathrm{NW}}}X_{q_{\mathrm{NE}}}$ (commuting with $z_{\mathrm{T}}, z_{\mathrm{B}}$) and $Z_{q_{\mathrm{NW}}}Z_{q_{\mathrm{SW}}}$ (commuting with $x_{\mathrm{L}}, x_{\mathrm{R}}$) of weight $2$. The eight right-isosceles incident triangles $\Delta(x, q, z)$ have side lengths $1, 1, \sqrt{2}$ and altitudes $1, 1, 1/\sqrt{2}$, pairing into four incidence diamonds $Q_{x,z}$ meeting at $q_{\mathrm{C}}$.
\end{proof}
\fi

\subsection{Step 1: Coordinate De-Crowding}
\label{subsec:step1-lll}

Each renormalization step consists of two geometric operations~\cite[Section~3]{BDL24}: a spatial dilation and coordinate de-crowding step on the coarse base complex (Step~1, following~\cite[Section~3.1 and Section~3.4]{BDL24} and building on subdivided manifold embeddings~\cite{gromov2012generalizations, freedman2021building, portnoy2023local}), followed by concentric shell subdivision and three-layer gadget replacement (Step~2). Step~1 is needed because when a fine gadget replaces a simplex of $\Sigma_{\mathrm{base}}(\cC)$, the newly created cells lie in a tubular neighborhood of that simplex; without rescaling and perturbing the coarse vertex coordinates at each scale, simplices whose affine spans pass near unit balls in $\R^3$ would crowd together across scales $m \to \infty$, eventually violating the unit-ball cell density $\rho_0$.

\begin{figure}[t]
\centering
\includegraphics[width=0.90\textwidth]{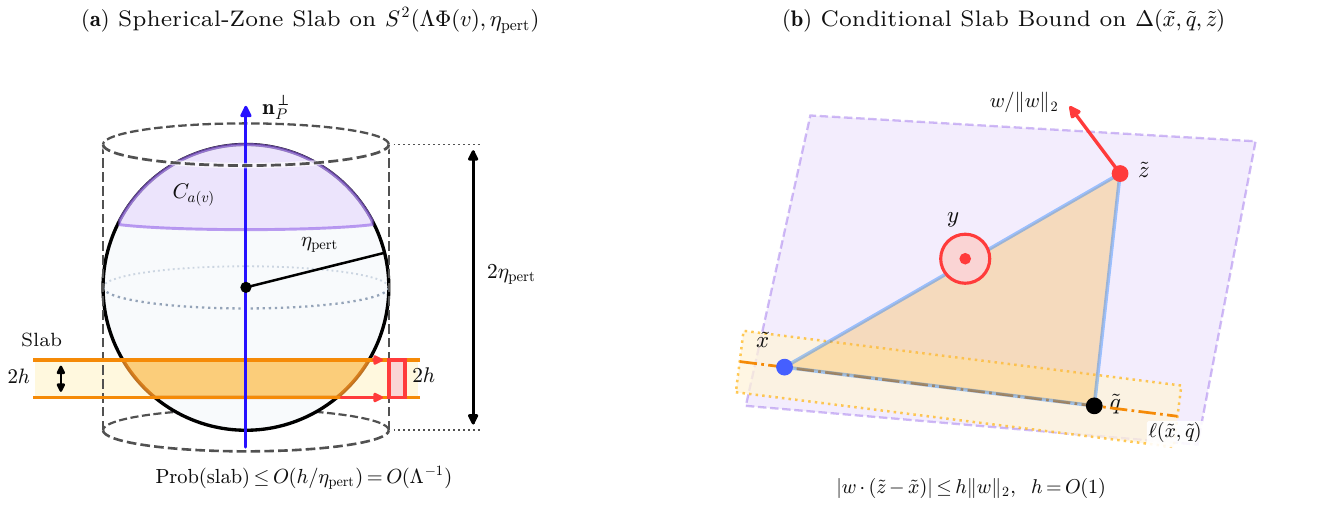}
\caption{Geometric ingredients of the Step~1 coordinate de-crowding theorem (Proposition~\ref{prop:step1-lll}). \textbf{(a)} On the perturbation sphere $S^2(\Lambda\Phi(v), \eta_{\mathrm{pert}})$ of radius $\eta_{\mathrm{pert}} = \alpha\Lambda$ (with normal axis $\mathbf{n}_P^\perp$, blue vertical arrow, and assigned spherical cap $C_{a(v)}$, violet shading), a horizontal planar slab of thickness $2h$ (light-yellow band with orange borders slicing the sphere in an amber zone) projects radially (red arrows) onto an equal-height band of height $2h$ (light-red shading) on the circumscribed cylinder of height $2\eta_{\mathrm{pert}}$, bounding the full-sphere area fraction by $2h/(2\eta_{\mathrm{pert}})$ and the cap-conditioned slab probability on $C_{a(v)}$ by $C_{\mathrm{cap}}h/\eta_{\mathrm{pert}} = O(\Lambda^{-1})$. \textbf{(b)} For a perturbed triangle $\Delta(\tilde{x}, \tilde{q}, \tilde{z})$ (amber shading with blue $X$-check $\tilde{x}$, black qubit $\tilde{q}$, and red $Z$-check $\tilde{z}$) to intersect a target ball $B(y, 1)$ (red circle around $y$), if $\dist(y, \ell(\tilde{x}, \tilde{q})) = \Theta(\Lambda)$ (outside a cylindrical tube around the edge line $\ell(\tilde{x}, \tilde{q})$), then $\tilde{z}$ must land in a thin violet dashed planar slab of thickness $2h = O(1)$ perpendicular to $w/\norm{w}_2$ for $w := (\tilde{q} - \tilde{x}) \times (y - \tilde{x})$; eliminating $y$ across simplices with distinct active random vertices bounds each amplified bad-event probability against the $O(1)$ dependency degree $\Delta_{\mathrm{dep}}$.}
\label{fig:archimedes-lll}
\end{figure}

Following Section~3.1 and Section~3.4 of~\cite{BDL24} and~\cite{gromov2012generalizations, portnoy2023local}, Step~1 first dilates the coarse embedding $\Phi \mapsto \Lambda\Phi$ by an integer scale factor $\Lambda \ge \Lambda_0 := 5.4 \times 10^{24}$ ($\Lambda = \Lambda_0$ in our formalization) and shifts each coarse vertex $v \in M_X \sqcup Q \sqcup M_Z$ independently to a perturbed position $\tilde{v} := \Phi'(v)$ on an assigned cap $C_{a(v)}$ of the sphere $S^2(\Lambda\Phi(v), \eta_{\mathrm{pert}})$ of radius $\eta_{\mathrm{pert}} = \alpha\Lambda$ ($\alpha = 1000$) and cap arc radius $\alpha_1\Lambda$ ($\alpha_1 = 139$), where $a = a(v) \in \{1, 2, 3\}$ indexes the sector $(M_X, Q, M_Z)$ along the standard basis axes $\Lambda\Phi(v) + \alpha\Lambda e_a$ (separated on $S^2(0, \alpha\Lambda)$ by arc distance $> 2\alpha_2\Lambda = 139\Lambda$ for $\alpha_2 := \frac{139}{2}$). Step~2 then subdivides each dilated simplex into $m_{\mathrm{sub}} = 2 D_{\mathrm{edge}}\Lambda = 4002\Lambda$ ($D_{\mathrm{edge}} := 2\alpha + 1 = 2001$) concentric shells, so every subdivided segment (including $(x, z)$ diagonals of length $\le 2 D_{\mathrm{edge}}\Lambda$) has length $\le \frac{4002\Lambda}{4002\Lambda} = 1$ and every fine cell has Euclidean interaction range $\le r_0 = 8$. Writing $\aff(\cdot)$ for affine hull and $\ell(u, v) := \aff(u, v)$ for the line through $u \neq v$, whenever a perturbed triangle $\Delta(\tilde{x}, \tilde{q}, \tilde{z})$ intersects a unit ball $B(y, 1) \subset \R^3$ with $y$ at distance $\Theta(\Lambda)$ from the edge line $\ell(\tilde{x}, \tilde{q})$ (outside a cylindrical tube around $\ell(\tilde{x}, \tilde{q})$), tetrahedron altitude geometry forces $\tilde{z}$ into a thin planar slab of thickness $2h = O(1)$ around $P := \aff(\tilde{x}, \tilde{q}, y)$ (Claim~3.6 of~\cite{BDL24}). Because orthogonal projection from $S^2(\Lambda\Phi(v), \eta_{\mathrm{pert}})$ onto the unit normal axis $\mathbf{n}_P^\perp$ pushes the normalized surface measure $\sigma_{S^2}$ forward to the uniform Lebesgue measure $\frac{1}{2\eta_{\mathrm{pert}}}\,dz$ on $[-\eta_{\mathrm{pert}}, \eta_{\mathrm{pert}}]$ (Figure~\ref{fig:archimedes-lll}(a)), any planar slab of half-width $h = O(1)$ has full-sphere measure at most $2h/(2\eta_{\mathrm{pert}})$ and cap-conditioned probability at most $C_{\mathrm{cap}}h/\eta_{\mathrm{pert}} = O(\Lambda^{-1})$ (where $C_{\mathrm{cap}} := 2/(1 - \cos(\alpha_1/\alpha))$ is the ratio of the full sphere area to the area of $C_{a(v)}$). Next, conditioning on two cell sectors at a time (cf.~\cite[Section~3.1]{BDL24} and~\cite[Theorem~7]{portnoy2023local}), bounded single-vertex simplex degree $\Delta_{\mathrm{simp}} \le 32$ greedily extracts large tuples of triangles with distinct active random vertices intersecting $B(y, 1)$ (Figure~\ref{fig:archimedes-lll}(b)) and eliminates $y$ via $4$-simplex determinant combinations so the resulting tube and coplanar-slab intersection events multiply conditionally across the active random vertices, enabling a clean application of the Lov\'asz Local Lemma (LLL)~\cite{erdos1975problems, moser2009constructive, moser2010constructive, chandrasekaran2013deterministic}.

In geometric terms, shifting each rescaled coarse vertex within its assigned spherical cap keeps every incident triangle uniformly thick while preventing more than a fixed constant number of coarse triangles from passing near any unit ball in $\R^3$. Proposition~\ref{prop:step1-lll} establishes the same spherical-cap perturbation invariants as Lemma~3.3 and Claims~3.4--3.6 of~\cite{BDL24} (building on~\cite[Theorem~7]{portnoy2023local}), but replaces the discrete $\alpha^3\Lambda^3$-ball net covering of~\cite[Section~3.1]{BDL24} with an algebraic Cramer-determinant elimination of the continuous ball center $y \in \R^3$ and explicit concentric-shell grid packing, matching the concrete $\Lambda$-independent bounds of~\cite[Section~3.4]{BDL24} (unit-ball triangle multiplicity $\rho_1' = t' = 5595$ and cell density $\rho_0 \le 6.6 \times 10^{10}$ for $\Lambda \ge \Lambda_0 = 5.4 \times 10^{24}$; realized in our formalization at fixed $\Lambda = \Lambda_0$ via $1$D rational latitude-circle arcs of $8.2 \times 10^{21} + 1$ points per cap and algebraic root separation, which also bounds coarse vertices per unit ball by $5595$) directly suited to formal verification.

\leanlink{prop:step1-lll}
\begin{proposition}[{Step 1: Spherical-Cap Coordinate De-Crowding~\cite[Sections~3.1 and~3.4, Lemma~3.3]{BDL24}}]
\label{prop:step1-lll}
Suppose $(\cC, \Phi)$ is a 3D CSS complex with $(8, r_0, \rho_0)$-bounded geometry ($d_X, d_Z \le 8$, $1 \le d_{QX}, d_{QZ} \le 4$, $r_0 \le 8$, $\rho_0 \le 6.6 \times 10^{10}$, so every cell is incident to at most $\Delta_{\mathrm{simp}} = \max(d_X d_{QZ}, d_{QX} d_{QZ}, d_Z d_{QX}) \le 32$ triangles of $\Sigma_{\mathrm{base}}(\cC)$). For $\Lambda_0 := 5.4 \times 10^{24}$ and any integer spatial dilation factor $\Lambda \ge \Lambda_0$, after dilating $\Phi \mapsto \Lambda\Phi$, there exists a spherical-cap perturbation $\Phi'$ of the coarse vertices on orthogonal-axis caps of $S^2(\Lambda\Phi(v), \alpha\Lambda)$ ($\alpha = 1000$, cap arc radius $\alpha_1\Lambda = 139\Lambda$) with $(\frac{137}{2}\Lambda, 2001\Lambda, \frac{7}{5}\Lambda, 5595)$-regular simplex geometry: every perturbed check-qubit active edge has length in $[\frac{137}{2}\Lambda, 2001\Lambda]$, every perturbed active triangle $\Delta$ has minimum altitude $\alt(\Delta) \ge \gamma\Lambda$ ($\gamma := \frac{7}{5}$), and every unit ball $B(p, 1) \subset \R^3$ intersects at most $\rho_1' = t' = 5595$ triangles of $\Phi'(\Sigma_{\mathrm{base}}(\cC))$. Subdividing each dilated quadrilateral $Q_{x,z}$ into $m_{\mathrm{sub}} = 2(2\alpha + 1)\Lambda = 4002\Lambda$ concentric shells in Step~2 then restores fine interaction range $r_0 = 8$ and uniform unit-ball cell density $\rho_0 \le 6.6 \times 10^{10}$ per sector, independent of the total number of cells in $\cC$ and of the recursion depth $m$.
\end{proposition}

\begin{proof}
Following Claims~3.4--3.6 of~\cite{BDL24} (and~\cite{gromov2012generalizations, portnoy2023local}), write $\tilde{v} := \Phi'(v)$ on the orthogonal-axis spherical caps centered at $\Lambda\Phi(v) + 1000\Lambda e_a$ ($a \in \{1, 2, 3\}$) of arc radius $139\Lambda$. Because coarse incident cells satisfy $\norm{\Phi(u) - \Phi(v)}_2 \le r_0 \le 8$ (so co-incident check pairs satisfy $\norm{\Phi(x) - \Phi(z)}_2 \le 2 r_0 \le 16$), each active difference vector lies within $(139 + 139 + 16)\Lambda = 294\Lambda$ of $1000\Lambda(e_a - e_b)$ ($a \neq b$, norm $1000\sqrt{2}\Lambda \approx 1414.2\Lambda$), deterministically keeping all perturbed active edge lengths inside $[1120\Lambda, 1709\Lambda] \subset [\frac{137}{2}\Lambda, 2001\Lambda]$ and all three altitudes of every perturbed active $2$-simplex $\Delta(\tilde{x}, \tilde{q}, \tilde{z})$ above $480\Lambda \ge \frac{7}{5}\Lambda$.

To bound the number of perturbed coarse $2$-simplices intersecting any unit ball $B(y, 1) \subset \R^3$, observe that only coarse simplices within Euclidean distance $O(\Lambda)$ of $y$ can reach $B(y, 1)$. Whenever a perturbed $2$-simplex $s_i = (A_i, B_i, C_i)$ intersects $B(y, 1)$, its normal vector $N_i = (B_i - A_i) \times (C_i - A_i)$ and plane offset $b_i = \langle N_i, A_i\rangle$ satisfy $\abs{\langle N_i, y\rangle - b_i} \le \norm{N_i}_2$, i.e., $\dist(y, \aff(A_i, B_i, C_i)) \le 1$. By the tetrahedron altitude identity $\dist(y, \ell(\tilde{x}, \tilde{q}))\,\dist(\tilde{z}, \aff(\tilde{x}, \tilde{q}, y)) = \dist(\tilde{z}, \ell(\tilde{x}, \tilde{q}))\,\dist(y, \aff(\tilde{x}, \tilde{q}, \tilde{z})) \le (2\alpha + 1)\Lambda$, if $\dist(y, \ell(\tilde{x}, \tilde{q})) = \Theta(\Lambda)$ then $\tilde{z}$ must fall within perpendicular distance $h = O(1)$ of $\aff(\tilde{x}, \tilde{q}, y)$ (Figure~\ref{fig:archimedes-lll}(b)). Because orthogonal projection onto a plane normal pushes the normalized spherical surface measure forward to the uniform Lebesgue measure on $[-\eta_{\mathrm{pert}}, \eta_{\mathrm{pert}}]$ (Figure~\ref{fig:archimedes-lll}(a); cf.~Claim~3.6 of~\cite{BDL24}), the cap-conditioned planar-slab probability of any fixed plane is bounded by
\begin{equation}
    \label{eq:archimedes-slab-bound}
    \Pr\bigl[\dist(\tilde{z}, P) \le h\bigr] \;\le\; \frac{C_{\mathrm{cap}}\,h}{\eta_{\mathrm{pert}}} \;=\; O(\Lambda^{-1}).
\end{equation}
To eliminate both the unknown ball center $y$ and shared-vertex dependencies while amplifying $O(\Lambda^{-1})$ past the polynomial entropy of local candidate tuples, suppose $> t' = 5595$ active simplices intersect $B(y, 1)$. Across three degeneracy stages (non-degenerate $4$-simplex slab, nearly coplanar normals, and edge-tube incidence, accounting for $(9 + 3 \times 54)\times 32 = 5472 \le 5595$ triangles), conditioning on two of $(M_X, Q, M_Z)$ leaves a single active random vertex ($\tilde{x}$, $\tilde{q}$, or $\tilde{z}$) per simplex; since each coarse cell belongs to at most $\Delta_{\mathrm{simp}} \le 32$ simplices, greedy extraction yields $\ge 55$ simplices with distinct active random vertices intersecting $B(y, 1)$. Selecting a dominant triple $(s_1^*, s_2^*, s_3^*)$ maximizing $\abs{\det(N_1, N_2, N_3)}$ (or a non-collinear pair), Cramer's rule eliminates $y$ across $K = 55$ simplices ($3$ anchors plus $52$ fourth simplices) via $\bigl(\sum_{i=1}^4 D_i b_i\bigr)^2 \le 4\sum_{i=1}^4 D_i^2 \norm{N_i}_2^2$ for the signed $3 \times 3$ cofactors $D_i := (-1)^i\det(N_j)_{j \neq i}$, simultaneously triggering independent $O(\Lambda^{-1})$ cap-slab or tube events of joint probability $O(\Lambda^{-52})$. Since each coarse cell has bounded simplex degree $\Delta_{\mathrm{simp}} \le 32$ and local density $\rho_0$, each amplified bad event depends on at most $\Delta_{\mathrm{dep}} = O(1)$ local neighborhoods, so for $\Lambda \ge \Lambda_0 = 5.4 \times 10^{24}$ the Lov\'asz Local Lemma~\cite{erdos1975problems, moser2009constructive, moser2010constructive, chandrasekaran2013deterministic} guarantees a realization $\Phi'$ in which every unit ball $B(y, 1)$ intersects at most $t' = 5595$ coarse $2$-simplices (the explicit bad-event indexing and numerical LLL verification at $\Lambda_0 = 5.4 \times 10^{24}$ are carried out in \leanfile[SphericalCapAndLLL.lean]{Proofs/Step1LLLDensity/SphericalCapAndLLL.lean}).

Finally, when each perturbed $2$-simplex $s = (x, q, z)$ is subdivided into $m_{\mathrm{sub}} = 2(2\alpha + 1)\Lambda = 4002\Lambda$ concentric shells in Step~2, the dilation factor $\Lambda$ cancels identically in the unit-ball grid-packing bounds: along any single radial shell $u$, two azimuthal grid points $(u, v_1), (u, v_2)$ inside $B(y, 1)$ are separated along the $(q, z)$-edge (of length $\ge (\alpha_2 - 1)\Lambda = \frac{137}{2}\Lambda$) by $\frac{\abs{v_1 - v_2}}{4002\Lambda}(\alpha_2 - 1)\Lambda = \frac{137}{8004}\abs{v_1 - v_2} \le 2$, giving $\abs{v_1 - v_2} \le 116$ ($\le 118$ azimuthal points per shell); across distinct shells $u_1, u_2$, the simplex altitude bound $\ge \gamma\Lambda = \frac{7}{5}\Lambda$ forces perpendicular separation $\frac{\abs{u_1 - u_2}}{4002\Lambda}\cdot\frac{7}{5}\Lambda = \frac{7}{20010}\abs{u_1 - u_2} \le 2$, giving $\abs{u_1 - u_2} \le 5717$ ($\le 5719$ radial shells per simplex). Each coarse $2$-simplex intersecting $B(y, 1)$ thus contributes at most $5719 \times 118 = 674{,}842$ grid points to $B(y, 1)$ (and each grid point hosts a bounded number of horizontal and vertical bridge cells across the three layers), restoring fine interaction range $\le r_0 = 8$ and a $\Lambda$-independent unit-ball cell density $\rho_0 \le 6.6 \times 10^{10}$ across all scales $m$.
\end{proof}

\subsection{Step 2: Cylinder Welding}
\label{subsec:step2-mapping-cone}

We now define the primal three-layer replacement functor $\widetilde{\cC} := \cR_X^{(3)}(\cC)$ (and its CSS dual $\cR_Z^{(3)}(\cC) := (\cR_X^{(3)}(\cC^\vee))^\vee$). Our construction directly adapts the homological quantum code embedding framework of Section~2.3 and Section~3.2 of~\cite{BDL24} (see also~\cite{williamson2023layer, yuan2026unified}), which formulates cellular replacement as a height-2 three-sector cochain diagram $\cS_X^\bullet \oplus \cS_Q^\bullet \oplus \cS_Z^\bullet$ (graded by degree $\bullet \in \{0, 1, 2\}$) along the directed acyclic graph (DAG) $X \to Q \to Z$, while modifying the local block gadgets of~\cite[Section~3.2.2]{BDL24} from two horizontal layers ($n_{\mathrm{lay}} = 2$) to three horizontal layers ($n_{\mathrm{lay}} = 3$). Specifically, whereas~\cite[Section~3.2.2]{BDL24} replaces each coarse $X$-check $x \in M_X$ by a two-layer cylinder $Y_x$, each coarse qubit $q \in Q$ by a degree-shifted two-layer star $Y_q$, and each coarse $Z$-check $z \in M_Z$ by a two-point check $Y_z$, our functor $\cR_X^{(3)}$ replaces each $x \in M_X$ by a subdivided cone complex replicated across three horizontal layers $b \in \{0, 1, 2\}$ and coupled vertically along its distinguished outer rim to reference layer $0$ (forming the hollow cylinder gadget $Y_x^{(3)}$), replaces each $q \in Q$ by a degree-shifted three-layer star interface $Y_q^{(3)}$, and replicates each $z \in M_Z$ into a three-layer column $Y_z^{(3)} = \{z\} \times \{0, 1, 2\}$.

Following~\cite[Section~2.3]{BDL24}, formulating $\cR_X^{(3)}(\cC)$ algebraically as an iterated cellular mapping cone $\Cone(\Psi)$ over the DAG $X \to Q \to Z$ decomposes the coboundary differential $\delta_{\widetilde{\cC}}$ into a block-diagonal internal differential $d_{\mathrm{blk}}$ on the disjoint blocks $\bigsqcup_x Y_x^{(3)} \sqcup \bigsqcup_q Y_q^{(3)} \sqcup \bigsqcup_z Y_z^{(3)}$ plus a strictly nilpotent off-diagonal welding operator $\Psi = \Psi_{XQ} + \Psi_{QZ}$ pointing exclusively along $X \to Q \to Z$ ($\Psi^3 = 0$). Refining~\cite[Sections~2.3.1--2.3.2 and~4.3]{BDL24} in Sections~\ref{sec:linear-tqo2-proof}--\ref{sec:two-stage-cleaner} below, this nilpotency $\Psi^3 = 0$ causes both the homological perturbation series for $\RCA(L^*, C)$ and the two-stage syndrome cleaner to terminate in at most two steps with zero feedback loops.

\begin{figure}[t]
\centering
\includegraphics[width=0.88\textwidth]{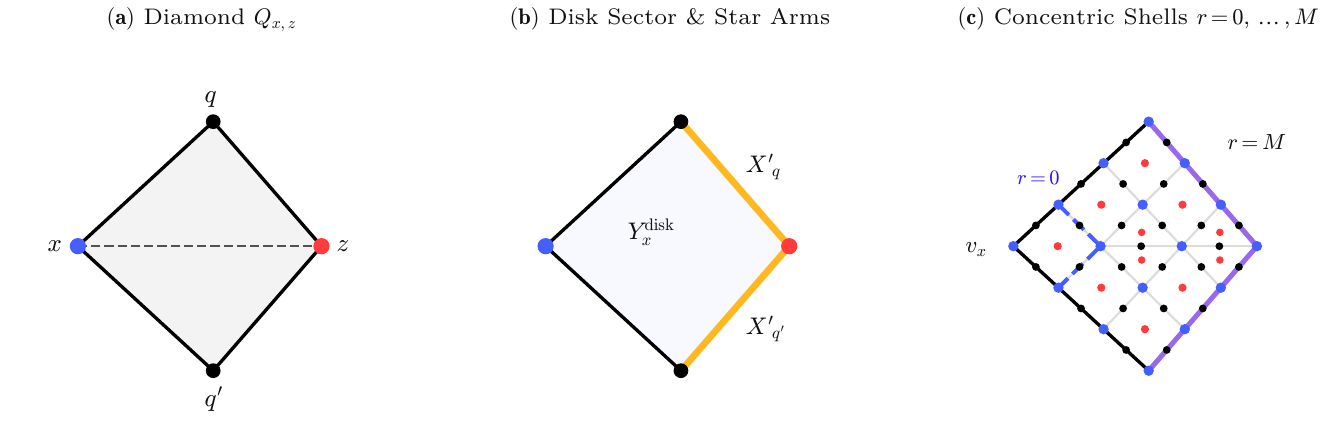}
\caption{Construction of the 2D cellular cone complex $Y_x^{\mathrm{disk}}$ for a coarse $X$-check $x \in M_X$. \textbf{(a)} Coarse diamond $Q_{x,z}$ formed by two gray coarse triangles $\Delta(x, q, z), \Delta(x, q', z)$ sharing the dashed diagonal $(x, z)$ (with blue $x$, black $q, q'$, and red $z$). \textbf{(b)} Diamond-sector decomposition into the cone disk $Y_x^{\mathrm{disk}}$ (blue shading) and thick amber star arms $X'_q, X'_{q'}$ (Definition~\ref{def:three-layer-mapping-cone}(2)). \textbf{(c)} Radial subdivision of $Y_x^{\mathrm{disk}}$ into shells $r = 0, \dots, M$ around $v_x$ (omitting the $r = 0$ spoke in $s_{x,z}^{(0)}$ while retaining the $r \ge 1$ midline $(x, z)$ spokes), with blue $0$-cells ($X$-checks), black $1$-cells (qubits), red $2$-cells ($Z$-checks), dashed blue $r = 0$ shell, and solid violet outer rim at $r = M$.}
\label{fig:barycentric-subdivision}
\end{figure}

Let us first specify the 2D cone complex $Y_x^{\mathrm{disk}}$ associated to each coarse $X$-check $x \in M_X$, following Section~3.2.2 of~\cite{BDL24}. Let $G_x$ be the bipartite incidence link graph between the qubits $q \in \partial_X(x)$ and the $Z$-checks $z \in M_Z$ co-incident with $x$ ($\partial_X(x) \cap \partial_Z(z) \neq \emptyset$). By Definition~\ref{def:3d-css-complex}(b), $G_x$ is a connected bipartite graph (with check-vertex degree $2$ and qubit-vertex degree at most $d_{QZ} \le 4$) whose edges $(q, z)$ correspond to the incident triangles $(x, q, z) \in \Sigma_{\mathrm{base}}(\cC)$. Coning $G_x$ over an apex vertex $v_x$ at $\Phi'(x)$ produces a 2D cone of triangles $(v_x, q, z)$, which we subdivide radially into $m_{\mathrm{sub}} = M + 1 = 4002\Lambda$ concentric shells $r \in \{0, \dots, M\}$ under a triangular-staircase shell rule (Figure~\ref{fig:barycentric-subdivision}). Specifically, coarse-qubit link vertices $v_{x,q}^{(r)}$ and coarse-check link vertices $v_{x,z}^{(r)}$ appear in every shell $r$; pairs of innermost triangles sharing an auxiliary diagonal $(x, z)$ at $r = 0$ are merged into $4$-edge corner squares $s_{x,z}^{(0)}$ by omitting the $r = 0$ radial $(x, z)$ spokes (implemented in the formalization by zeroing the duplicate non-canonical $r = 0$ row, whose trivial stabilizer $I$ contributes $\frac{1}{2}(I - I) = 0$ to $H_{0,k}$); and the $j$-th interior subdivided link vertex $v_{x,q,z,j+1}^{(r)}$ ($0 \le j < M$) enters at shells $r > j$ with diagonal radial parent $v_{x,z}^{(r-1)}$ at shell $r - 1 = j$. Fixing a canonical qubit $q_{\mathrm{can}}(x,z) \in \partial_X(x) \cap \partial_Z(z)$ for each co-incident $z$ and adjoining its $r = 0$ corner azimuthal segment $e_{\mathrm{can}}(x,z) = (v_{x,q_{\mathrm{can}}(x,z)}^{(0)}, v_{x,z}^{(0)})$ to the radial-parent edges forms a canonical spanning tree $T_x^{\mathrm{rad}} \subset Y_x^{\mathrm{disk}}$ rooted at $v_x$, converting each sector into a grid of cells of Euclidean diameter $\le r_0 = 8$ with $0$-cells $C^0(Y_x^{\mathrm{disk}})$, $1$-cells $C^1(Y_x^{\mathrm{disk}})$, and $2$-cells $C^2(Y_x^{\mathrm{disk}})$. Along the distinguished outer rim $\partial Y_x^{\mathrm{disk}}$ at shell $r = M$ (the subdivided link graph of $G_x$), each coarse qubit $q \in \partial_X(x)$ is represented by a corner $0$-cell $v_{x,q} := v_{x,q}^{(M)} \in C^0(\partial Y_x^{\mathrm{disk}})$, and each co-incident check $z \in M_Z$ ($\abs{\partial_X(x) \cap \partial_Z(z)} = 2$) is represented by a shared terminal corner $0$-cell $v_{x,z} := v_{x,z}^{(M)} \in C^0(\partial Y_x^{\mathrm{disk}})$ where the two incident arms $(q, z)$ and $(q', z)$ for $\{q, q'\} = \partial_X(x) \cap \partial_Z(z)$ meet. Each incident pair $(q, z)$ is represented by $M + 1$ subdivided outer-rim $1$-cells $e_{x,q,z,k} \in C^1(\partial Y_x^{\mathrm{disk}})$ ($0 \le k \le M$) joining outer-rim $0$-cells $v_{x,q,z,0} := v_{x,q}$ and $v_{x,q,z,k} := v_{x,q,z,k}^{(M)}$ ($1 \le k \le M$) to the shared terminal corner $v_{x,z} =: v_{x,q,z,M+1}$.

\begin{figure}[t]
\centering
\includegraphics[width=0.88\textwidth]{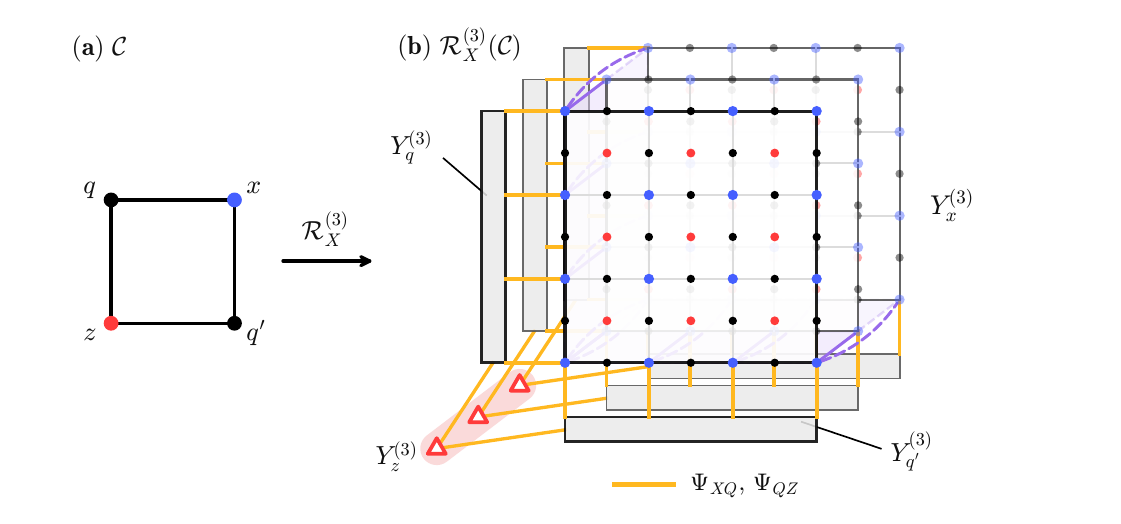}
\caption{Three-layer replacement blocks of $\cR_X^{(3)}(\cC)$ (adapted from~\cite[Section~3.2]{BDL24}). \textbf{(a)} Coarse diamond cell $Q_{x,z}$ in $\cC$ ($X$-check $x$ in blue, qubits $q, q'$ in black, and $Z$-check $z$ in red). \textbf{(b)} Three-dimensional oblique projection of $\cR_X^{(3)}(\cC)$ showing the three stacked horizontal sheets $b \in \{0, 1, 2\}$ of the cylinder $Y_x^{(3)}$ joined at the outer rim by violet vertical bridge collars $[0,1], [0,2]$ to layer $0$, the gray three-layer star blocks $Y_q^{(3)}, Y_{q'}^{(3)}$, the pink capsule $Y_z^{(3)}$ containing three replicated $Z$-check $2$-cells $(z,0), (z,1), (z,2)$ (red $\triangle$ markers), and the amber welding maps $\Psi_{XQ}, \Psi_{QZ}$ across adjacent blocks.}
\label{fig:three-layer-complexes}
\end{figure}

To assemble the three-layer replacement complex from these planar disks, we stack three parallel horizontal copies $b \in \{0, 1, 2\}$ of each subdivided disk and star tree, attach vertical bridge cells along their outer rims connecting layers $1$ and $2$ directly to reference layer $0$, and weld the cylinder rims through the star arms into three replicated copies of each coarse $Z$-check (Figure~\ref{fig:three-layer-complexes}).

\leanlink{def:three-layer-mapping-cone}
\begin{definition}[{Three-Layer Cylinder $Y_x^{(3)}$, Star Interface $Y_q^{(3)}$, and Mapping Cone $\cR_X^{(3)}(\cC)$, adapted from~\cite[Section~2.3 and Section~3.2]{BDL24}}]
\label{def:three-layer-mapping-cone}
Given a 3D CSS complex $(\cC, \Phi)$, the \emph{primal three-layer replacement complex} $\widetilde{\cC} = \cR_X^{(3)}(\cC)$ is the cellular cochain complex constructed from three families of local block complexes indexed by $x \in M_X$, $q \in Q$, and $z \in M_Z$ (writing $C^k(Y)$ interchangeably for a complex $Y$'s $k$-cell basis and its $\F$-cochain space $\F^{C^k(Y)}$):
\begin{enumerate}[label=\textbf{(\arabic*)}, leftmargin=2.2em, itemsep=3pt]
    \item \textbf{Three-layer cylinder blocks $Y_x^{(3)}$ ($x \in M_X$)}: Let $I^{(3)}$ be the $1$-dimensional interval complex with three $0$-cells $\{[0], [1], [2]\}$ (horizontal layers $b \in \{0, 1, 2\}$) and two $1$-cells $\{[0, 1], [0, 2]\}$ (vertical bridges $w \in \{1, 2\}$ joining layer $0$ to layer $w$), with coboundary $\delta_I^0([0]) = [0,1] + [0,2]$ and $\delta_I^0([w]) = [0,w]$ ($w \in \{1,2\}$). Set $Y_x^{(3)}$ to be the three-layer cylinder obtained by triplicating $Y_x^{\mathrm{disk}}$ across $b \in \{0, 1, 2\}$ and attaching vertical bridges along the outer boundary rim $\partial Y_x^{\mathrm{disk}}$:
    \begin{align}
        C^0(Y_x^{(3)}) &= C^0(Y_x^{\mathrm{disk}}) \times \{0, 1, 2\}, \nonumber\\
        C^1(Y_x^{(3)}) &= \bigl(C^1(Y_x^{\mathrm{disk}}) \times \{0, 1, 2\}\bigr) \;\sqcup\; \bigl(C^0(\partial Y_x^{\mathrm{disk}}) \times \{[0,1], [0,2]\}\bigr), \label{eq:Yx3-cells}\\
        C^2(Y_x^{(3)}) &= \bigl(C^2(Y_x^{\mathrm{disk}}) \times \{0, 1, 2\}\bigr) \;\sqcup\; \bigl(C^1(\partial Y_x^{\mathrm{disk}}) \times \{[0,1], [0,2]\}\bigr), \nonumber
    \end{align}
    with internal coboundary operators (for the planar coboundaries $\delta_{\mathrm{disk}}^0, \delta_{\mathrm{disk}}^1$ on $Y_x^{\mathrm{disk}}$ and $\delta_{\partial}^0$ on $\partial Y_x^{\mathrm{disk}}$)
    \begin{align}
        d_x^0(u, b) &= (\delta_{\mathrm{disk}}^0 u, b) + \mathbf{1}_{\{u \in \partial Y_x^{\mathrm{disk}}\}}(u, \delta_I^0 b), \nonumber\\
        d_x^1\bigl((e, b) + (u, [0,w])\bigr) &= (\delta_{\mathrm{disk}}^1 e, b) + (\delta_{\partial}^0 u, [0,w]) + \mathbf{1}_{\{e \in \partial Y_x^{\mathrm{disk}}\}}\sum_{w'=1}^2 \mathbf{1}_{\{b \in \{0,w'\}\}}(e, [0,w']). \label{eq:dx-coboundary}
    \end{align}
    \item \textbf{Three-layer star interface blocks $Y_q^{(3)}$ ($q \in Q$)}: Following~\cite[Section~3.2.2]{BDL24}, let $X'_q$ be the subdivided $1$-dimensional star tree with central hub vertex $v_q$ and radial arms indexed by the incident coarse $Z$-checks $z \in \partial_Z^\top(q)$, where each arm $(q, z)$ is subdivided into $M + 1$ edges $e_{q,z,k} \in C^1(X'_q)$ ($0 \le k \le M$) and vertices $v_{q,z,k} \in C^0(X'_q)$ ($1 \le k \le M+1$, with $v_{q,z,0} := v_q$) ending at the terminal boundary tip $v_{q,z} := v_{q,z,M+1} \in C^0(\partial X'_q)$. Triplicating $X'_q$ across $b \in \{0, 1, 2\}$, attaching vertical bridges at the terminal boundary tips $\partial X'_q$, and shifting cohomological degree by $[-1]$ (so that $0$-cells of $X'_q$ are physical qubits in degree $1$ and $1$-cells of $X'_q$ are $Z$-checks in degree $2$) gives
    \begin{align}
        C^0(Y_q^{(3)}) &= 0, \nonumber\\
        C^1(Y_q^{(3)}) &= C^0(X'_q) \times \{0, 1, 2\}, \label{eq:Yq3-cells}\\
        C^2(Y_q^{(3)}) &= \bigl(C^1(X'_q) \times \{0, 1, 2\}\bigr) \;\sqcup\; \bigl(C^0(\partial X'_q) \times \{[0,1], [0,2]\}\bigr), \nonumber
    \end{align}
    with internal coboundary operators (for the edge coboundary $\delta_{\mathrm{star}}^0$ on $X'_q$)
    \begin{equation}
        \label{eq:dq-coboundary}
        d_q^0 \;=\; 0, \qquad
        d_q^1(u, b) \;=\; (\delta_{\mathrm{star}}^0 u, b) + \mathbf{1}_{\{u \in \partial X'_q\}}\sum_{w=1}^2 \mathbf{1}_{\{b \in \{0,w\}\}}(u, [0,w]).
    \end{equation}
    \item \textbf{Three-layer coarse $Z$-check blocks $Y_z^{(3)}$ ($z \in M_Z$)}: Each coarse $Z$-check $z \in M_Z$ is replaced by three $2$-cells $C^2(Y_z^{(3)}) := \{(z, 0), (z, 1), (z, 2)\}$ concentrated in degree $2$ ($C^0(Y_z^{(3)}) = C^1(Y_z^{(3)}) = 0$, $d_z = 0$).
    \item \textbf{DAG coupling operators $\Psi_{XQ}$ and $\Psi_{QZ}$}: For each degree $n \in \{0, 1, 2\}$, the total cochain space $\widetilde{\cC} = \cR_X^{(3)}(\cC)$ is the three-sector direct sum $C^n(\widetilde{\cC}) =: \cS_X^n \oplus \cS_Q^n \oplus \cS_Z^n$, where
    \begin{equation}
        \label{eq:cone-total-space}
        \cS_X^n := \bigoplus_{x \in M_X} C^n(Y_x^{(3)}), \;\; \cS_Q^n := \bigoplus_{q \in Q} C^n(Y_q^{(3)}), \;\; \cS_Z^n := \bigoplus_{z \in M_Z} C^n(Y_z^{(3)}), \quad
        \delta_{\widetilde{\cC}} := d_{\mathrm{blk}} + \Psi_{XQ} + \Psi_{QZ},
    \end{equation}
    where $d_{\mathrm{blk}} = \bigoplus_x d_x \oplus \bigoplus_q d_q \oplus 0$, the cylinder-to-star welding map $\Psi_{XQ} : \bigoplus_x C^\bullet(Y_x^{(3)}) \to \bigoplus_q C^{\bullet+1}(Y_q^{(3)})$ (denoted $g_{XQ}$ in~\cite[Section~2.3]{BDL24}) identifies the outer-rim boundary cells of $Y_x^{(3)}$ with the corresponding star cells of $Y_q^{(3)}$ by
    \begin{align}
        \Psi_{XQ}^0(v_{x,q,z,k}, b) &= (v_{q,z,k}, b) \quad (0 \le k \le M), \qquad
        \Psi_{XQ}^0(v_{x,z}, b) = \sum_{q \in \partial_X(x) \cap \partial_Z(z)} (v_{q,z}, b), \label{eq:Psi-XQ-def}\\
        \Psi_{XQ}^1(e_{x,q,z,k}, b) &= (e_{q,z,k}, b) \quad (0 \le k \le M), \qquad
        \Psi_{XQ}^1(v_{x,z}, [0,w]) = \sum_{q \in \partial_X(x) \cap \partial_Z(z)} (v_{q,z}, [0,w]), \nonumber
    \end{align}
    (and $\Psi_{XQ}^1(u, [0,w]) = 0$ for non-terminal rim $0$-cells $u \in C^0(\partial Y_x^{\mathrm{disk}})$), and the star-to-$Z$ welding map $\Psi_{QZ} : \bigoplus_q C^1(Y_q^{(3)}) \to \bigoplus_z C^2(Y_z^{(3)})$ (denoted $g_{QZ}$ in~\cite[Section~2.3]{BDL24}) acts on the terminal star tip qubits $(v_{q,z}, b)$ ($z \in \partial_Z^\top(q)$) by
    \begin{equation}
        \label{eq:Psi-QZ-def}
        \Psi_{QZ}^1(v_{q,z}, b) = (z, b) \quad (b \in \{0, 1, 2\}),
    \end{equation}
    and vanishes on all non-terminal vertices of $X'_q$. Each fine cell (across all layers $b \in \{0,1,2\}$ and bridges $[0,w]$; drawn separated vertically in Figures~\ref{fig:three-layer-complexes}--\ref{fig:full-welded-replacement-3d}) is assigned by the fine embedding $\widetilde{\Phi}$ to its subdivided grid coordinate on $\Phi'(\Sigma_{\mathrm{base}}(\cC)) \subset \R^3$.
\end{enumerate}
\end{definition}

\begin{figure}[t]
\centering
\includegraphics[width=0.90\textwidth]{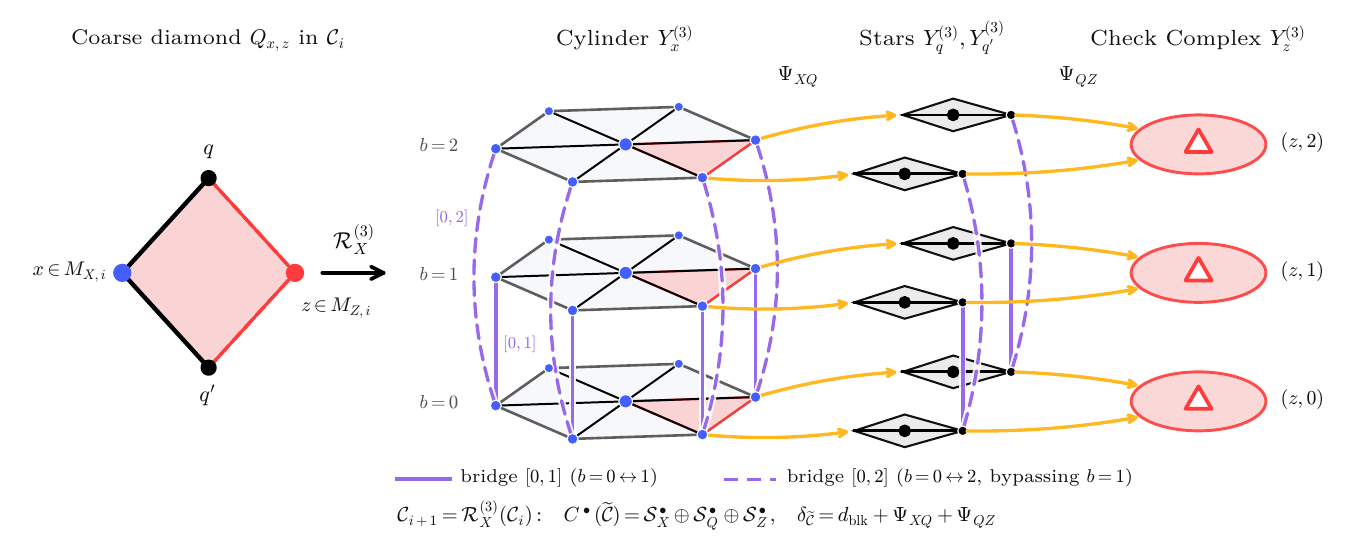}
\caption{Three-dimensional view of the mapping-cone transition $\cC_i \to \cC_{i+1} = \cR_X^{(3)}(\cC_i)$ (even $i$) along the directed acyclic graph $X \to Q \to Z$. From left to right, a coarse diamond $Q_{x,z}$ ($x \in M_{X,i}$, $q, q' \in Q_i$, $z \in M_{Z,i}$) is replaced by the three-layer cylinder $Y_x^{(3)}$ (three stacked hexagonal disk layers with blue $0$-cells, black spokes, one pink-shaded sector per layer, and violet outer-rim vertical bridges $[0,1], [0,2]$), the three-layer star interfaces $Y_q^{(3)}, Y_{q'}^{(3)}$ (three layers of gray diamond blocks with black hub/spoke $1$-cells and violet tip bridges), and the replicated check complex $Y_z^{(3)}$ (three red $2$-cells $(z,0), (z,1), (z,2)$ with open red triangles $\triangle$ marking active $Z$-syndrome defects), glued along $X \to Q \to Z$ by the curved amber welding arrows $\Psi_{XQ}$ and $\Psi_{QZ}$ ($\delta_{\widetilde{\cC}} = d_{\mathrm{blk}} + \Psi_{XQ} + \Psi_{QZ}$).}
\label{fig:full-welded-replacement-3d}
\end{figure}

Because the welding maps point strictly from cylinder rims to star arms and from star tips to replicated checks, two consequences follow directly, adapting~\cite[Section~2.3 and Section~3.2.7, Lemma~3.14]{BDL24} to $\cR_X^{(3)}(\cC)$ (Figure~\ref{fig:full-welded-replacement-3d}). First, as in~\cite[Section~2.3]{BDL24}, any two-step composition across the welding interfaces cancels mod~$2$ wherever a coarse $X$-check and $Z$-check share an even number of qubits, ensuring that all fine checks commute. Second, whereas the two-layer replacement of~\cite[Section~3.2.7, Lemma~3.14]{BDL24} attaches one vertical bridge (check degree $\le 7$, qubit degree $\le 4$), attaching two vertical bridges $[0,1], [0,2]$ to layer $0$ adds at most one bridge incidence on the layer-$0$ cylinder rim; under our triangular-staircase shell subdivision, check weights reset to $\le 8$ and qubit degrees to $\le 4$ at every step rather than growing with recursion depth.

\leanlink{prop:mapping-cone-invariants}
\begin{proposition}[{Welding Differential Nilpotency and Bounded Geometry, adapted from~\cite[Section~2.3 and Section~3.2]{BDL24}}]
\label{prop:mapping-cone-invariants}
For any 3D CSS complex $(\cC, \Phi)$ satisfying Definition~\ref{def:3d-css-complex} with $d_X, d_Z \le 8$ and $1 \le d_{QX}, d_{QZ} \le 4$:
\begin{enumerate}[label=\textbf{(\roman*)}, leftmargin=2.2em, itemsep=2.5pt]
    \item The off-diagonal operator $\Psi = \Psi_{XQ} + \Psi_{QZ}$ satisfies $\Psi^3 = 0$, $d_{\mathrm{blk}}\Psi_{XQ} = \Psi_{XQ} d_{\mathrm{blk}}$, and $\Psi_{QZ} d_{\mathrm{blk}} + \Psi_{QZ}\Psi_{XQ} = 0$, so $\delta_{\widetilde{\cC}}^1 \circ \delta_{\widetilde{\cC}}^0 = 0$.
    \item In $\widetilde{\cC} = \cR_X^{(3)}(\cC)$, every $X$-check has weight $\le \max(d_X(\cC), d_{QZ}(\cC) + 4, 7) \le 8$, every $Z$-check has weight $\le 8$, every qubit has $d_{QX}(\widetilde{\cC}) \le 4$ and $d_{QZ}(\widetilde{\cC}) \le 4$, every check pair shares $0$ or $2$ qubits, and every check link $G_{\tilde{x}}$ or $G_{\tilde{z}}$ is connected with check-vertex degree $2$ and qubit-vertex degree at most $4$. Consequently, setting $(\cC_{2j+1}, \Phi_{2j+1}) := \cR_X^{(3)}(\cC_{2j}, \Phi'_{2j})$ and $(\cC_{2j+2}, \Phi_{2j+2}) := \cR_Z^{(3)}(\cC_{2j+1}, \Phi'_{2j+1})$ via Step~1 perturbations $\Phi'_i$ (Proposition~\ref{prop:step1-lll}) maintains $(8, r_0, \rho_0)$-bounded geometry on $(\cC_m, \Phi_m)$ for all $m \ge 0$ with $r_0 = 8$ and $\rho_0 \le 6.6 \times 10^{10}$.
\end{enumerate}
\end{proposition}

\begin{proof}
For (i), grading $C^\bullet(\widetilde{\cC}) = \cS_X^\bullet \oplus \cS_Q^\bullet \oplus \cS_Z^\bullet$ by the three sectors $(X, Q, Z)$ makes $\Psi$ strictly upper-triangular along $X \to Q \to Z$, so $\Psi^3 = 0$. On every outer-rim $0$-cell $u \in C^0(\partial Y_x^{\mathrm{disk}})$ and layer $b \in \{0, 1, 2\}$, the cellular boundary identification~\eqref{eq:Psi-XQ-def} intertwines the horizontal edge coboundary $\delta_{\partial}^0$ with $\delta_{\mathrm{star}}^0$ and the terminal vertical bridge coboundary $\delta_I^0$ at the shared corner $v_{x,z}$, giving $\bigoplus_q d_q^1(\Psi_{XQ}^0(u, b)) = \Psi_{XQ}^1(d_x^0(u, b))$ over $\F$. Likewise, $d_q^0 = 0$, while on every non-terminal $0$-cell of $Y_x^{(3)}$ the image of $\Psi_{XQ}^0$ has no terminal star tips so $\Psi_{QZ}^1 \circ \Psi_{XQ}^0$ vanishes trivially, and on each shared terminal corner $0$-cell $(v_{x,z}, b) \in C^0(Y_x^{(3)})$,~\eqref{eq:Psi-XQ-def}--\eqref{eq:Psi-QZ-def} give $\Psi_{QZ}^1(\Psi_{XQ}^0(v_{x,z}, b)) = \sum_{q \in \partial_X(x) \cap \partial_Z(z)} (z, b) = 0 \pmod 2$ since $\abs{\partial_X(x) \cap \partial_Z(z)} \in \{0, 2\}$. Thus $\Psi_{QZ}^1 \circ \Psi_{XQ}^0 = 0$ on every basis $0$-cell and $\delta_{\widetilde{\cC}}^1 \circ \delta_{\widetilde{\cC}}^0 = 0$.

For (ii), all $X$-checks of $\widetilde{\cC}$ lie in $\bigoplus_x C^0(Y_x^{(3)})$. Under the triangular-staircase shell rule, the central apex $0$-cell $(v_x, b)$ is incident to one radial spoke per coarse qubit $q \in \partial_X(x)$ and $0$ vertical bridges (weight $\abs{\partial_X(x)} \le d_X(\cC) \le 8$). An interior shell $0$-cell ($0 \le r < M$) has at most $d_{QZ}(\cC) \le 4$ azimuthal segments and $2$ radial edges over a coarse qubit $q \in \partial_X(x)$ (weight $\le 6$), $2$ azimuthal segments and $2$ radial edges over an interior subdivided arm vertex $v_{x,q,z,j+1}$ (weight $\le 4$), or $2$ azimuthal segments, at most $2$ radial edges, and $2$ diagonal staircase children over a coarse $Z$-check $z_0$ (weight $\le 6$). An outer-rim $0$-cell at $r = M$ has at most $d_{QZ}(\cC) \le 4$ azimuthal segments, $1$ inward radial edge, at most $2$ vertical bridges (at $b = 0$), and $1$ welded star hub qubit $(v_q, b)$ over $q \in \partial_X(x)$ (weight $\le d_{QZ}(\cC) + 4 \le 8$), at most $6$ incident qubits over $v_{x,q,z,j+1}$ ($2$ azimuthal segments, $1$ inward radial edge, at most $2$ vertical bridges, and $1$ welded star arm qubit), and at most $7$ incident qubits at each shared terminal corner $(v_{x,z}, b)$ ($2$ azimuthal segments, $1$ inward radial edge, at most $2$ vertical bridges, and the $2$ welded terminal star tip qubits $(v_{q,z}, b), (v_{q',z}, b)$ for $\{q, q'\} = \partial_X(x) \cap \partial_Z(z)$), giving $d_X(\widetilde{\cC}) \le 8$. For $Z$-checks, horizontal disk plaquettes in $Y_x^{(3)}$ have weight $3$ or $4$, outer vertical plaquettes $(e_{x,q,z,k}, [0,w])$ have weight $4$, horizontal star arm checks $(e_{q,z,k}, b)$ have $2$ star qubits plus $\abs{\partial_X^\top(q)} \le d_{QX}(\cC) \le 4$ welded cylinder rim qubits (weight $2 + \abs{\partial_X^\top(q)} \in [3, 6] \le 8$), terminal star bridge checks $(v_{q,z}, [0,w])$ have $2$ star tip qubits plus $\abs{\partial_X^\top(q)} \le d_{QX}(\cC) \le 4$ welded cylinder bridge qubits (weight $2 + \abs{\partial_X^\top(q)} \in [3, 6] \le 8$), and replicated coarse checks $(z, b)$ have $\abs{\partial_Z(z)} \le 8$ incident star tip qubits $(v_{q,z}, b)$, giving $d_Z(\widetilde{\cC}) \le 8$. Every qubit in $Y_x^{(3)}$ lies in at most $2$ $X$-checks and $4$ $Z$-checks (attained at a layer-$0$ outer-rim edge $(e_{x,q,z,k}, 0)$, which meets $1$ horizontal plaquette, $2$ vertical bridge plaquettes $[0,1],[0,2]$, and $1$ welded star-arm check), while a star qubit in $Y_q^{(3)}$ lies in at most $\abs{\partial_X^\top(q)} \le d_{QX}(\cC) \le 4$ welded cylinder $X$-checks and at most $4$ $Z$-checks (at most $2$ along an interior arm vertex, $\abs{\partial_Z^\top(q)} \le 4$ at the hub $(v_q, b)$, $3$ at a layer-$w$ terminal tip $(v_{q,z}, w)$ for $w \in \{1, 2\}$, and $4$ at a layer-$0$ terminal tip $(v_{q,z}, 0)$), giving $d_{QX}(\widetilde{\cC}) \le 4$ and $d_{QZ}(\widetilde{\cC}) \le 4$. Finally, connectivity of $G_x$ and $G_z$ transfers to every fine check link $G_{\tilde{x}}$ and $G_{\tilde{z}}$. Dualizing gives $d_X(\cR_Z^{(3)}(\widetilde{\cC})) \le 8$, $d_Z(\cR_Z^{(3)}(\widetilde{\cC})) \le 8$, and $d_{QX}, d_{QZ} \le 4$, so all degree bounds reset without accumulating across $m$.
\end{proof}

\section{Relative Cone Acyclicity}
\label{sec:linear-tqo2-proof}

\subsection{Local Topological Quantum Order}
\label{subsec:rca-trinity}

Both the finite-temperature free-energy barrier and the zero-temperature perturbative spectral gap rest on a single local filling property: any syndrome-free error chain supported inside a ball of radius $r$ up to the macroscopic code scale should be generated by parity checks supported inside a linearly larger ball of radius $\ceil{C r}$. Whereas~\cite[Section~3.2]{BDL24} established global cohomology preservation $H^1(\cC_m) \cong H^1(\cC_0)$ for the welded code hierarchy while leaving local topological quantum order ($\mathrm{TQO\text{-}2}$) to future work~\cite[Section~1.1]{BDL24}, we formulate and prove this local exactness condition across scales as \emph{Relative Cone Acyclicity} $\RCA(R_0, C)$.

\leanlink{def:rca}
\begin{definition}[Relative Cone Acyclicity $\RCA(R_0, C)$]
\label{def:rca}
\label{def:rca-tqo}
Let $(\cC, \Phi)$ be a 3D CSS complex. For any Euclidean ball $B \subset \R^3$, let $Q(B) := \{q \in Q : \Phi(q) \in B\}$ denote the set of physical qubits in $B$, and let
\begin{equation}
    \label{eq:local-check-sets}
    M_X(B) := \{x \in M_X : \Phi(\partial_X(x)) \subset B\}, \qquad
    M_Z(B) := \{z \in M_Z : \Phi(\partial_Z(z)) \subset B\}
\end{equation}
denote the sets of $X$-checks and $Z$-checks whose qubit supports lie entirely inside $B$. For a cutoff radius $R_0 \ge 1$ and a linear collar factor $C \ge 1$, we say that $(\cC, \Phi)$ satisfies \emph{Relative Cone Acyclicity} $\RCA(R_0, C)$ if for every center $x_0 \in \R^3$ and radius $1 \le r \le R_0$, putting $B := B(x_0, r)$ and $B^+ := B(x_0, \ceil{C r})$, every homology $1$-cycle (primal cycle) $c_X \in \F^{Q(B)} \cap \ker \partial_1$ ($Z$-Pauli sector, zero $X$-syndrome) bounds in $\F^{M_Z(B^+)}$, and every cohomology $1$-cocycle (dual cocycle) $c_Z \in \F^{Q(B)} \cap \ker \delta^1$ ($X$-Pauli sector, zero $Z$-syndrome) cobounds in $\F^{M_X(B^+)}$:
\begin{equation}
    \label{eq:rca-def}
    \F^{Q(B)} \cap \ker \partial_1 \;\subseteq\; \partial_2\bigl(\F^{M_Z(B^+)}\bigr), \qquad
    \F^{Q(B)} \cap \ker \delta^1 \;\subseteq\; \delta^0\bigl(\F^{M_X(B^+)}\bigr).
\end{equation}
\end{definition}

On the stabilizer Hilbert space, local chain filling translates into a pair of operator-algebraic local topological quantum order conditions relaxing those in the perturbative stability theory of Bravyi, Hastings, and Michalakis~\cite{BHM10, BH11}. The first condition ($\mathrm{TQO\text{-}1}$) is unchanged from~\cite{BHM10, BH11}: observables supported on balls up to the cutoff radius act as scalars on the global ground space, so no local measurement can distinguish orthogonal encoded states. For the second condition, however,~\cite{BHM10, BH11} require a \emph{constant additive} collar $B(x_0, r + O(1))$ ($\mathrm{TQO\text{-}2}$): any operator on $B(x_0, r)$ that annihilates the global ground space must already annihilate the local ground space on $B(x_0, r + O(1))$. Multiscale welded codes violate that stronger condition whenever a stabilizer in $B(x_0, r)$ straddles a welding interface at scale $\Theta(r)$, since generating it from checks requires sweeping across adjacent cylinders of diameter $\Theta(r)$. Definition~\ref{def:rca} instead yields a relaxed condition, $\mathrm{Linear\ TQO\text{-}2}(C)$, in which the additive buffer $r + O(1)$ is weakened to a \emph{linearly dilated} collar $B(x_0, \ceil{C r})$---a relaxation that still suffices for quasi-adiabatic stability under superpolynomially decaying perturbations in Sections~\ref{sec:qac-banach}--\ref{sec:telescoping-main} (cf.~\cite{MZ13, NSY22}).

\leanlink{def:tqo}
\begin{definition}[Topological Quantum Order $\mathrm{TQO\text{-}1}(R_0)$ and $\mathrm{Linear\ TQO\text{-}2}(C)$, relaxing~\cite{BHM10, BH11}]
\label{def:tqo}
Let $(\cC, \Phi)$ be a 3D CSS complex with commuting stabilizer Hamiltonian $H_0$, global ground-space projector $P_0$, and local ground-space projector $P_B$ on a ball $B \subset \R^3$ given by
\begin{align}
    H_0 &\;=\; \sum_{x \in M_X} \frac{I - A_x}{2} + \sum_{z \in M_Z} \frac{I - B_z}{2}, \nonumber\\
    P_0 &\;=\; \prod_{x \in M_X} \frac{I + A_x}{2}\prod_{z \in M_Z} \frac{I + B_z}{2}, \qquad
    P_B \;:=\; \prod_{x \in M_X(B)} \frac{I + A_x}{2}\prod_{z \in M_Z(B)} \frac{I + B_z}{2}.
    \label{eq:H0-P0-PB-def}
\end{align}
For a cutoff radius $R_0 \ge 1$ and a linear collar factor $C \ge 1$, define:
\begin{enumerate}[label=\textbf{(\alph*)}, leftmargin=2.2em, itemsep=2.5pt]
    \item \textbf{Topological Quantum Order $\mathrm{TQO\text{-}1}(R_0)$}: For every ball $B = B(x_0, r)$ of radius $1 \le r \le R_0$ and every operator $O_B$ supported on $Q(B)$, there exists a scalar $c(O_B) \in \C$ such that $P_0 O_B P_0 = c(O_B) P_0$.
    \item \textbf{Linear Topological Quantum Order $\mathrm{Linear\ TQO\text{-}2}(C)$}: For every ball $B = B(x_0, r)$ of radius $1 \le r \le R_0$ and every operator $O_B$ supported on $Q(B)$, if $O_B P_0 = 0$, then $O_B P_{B^+} = 0$ where $B^+ = B(x_0, \ceil{C r})$.
\end{enumerate}
\end{definition}

\begin{figure}[t]
\centering
\includegraphics[width=0.88\textwidth]{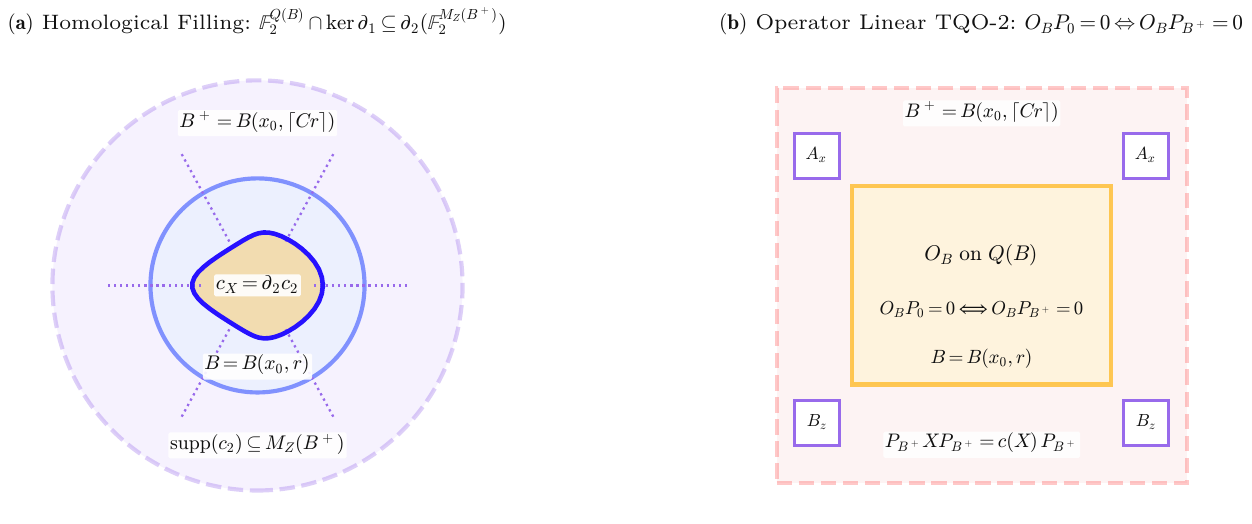}
\caption{Homological and operator-algebraic sides of Proposition~\ref{prop:trinity-equivalence}. \textbf{(a)} Homological cycle filling on $(B, B^+)$ (Definition~\ref{def:rca}), showing a blue cycle loop $c_X = \partial_2 c_2$ bounding an amber surface inside the inner blue ball $B = B(x_0, r)$, enclosed inside the outer dashed violet ball $B^+ = B(x_0, \ceil{C r})$ with $\supp(c_2) \subseteq M_Z(B^+)$. \textbf{(b)} Operator $\mathrm{Linear\ TQO\text{-}2}(C)$, showing an inner amber region supporting $O_B$ on $Q(B)$ inside an outer dashed light-red collar $B^+ = B(x_0, \ceil{C r})$ containing local checks $A_x, B_z$ (violet borders), where local indistinguishability gives $P_{B^+} X P_{B^+} = c(X)\,P_{B^+}$ (for any operator $X$ on $Q(B)$ with $\mathrm{TQO\text{-}1}$ scalar $c(X)$) and $O_B P_0 = 0 \Longleftrightarrow O_B P_{B^+} = 0$.}
\label{fig:fredholm-cone-duality}
\end{figure}

Relative Cone Acyclicity captures the conjunction of these two operator conditions through linear duality over $\F$ (Figure~\ref{fig:fredholm-cone-duality}). Taking orthogonal complements converts primal cycle filling on $(B, B^+)$ into \emph{local cocycle extension}: any error pattern whose syndrome vanishes inside the dilated ball $B^+$ agrees on $Q(B)$ with a global stabilizer coboundary. Combining the primal and dual sectors yields \emph{Local Check Generation} at $(R_0, C)$: for every $x_0 \in \R^3$ and $1 \le r \le R_0$, putting $B := B(x_0, r)$ and $B^+ := B(x_0, \ceil{C r})$, a Hermitian Pauli operator supported on $Q(B)$ commutes with all checks in $M_X(B^+) \sqcup M_Z(B^+)$ if and only if it is a signed product of checks in $M_X(B^+) \sqcup M_Z(B^+)$. Neither operator condition suffices by itself: local indistinguishability rules out small logical operators inside $B$, while linear local-to-global consistency forces globally trivial stabilizers on $B$ to be generated locally inside $B^+$. Definition~\ref{def:rca}, therefore, doubles as a local homological immunity condition for thermal decoding and an operator-algebraic local topological order condition for perturbation theory.

\leanlink{prop:trinity-equivalence}
\begin{proposition}[Homological--Operator Equivalence]
\label{prop:trinity-equivalence}
\label{prop:triple-equivalence}
For any 3D CSS complex $(\cC, \Phi)$ of locality range $r_0 > 0$ (Definition~\ref{def:3d-css-complex}(c)) with $P_0 \neq 0$ and any $R_0, C \ge 1$ with $\ceil{C r} \ge r + 2 r_0$ for all $1 \le r \le R_0$, the following three statements are equivalent:
\begin{equation}
    \label{eq:trinity-equiv}
    \RCA(R_0, C) \;\Longleftrightarrow\;
    \text{Local Check Generation at }(R_0, C) \;\Longleftrightarrow\;
    \mathrm{TQO\text{-}1}(R_0) \;\wedge\; \mathrm{Linear\ TQO\text{-}2}(C).
\end{equation}
In particular, under $\RCA(R_0, C)$, every nontrivial logical cocycle $c \in \ker \delta^1 \setminus \im \delta^0$ (or cycle in $\ker \partial_1 \setminus \im \partial_2$) has Euclidean support diameter greater than $R_0$.
\end{proposition}

\begin{proof}
Consider the linear restriction map $\pi_B : \F^Q \to \F^{Q(B)}$ and its transpose inclusion $\iota_B = \pi_B^\top : \F^{Q(B)} \hookrightarrow \F^Q$. Since $\partial_1 = (\delta^0)^\top$, taking $\F$-orthogonal complements gives $(\F^{Q(B)} \cap \ker \partial_1)^\perp = \F^{Q \setminus Q(B)} + (\ker \partial_1)^\perp = \ker \pi_B + \im \delta^0$, while $(\partial_2(\F^{M_Z(B^+)}))^\perp = \ker(\delta_{B^+}^1)$, where $\delta_{B^+}^1 : \F^Q \to \F^{M_Z(B^+)}$ is the restricted $Z$-syndrome map. Thus the first inclusion of~\eqref{eq:rca-def} is equivalent to local cocycle extension $\ker(\delta_{B^+}^1) \subseteq \ker \pi_B + \im \delta^0$: any cochain $u \in \F^Q$ satisfying $\delta_{B^+}^1(u) = 0$ agrees on $Q(B)$ with a global stabilizer coboundary $\tilde{u} \in \im \delta^0$ (ruling out any nontrivial logical operator on $Q(B)$; in particular, if a nonzero cocycle or cycle $c$ had $\diam(\Phi(\supp(c))) \le R_0$, choosing $q_0 \in \supp(c)$ would place $\supp(c) \subseteq Q(B(\Phi(q_0), R_0))$, forcing $c \in \im \delta^0$ or $\im \partial_2$). Moreover, since $\ceil{C r} \ge r + 2 r_0$, every $Z$-check intersecting $Q(B)$ lies inside $M_Z(B^+)$, so for $u \in \F^{Q(B)}$ having $\delta_{B^+}^1(u) = 0$ implies $\delta^1(u) = 0$ globally, and the second inclusion of~\eqref{eq:rca-def} gives $u \in \delta^0(\F^{M_X(B^+)})$. Dualizing $X \leftrightarrow Z$ shows that a Hermitian Pauli operator $P(u, v) := \prod_{q \in Q} i^{u_q v_q} X_q^{u_q} Z_q^{v_q}$ ($u, v \in \F^{Q(B)}$) commutes with all local checks in $B^+$ if and only if $P(u, v)$ is a signed product of checks in $M_X(B^+) \sqcup M_Z(B^+)$.

Expanding any operator $O_B = \sum_{u,v} \alpha_{u,v} P(u,v)$ in the Pauli basis on $Q(B)$, since $\ceil{C r} \ge r + 2 r_0$, every global check overlapping $Q(B)$ already lies in $M_X(B^+) \sqcup M_Z(B^+)$. Consequently, both $P_{B^+} P(u,v) P_{B^+}$ and $P_0 P(u,v) P_0$ vanish unless $P(u,v)$ commutes with all checks in $B^+$, in which case Local Check Generation implies $P(u,v)$ is a signed product of checks in $M_X(B^+) \sqcup M_Z(B^+)$ and hence $P(u,v) P_{B^+} = \pm P_{B^+}$ and $P(u,v) P_0 = \pm P_0$ with the same sign. By linearity, every operator $X$ supported on $Q(B)$ satisfies the common-scalar sandwich $P_{B^+} X P_{B^+} = c(X) P_{B^+}$ and $P_0 X P_0 = c(X) P_0$, proving $\mathrm{TQO\text{-}1}(R_0)$. Moreover, if $O_B P_0 = 0$, applying this common-scalar identity to $X := O_B^\dagger O_B$ (which is also supported on $Q(B)$) yields $c(O_B^\dagger O_B) P_0 = P_0 O_B^\dagger O_B P_0 = 0$; since $P_0 \neq 0$, $c(O_B^\dagger O_B) = 0$, so $(O_B P_{B^+})^\dagger (O_B P_{B^+}) = P_{B^+} O_B^\dagger O_B P_{B^+} = 0$ and thus $O_B P_{B^+} = 0$, proving $\mathrm{Linear\ TQO\text{-}2}(C)$. Conversely, if $\mathrm{TQO\text{-}1}(R_0) \wedge \mathrm{Linear\ TQO\text{-}2}(C)$ holds and $P(u,v)$ on $Q(B)$ commutes with all checks in $B^+$ (hence all global checks), then $P_0 P(u,v) P_0 = c P_0$ with $c \in \{\pm 1\}$, so $(I - c P(u,v))P_0 = 0$; applying $\mathrm{Linear\ TQO\text{-}2}(C)$ gives $(I - c P(u,v))P_{B^+} = 0$, so $P(u,v)$ is generated by checks in $M_X(B^+) \sqcup M_Z(B^+)$.
\end{proof}

\begin{remark}[Transfer from Euclidean Balls to $\ell_\infty$ Lattice Cubes]
\label{rem:cube-transfer}
Because $\norm{y}_\infty \le \norm{y}_2 \le \sqrt{3}\,\norm{y}_\infty$ on $\R^3$, Euclidean $\RCA(R_0, C_{\mathrm{euc}})$ immediately implies $\RCA$ (and hence $\mathrm{TQO\text{-}1} \wedge \mathrm{Linear\ TQO\text{-}2}$, with single-qubit $Z$ pins on unused site slots of the cubic lattice $\Gamma_k$ included; see Definition~\ref{def:safe-decoded-observables}(a)) on boundary-clamped $\ell_\infty$ cubes of $\Gamma_k$ of side $1 \le r \le L_k^* \le \floor{R_0/\sqrt{3}}$ into concentric collar cubes of side $\ceil{C r}$ (written $\RCA(L_k^*, C)$, $\mathrm{TQO\text{-}1}(L_k^*)$, and $\mathrm{Linear\ TQO\text{-}2}(C)$ on $\Gamma_k$, with $C = O(C_{\mathrm{euc}})$), which is the geometry used in Sections~\ref{sec:qac-banach}--\ref{sec:telescoping-main}. (In the formalization, continuum $\RCA$ uses check-center membership $\Phi(m) \in B^+$, shifting radii by at most $r_0 = 8$, before transferring to full check-support containment $\supp(Q_i) \subseteq B^+$ on $\ell_\infty$ balls and clamped cubes of $\Gamma_k$ in the forward direction needed downstream.)
\end{remark}

\subsection{Homological Perturbation Lemma}
\label{subsec:nilpotent-hpl}

To relate the global cohomology and local filling properties of a fine welded code to those of its coarse predecessor, we follow the homological perturbation strategy of Section~2.3.1, Section~3.2.3, and Appendix~A of~\cite{BDL24}: we first contract each uncoupled cylinder, star, and check gadget onto its coarse representative via a strong deformation retraction, and then turn on the inter-block welding couplings as a perturbation of the differential~\cite{brown1965twisted, gugenheim1972chain, lambe1987applications, crainic2004perturbation}. Whenever the coupling series terminates at finite order, this procedure produces explicit algebraic formulas for lifting coarse chains, projecting fine chains, and contracting residual boundaries.

\begin{lemma}[{Classical Homological Perturbation Lemma~\cite{brown1965twisted, gugenheim1972chain, lambe1987applications}, \cite[Sections~2.3--2.4 and~3.2]{crainic2004perturbation}; see also~\cite[Section~2.3.1 and Appendix~A]{BDL24}}]
\label{lem:general-hpl}
Let $(M^\bullet, d_M)$ and $(N^\bullet, d_N)$ be cochain complexes over a commutative ring (here $\F$), equipped with chain maps $\iota : N^\bullet \to M^\bullet$ and $\pi : M^\bullet \to N^\bullet$ of degree $0$ and a contracting homotopy $h : M^\bullet \to M^{\bullet - 1}$ of degree $-1$ satisfying the strong deformation retraction (SDR) identities and side conditions
\begin{equation}
    \label{eq:general-sdr-conditions}
    \pi\,\iota = \id_N, \qquad d_M h + h d_M = \id_M - \iota\,\pi, \qquad h^2 = 0, \qquad \pi h = 0, \qquad h\iota = 0.
\end{equation}
Suppose $\Psi : M^\bullet \to M^{\bullet + 1}$ is a perturbation of the differential such that $(d_M + \Psi)^2 = 0$ and $\id_M + \Psi h$ is invertible, with perturbation series $\Theta := (\id_M + \Psi h)^{-1}\Psi = \sum_{n=0}^\infty (-1)^n \Psi(h\Psi)^n$ (which converges whenever $h\Psi$ or $\Psi h$ is nilpotent). Then the perturbed operators
\begin{equation}
    \label{eq:general-hpl-formulas}
    \widetilde{\iota} := \iota - h\Theta\iota, \qquad
    \widetilde{\pi} := \pi - \pi\Theta h, \qquad
    \Sigma := h - h\Theta h, \qquad
    \widetilde{d}_N := d_N + \pi\Theta\iota
\end{equation}
define a strong deformation retraction of $(M^\bullet, d_M + \Psi)$ onto $(N^\bullet, \widetilde{d}_N)$ satisfying $\widetilde{d}_N^2 = 0$, $\widetilde{\pi}\,\widetilde{\iota} = \id_N$, $(d_M + \Psi)\widetilde{\iota} = \widetilde{\iota}\,\widetilde{d}_N$, $\widetilde{\pi}(d_M + \Psi) = \widetilde{d}_N\widetilde{\pi}$, $(d_M + \Psi)\Sigma + \Sigma(d_M + \Psi) = \id_M - \widetilde{\iota}\,\widetilde{\pi}$, $\Sigma^2 = 0$, $\widetilde{\pi}\Sigma = 0$, and $\Sigma\widetilde{\iota} = 0$ (where $\widetilde{d}_N^2 = 0$, both chain-map relations, and the homotopy identity hold from the first two homotopy-equivalence equations of~\eqref{eq:general-sdr-conditions} alone~\cite[Section~2.4]{crainic2004perturbation}).
\end{lemma}

\begin{figure}[t]
\centering
\includegraphics[width=0.90\textwidth]{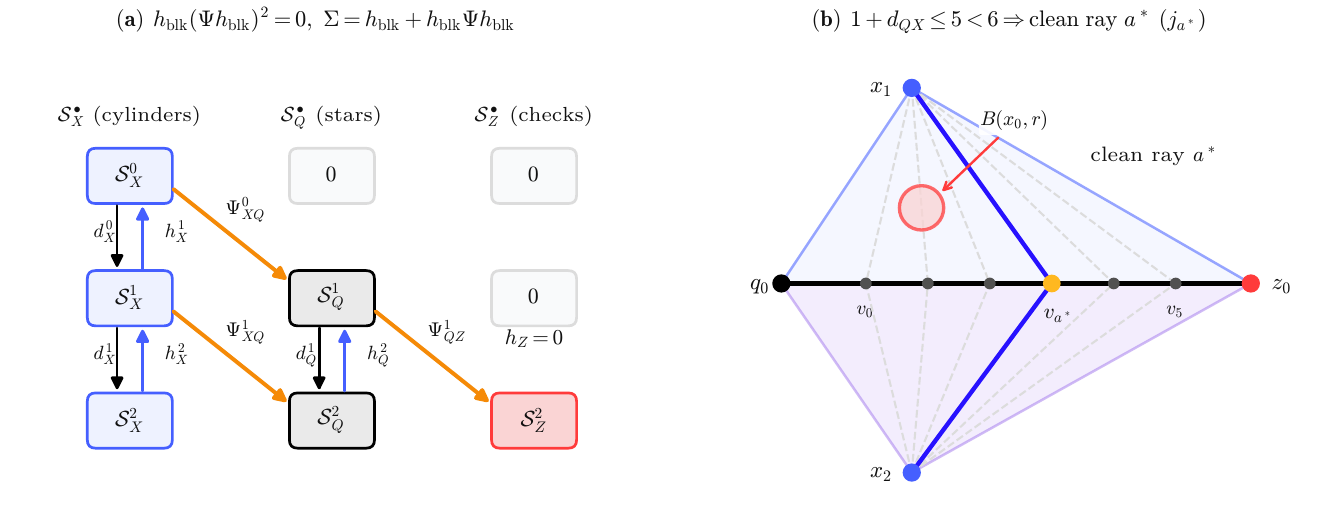}
\caption{Two-step nilpotent HPL and six-checkpoint clean-ray selection on $\widetilde{\cC} = \cR_X^{(3)}(\cC)$. \textbf{(a)} In the three-tier sector grading $q_{\mathrm{sec}} \in \{0, 1, 2\}$ ($\cS_X^\bullet$ in blue boxes, $\cS_Q^\bullet$ in gray boxes, $\cS_Z^\bullet$ in pink boxes), the block differentials $d_X^n := \bigoplus_x d_x^n,\, d_Q^1 := \bigoplus_q d_q^1$ (downward black arrows) and contracting homotopies $h_X^n := \bigoplus_x h_x^n,\, h_Q^2 := \bigoplus_q h_q^2$ (upward blue arrows, with $h_Z = 0$) preserve each column, while the welding maps $\Psi_{XQ}^n$ and $\Psi_{QZ}^1$ (diagonal orange arrows) advance strictly rightwards ($\Delta q_{\mathrm{sec}} := q_{\mathrm{sec}}(\text{target}) - q_{\mathrm{sec}}(\text{source}) = +1$), forcing $h_{\mathrm{blk}}(\Psi h_{\mathrm{blk}})^2 = 0$ and $\Sigma = h_{\mathrm{blk}} + h_{\mathrm{blk}}\Psi h_{\mathrm{blk}}$. \textbf{(b)} Previewing Subsection~\ref{subsec:scale-matched-rca} (Lemma~\ref{lem:clean-shell-subcylinder}), around a star hub $q_0$ (black) and $Z$-check $z_0$ (red) with six arm checkpoints $v_a := (v_{q_0, z_0, j_a}, 0)$ at spaced shells $1 \le j_0 < \dots < j_5 \le M$ ($0 \le a \le 5$) and incident $X$-checks $x_1, x_2$ (blue), a ball $B(x_0, r)$ (red circle) can intersect at most $1 + d_{QX} \le 5$ checkpoint transition zones along $(q_0, z_0)$, leaving at least one clean ray $a^*$ at checkpoint shell $j_{a^*}$ (solid blue ray through the amber star-arm vertex $v_{a^*}$).}
\label{fig:dag-nilpotent-hpl}
\end{figure}

Refining the bounded spectral-sequence filtration argument of~\cite[Sections~2.3.1--2.3.2]{BDL24}, because $\Psi = \Psi_{XQ} + \Psi_{QZ}$ points strictly along the two-step DAG $X \to Q \to Z$ (sector grade $q_{\mathrm{sec}} \in \{0, 1, 2\}$) and the block homotopy $h_{\mathrm{blk}} = \bigoplus_x h_x \oplus \bigoplus_q h_q \oplus 0 =: h_X \oplus h_Q \oplus h_Z$ (Lemma~\ref{lem:block-sdr} below) preserves each sector $(X, Q, Z)$ while vanishing on $Z$, any two steps of $\Psi h_{\mathrm{blk}}$ land in $Z$ where $h_Z = 0$, so $h_{\mathrm{blk}}(\Psi h_{\mathrm{blk}})^2 = 0$ and $(\Psi h_{\mathrm{blk}})^3 = 0$ (Figure~\ref{fig:dag-nilpotent-hpl}(a)). Thus the perturbation series $\Sigma = h_{\mathrm{blk}} - h_{\mathrm{blk}}\Theta h_{\mathrm{blk}} = \sum_{n=0}^\infty (-1)^n h_{\mathrm{blk}}(\Psi h_{\mathrm{blk}})^n$ in Lemma~\ref{lem:general-hpl} terminates at $n = 1$ as $\Sigma = h_{\mathrm{blk}} + h_{\mathrm{blk}}\Psi h_{\mathrm{blk}}$ over $\F$.

Before turning on the welding maps, we construct explicit strong deformation retractions on each uncoupled three-layer gadget along spanning trees on its horizontal sheets and vertical bridges. Lemma~\ref{lem:block-sdr} adapts the local cohomology calculations of Lemmas~3.8 and~3.11 of~\cite{BDL24} in two ways: lifting the gadgets from two layers to three triples coarse check cohomology from $\F^2$ to $H^2(Y_z^{(3)}) \cong \F^3$, and writing down concrete spanning-tree homotopies $h_x, h_q$ rather than only abstract cohomology groups gives us chain-level control over spatial supports in Subsection~\ref{subsec:scale-matched-rca}. Specifically, each hollow cylinder has trivial first cohomology and retracts in degree $0$ onto its uniform all-ones check cochain; each degree-shifted star retracts in degree $1$ onto its uniform all-ones hub-to-tip qubit cochain; and each coarse $Z$-block carries three replicated check cells in degree $2$.

\begin{lemma}[{Block SDRs on $Y_x^{(3)}$, $Y_q^{(3)}$, and $Y_z^{(3)}$, adapted from~\cite[Section~3.2, Lemmas~3.8 and~3.11]{BDL24}}]
\label{lem:block-sdr}
For each coarse check $x \in M_X$, qubit $q \in Q$, and check $z \in M_Z$, the local blocks of Definition~\ref{def:three-layer-mapping-cone} admit strong deformation retractions onto their block cohomology spaces:
\begin{enumerate}[label=\textbf{(\alph*)}, leftmargin=2.2em, itemsep=2.5pt]
    \item \emph{Cylinder block $Y_x^{(3)}$}: Here $H^0(Y_x^{(3)}) \cong \F \cdot \mathbf{1}_x$, $H^1(Y_x^{(3)}) = 0$, and $H^2(Y_x^{(3)}) \cong H^1(G_x) \oplus H^1(G_x)$. In degree $0$, define $\iota_X^0(\mathbf{1}_x) := \sum_{u \in C^0(Y_x^{\mathrm{disk}}),\, b \in \{0,1,2\}} (u, b)$ (the all-ones $0$-cochain on $Y_x^{(3)}$) and $\pi_X^0(\alpha) := \alpha(v_x, 0)$ (evaluation at the layer-$0$ apex $(v_x, 0)$). Fixing a spanning tree $T_\partial \subset \partial Y_x^{\mathrm{disk}} \simeq G_x$ with $b_1(G_x) := \dim_{\F} H^1(G_x)$ complement chord edges $E_{\mathrm{chord}}(G_x) := C^1(\partial Y_x^{\mathrm{disk}}) \setminus T_\partial$, relative cohomology along the outer-rim bridges $\partial Y_x^{\mathrm{disk}} \times [0, w]$ ($w \in \{1, 2\}$) gives $H^1(Y_x^{(3)}) = 0$ and $H^2(Y_x^{(3)}) \cong H^1(\partial Y_x^{\mathrm{disk}})^{\oplus 2} \cong H^1(G_x)^{\oplus 2}$ of dimension $2 b_1(G_x)$, represented by $W_x := \mathrm{span}\{(e, [0,w]) : e \in E_{\mathrm{chord}}(G_x),\, w \in \{1, 2\}\} \subset C^2(Y_x^{(3)})$. Choosing a spanning tree $T_x \subset Y_x^{\mathrm{disk}}$ rooted at $v_x$ and a dual spanning tree $T_x^* \subset (Y_x^{\mathrm{disk}})^\vee$ of horizontal faces, let $h_{\mathrm{disk}}^1, h_{\mathrm{disk}}^2$ be the tree contraction homotopies on $Y_x^{\mathrm{disk}}$ and let $h_I^1([0,w]) := [w]$ ($w \in \{1,2\}$) retract vertical bridges onto layer $0$ along $C^0(\partial Y_x^{\mathrm{disk}}) \cup T_\partial$. This decomposes $C^2(Y_x^{(3)}) = \im(d_x^1) \oplus W_x$ with projection $\pi_{W_x}$ onto $W_x$ along $\ker \pi_{W_x} = \im(d_x^1)$ and $\ker h_x^2 = W_x$, so that $h_x^1 d_x^0 = \id_{C^0(Y_x^{(3)})} - \iota_X^0 \pi_X^0$, $d_x^0 h_x^1 + h_x^2 d_x^1 = \id_{C^1(Y_x^{(3)})}$, and $d_x^1 h_x^2 = \id_{C^2(Y_x^{(3)})} - \iota_{W_x}\pi_{W_x}$.
    \item \emph{Star block $Y_q^{(3)}$ and coarse $Z$-block $Y_z^{(3)}$}: On $Y_q^{(3)}$, where $C^0(Y_q^{(3)}) = 0$, we have $H^0(Y_q^{(3)}) = 0$, $H^1(Y_q^{(3)}) \cong \F \cdot \bar{\gamma}_q$, and $\dim_{\F} H^2(Y_q^{(3)}) = 2(\abs{\partial_Z^\top(q)} - 1)$, where $\bar{\gamma}_q := \sum_{u \in C^0(X'_q),\, b \in \{0,1,2\}}(u, b)$ is the all-ones horizontal star $1$-cocycle (satisfying $d_q^1(\bar{\gamma}_q) = 0$ since each horizontal arm edge has two endpoints and each terminal vertical bridge $(v_{q,z}, [0,w])$ receives two cancelling contributions from $(v_{q,z}, 0)$ and $(v_{q,z}, w)$ over $\F$). Fixing a reference incident check $z_0 \in \partial_Z^\top(q)$, let $W_q \subset C^2(Y_q^{(3)})$ be the subspace of dimension $2(\abs{\partial_Z^\top(q)} - 1)$ spanned by the non-reference terminal vertical bridges $(v_{q,z}, [0,w])$ ($z \in \partial_Z^\top(q) \setminus \{z_0\}$, $w \in \{1, 2\}$). Defining $\iota_Q^1(\mathbf{1}_q) := \bar{\gamma}_q$, $\pi_Q^1(\beta) := \beta(v_q, 0)$ (evaluation at the layer-$0$ hub $(v_q, 0)$), and the star-tree homotopy $h_q^2 : C^2(Y_q^{(3)}) \to C^1(Y_q^{(3)})$ that vanishes on $W_q$ and integrates across $(v_{q,z_0}, [0,w])$ and along $X'_q \times \{b\}$, we have $C^2(Y_q^{(3)}) = \im(d_q^1) \oplus W_q$ (with projection $\pi_{W_q}$ along $\ker \pi_{W_q} = \im(d_q^1)$), $h_q^2 d_q^1 = \id_{C^1(Y_q^{(3)})} - \iota_Q^1\pi_Q^1$, and $d_q^1 h_q^2 = \id_{C^2(Y_q^{(3)})} - \iota_{W_q}\pi_{W_q}$. Finally, on $Y_z^{(3)}$ where $C^0(Y_z^{(3)}) = C^1(Y_z^{(3)}) = 0$ and $d_z = 0$, the exact block cohomology is $H^2(Y_z^{(3)}) = C^2(Y_z^{(3)}) = \bigoplus_{b=0}^2 \F \cdot (z, b) \cong \F^3$ with $\iota_z^2 = \pi_z^2 = \id_{\F^3}$ and $h_z = 0$, so putting $W_X := \bigoplus_{x \in M_X} W_x$ and $W_Q := \bigoplus_{q \in Q} W_q$, the combined block SDR $(\iota_{\mathrm{blk}}, \pi_{\mathrm{blk}}, h_{\mathrm{blk}})$ onto $N^\bullet = \bigl(\F^{M_X},\, \F^Q,\, W_X \oplus W_Q \oplus (\F^{M_Z})^{\oplus 3}\bigr)$ satisfies $d_{\mathrm{blk}} h_{\mathrm{blk}} + h_{\mathrm{blk}} d_{\mathrm{blk}} = \id - \iota_{\mathrm{blk}}\pi_{\mathrm{blk}}$.
\end{enumerate}
\end{lemma}

\begin{proof}
On $I^{(3)}$, $\pi_I^0(\sum_b c_b [b]) = c_0$, $\iota_I^0(1) = [0] + [1] + [2]$, and $h_I^1([0,w]) = [w]$ ($w \in \{1,2\}$) satisfy $(h_I^1 \delta_I^0)([0]) = [1] + [2] = [0] - \iota_I^0\pi_I^0([0])$ and $(h_I^1 \delta_I^0)([w]) = [w] = [w] - \iota_I^0\pi_I^0([w])$ over $\F$, while $\delta_I^0 h_I^1([0,w]) = [0,w]$. Combining the disk SDR $(Y_x^{\mathrm{disk}}, \iota_{\mathrm{disk}}, \pi_{\mathrm{disk}}, h_{\mathrm{disk}})$ with the outer-rim vertical bridge retraction $(I^{(3)}, \iota_I, \pi_I, h_I)$ along $C^0(\partial Y_x^{\mathrm{disk}}) \cup T_\partial$ leaves the complementary chord bridge space $W_x$, yielding the cylinder SDR in (a). Similarly, on the degree-shifted star block $Y_q^{(3)}$, integrating along the star tree $X'_q \times \{0\}$ on layer $0$ from $(v_q, 0)$ to the reference terminal tip $(v_{q,z_0}, 0) \in \partial X'_q \times \{0\}$, crossing the reference terminal vertical bridge $(v_{q,z_0}, [0,w])$ via $h_I^1([0,w]) = [w]$ to $(v_{q,z_0}, w)$, and integrating along $X'_q \times \{w\}$ on layer $w \in \{1, 2\}$ (while setting $h_q^2|_{W_q} = 0$) defines $h_q^2 : C^2(Y_q^{(3)}) \to C^1(Y_q^{(3)})$ satisfying $d_q^1 h_q^2 = \id_{C^2(Y_q^{(3)})} - \iota_{W_q}\pi_{W_q}$ and $h_q^2 d_q^1(\beta) = \beta - \beta(v_q, 0)\bar{\gamma}_q = \beta - \iota_Q^1\pi_Q^1(\beta)$ for every $\beta \in C^1(Y_q^{(3)})$.
\end{proof}

Plugging these block retractions into Lemma~\ref{lem:general-hpl} with the one-way welding differential $\Psi = \Psi_{XQ} + \Psi_{QZ}$ deforms the retraction operators by at most one step across adjacent gadgets. Whereas Sections~3.2.4 and~3.2.6 (Lemmas~3.10 and~3.13) of~\cite{BDL24} (via the HPL cochain equivalence of~\cite[Sections~2.3.1--2.3.2, Lemma~2.1]{BDL24}) and Theorem~1 of~\cite{yuan2026unified} (via exact sequences of code embeddings) establish global logical qubit preservation $H^1(\cC_{i+1}) \cong H^1(\cC_i)$ (with diagonal check doubling $b \mapsto (b, b)$ in two layers), Proposition~\ref{prop:classical-hpl} produces explicit closed-form chain maps $\widetilde{\iota}, \widetilde{\pi}$ and a two-term contracting homotopy $\Sigma = h_{\mathrm{blk}} + h_{\mathrm{blk}}\Psi h_{\mathrm{blk}}$ (with diagonal tripling $\Delta_3(b) = (b, b, b)$ in three layers) whose Euclidean support growth is bounded by $c_0\Lambda$ for $c_0 := 3 D_{\mathrm{edge}} + \alpha = 7003$.

\leanlink{prop:classical-hpl}
\begin{proposition}[{Two-Step Nilpotent HPL and Degree-$1$ Cohomology Isomorphism~\cite{BDL24, yuan2026unified}}]
\label{prop:classical-hpl}
Let $(\iota_{\mathrm{blk}}, \pi_{\mathrm{blk}}, h_{\mathrm{blk}})$ be the block SDR of Lemma~\ref{lem:block-sdr} and turn on the DAG welding differential $\Psi = \Psi_{XQ} + \Psi_{QZ}$. Because $h_{\mathrm{blk}}$ preserves the sector grading $(X, Q, Z)$ while $\Psi$ maps $X \to Q \to Z$ and $h_Z = 0$, the composite operator $\Psi h_{\mathrm{blk}}$ satisfies $h_{\mathrm{blk}}(\Psi h_{\mathrm{blk}})^2 = 0$ and $(\Psi h_{\mathrm{blk}})^3 = 0$ on $C^\bullet(\widetilde{\cC})$. Consequently, the HPL perturbation series~\eqref{eq:general-hpl-formulas} of Lemma~\ref{lem:general-hpl} terminates at finite order over $\F$:
\begin{align}
    \widetilde{\iota} &= (\id + h_{\mathrm{blk}}\Psi)\,\iota_{\mathrm{blk}}, \qquad
    \widetilde{\pi} = \pi_{\mathrm{blk}}\,\bigl(\id + \Psi h_{\mathrm{blk}} + (\Psi h_{\mathrm{blk}})^2\bigr), \qquad
    \Sigma = h_{\mathrm{blk}} + h_{\mathrm{blk}}\Psi h_{\mathrm{blk}}, \label{eq:hpl-two-term}\\
    \widetilde{\delta} &= \pi_{\mathrm{blk}}\Psi(\id + h_{\mathrm{blk}}\Psi)\iota_{\mathrm{blk}}, \qquad
    \delta_{\widetilde{\cC}}\Sigma + \Sigma\delta_{\widetilde{\cC}} = \id_{\widetilde{\cC}} - \widetilde{\iota}\,\widetilde{\pi}, \label{eq:hpl-sdr-identities}
\end{align}
where $\widetilde{\pi}^k = \pi_{\mathrm{blk}}^k(\id + \Psi h_{\mathrm{blk}})$ in degrees $k \in \{0, 1\}$. In degree $0 \to 1$, the induced differential is $\widetilde{\delta}^0 = \delta_{\cC}^0 : \F^{M_X} \to \F^Q$, while on $c \in \F^Q$ the degree-$1$ lift $\iota_{\mathrm{lift}}^1(c) := \widetilde{\iota}^1(c) = (\id + h_{\mathrm{blk}}\Psi)\iota_Q^1(c)$ has zero coboundary in $W_X \oplus W_Q$ and satisfies
\begin{equation}
    \label{eq:degree1-lift-intertwining}
    \delta_{\widetilde{\cC}}^1\,\iota_{\mathrm{lift}}^1(c) \;=\; \Delta_3\bigl(\delta_{\cC}^1 c\bigr) \;:=\; \sum_{z \in M_Z} (\delta_{\cC}^1 c)_z \sum_{b=0}^2 (z, b), \qquad \forall\, c \in \F^Q,
\end{equation}
so that composing with the layer-$0$ evaluation $\pi_{Z,0}^2(\sum_{z,b}\zeta_{z,b}(z,b)) := \sum_z \zeta_{z,0}\mathbf{1}_z$ recovers $\pi_{Z,0}^2 \delta_{\widetilde{\cC}}^1 \iota_{\mathrm{lift}}^1 = \delta_{\cC}^1$. In particular, $\widetilde{\iota}^1$ and $\widetilde{\pi}^1$ (and their transpose chain maps) induce canonical isomorphisms $H^1(\widetilde{\cC}) \cong H^1(\cC)$ and $H_1(\widetilde{\cC}) \cong H_1(\cC)$, so $\dim_{\F} H^1(\cC_m) = \dim_{\F} H_1(\cC_m) = 1$ for all $m \ge 0$. Moreover, each application of $\widetilde{\iota}, \widetilde{\pi}, \Sigma$ expands Euclidean chain supports by at most $c_0 \Lambda$ ($c_0 = 7003$).
\end{proposition}

\begin{proof}
Since $h_{\mathrm{blk}} = h_X \oplus h_Q \oplus 0$ vanishes on the $Z$-sector $\cS_Z$ and $\Psi$ strictly increases the sector grade along $X \to Q \to Z$, any two-step product $(\Psi h_{\mathrm{blk}})^2$ maps $\cS_X \to \cS_Q \to \cS_Z$ and annihilates $\cS_Q \oplus \cS_Z$. Since $h_{\mathrm{blk}}|_{\cS_Z} = 0$, we have $h_{\mathrm{blk}}(\Psi h_{\mathrm{blk}})^2 = 0$, $(h_{\mathrm{blk}}\Psi)^2 = h_{\mathrm{blk}}(\Psi h_{\mathrm{blk}})\Psi = 0$, and $(\Psi h_{\mathrm{blk}})^3 = \Psi(h_{\mathrm{blk}}(\Psi h_{\mathrm{blk}})^2) = 0$; moreover, on degree-$1$ cochains, $h_Q^1 = 0$ because $C^0(Y_q^{(3)}) = 0$, so $(\Psi h_{\mathrm{blk}})^2 = 0$ on $C^0(\widetilde{\cC}) \oplus C^1(\widetilde{\cC})$. Substituting this nilpotency into Lemma~\ref{lem:general-hpl} (with $-1 \equiv +1$ over $\F$, and noting $\widetilde{\pi}^1\widetilde{\iota}^1 = \pi_Q^1\iota_Q^1 = \id_{\F^Q}$ directly since $h_Q^1 = 0$ and $h_Z = 0$) yields~\eqref{eq:hpl-two-term}--\eqref{eq:hpl-sdr-identities}. Checking the induced coarse differential on basis elements in degrees $0$ and $1$: for $x \in M_X$, $\Psi_{XQ}^0\iota_X^0(\mathbf{1}_x)$ evaluates at each incident hub $(v_q, 0)$ ($q \in \partial_X(x)$) to $1$, projecting under $\pi_Q^1$ to $\sum_{q \in \partial_X(x)} \mathbf{1}_q = \delta_{\cC}^0(\mathbf{1}_x)$; for $q \in Q$, $\iota_{\mathrm{lift}}^1(\mathbf{1}_q) = \bar{\gamma}_q$ satisfies $d_q^1(\bar{\gamma}_q) = 0$ on $Y_q^{(3)}$ and maps under $\Psi_{QZ}^1$ to $\sum_{z \in \partial_Z^\top(q)} \sum_{b=0}^2 (z, b) = \Delta_3(\delta_{\cC}^1 \mathbf{1}_q)$, proving~\eqref{eq:degree1-lift-intertwining}. Since $\Delta_3 : \F^{M_Z} \hookrightarrow (\F^{M_Z})^{\oplus 3}$ is injective with left inverse $\pi_{Z,0}^2$, a coarse $1$-cochain $c \in \F^Q$ satisfies $\delta_{\cC}^1 c = 0$ if and only if $\delta_{\widetilde{\cC}}^1 \iota_{\mathrm{lift}}^1(c) = 0$, and for any fine $1$-cocycle $\tilde{c} \in \ker \delta_{\widetilde{\cC}}^1$, the SDR identity~\eqref{eq:hpl-sdr-identities} gives $\tilde{c} = \widetilde{\iota}^1(\widetilde{\pi}^1 \tilde{c}) + \delta_{\widetilde{\cC}}^0(\Sigma^1 \tilde{c})$, establishing $H^1(\widetilde{\cC}) \cong H^1(\cC)$ and by transposition $H_1(\widetilde{\cC}) \cong H_1(\cC)$. Since $\Sigma = h_{\mathrm{blk}} + h_{\mathrm{blk}}\Psi h_{\mathrm{blk}}$ crosses at most adjacent cylinder and star blocks (three active edges of length $\le D_{\mathrm{edge}}\Lambda = 2001\Lambda$ plus the $\alpha\Lambda = 1000\Lambda$ spherical-cap shift), $\widetilde{\iota}, \widetilde{\pi}, \Sigma$ expand Euclidean supports by at most $(3 D_{\mathrm{edge}} + \alpha)\Lambda = c_0\Lambda$.
\end{proof}

\subsection{Scale-Matched Coarse-Graining}
\label{subsec:scale-matched-rca}

Iterating the coarse projection $\widetilde{\pi}$ of Proposition~\ref{prop:classical-hpl} all the way down from level $m$ to $\cC_0$ does not by itself establish $\RCA(L^*, C)$, for two geometric reasons:
\begin{enumerate}[label=\textbf{(\arabic*)}, leftmargin=2.2em, itemsep=2.5pt]
    \item \emph{Affine radius stalling at deep scales}: Each coarse pullback step shrinks Euclidean radii by $\Lambda^{-1}$ while adding an $O(1)$ block-collar overhead $c_0$: $r_{l+1} \le r_l/\Lambda + c_0$. As $l$ increases, $r_l$ stalls at the fixed point $r_{\mathrm{fix}} = c_0\Lambda/(\Lambda - 1) = O(1)$, and lifting an $O(1)$ chain from $\cC_0$ all the way back through $m$ levels inflates its radius to $\Theta(\Lambda^m) \gg C r$. Thus we stop coarse-graining a ball $B(x_0, r)$ at the scale-matched depth $s_*(r) := \ceil{\log_\Lambda r}$, where the rescaled radius first enters $[1, R_{\mathrm{stop}}]$ for an $O(1)$ threshold $R_{\mathrm{stop}}$ on an intermediate macroscopic code $\cC_{m - s_*(r)}$.
    \item \emph{Sub-cylinder filling and alternating $\cR_X^{(3)} / \cR_Z^{(3)}$ duality}: At the stopping level $m - s_*$ ($s_* := s_*(r)$), the rescaled ball $B(\Lambda^{-s_*} x_0, r_{s_*})$ ($r_{s_*} \le R_{\mathrm{stop}}$) is much smaller than the level-$(m - s_*)$ gadget scale $\Lambda$, while $\cC_{m - s_*}$ itself has nontrivial global homology $H_1 \cong H^1 \cong \F$; moreover, because $\cC_m$ alternates primal $\cR_X^{(3)}$ and dual $\cR_Z^{(3)}$ steps, a single $\cR_X^{(3)}$ step must transfer both dual cocycle filling and primal cycle filling across scales without assuming disjoint simplices are separated in $\R^3$.
\end{enumerate}

\begin{figure}[t]
\centering
\includegraphics[width=0.88\textwidth]{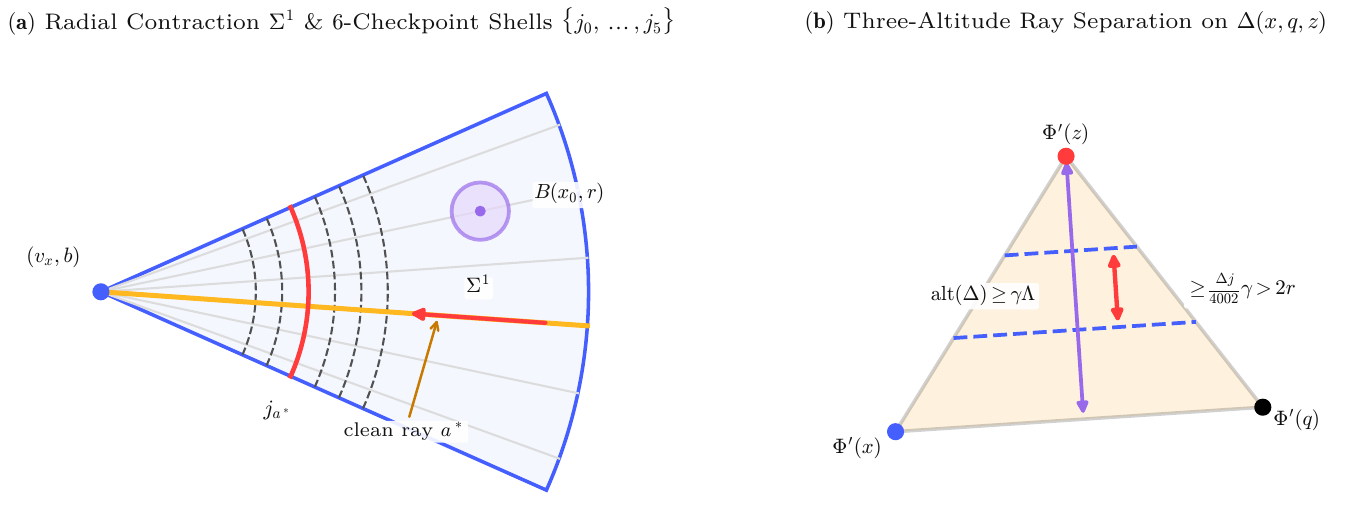}
\caption{Six-checkpoint pigeonhole ray selection and three-altitude ray separation on a single replacement step $\widetilde{\cC} = \cR_X^{(3)}(\cC)$ (Lemma~\ref{lem:clean-shell-subcylinder}). \textbf{(a)} Radial contraction $\Sigma^1$ and 6-checkpoint selection $\{j_0, \dots, j_5\}$ ($j_a = (a+1)\Delta j$, spacing $\Delta j = 5 \times 10^7$) on a cylinder sector rooted at the blue apex $(v_x, b)$, showing 6 concentric checkpoint shells (with the clean shell $j_{a^*}$ in solid red and the others in dashed gray), radial rays (with the clean ray $a^*$ in solid amber), a violet ball $B(x_0, r)$, and an inward red radial arrow $\Sigma^1$. \textbf{(b)} Single-triangle altitude ray separation on a coarse triangle $\Delta(x, q, z)$ (amber shading) with vertices $\Phi'(x)$ (blue), $\Phi'(q)$ (black), $\Phi'(z)$ (red), a violet double-headed altitude arrow $\alt(\Delta) \ge \gamma\Lambda$, and a red double-headed ray-separation arrow $\ge \frac{\Delta j}{4002}\gamma > 2r$ between two candidate rays.}
\label{fig:subcylinder-homotopy}
\end{figure}

\begin{figure}[t]
\centering
\includegraphics[width=0.88\textwidth]{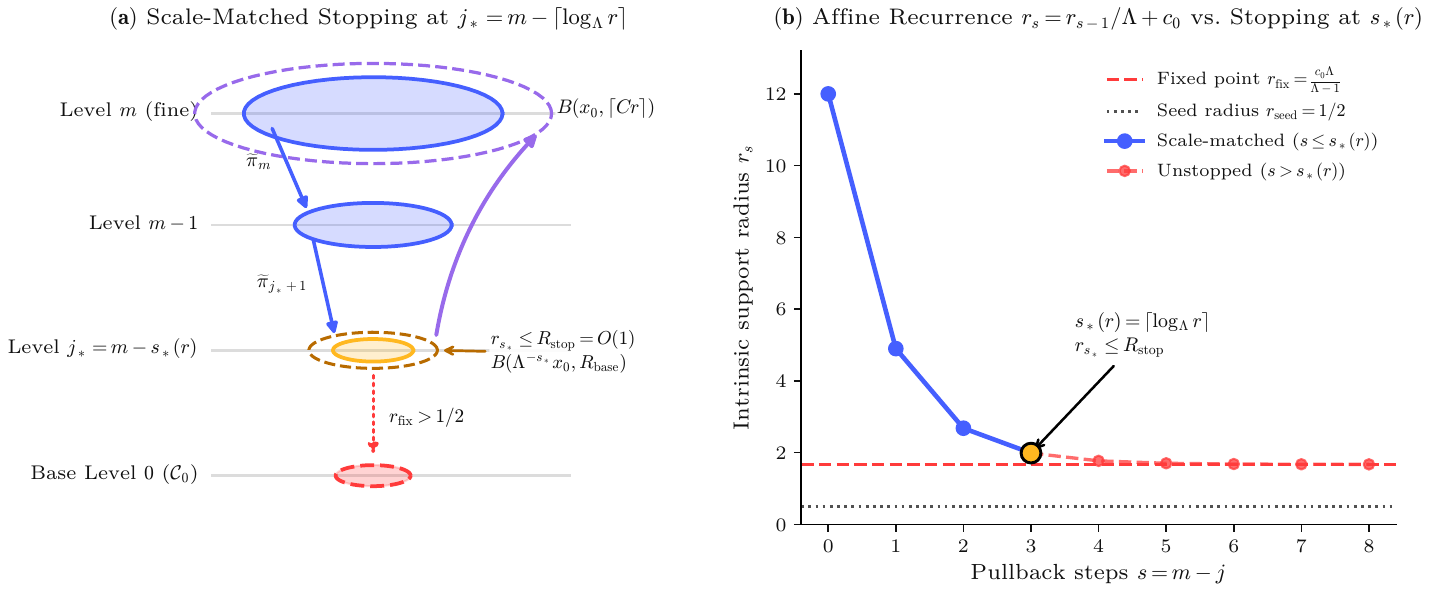}
\caption{Resolution of the affine pullback stalling obstacle via scale-matched stopping (Theorem~\ref{thm:rca-welded-hierarchy}). \textbf{(a)} Scale-matched stopping across hierarchy levels ($\text{Level } m$, $\text{Level } m-1$, $\text{Level } j_* = m - s_*(r)$, and $\text{Base Level } 0$), showing downward blue pullback arrows $\widetilde{\pi}_m, \widetilde{\pi}_{j_*+1}$ stopping at the amber ellipse ($r_{s_*} \le R_{\mathrm{stop}} = O(1)$, $s_* := s_*(r)$) with dashed amber collar $B(\Lambda^{-s_*} x_0, R_{\mathrm{base}})$ and lifting back to level $m$ via a curved violet arrow into the dashed violet collar $B(x_0, \ceil{C r})$ (avoiding the dotted red continuation to $r_{\mathrm{fix}} > 1/2$ at level $0$). \textbf{(b)} Rescaled ball radius $r_s$ versus pullback step $s$, comparing the solid blue scale-matched trajectory stopping at the amber circle $s_*(r) = \ceil{\log_\Lambda r}$ (for $r \le \Lambda^{m - i_{\min}}$) against the dashed red unstopped trajectory stalling at the horizontal asymptote $r_{\mathrm{fix}} = \frac{c_0\Lambda}{\Lambda - 1}$ above the dotted gray seed single-cell clearance radius $r_{\mathrm{seed}} = 1/2$.}
\label{fig:scale-matched-stopping}
\end{figure}

The idea behind resolving both obstacles is simple, requiring only single-triangle altitude regularity $\ge \gamma\Lambda \ge \frac{7}{5}\Lambda$, bounded qubit degree $d_{QX} \le 4$, and $\F$-orthogonal cocycle-extension duality (Figure~\ref{fig:subcylinder-homotopy} and Figure~\ref{fig:scale-matched-stopping}).

At the stopping scale, any ball of bounded radius is tiny compared to the surrounding $\Theta(\Lambda)$-sized cylinders, so its coarse projection could only be nonzero if the ball blocked every path from a star hub out to its incident cylinder apices. Because each coarse triangle is uniformly thick and each qubit meets at most four coarse $X$-checks, choosing six well-spaced radial paths along an incident triangle guarantees by the pigeonhole principle that at least one ray misses the ball entirely, allowing local cocycles and cycles to be filled across two consecutive replacement steps.

\leanlink{lem:clean-shell-subcylinder}
\begin{lemma}[Clean-Ray Selection and Two-Step Sub-Cylinder Filling]
\label{lem:clean-shell-subcylinder}
Let $(\cC, \Phi)$ be a 3D CSS complex with $(8, r_0, \rho_0)$-bounded geometry ($1 \le d_{QX}, d_{QZ} \le 4$, $r_0 \le 8$, $\rho_0 \le 6.6 \times 10^{10}$), and let $\widetilde{\cC} = \cR_X^{(3)}(\cC)$ at scale $\Lambda \ge \Lambda_0 = 5.4 \times 10^{24}$ after the Step~1 perturbation $\Phi'$ (Proposition~\ref{prop:step1-lll}). Set $R_{\mathrm{stop}} := 7015$, $c_0 := 7003$, $c_{\mathrm{base}} := 10005$, and $C_{\mathrm{diam}} := 13006$:
\begin{enumerate}[label=\textbf{(\roman*)}, leftmargin=2.2em, itemsep=2.5pt]
    \item For every center $x_0 \in \R^3$ and radius $1 \le r \le R_{\mathrm{stop}}$, every dual cocycle $c_Z \in \F^{Q(\widetilde{\cC})(B(x_0, r))} \cap \ker \delta_{\widetilde{\cC}}^1$ lies in $\delta_{\widetilde{\cC}}^0\bigl(\F^{M_X(\widetilde{\cC})(B(x_0, c_{\mathrm{base}}\Lambda))}\bigr)$. By CSS duality, on $\cR_Z^{(3)}(\cC)$ every primal cycle $c_X \in \F^{Q(B(x_0, r))} \cap \ker \partial_1$ lies in $\partial_2\bigl(\F^{M_Z(B(x_0, c_{\mathrm{base}}\Lambda))}\bigr)$.
    \item If $(\cC, \Phi)$ fills primal cycles (resp.\ dual cocycles) from inner radius $r_{\mathrm{in}} := r/\Lambda + c_0$ into outer radius $R_{\mathrm{coarse}}$, then $\widetilde{\cC} = \cR_X^{(3)}(\cC)$ (and dually $\cR_Z^{(3)}(\cC)$) fills primal cycles (resp.\ dual cocycles) from radius $r$ into radius $\Lambda(R_{\mathrm{coarse}} + C_{\mathrm{diam}})$. Consequently, across two consecutive steps $\cC'' := \cR_Z^{(3)}(\cR_X^{(3)}(\cC))$, every ball $B(x_0, r)$ with $1 \le r \le R_{\mathrm{stop}}$ satisfies both primal and dual filling inside $B(x_0, R_{\mathrm{base}})$ with $R_{\mathrm{base}} := \Lambda(c_{\mathrm{base}}\Lambda + C_{\mathrm{diam}})$.
\end{enumerate}
\end{lemma}

\begin{proof}
For (i), let $c_Z \in \F^{Q(\widetilde{\cC})(B(x_0, r))} \cap \ker \delta_{\widetilde{\cC}}^1$ with $r \le R_{\mathrm{stop}} = 7015$. By Proposition~\ref{prop:classical-hpl}, $c_Z = \delta_{\widetilde{\cC}}^0(\Sigma^1 c_Z) + \widetilde{\iota}^1(\bar{c})$ where $\bar{c} := \widetilde{\pi}^1(c_Z) \in \ker \delta_{\cC}^1$ and $\Sigma^1 c_Z$ is supported within $r + 3 D_{\mathrm{edge}}\Lambda \le 7000\Lambda$ of $x_0$. Define a coarse $0$-cochain $c_{\mathrm{apex}} \in \F^{M_X}$ by splitting coarse apices at radius $\frac{\gamma}{2}\Lambda \ge \frac{7}{10}\Lambda \gg R_{\mathrm{stop}}$: if $\norm{\Phi'(x) - x_0}_2 \le \frac{\gamma}{2}\Lambda$ (\emph{near apex}), every outer-rim vertex of $Y_x^{(3)}$ lies at distance $\ge \gamma\Lambda$ from $\Phi'(x)$ by the triangle altitude bound and hence at distance $\ge \frac{\gamma}{2}\Lambda > r$ from $x_0$, so $c_Z$ vanishes identically on $\partial Y_x^{(3)}$; by connectivity of the link $G_x$, the layer-$0$ radial spoke sum of $c_Z$ from $(v_x, 0)$ to $\partial Y_x^{\mathrm{disk}}$ takes a ray-independent value $c_{\mathrm{apex}}(x) \in \F$. For every \emph{far apex} ($\norm{\Phi'(x) - x_0}_2 > \frac{\gamma}{2}\Lambda$), set $c_{\mathrm{apex}}(x) := 0$.

Now fix a coarse qubit $q_0 \in Q$ and pick an incident coarse $Z$-check $z_0 \in \partial_Z^\top(q_0)$ (no disjoint-simplex separation is needed). Along the $(q_0, z_0)$ arm, consider the $6$ checkpoint shell indices $j_a = (a+1)\Delta j$ ($a \in \{0, \dots, 5\}$, $\Delta j = 5 \times 10^7 \ll m_{\mathrm{sub}} = 4002\Lambda$). For any far incident check $x \in \partial_X^\top(q_0)$, inner cylinder shells $j \le j_5 = 3 \times 10^8$ (radial step $u = j + 1 \le j_5 + 1$) lie within $\frac{j_5 + 1}{4002\Lambda}\cdot 4002\Lambda = 3 \times 10^8 + 1 < \frac{\gamma}{2}\Lambda - R_{\mathrm{stop}}$ of $\Phi'(x)$ and miss $B(x_0, r)$ altogether, whereas on outer steps $u \ge j_a + 1$ ($j \ge j_a$) of the single triangle $\Delta(\Phi'(x), \Phi'(q_0), \Phi'(z_0))$, the altitude bound $\ge \gamma\Lambda \ge \frac{7}{5}\Lambda$ separates distinct candidate rays $a \neq b$ in $\R^3$ by at least
\begin{equation}
    \label{eq:six-checkpoint-ray-separation}
    \frac{\abs{j_b - j_a}}{4002\Lambda}\,\gamma\Lambda \;=\; \frac{\abs{j_b - j_a}}{4002}\,\gamma \;\ge\; \frac{\Delta j}{4002}\cdot\frac{7}{5} \;\ge\; 17490 \;>\; 2 R_{\mathrm{stop}} \;\ge\; 2 r.
\end{equation}
Hence $B(x_0, r)$ intersects at most $1$ candidate ray per far check $x \in \partial_X^\top(q_0)$ and at most $1$ star-arm checkpoint $(v_{q_0, z_0, j_a}, 0)$. Since $\abs{\partial_X^\top(q_0)} \le d_{QX} \le 4$, at most $1 + 4 = 5 < 6$ indices $a$ are hit, leaving a clean checkpoint $a^*$ (Figure~\ref{fig:dag-nilpotent-hpl}(b)) at which $c_Z$ vanishes at $(v_{q_0, z_0, j_{a^*}}, 0)$ and along every far ray, while each near ray contributes $c_{\mathrm{apex}}(x)$. Telescoping $(\delta_{\widetilde{\cC}}^1 c_Z)(e_{q_0, z_0, k}, 0) = 0$ ($0 \le k < j_{a^*}$) from $(v_{q_0}, 0)$ to $(v_{q_0, z_0, j_{a^*}}, 0)$ and across disk plaquettes to ray $a^*$ gives $\bar{c}(q_0) = \sum_{x \in \partial_X^\top(q_0)} c_{\mathrm{apex}}(x) = (\delta_{\cC}^0 c_{\mathrm{apex}})(q_0)$. Since $q_0 \in Q$ was arbitrary, $\bar{c} = \delta_{\cC}^0 c_{\mathrm{apex}}$, so $c_Z = \delta_{\widetilde{\cC}}^0(\Sigma^1 c_Z + \widetilde{\iota}^0 c_{\mathrm{apex}})$ with support inside $B(x_0, c_{\mathrm{base}}\Lambda)$.

For (ii), dual cocycle filling transfers from $(r/\Lambda + c_0,\, R_{\mathrm{coarse}})$ on $\cC$ to $(r,\, \Lambda(R_{\mathrm{coarse}} + C_{\mathrm{diam}}))$ on $\widetilde{\cC} = \cR_X^{(3)}(\cC)$ directly via $c_Z = \widetilde{\iota}^1(\widetilde{\pi}^1 c_Z) + \delta_{\widetilde{\cC}}^0(\Sigma^1 c_Z)$ and $\widetilde{\iota}^1\delta_{\cC}^0 = \delta_{\widetilde{\cC}}^0\widetilde{\iota}^0$. Simultaneously, as proved in Proposition~\ref{prop:trinity-equivalence}, primal cycle filling $\F^{Q(B_{\mathrm{in}})} \cap \ker \partial_1 \subseteq \partial_2(\F^{M_Z(B_{\mathrm{out}})})$ is equivalent by $\F$-orthogonal complements to \emph{local cocycle extension} $\ker \delta_{B_{\mathrm{out}}}^1 \subseteq \ker \pi_{B_{\mathrm{in}}} + \im \delta^0$ in degree-$1$ cohomology. Applying the same local cylinder and star HPL gauge-fixing on blocks inside $B_{\mathrm{out}}$ reduces fine local cocycle extension on $\widetilde{\cC}$ to coarse local cocycle extension on $\cC$ at the identical radii $(r/\Lambda + c_0,\, R_{\mathrm{coarse}}) \mapsto (r,\, \Lambda(R_{\mathrm{coarse}} + C_{\mathrm{diam}}))$. Dualizing $X \leftrightarrow Z$ transfers both sectors across $\cR_Z^{(3)}$ as well. Finally, on $\cC'' = \cR_Z^{(3)}(\cR_X^{(3)}(\cC))$, dualize part~(i) to fill primal cycles inside $c_{\mathrm{base}}\Lambda \le R_{\mathrm{base}}$ directly at the outer $\cR_Z^{(3)}$ step, and apply the $1$-step transfer across $\cR_Z^{(3)}$ to the base cocycle filling $c_{\mathrm{base}}\Lambda$ of $\cR_X^{(3)}(\cC)$ (since $r/\Lambda + c_0 \le R_{\mathrm{stop}}/\Lambda + c_0 \le 7015 = R_{\mathrm{stop}}$) to fill dual cocycles inside $\Lambda(c_{\mathrm{base}}\Lambda + C_{\mathrm{diam}}) = R_{\mathrm{base}}$.
\end{proof}

Now consider a ball $B(x_0, r)$ at top level $m = 2k + 2 \ge i_{\min} := 2$ of arbitrary radius $1 \le r \le 11 L_k^*$ with $L_k^* := \Lambda^{m - i_{\min}} = \Lambda^{2k}$. Instead of pulling a syndrome-free chain $c^{(m)}$ all the way down to $\cC_0$, we pull it down for
\begin{equation}
    \label{eq:scale-matched-stopping-depth}
    s_*(r) \;:=\; \min\bigl(m - i_{\min},\, \ceil{\log_\Lambda r}\bigr) \;\in\; \{0, 1, \dots, m - i_{\min}\}
\end{equation}
renormalization steps to level $m - s_*(r) \ge 2$ (which has at least two replacement steps above $\cC_0$), where $r \Lambda^{-s_*(r)} \le 11$ so its rescaled Euclidean radius $r_{s_*(r)} \le 11 + r_{\mathrm{fix}} \le R_{\mathrm{stop}} = 7015$ falls inside the two-step base-filling range of Lemma~\ref{lem:clean-shell-subcylinder}(ii) (Figure~\ref{fig:scale-matched-stopping}).

Stopping the coarse pullback at this scale-matched intermediate level, filling the rescaled chain locally across two consecutive replacement steps, and lifting the resulting filling back to the physical level bounds the outer support radius linearly in $r$ across all even hierarchy depths.

\leanlink{thm:rca-welded-hierarchy}
\begin{theorem}[Relative Cone Acyclicity $\RCA(L_k^*, C)$ for the Welded Hierarchy $\cC_m$]
\label{thm:rca-welded-hierarchy}
There exists a universal linear collar constant $C = C(\Lambda) \ge 1$ independent of $m$ such that for every $m = 2k + 2 \ge 2$ ($k \ge 0$), the welded code $\cC_m := (\cR_Z^{(3)} \circ \cR_X^{(3)})^{\circ(k+1)}(\cC_0)$ (with Step~1--2 embedding $\Phi_m$) satisfies Euclidean Relative Cone Acyclicity $\RCA(11 L_k^*, C_{\mathrm{euc}})$, and hence $\ell_\infty$-cube $\RCA(L_k^*, C)$, $\mathrm{TQO\text{-}1}(L_k^*)$, and $\mathrm{Linear\ TQO\text{-}2}(C)$ on $\Gamma_k$ (Definition~\ref{def:safe-decoded-observables}) by Proposition~\ref{prop:trinity-equivalence}, up to the macroscopic cutoff scale $L_k^* = \Lambda^{m - i_{\min}} = \Lambda^{2k}$ with $i_{\min} = 2$.
\end{theorem}

\begin{proof}
Let $1 \le r \le 11 L_k^* = 11\Lambda^{m-2}$ and set $s := s_*(r)$ as in~\eqref{eq:scale-matched-stopping-depth}, so that $0 \le s \le m - 2$, $r \Lambda^{-s} \le 11$, and $\Lambda^s \le \Lambda r$. At each downward step $m - l$ ($0 \le l < s$), whether $\cC_{m-l}$ is a primal $\cR_X^{(3)}$ step or a dual $\cR_Z^{(3)}$ step, the HPL decomposition (in cohomology or homology, paired with $\F$-orthogonal cocycle extension as in Lemma~\ref{lem:clean-shell-subcylinder}(ii)) takes the form
\begin{equation}
    \label{eq:telescoping-sdr-chain}
    c^{(m-l)} \;=\; \widetilde{\iota}_{m-l}\bigl(c^{(m-l-1)}\bigr) + \delta_{\cC_{m-l}}^0\bigl(b^{(m-l)}\bigr)
\end{equation}
(or its $\partial_2$-transpose), reducing filling at radius $r_l$ on $\cC_{m-l}$ to filling at radius $r_{l+1} = r_l/\Lambda + c_0$ on $\cC_{m-l-1}$ with $c_0 = 3 D_{\mathrm{edge}} + \alpha = 3 \times 2001 + 1000 = 7003$. Iterating $s$ steps yields
\begin{equation}
    \label{eq:affine-radius-bound}
    r_s \;\le\; \frac{r}{\Lambda^s} + c_0 \sum_{j=0}^{s-1} \Lambda^{-j} \;\le\; 11 + \frac{c_0 \Lambda}{\Lambda - 1} \;\le\; 7015 \;=\; R_{\mathrm{stop}}.
\end{equation}
Because $m - s \ge 2$ is at least two replacement steps above $\cC_0$ (of the form $\cR_Z^{(3)} \circ \cR_X^{(3)}$ when $s$ is even, or $\cR_X^{(3)} \circ \cR_Z^{(3)}$ when $s$ is odd), Lemma~\ref{lem:clean-shell-subcylinder}(ii) fills both primal cycles and dual cocycles of radius $r_s \le R_{\mathrm{stop}}$ on $\cC_{m-s}$ inside radius $R_{\mathrm{base}} = \Lambda(c_{\mathrm{base}}\Lambda + C_{\mathrm{diam}})$. Applying the two-sector scale-up transfer $R_{j+1} = \Lambda(R_j + C_{\mathrm{diam}})$ ($C_{\mathrm{diam}} = 13006$) of Lemma~\ref{lem:clean-shell-subcylinder}(ii) upward through the $s$ steps reconstructs the global coboundary (and dually the global $\partial_2$-boundary)
\begin{equation}
    \label{eq:global-coboundary-reconstruction}
    c^{(m)} \;=\; \delta_{\cC_m}^0\!\Biggl( \widetilde{\iota}_{m} \cdots \widetilde{\iota}_{m-s+1}\bigl(a^{(m-s)}\bigr) \;+\; \sum_{l=0}^{s-1} \widetilde{\iota}_{m} \cdots \widetilde{\iota}_{m-l+1}\bigl(b^{(m-l)}\bigr) \Biggr)
\end{equation}
supported inside Euclidean radius
\begin{equation}
    \label{eq:linear-collar-radius-final}
    R_{\mathrm{final}}(r) \;\le\; \Lambda^s \Bigl(R_{\mathrm{base}} + \frac{C_{\mathrm{diam}} \Lambda}{\Lambda - 1}\Bigr) \;\le\; \Lambda \Bigl(R_{\mathrm{base}} + \frac{C_{\mathrm{diam}} \Lambda}{\Lambda - 1}\Bigr) r \;:=\; C_{\mathrm{euc}}\, r.
\end{equation}
By Proposition~\ref{prop:trinity-equivalence} and $\norm{y}_2 \le \sqrt{3}\,\norm{y}_\infty$, Euclidean $\RCA(11\Lambda^{2k}, C_{\mathrm{euc}})$ implies $\ell_\infty$-cube $\RCA(\Lambda^{2k}, C)$, $\mathrm{TQO\text{-}1}(L_k^*)$, and $\mathrm{Linear\ TQO\text{-}2}(C)$ on $\Gamma_k$ up to $L_k^* = \Lambda^{2k}$.
\end{proof}

\section{Two-Stage Syndrome Cleaning}
\label{sec:two-stage-cleaner}

\subsection{Cleaner Architecture}
\label{subsec:cleaner-architecture}

We now turn to the finite-temperature side of the problem (the left branch of Figure~\ref{fig:proof-roadmap}). Suppose a Pauli $X$-error $e \in C^1(\widetilde{\cC})$ on the fine welded complex $\widetilde{\cC} = \cR_X^{(3)}(\cC)$ excites a realizable $Z$-check syndrome $\sigma = \delta_{\widetilde{\cC}}^1 e \in \im(\delta_{\widetilde{\cC}}^1) \subset C^2(\widetilde{\cC})$. Following the two-stage syndrome cleaning strategy introduced in Section~4.3 of~\cite{BDL24} (Theorem~4.7, Lemma~4.8, and Lemma~4.12 of~\cite{BDL24}), we locally sweep $\sigma$ out of the cylinder gadgets $Y_x^{(3)}$ and star interfaces $Y_q^{(3)}$ onto the replicated coarse $Z$-checks $Y_z^{(3)}$. Whereas the two-layer cleaner of~\cite[Section~4.3]{BDL24} ($n_{\mathrm{lay}} = 2$) relies on sector-subregion string pairing and $(1,1)$-versus-$(0,0)$ arm pairing to avoid $1$-versus-$1$ layer ties, our three-layer geometry $Y_x^{(3)}, Y_q^{(3)}, Y_z^{(3)}$ ($n_{\mathrm{lay}} = 3$) ensures that each cleaning stage expands syndrome weight by at most $\frac{4}{3}$, so that dividing by the $3$-fold coarse check replication factor contracts the coarse syndrome $\sigma_{\mathrm{coarse}} \in \im(\delta_{\cC}^1)$ in Hamming weight by a strict factor $\frac{1}{3}\times\frac{4}{3}\times\frac{4}{3} = \frac{16}{27} < 1$.

\begin{figure}[t]
\centering
\includegraphics[width=0.92\textwidth]{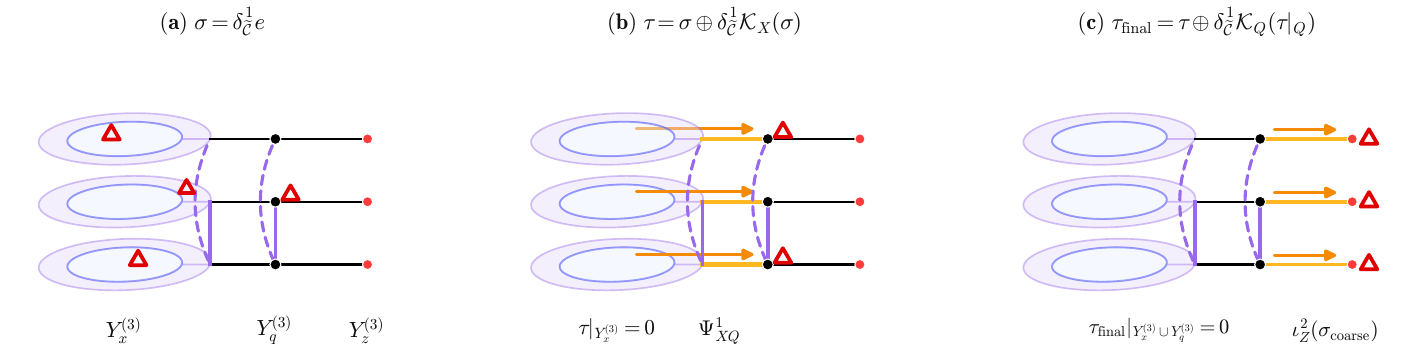}
\caption{Two-stage local syndrome cleaning along the DAG $X \to Q \to Z$: \textbf{(a)} raw syndrome $\sigma = \delta_{\widetilde{\cC}}^1 e$, \textbf{(b)} Stage-1 intermediate syndrome $\tau = \sigma \oplus \delta_{\widetilde{\cC}}^1 \cK_X(\sigma)$ with $\tau|_{Y_x^{(3)}} = 0$, and \textbf{(c)} Stage-2 cleaned syndrome $\tau_{\mathrm{final}} = \tau \oplus \delta_{\widetilde{\cC}}^1 \cK_Q(\tau|_Q) = \iota_Z^2(\sigma_{\mathrm{coarse}})$, showing the 3-layer cylinder $Y_x^{(3)}$ (blue disks with violet outer collars) connected to the 3-layer star $Y_q^{(3)}$ (black hubs with violet terminal vertical bridges) as open red triangle defects ($\triangle$) are swept out of $Y_x^{(3)}$ along orange arrows and amber links $\Psi_{XQ}^1$ and aligned across all three layers.}
\label{fig:two-stage-cleaner}
\end{figure}

As in~\cite[Section~4.3]{BDL24}, because the mapping cone differential $\delta_{\widetilde{\cC}} = d_{\mathrm{blk}} + \Psi_{XQ} + \Psi_{QZ}$ points strictly along the directed acyclic graph $X \to Q \to Z$ (Definition~\ref{def:three-layer-mapping-cone}), we clean $\sigma$ in two block-parallel stages with zero feedback from $Q$ or $Z$ back to $X$ (Figure~\ref{fig:two-stage-cleaner}):
\begin{enumerate}[label=\textbf{Stage \arabic*:}, leftmargin=4.2em, itemsep=3pt]
    \item \textbf{Cylinder Cleaning ($\cK_X$ on $\bigsqcup_{x \in M_X} Y_x^{(3)}$; cf.~\cite[Section~4.3.1, Lemma~4.8]{BDL24})}: Since there are no edges in the DAG pointing into the $X$-sector $\cS_X$, the restriction $\sigma_x := \sigma|_{Y_x^{(3)}}$ of any globally realizable syndrome $\sigma = \delta_{\widetilde{\cC}}^1 e$ to each cylinder block $Y_x^{(3)}$ already satisfies the relative compatibility condition $\sigma_x \in \im(d_x^1)$. Independently on each cylinder $Y_x^{(3)}$, we choose a minimum-export local $1$-chain $c_x = \cK_{X,x}(\sigma_x) \in C^1(Y_x^{(3)})$ satisfying $d_x^1 c_x = \sigma_x$. Putting $\cK_X(\sigma) := \sum_{x \in M_X} c_x \in \cS_X^1$ and subtracting $\delta_{\widetilde{\cC}}^1\cK_X(\sigma)$ from $\sigma$ produces the intermediate syndrome (writing $\oplus$ for binary $\F$-cochain addition alongside $\sum$ for disjoint-support sums, while keeping $\oplus$ between subspaces for direct sum)
    \begin{equation}
        \label{eq:tau-intermediate-def}
        \tau \;:=\; \sigma \oplus \delta_{\widetilde{\cC}}^1\cK_X(\sigma) \;\in\; \cS_Q^2 \oplus \cS_Z^2,
    \end{equation}
    which vanishes identically on all cylinder blocks $Y_x^{(3)}$.
    \item \textbf{Star Interface Cleaning ($\cK_Q$ on $\bigsqcup_{q \in Q} Y_q^{(3)}$; cf.~\cite[Section~4.3.3, Lemma~4.12]{BDL24})}: Since $\tau$ vanishes on $\cS_X$ and remains in $\im(\delta_{\widetilde{\cC}}^1)$, its restriction $\tau_q := \tau|_{Y_q^{(3)}}$ to each star block $Y_q^{(3)}$ lies in $\im(d_q^1)$ (as $H^1(Y_x^{(3)}) = 0$ by Lemma~\ref{lem:block-sdr}(a) writes the cylinder preimage component as $d_x^0 \alpha_X$, and $d_q^1\Psi_{XQ}^0 = \Psi_{XQ}^1 d_x^0$ absorbs its star export into $\im(d_q^1)$). Independently on each star $Y_q^{(3)}$, we choose between the two coset candidate $1$-cochains $c_q^{(0)}, c_q^{(1)} \in C^1(Y_q^{(3)})$ satisfying $d_q^1 c_q^{(u)} = \tau_q$ ($u \in \{0, 1\}$, differing by the local star cocycle $c_q^{(1)} = c_q^{(0)} \oplus \bar{\gamma}_q$ with $\bar{\gamma}_q \in \ker d_q^1$) to minimize the exported weight onto the adjacent coarse $Z$-checks $Y_z^{(3)}$. Putting $\tau|_Q := \sum_{q \in Q}\tau_q \in \cS_Q^2$ and $\cK_Q(\tau|_Q) := \sum_{q \in Q} c_q \in \cS_Q^1$, subtracting $\delta_{\widetilde{\cC}}^1\cK_Q(\tau|_Q)$ from $\tau$ leaves a residual syndrome
    \begin{equation}
        \label{eq:tau-final-def}
        \tau_{\mathrm{final}} \;:=\; \tau \oplus \delta_{\widetilde{\cC}}^1\cK_Q(\tau|_Q) \;\in\; \cS_Z^2 \cap \im(\delta_{\widetilde{\cC}}^1),
    \end{equation}
    which is supported entirely on the coarse $Z$-blocks $Y_z^{(3)}$; writing $\iota_Z^2 := \Delta_3$ for the $3$-layer check replication map~\eqref{eq:degree1-lift-intertwining}, we show in Lemma~\ref{lem:stage2-star-cleaner}(ii) that $\tau_{\mathrm{final}} = \iota_Z^2(\sigma_{\mathrm{coarse}})$ for $\sigma_{\mathrm{coarse}} := \cR_X^{\syn}(\sigma) := \pi_{Z,0}^2(\tau_{\mathrm{final}}) \in \im(\delta_{\cC}^1)$.
\end{enumerate}

\begin{figure}[t]
\centering
\includegraphics[width=0.90\textwidth]{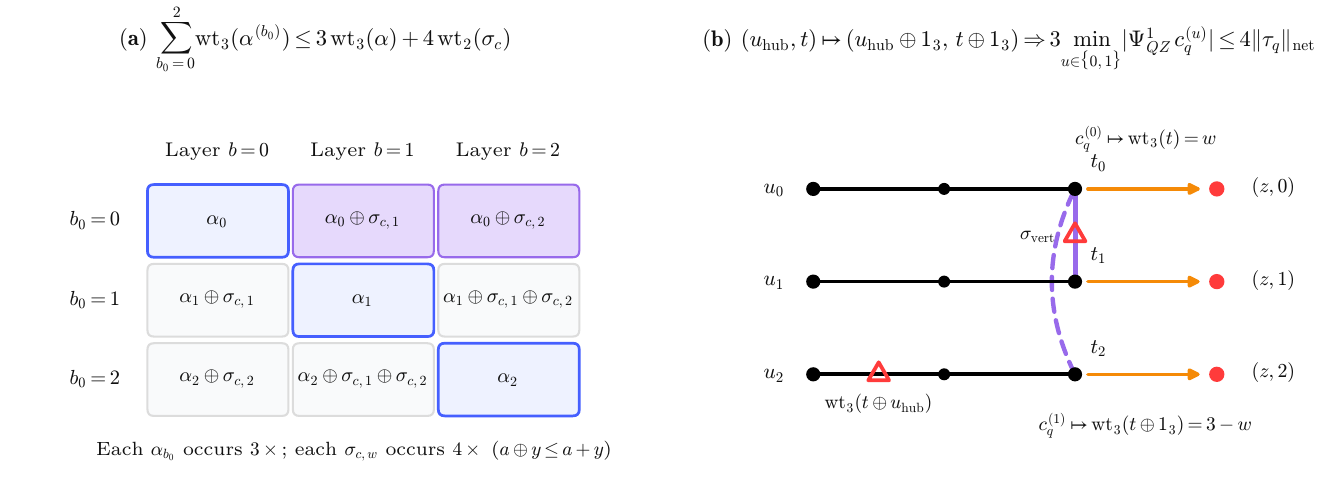}
\caption{Algebraic mechanisms behind the Stage~1 and Stage~2 $\frac{4}{3}$ export bounds. \textbf{(a)} In Stage~1 (Lemma~\ref{lem:stage1-cylinder-cleaner}), the $3 \times 3$ table across reference layers $b_0 \in \{0, 1, 2\}$ (rows) and horizontal layers $b \in \{0, 1, 2\}$ (columns) shows that for candidate export vectors $\alpha^{(b_0)} \in \F^3$ built from an arm's horizontal parity $\alpha = (\alpha_0, \alpha_1, \alpha_2) \in \F^3$ and bridge parity $\sigma_c = (\sigma_{c,1}, \sigma_{c,2}) \in \F^2$ (Section~\ref{subsec:stage1-cylinder}), each $\alpha_b$ (blue diagonal cells) appears $3$ times and each $\sigma_{c,w}$ ($w \in \{1, 2\}$, violet/gray off-diagonal cells) appears $4$ times, bounding the summed export of the three candidates by $4\abs{\sigma_x}$ via $\Z_{\ge 0}$-subadditivity. \textbf{(b)} In Stage~2 (Lemma~\ref{lem:stage2-star-cleaner}), along the three horizontal layers $b \in \{0, 1, 2\}$ of a star arm (with hub and exported tip vectors $u_{\mathrm{hub}} = (u_b)_{b=0}^2, t = (t_b)_{b=0}^2 \in \F^3$, arm parity $t \oplus u_{\mathrm{hub}}$, purple bridge checks, and red $\triangle$ defects; Section~\ref{subsec:stage2-star}), $c_q^{(0)}$ (exporting $t$ of weight $w := \wt_3(t)$ along orange arrows to $(z,b)$) and $c_q^{(1)} = c_q^{(0)} \oplus \bar{\gamma}_q$ (exporting $t \oplus \mathbf{1}_3$ of weight $3 - w$) are bitwise complements, giving $\min_{u \in \{0,1\}}\abs{\Psi_{QZ}^1 c_q^{(u)}} \le \frac{4}{3}\abs{\tau_q}$.}
\label{fig:subadditivity-complement}
\end{figure}

To measure intermediate syndromes exported onto $Y_q^{(3)}$, write $\tau_{q,z}$ for the restriction of $\tau_q$ to the checks along arm $z \in \partial_Z^\top(q)$, $\sigma_{\mathrm{arm}}(q, z, b) := \sum_{k=0}^M \tau_q(e_{q,z,k}, b) \in \F$ ($b \in \{0, 1, 2\}$) for the net horizontal arm parity, and $\sigma_{\mathrm{vert}}(q, z, w) := \tau_q(v_{q,z}, [0,w]) \in \F$ ($w \in \{1, 2\}$) for the terminal bridge syndrome along $z \in \partial_Z^\top(q)$, and define the \emph{collapsed star syndrome seminorm} (extended additively over $Q$ via $\norm{\tau|_Q}_{\mathrm{net}} := \sum_{q \in Q}\norm{\tau_q}_{\mathrm{net}}$)
\begin{equation}
    \label{eq:tau-q-net-def}
    \norm{\tau_q}_{\mathrm{net}} \;:=\; \sum_{z \in \partial_Z^\top(q)} \norm{\tau_{q,z}}_{\mathrm{net}}, \qquad
    \norm{\tau_{q,z}}_{\mathrm{net}} \;:=\; \sum_{b=0}^2 \abs{\sigma_{\mathrm{arm}}(q, z, b)} \;+\; \sum_{w=1}^2 \abs{\sigma_{\mathrm{vert}}(q, z, w)} \;\le\; \abs{\tau_{q,z}}.
\end{equation}
Figure~\ref{fig:subadditivity-complement} summarizes the two structural mechanisms that yield the $\frac{4}{3}$ bound in Stage~1 and Stage~2 without case-by-case enumeration: in Stage~1, $\Z_{\ge 0}$-subadditivity of the radial-spoke disk solver $E$ (defined in Section~\ref{subsec:stage1-cylinder}) combined with outer-rim vertical bridge gauge shifts across the three reference layers $b_0 \in \{0, 1, 2\}$ bounds the minimum export by $\frac{1}{3}\cdot 4 = \frac{4}{3}$ of the internal syndrome weight; in Stage~2, layer-complement symmetry $c_q^{(1)} = c_q^{(0)} \oplus \bar{\gamma}_q$ bounds an asymmetric $2:1$ combination ($2 W_0 + W_1$ or $W_0 + 2 W_1$, where $W_u := \abs{\Psi_{QZ}^1 c_q^{(u)}}$) on mixed-hub cosets---and $3 W_u$ on uniform-hub cosets---by $4\norm{\tau_q}_{\mathrm{net}} \le 4\abs{\tau_q}$, so the better candidate exports at most $\frac{4}{3}\abs{\tau_q}$ in the tie-free three-layer geometry.

\subsection{Stage 1: Cylinder Subadditivity}
\label{subsec:stage1-cylinder}

Consider a single cylinder block $Y_x^{(3)}$ ($x \in M_X$, Definition~\ref{def:three-layer-mapping-cone}(1)), whose internal $2$-cells decompose into horizontal plaquettes $C_H^2(Y_x^{(3)}) = C^2(Y_x^{\mathrm{disk}}) \times \{0, 1, 2\}$ (producing syndromes $\sigma_H = (\sigma_{H,0}, \sigma_{H,1}, \sigma_{H,2})$) and vertical bridge plaquettes $C_V^2(Y_x^{(3)}) = C^1(\partial Y_x^{\mathrm{disk}}) \times \{[0,1], [0,2]\}$ (producing syndromes $\sigma_V = (\sigma_{V,1}, \sigma_{V,2})$). When a local chain $c_x \in C^1(Y_x^{(3)})$ cleans the internal cylinder syndrome $d_x^1 c_x = \sigma_x := \sigma|_{Y_x^{(3)}}$, its outer-rim values export a net residual syndrome $\norm{\Psi_{XQ}^1 c_x}_{\mathrm{net}}$ across $\partial Y_x^{(3)}$ onto the adjacent star arms in $Y_q^{(3)}$ (with $\Psi_{QZ}|_{Y_x^{(3)}} = 0$).

\begin{figure}[t]
\centering
\includegraphics[width=0.90\textwidth]{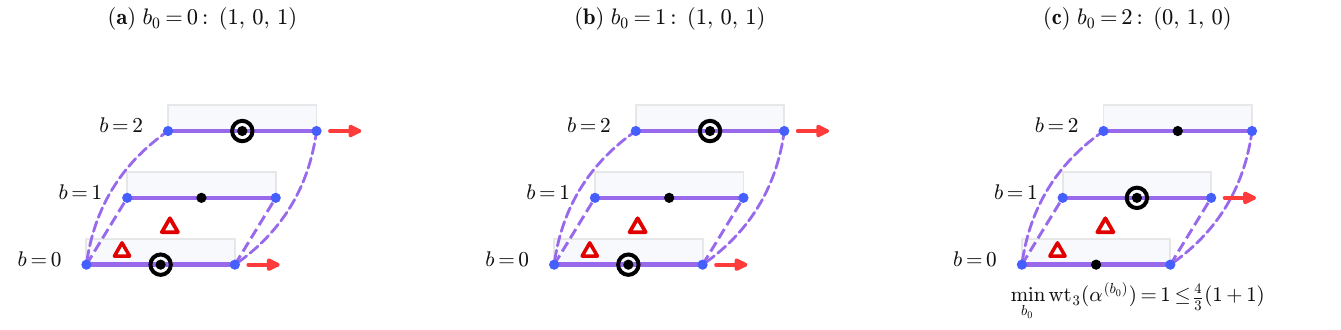}
\caption{The three reference-layer candidate chains $c_x^{(b_0)}$ ($b_0 \in \{0, 1, 2\}$) in Stage~1 cylinder cleaning (Lemma~\ref{lem:stage1-cylinder-cleaner}) for a boundary arm with horizontal parities $(\alpha_0, \alpha_1, \alpha_2) = (1, 0, 0)$ and bridge parities $(\sigma_{c,1}, \sigma_{c,2}) = (1, 0)$, giving export vectors $\alpha^{(b_0)}$: \textbf{(a)} $b_0 = 0 : (1, 0, 1)$, \textbf{(b)} $b_0 = 1 : (1, 0, 1)$, and \textbf{(c)} $b_0 = 2 : (0, 1, 0)$ across the three outer-rim strips $b \in \{0, 1, 2\}$ (violet boundary edges with blue endpoints, central black qubit, dashed violet vertical bridges, and active red syndrome triangles $\triangle$ on strip $b = 0$ and bridge $[0,1]$), where active boundary qubits are circled by open black rings ($\circ$) with red outward export arrows and $b_0 = 2$ achieves $\min_{b_0}\wt_3(\alpha^{(b_0)}) = 1 \le \frac{4}{3}(1 + 1)$.}
\label{fig:stage1-collar-candidates}
\end{figure}

For each horizontal layer $b \in \{0, 1, 2\}$, define the \emph{canonical radial-spoke gauge 2D disk solver} $E : \im(\delta_{\mathrm{disk}}^1) \to C^1(Y_x^{\mathrm{disk}})$ that maps any horizontal $2$-cochain $s \in \im(\delta_{\mathrm{disk}}^1)$ on $Y_x^{\mathrm{disk}}$ to the unique $1$-cochain $E(s) \in C^1(Y_x^{\mathrm{disk}})$ vanishing on the canonical spanning tree $T_x^{\mathrm{rad}} \subset Y_x^{\mathrm{disk}}$ (that is, vanishing on all radial-parent spokes and on the canonical $r = 0$ corner azimuthal segment $e_{\mathrm{can}}(x,z)$ for each co-incident check $z$) and satisfying $\delta_{\mathrm{disk}}^1(E(s)) = s$ by integrating outward in each sector $(q, z)$. On each boundary arm $(q, z)$ along $\partial Y_x^{\mathrm{disk}}$, the collapsed net parity $\alpha_s(q,z) := \sum_{k=0}^M E(s)(e_{x,q,z,k}) \in \F$ across the $M + 1$ subdivided rim segments equals the modulo-$2$ sum of the $r \ge 1$ sector plaquettes of $s$ in $(q, z)$ when $q = q_{\mathrm{can}}(x,z)$, plus the merged $r = 0$ corner square $s_{x,z}^{(0)}$ when $q \neq q_{\mathrm{can}}(x,z)$. Because these plaquette sets are pairwise disjoint across $(q, z)$, the collapsed boundary export weight satisfies $\norm{E(s)}_{\partial} := \sum_{(q,z)}\abs{\alpha_s(q,z)} \le \abs{s}$, which is the exact quantity entering Stage~2. Since $\F$-addition lifted to $\Z_{\ge 0}$ via $0 \mapsto 0, 1 \mapsto 1$ is subadditive ($\abs{a \oplus b} \le \abs{a} + \abs{b}$), $s \mapsto \norm{E(s)}_{\partial}$ is $\Z_{\ge 0}$-subadditive:
\begin{equation}
    \label{eq:E-subadditivity}
    \norm{E(s_1 \oplus s_2)}_{\partial} \;\le\; \norm{E(s_1)}_{\partial} + \norm{E(s_2)}_{\partial}, \qquad
    \norm{E(s)}_{\partial} \;\le\; \abs{s}.
\end{equation}
Since $\sigma_x = (\sigma_H, \sigma_V) \in \im(d_x^1)$, there exists a unique baseline $1$-cochain $\tilde{e}_x = (E(\sigma_{H,b}), \tilde{u}_w) \in C^1(Y_x^{(3)})$ with horizontal components $E(\sigma_{H,b})$ and vertical bridge $1$-cochains $\tilde{u}_w \in C^0(\partial Y_x^{\mathrm{disk}})$ ($w \in \{1, 2\}$, normalized by $\tilde{u}_w(v_{\mathrm{base}}) = 0$ at a chosen rim basepoint $v_{\mathrm{base}} \in \partial Y_x^{\mathrm{disk}}$) satisfying $d_x^1 \tilde{e}_x = \sigma_x$. Fixing a reference layer $b_0 \in \{0, 1, 2\}$, we apply the outer-rim $0$-cochain gauge transformation $c_x^{(b_0)} := \tilde{e}_x \oplus d_x^0 \phi^{(b_0)}$, where $\phi^{(b_0)}$ is supported on $\partial Y_x^{\mathrm{disk}} \times \{0, 1, 2\}$ with $\phi_{b_0}^{(b_0)} := 0$ and $\phi_w^{(b_0)} := \phi_0^{(b_0)} \oplus \tilde{u}_w$ ($w \in \{1, 2\}$) so that all vertical bridge $1$-cells vanish ($c_{V,w}^{(b_0)} = 0$). Under this gauge shift, the exported boundary chain on layer $b \neq b_0$ equals that of reference layer $b_0$ shifted by $\sigma_{V, b_0 \to b} := \sigma_{V,b} \oplus \sigma_{V,b_0}$ (setting $\sigma_{V,0} := 0$), with arm parities $\alpha_b(q,z) := \alpha_{\sigma_{H,b}}(q,z)$ and $\sigma_{c,w}(q,z) := \sum_{k=0}^M \sigma_{V,w}(e_{x,q,z,k}) \in \F$ (Figure~\ref{fig:stage1-collar-candidates}).

Summing the boundary exports across all three choices of reference layer counts each horizontal sheet syndrome three times and each outer-rim vertical bridge defect at most four times, so choosing the smallest-export candidate guarantees at most a $4/3$ expansion of the internal cylinder syndrome. In two layers, the corresponding cylinder cleaner of~\cite[Section~4.3.1, Lemma~4.8]{BDL24} achieves export at most $\abs{\sigma_x}$ by solving a link-graph parity coboundary on the sparser sheet and pairing the remaining defects by open strings across four subregions per sector; in three layers, our canonical radial-spoke solver $E$ and outer-rim gauge shifts replace sector-by-sector string pairing with a single subadditivity average across the three reference layers, trading a factor of $4/3$ here for the $3$-fold check replication in Stage~2.

\leanlink{lem:stage1-cylinder-cleaner}
\begin{lemma}[{Stage 1 Cylinder Cleaner $\cK_X$ and the $\frac{4}{3}$ Export Bound, adapted from~\cite[Section~4.3.1, Lemma~4.8]{BDL24}}]
\label{lem:stage1-cylinder-cleaner}
For every coarse $X$-check $x \in M_X$ and every local cylinder syndrome $\sigma_x = (\sigma_H, \sigma_V) \in \im(d_x^1) \subset C^2(Y_x^{(3)})$, let $b_0^* := \argmin_{b_0 \in \{0,1,2\}} \norm{\Psi_{XQ}^1 c_x^{(b_0)}}_{\mathrm{net}}$ (breaking ties by choosing the smallest index $b_0 \in \{0, 1, 2\}$, so $\cK_{X,x}(0) = 0$) and define $\cK_{X,x}(\sigma_x) := c_x^{(b_0^*)} \in C^1(Y_x^{(3)})$ to be the minimum-export candidate among the three reference-layer solutions $c_x^{(0)}, c_x^{(1)}, c_x^{(2)}$. Then $d_x^1(\cK_{X,x}(\sigma_x)) = \sigma_x$, $\cK_{X,x}(\sigma_x)$ vanishes on all vertical bridge $1$-cells, and its net exported syndrome weight onto the adjacent star arms satisfies
\begin{equation}
    \label{eq:stage1-4-3-bound}
    \norm{\Psi_{XQ}^1\,\cK_{X,x}(\sigma_x)}_{\mathrm{net}} \;\le\; \frac{4}{3}\,\abs{\sigma_x}.
\end{equation}
\end{lemma}

\begin{proof}
Since $\sigma_x = (\sigma_H, \sigma_V) \in \im(d_x^1)$, choose any preimage $e_x = (e_{H,b}, u_w) \in C^1(Y_x^{(3)})$ with $d_x^1 e_x = \sigma_x$. On each horizontal layer $b \in \{0, 1, 2\}$, $\delta_{\mathrm{disk}}^1(e_{H,b} \oplus E(\sigma_{H,b})) = \sigma_{H,b} \oplus \sigma_{H,b} = 0$, and since $Y_x^{\mathrm{disk}}$ is a contractible cone ($H^1(Y_x^{\mathrm{disk}}) = 0$), there exists a $0$-cochain $\xi_b \in C^0(Y_x^{\mathrm{disk}})$ (fixing the global constant of $\xi_w$ for $w \in \{1, 2\}$ by $\xi_w(v_{\mathrm{base}}) = \xi_0(v_{\mathrm{base}}) \oplus u_w(v_{\mathrm{base}})$) such that $e_{H,b} = E(\sigma_{H,b}) \oplus \delta_{\mathrm{disk}}^0 \xi_b$. Subtracting $d_x^0(\xi_0, \xi_1, \xi_2)$ from $e_x$ replaces $e_{H,b}$ by $E(\sigma_{H,b})$ and updates the vertical bridge $1$-cochains to $\tilde{u}_w := u_w \oplus \xi_0|_{\partial Y_x^{\mathrm{disk}}} \oplus \xi_w|_{\partial Y_x^{\mathrm{disk}}} \in C^0(\partial Y_x^{\mathrm{disk}})$ ($\tilde{u}_w(v_{\mathrm{base}}) = 0$ and $\delta_\partial^0$ injective off $v_{\mathrm{base}}$ on connected $\partial Y_x^{\mathrm{disk}}$), so $\tilde{e}_x = (E(\sigma_{H,b}), \tilde{u}_w)$ is uniquely determined by $\sigma_x$ and satisfies $d_x^1 \tilde{e}_x = \sigma_x$, i.e.,
\[
    E(\sigma_{H,0})\big|_{\partial Y_x^{\mathrm{disk}}} \oplus E(\sigma_{H,w})\big|_{\partial Y_x^{\mathrm{disk}}} \;=\; \sigma_{V,w} \oplus \delta_{\partial}^0 \tilde{u}_w \qquad (w \in \{1, 2\}).
\]
For each reference layer $b_0 \in \{0, 1, 2\}$, define the outer-rim $0$-cochain $\phi^{(b_0)} \in C^0(Y_x^{(3)})$ supported on $\partial Y_x^{\mathrm{disk}} \times \{0, 1, 2\}$ (extended by $0$ on all interior vertices of $Y_x^{\mathrm{disk}}$) by
\[
    \phi^{(0)} := (0,\, \tilde{u}_1,\, \tilde{u}_2), \qquad
    \phi^{(1)} := (\tilde{u}_1,\, 0,\, \tilde{u}_1 \oplus \tilde{u}_2), \qquad
    \phi^{(2)} := (\tilde{u}_2,\, \tilde{u}_1 \oplus \tilde{u}_2,\, 0),
\]
and set $c_x^{(b_0)} := \tilde{e}_x \oplus d_x^0 \phi^{(b_0)}$. By construction, $d_x^1 c_x^{(b_0)} = d_x^1 \tilde{e}_x = \sigma_x$, all vertical bridge $1$-cells of $c_x^{(b_0)}$ vanish ($c_{V,w}^{(b_0)} = \tilde{u}_w \oplus \phi_0^{(b_0)} \oplus \phi_w^{(b_0)} = 0$), and on each boundary arm $(q, z)$ along $\partial Y_x^{\mathrm{disk}}$ the restriction $c_{H,b}^{(b_0)}|_{\partial Y_x^{\mathrm{disk}}} = E(\sigma_{H,b_0})|_{\partial Y_x^{\mathrm{disk}}} \oplus \sigma_{V, b_0 \to b}$ has collapsed boundary export parity
\begin{equation}
    \label{eq:candidate-b0-formula}
    \alpha_b^{(b_0)} \;:=\; \alpha_{b_0} \;\oplus\; \sigma_{c, b_0 \to b}, \qquad
    \sigma_{c, b_0 \to b} \;:=\; \begin{cases}
        0 & \text{if } b = b_0,\\
        \sigma_{c, b} & \text{if } b_0 = 0, b \in \{1,2\},\\
        \sigma_{c, b_0} & \text{if } b_0 \in \{1,2\}, b = 0,\\
        \sigma_{c, b_0} \oplus \sigma_{c, b} & \text{if } \{b_0, b\} = \{1, 2\},
    \end{cases}
\end{equation}
where $\alpha_b(q, z) \in \F$ and $\sigma_{c,w}(q, z) \in \F$ denote the arm parities of $E(\sigma_{H,b})|_{\partial Y_x^{\mathrm{disk}}}$ and $\sigma_{V,w}$ on $(q, z)$. Using $\norm{E(\sigma_{H, b_0})}_{\partial} = \sum_{(q,z)}\abs{\alpha_{b_0}(q,z)} \le \abs{\sigma_{H, b_0}}$, $\sum_{(q,z)}\abs{\sigma_{c,w}(q,z)} \le \abs{\sigma_{V,w}}$, and $\Z_{\ge 0}$-subadditivity of~\eqref{eq:candidate-b0-formula} across the three layers $b \in \{0, 1, 2\}$ gives
\begin{align}
    \norm{\Psi_{XQ}^1 c_x^{(0)}}_{\mathrm{net}} &\;\le\; 3\norm{E(\sigma_{H,0})}_{\partial} + \abs{\sigma_{V,1}} + \abs{\sigma_{V,2}}, \nonumber\\
    \norm{\Psi_{XQ}^1 c_x^{(1)}}_{\mathrm{net}} &\;\le\; 3\norm{E(\sigma_{H,1})}_{\partial} + 2\abs{\sigma_{V,1}} + \abs{\sigma_{V,2}}, \label{eq:three-candidates-export}\\
    \norm{\Psi_{XQ}^1 c_x^{(2)}}_{\mathrm{net}} &\;\le\; 3\norm{E(\sigma_{H,2})}_{\partial} + \abs{\sigma_{V,1}} + 2\abs{\sigma_{V,2}}. \nonumber
\end{align}
In the $3 \times 3$ table of candidate arm exports~\eqref{eq:candidate-b0-formula} across $b_0, b \in \{0, 1, 2\}$ (Figure~\ref{fig:subadditivity-complement}(a)), each $\alpha_{b_0}$ appears three times and each bridge term $\sigma_{c,w}$ appears four times, so summing~\eqref{eq:three-candidates-export} over $b_0 \in \{0, 1, 2\}$ gives
\begin{equation}
    \label{eq:sum-three-candidates}
    \sum_{b_0 = 0}^2 \norm{\Psi_{XQ}^1 c_x^{(b_0)}}_{\mathrm{net}} \;\le\; 3\bigl(\abs{\sigma_{H,0}} + \abs{\sigma_{H,1}} + \abs{\sigma_{H,2}}\bigr) + 4\bigl(\abs{\sigma_{V,1}} + \abs{\sigma_{V,2}}\bigr) \;\le\; 4\abs{\sigma_x}.
\end{equation}
Taking the minimum over the $3$ candidates $b_0 \in \{0, 1, 2\}$ yields $\min_{b_0} \norm{\Psi_{XQ}^1 c_x^{(b_0)}}_{\mathrm{net}} \le \frac{4}{3}\abs{\sigma_x}$.
\end{proof}

\subsection{Stage 2: Star Complement Symmetry}
\label{subsec:stage2-star}

After Stage~1 subtracts $\delta_{\widetilde{\cC}}^1\cK_X(\sigma)$, the intermediate syndrome $\tau$~\eqref{eq:tau-intermediate-def} is supported purely on $\cS_Q^2 \oplus \cS_Z^2$. Because $\tau \in \im(\delta_{\widetilde{\cC}}^1)$ and $\tau|_{\cS_X} = 0$, $H^1(Y_x^{(3)}) = 0$ (Lemma~\ref{lem:block-sdr}(a)) and $d_q^1\Psi_{XQ}^0 = \Psi_{XQ}^1 d_x^0$ ensure that on each star block $Y_q^{(3)}$ ($q \in Q$) the restriction $\tau_q := \tau|_{Y_q^{(3)}}$ lies in $\im(d_q^1)$, so we can describe the affine solution space to $d_q^1(c_q) = \tau_q$ explicitly.

\begin{figure}[t]
\centering
\includegraphics[width=0.88\textwidth]{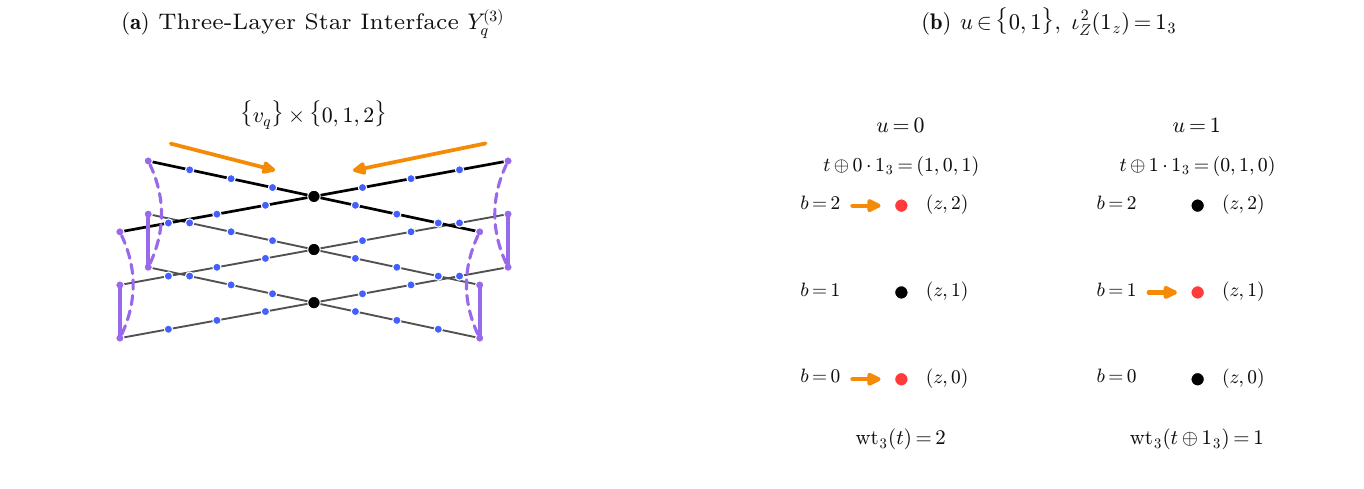}
\caption{Stage~2 star interface cleaning and complement majority selection on $Y_q^{(3)}$ (Lemma~\ref{lem:stage2-star-cleaner}). \textbf{(a)} Three-layer star interface $Y_q^{(3)}$ consisting of three horizontal copies $X'_q \times \{0, 1, 2\}$ with central black hubs $\{v_q\} \times \{0, 1, 2\}$, four radial arms per layer (blue interior nodes, violet tip nodes), and violet terminal vertical bridges $[0,1], [0,2]$ at the boundary tips $\partial X'_q$, with orange arrows pulling arm parities inward to fix the hub coset. \textbf{(b)} Comparison of the two hub candidates $u = 0$ ($t \oplus 0\cdot\mathbf{1}_3 = (1, 0, 1)$ for $t := (t_{q,z,b})_{b=0}^2 \in \F^3$, $\wt_3(t) = 2$, two red active checks and horizontal orange export arrows) and $u = 1$ ($t \oplus 1\cdot\mathbf{1}_3 = (0, 1, 0)$, $\wt_3(t \oplus \mathbf{1}_3) = 1$, one red active check) across layers $b \in \{0, 1, 2\}$.}
\label{fig:stage2-star-majority}
\end{figure}

Recall from Definition~\ref{def:three-layer-mapping-cone}(2) and Lemma~\ref{lem:block-sdr}(b) that $C^0(Y_q^{(3)}) = 0$ ($d_q^0 = 0$) and $\ker(d_q^1) = \F \cdot \bar{\gamma}_q$, where $\bar{\gamma}_q = \sum_{u \in C^0(X'_q),\, b \in \{0,1,2\}}(u, b)$ is the fundamental all-ones horizontal star $1$-cocycle (cf.~\cite[Section~4.3.3, Lemma~4.12]{BDL24}). With $\sigma_{\mathrm{arm}}(q, z, b)$ and $\sigma_{\mathrm{vert}}(q, z, w)$ as in~\eqref{eq:tau-q-net-def} (for the bridge-free candidate above, $\sigma_{\mathrm{vert}}(q, z, w) = \sigma(v_{q,z}, [0,w])$, though~\eqref{eq:stage2-4-3-bound} holds for any Stage-1 cleaner obeying~\eqref{eq:stage1-4-3-bound}), any $1$-cochain $c_q \in C^1(Y_q^{(3)})$ satisfying the horizontal arm equations of $d_q^1(c_q) = \tau_q$ is uniquely determined by its $3$-layer hub value $u_{\mathrm{hub}} = (u_0, u_1, u_2) \in \F^3$ at $(v_q, b)$: integrating outward along arm $(q, z)$ sets the terminal tip values and terminal vertical bridge compatibility relations to
\begin{equation}
    \label{eq:star-hub-compatibility}
    t_{q,z,b} \;=\; u_b \oplus \sigma_{\mathrm{arm}}(q, z, b), \qquad
    u_w \oplus u_0 \;=\; \sigma_{\mathrm{arm}}(q, z, w) \oplus \sigma_{\mathrm{arm}}(q, z, 0) \oplus \sigma_{\mathrm{vert}}(q, z, w) \quad (w \in \{1, 2\}).
\end{equation}
Thus there are exactly two solutions $c_q^{(0)}$ (with $u_0 = 0$) and $c_q^{(1)} := c_q^{(0)} \oplus \bar{\gamma}_q$ (with $u_0 = 1$), whose terminal tip values exported by $\Psi_{QZ}^1$ to $(z, b)$ are
\begin{align}
    \bigl(\Psi_{QZ}^1 c_q^{(0)}\bigr)(z, b) &\;=\; u_b^{(0)} \oplus \sigma_{\mathrm{arm}}(q, z, b), \nonumber\\
    \bigl(\Psi_{QZ}^1 c_q^{(1)}\bigr)(z, b) &\;=\; 1 \oplus u_b^{(0)} \oplus \sigma_{\mathrm{arm}}(q, z, b) \qquad (z \in \partial_Z^\top(q),\, b \in \{0, 1, 2\}), \label{eq:star-two-candidates}
\end{align}
so on each arm $z \in \partial_Z^\top(q)$ the exported 3-layer vector of $c_q^{(1)}$ is the exact bitwise complement in $\F^3$ across layers $\{0, 1, 2\}$ of that of $c_q^{(0)}$ (Figure~\ref{fig:stage2-star-majority}).

Because the two candidate star corrections export bitwise-complementary patterns across the three layers of every incident coarse check, choosing the lighter candidate prevents single-bridge ties and limits the exported weight to at most $4/3$ of the collapsed incoming star syndrome. In the two-layer star cleaner of~\cite[Section~4.3.3, Lemma~4.12]{BDL24}, toggling the all-ones star cocycle swaps $(0, 0)$ and $(1, 1)$ tip pairs on bridge-free arms and pairs each remaining $(1, 1)$ arm against a $(0, 0)$ arm to bound export by the full star weight $\abs{\tau_q}$; in three layers, complement symmetry ($w_z^{(1)} = 3 - w_z^{(0)}$ for exported arm weights $w_z^{(u)}$) bounds an asymmetric $2 : 1$ combination arm by arm on mixed-hub cosets (and $3 w_z^{(u)}$ on uniform-hub cosets) against the collapsed arm-and-bridge weight $\norm{\tau_q}_{\mathrm{net}}$ coming out of Stage~1. Once all cylinder and star defects are cleared, absorbing the residual cylinder $1$-cocycle via $H^1(Y_x^{(3)}) = 0$ (as in~\cite[Sections~4.3.2 and~4.3.4, Lemmas~4.10 and~4.14]{BDL24}) replicates every remaining coarse check defect across all three horizontal layers rather than two, so its coarse Hamming weight is one third of the residual fine syndrome weight.

\leanlink{lem:stage2-star-cleaner}
\begin{lemma}[{Stage 2 Star Cleaner $\cK_Q$ and the $\frac{4}{3}$ Majority Bound, adapted from~\cite[Sections~4.3.2--4.3.4, Lemmas~4.10, 4.12, and~4.14]{BDL24}}]
\label{lem:stage2-star-cleaner}
For every coarse qubit $q \in Q$ and realizable star syndrome $\tau_q \in \im(d_q^1) \subset C^2(Y_q^{(3)})$, let $u^* := \argmin_{u \in \{0,1\}} \abs{\Psi_{QZ}^1 c_q^{(u)}}$ (breaking ties by choosing $u = 0$, so $\cK_{Q,q}(0) = 0$) and define $\cK_{Q,q}(\tau_q) := c_q^{(u^*)} \in C^1(Y_q^{(3)})$ to be the smaller-export candidate between $c_q^{(0)}$ and $c_q^{(1)} = c_q^{(0)} \oplus \bar{\gamma}_q$. Then:
\begin{enumerate}[label=\textbf{(\roman*)}, leftmargin=2.2em, itemsep=2.5pt]
    \item $d_q^1(\cK_{Q,q}(\tau_q)) = \tau_q$, and for the collapsed star syndrome seminorm $\norm{\tau_q}_{\mathrm{net}}$ of~\eqref{eq:tau-q-net-def},
    \begin{equation}
        \label{eq:stage2-4-3-bound}
        \abs{\Psi_{QZ}^1\,\cK_{Q,q}(\tau_q)} \;\le\; \frac{4}{3}\,\norm{\tau_q}_{\mathrm{net}} \;\le\; \frac{4}{3}\,\abs{\tau_q}.
    \end{equation}
    \item Let $c_{\mathrm{clean}}(\sigma) := \sum_{x \in M_X} \cK_{X,x}(\sigma|_{Y_x^{(3)}}) \oplus \sum_{q \in Q} \cK_{Q,q}(\tau_q) \in C^1(\widetilde{\cC})$. For any fine chain $e \in C^1(\widetilde{\cC})$ with $\delta_{\widetilde{\cC}}^1 e = \sigma$, its cleaned coarse chain projection $e_{\mathrm{coarse}} := \widetilde{\pi}^1(e \oplus c_{\mathrm{clean}}(\sigma)) \in C^1(\cC)$ satisfies $\delta_{\cC}^1(e_{\mathrm{coarse}}) = \cR_X^{\syn}(\sigma)$, and the final residual syndrome $\tau_{\mathrm{final}} = \sigma \oplus \delta_{\widetilde{\cC}}^1(c_{\mathrm{clean}}(\sigma))$ is replicated identically across all three horizontal layers $b \in \{0, 1, 2\}$:
    \begin{equation}
        \label{eq:exact-three-layer-replication}
        \tau_{\mathrm{final}} \;=\; \iota_Z^2\bigl(\cR_X^{\syn}(\sigma)\bigr) \;=\; \sum_{z \in M_Z} \cR_X^{\syn}(\sigma)(z) \sum_{b=0}^2 (z, b), \qquad
        \bigl\lvert\cR_X^{\syn}(\sigma)\bigr\rvert \;=\; \frac{1}{3}\,\abs{\tau_{\mathrm{final}}}.
    \end{equation}
\end{enumerate}
\end{lemma}

\begin{proof}
For (i), since $d_q^1(\bar{\gamma}_q) = 0$ and $d_q^1(c_q^{(0)}) = \tau_q$, both $c_q^{(0)}$ and $c_q^{(1)}$ satisfy $d_q^1(c_q^{(u)}) = \tau_q$. To bound the exported weight $W_u := \abs{\Psi_{QZ}^1 c_q^{(u)}} = \sum_{z \in \partial_Z^\top(q)} w_z^{(u)}$, note that on each arm $z \in \partial_Z^\top(q)$, $c_q^{(0)}$ exports weight $w_z^{(0)} := \sum_{b=0}^2 \abs{u_b^{(0)} \oplus \sigma_{\mathrm{arm}}(q,z,b)} \in \{0, 1, 2, 3\}$ while $c_q^{(1)} = c_q^{(0)} \oplus \bar{\gamma}_q$ exports the complementary weight $w_z^{(1)} = 3 - w_z^{(0)}$, and the terminal bridge syndrome satisfies $\sigma_{\mathrm{vert}}(q, z, w) = (u_w^{(0)} \oplus \sigma_{\mathrm{arm}}(q, z, w)) \oplus (u_0^{(0)} \oplus \sigma_{\mathrm{arm}}(q, z, 0))$ for $w \in \{1, 2\}$. When $u^{(0)} = (0, 0, 0)$, we have $w_z^{(0)} = \sum_{b=0}^2 \abs{\sigma_{\mathrm{arm}}(q,z,b)} \le \norm{\tau_{q,z}}_{\mathrm{net}}$ on every arm, so $3 W_0 \le 3\norm{\tau_q}_{\mathrm{net}} \le 4\norm{\tau_q}_{\mathrm{net}}$. When $\wt_3(u^{(0)}) = 1$, a direct check over the $8$ vectors $(\sigma_{\mathrm{arm}}(q, z, b))_{b=0}^2 \in \F^3$ shows that every arm $z \in \partial_Z^\top(q)$ satisfies the asymmetric linear inequality $2 w_z^{(0)} + w_z^{(1)} \le 4\norm{\tau_{q,z}}_{\mathrm{net}}$, so summing over $z \in \partial_Z^\top(q)$ gives $3\min(W_0, W_1) \le 2 W_0 + W_1 \le 4\norm{\tau_q}_{\mathrm{net}}$. Symmetrically, when $\wt_3(u^{(0)}) \ge 2$, the complementary hub vector $u^{(1)} = \mathbf{1}_3 \oplus u^{(0)}$ has $\wt_3(u^{(1)}) \le 1$, so the same two subcases with $W_0$ and $W_1$ interchanged yield $3 W_1 \le 3\norm{\tau_q}_{\mathrm{net}}$ (if $\wt_3(u^{(1)}) = 0$) or $3\min(W_0, W_1) \le W_0 + 2 W_1 \le 4\norm{\tau_q}_{\mathrm{net}}$ (if $\wt_3(u^{(1)}) = 1$). In every case, $3\min(W_0, W_1) \le 4\norm{\tau_q}_{\mathrm{net}} \le 4\abs{\tau_q}$, proving~\eqref{eq:stage2-4-3-bound}.

For (ii), $\tilde{e} := e \oplus c_{\mathrm{clean}}(\sigma)$ satisfies $\delta_{\widetilde{\cC}}^1(\tilde{e}) = \tau_{\mathrm{final}}$, which vanishes identically on $\bigoplus_x C^2(Y_x^{(3)}) \oplus \bigoplus_q C^2(Y_q^{(3)})$. Because $d_x^1\tilde{e}_X = 0$ on every cylinder $Y_x^{(3)}$ and $H^1(Y_x^{(3)}) = 0$ (Lemma~\ref{lem:block-sdr}(a)), $\tilde{e}_X = d_x^0(\alpha_X)$ for some $0$-cochain $\alpha_X$. Replacing $\tilde{e}$ by $\tilde{e}' := \tilde{e} \oplus \delta_{\widetilde{\cC}}^0(\alpha_X)$ preserves both $\delta_{\widetilde{\cC}}^1(\tilde{e}') = \tau_{\mathrm{final}}$ and the coarse cohomology projection $\widetilde{\pi}^1(\tilde{e}') = \widetilde{\pi}^1(\tilde{e}) \oplus \delta_{\cC}^0(\widetilde{\pi}^0 \alpha_X)$, while reducing $\tilde{e}'$ to $\bigoplus_q C^1(Y_q^{(3)})$ with $d_q^1(\tilde{e}'_q) = 0$ on every star $Y_q^{(3)}$. By Lemma~\ref{lem:block-sdr}(b), $\ker(d_q^1) = \F \cdot \bar{\gamma}_q$, where $\widetilde{\pi}^1(\bar{\gamma}_q) = \mathbf{1}_q$ and
\[
    \delta_{\widetilde{\cC}}^1(\bar{\gamma}_q) \;=\; \Psi_{QZ}^1(\bar{\gamma}_q) \;=\; \sum_{z \in \partial_Z^\top(q)}\sum_{b=0}^2 (z, b) \;=\; \iota_Z^2(\delta_{\cC}^1 \mathbf{1}_q).
\]
Hence $\tau_{\mathrm{final}} = \iota_Z^2(\delta_{\cC}^1 e_{\mathrm{coarse}})$ with $\cR_X^{\syn}(\sigma) = \delta_{\cC}^1 e_{\mathrm{coarse}}$, establishing~\eqref{eq:exact-three-layer-replication}.
\end{proof}

\begin{figure}[t]
\centering
\includegraphics[width=0.86\textwidth]{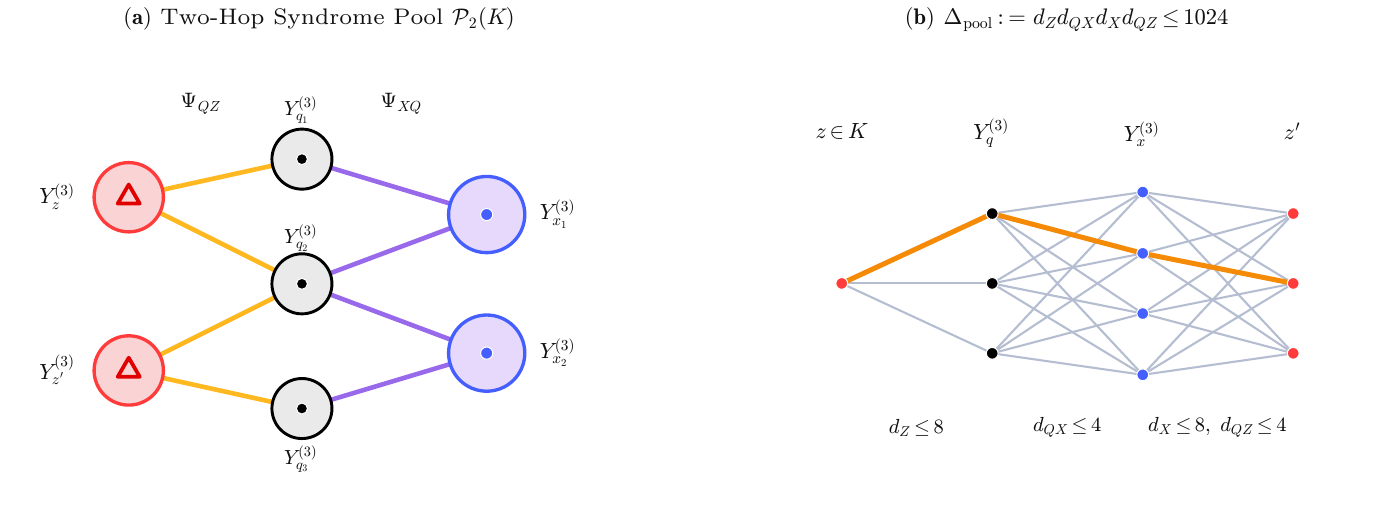}
\caption{The fine two-hop syndrome pool $\cP_2(K)$ and degree-bounding graph (Theorem~\ref{thm:16-27-contraction}). \textbf{(a)} Two-hop syndrome pool $\cP_2(K)$ (eq.~\eqref{eq:two-hop-pool-def}) showing coarse check blocks $Y_z^{(3)}, Y_{z'}^{(3)}$ (pink circles with $\triangle$ markers) connected via amber links $\Psi_{QZ}$ to star blocks $Y_{q_1}^{(3)}, Y_{q_2}^{(3)}, Y_{q_3}^{(3)}$ (gray circles with black centers) and via violet links $\Psi_{XQ}$ to cylinder blocks $Y_{x_1}^{(3)}, Y_{x_2}^{(3)}$ (violet circles with blue centers). \textbf{(b)} Degree-bounding paths in the incidence graph from $z \in K$ (red) through $Y_q^{(3)}$ (black, $d_Z \le 8$) and $Y_x^{(3)}$ (blue, $d_{QX} \le 4$), spanning at most $1 + d_Z + d_Z d_{QX} \le 41$ coarse gadget blocks per check $z \in K$, and onward along four-hop walks (highlighted in orange, via $q' \in \partial_X(x)$ to $z' \in \partial_Z^\top(q')$) to at most $\Delta_{\mathrm{pool}} := d_Z d_{QX} d_X d_{QZ} \le 1024$ pool-overlapping coarse checks $z'$ (red).}
\label{fig:two-hop-pool}
\end{figure}

Because each coarse check receives exported defects only from its incident stars and the cylinders adjacent to those stars, the two-stage cleaner acts locally within two-hop gadget neighborhoods (Figure~\ref{fig:two-hop-pool}). Composing the two $4/3$ export bounds and dividing by the three-fold layer replication contracts the coarse syndrome weight by $\frac{4}{3}\times\frac{4}{3}\times\frac{1}{3} = \frac{16}{27} < 1$, playing the role of the two-layer $1/2$ contraction in~\cite[Section~4.3, Theorem~4.7]{BDL24}. Moreover, whereas~\cite[Section~5.2, Claim~5.5 and Lemma~5.6]{BDL24} tracks locality along connected components of a global multi-level decoding graph, part~(ii) formulates the contraction directly on the fine two-hop $Z$-check cell pool $\cP_2(K)$ of any subset of coarse checks $K \subseteq M_Z$ while keeping two-hop separated subsets uncoupled, supplying the subset-level input used by the syndrome-only witness decoder in Section~\ref{sec:cluster-decoder-peierls}.

\leanlink{thm:16-27-contraction}
\begin{theorem}[{The $\frac{16}{27}$ Bipartite Syndrome Contraction Theorem, adapted from~\cite[Section~4.3, Theorem~4.7, and Section~5.2, Lemma~5.6]{BDL24}}]
\label{thm:16-27-contraction}
Let $\widetilde{\cC} = \cR_X^{(3)}(\cC)$ on a $(8, r_0, \rho_0)$-bounded-geometry CSS complex $\cC$ with $1 \le d_{QX}, d_{QZ} \le 4$, and let $\cR_X^{\syn} : \im(\delta_{\widetilde{\cC}}^1) \to \im(\delta_{\cC}^1)$ be the two-stage syndrome renormalization map defined by $\iota_Z^2(\cR_X^{\syn}(\sigma)) = \tau_{\mathrm{final}}$. Then:
\begin{enumerate}[label=\textbf{(\roman*)}, leftmargin=2.2em, itemsep=2.5pt]
    \item For every realizable fine syndrome $\sigma \in \im(\delta_{\widetilde{\cC}}^1)$, the coarse syndrome $\sigma_{\mathrm{coarse}} = \cR_X^{\syn}(\sigma)$ satisfies
    \begin{equation}
        \label{eq:global-16-27}
        \bigl\lvert\cR_X^{\syn}(\sigma)\bigr\rvert \;\le\; \frac{16}{27}\,\abs{\sigma}, \qquad\text{or equivalently}\qquad
        \abs{\sigma} \;\ge\; \frac{27}{16}\,\bigl\lvert\cR_X^{\syn}(\sigma)\bigr\rvert.
    \end{equation}
    \item For any subset of coarse checks $K \subseteq M_Z$, writing $\partial_Z(K) := \bigcup_{z \in K}\partial_Z(z)$ and $\partial_X^\top(S) := \bigcup_{q \in S}\partial_X^\top(q)$, define its fine two-hop $Z$-check cell pool
    \begin{equation}
        \label{eq:two-hop-pool-def}
        \cP_2(K) \;:=\; \bigsqcup_{z \in K} C^2(Y_z^{(3)}) \;\sqcup\; \bigsqcup_{q \in \partial_Z(K)} C^2(Y_q^{(3)}) \;\sqcup\; \bigsqcup_{x \in \partial_X^\top(\partial_Z(K))} C^2(Y_x^{(3)}).
    \end{equation}
    Let $\mu_{\mathrm{step}} = O(\Lambda^2)$ be the maximum number of fine checks in a single gadget block $Y_v^{(3)}$ ($v \in M_X \sqcup Q \sqcup M_Z$). Then $\abs{\cP_2(K)} \le 41\mu_{\mathrm{step}}\abs{K}$, $\cR_X^{\syn}(\sigma)|_K$ depends only on $\sigma|_{\cP_2(K)}$, and for every realizable fine syndrome $\sigma \in \im(\delta_{\widetilde{\cC}}^1)$,
    \begin{equation}
        \label{eq:local-16-27}
        \bigl\lvert\cR_X^{\syn}(\sigma)\big|_K\bigr\rvert \;\le\; \frac{16}{27}\,\bigl\lvert\sigma\big|_{\cP_2(K)}\bigr\rvert, \qquad\text{so}\qquad
        \bigl\lvert\sigma\big|_{\cP_2(K)}\bigr\rvert \;\ge\; \Bigl\lceil \frac{27}{16}\,\bigl\lvert\cR_X^{\syn}(\sigma)\big|_K\bigr\rvert \Bigr\rceil.
    \end{equation}
    In particular, if $K_1, \dots, K_r \subseteq M_Z$ are two-hop separated ($\partial_X^\top(\partial_Z(K_a)) \cap \partial_X^\top(\partial_Z(K_b)) = \emptyset$ for $a \neq b$), their fine pools $\cP_2(K_a)$ are pairwise disjoint.
\end{enumerate}
\end{theorem}

\begin{proof}
Each coarse check $z \in K$ is incident to at most $d_Z \le 8$ qubits $q \in \partial_Z(z)$, each of which is incident to at most $d_{QX} \le 4$ checks $x \in \partial_X^\top(q)$, so $K \sqcup \partial_Z(K) \sqcup \partial_X^\top(\partial_Z(K))$ comprises at most $(1 + 8 + 8 \cdot 4)\abs{K} = 41\abs{K}$ coarse gadget blocks and hence $\abs{\cP_2(K)} \le 41\mu_{\mathrm{step}}\abs{K}$ fine checks. Decompose $\sigma|_{\cP_2(K)} = \sigma_X \oplus \sigma_Q \oplus \sigma_Z$ into its components on the cylinders $Y_x^{(3)}$ ($x \in \partial_X^\top(\partial_Z(K))$), stars $Y_q^{(3)}$ ($q \in \partial_Z(K)$), and coarse checks $Y_z^{(3)}$ ($z \in K$), so that $\abs{\sigma|_{\cP_2(K)}} = \abs{\sigma_X} + \abs{\sigma_Q} + \abs{\sigma_Z}$. By Lemma~\ref{lem:stage1-cylinder-cleaner}, Stage~1 cleaning on each $Y_x^{(3)}$ exports net star weight $\le \frac{4}{3}\abs{\sigma_X}$ to the adjacent stars (with zero direct export to $Y_z^{(3)}$), so writing $\tau_Q := \tau|_{\partial_Z(K)}$ (with $\norm{\tau_Q}_{\mathrm{net}} := \sum_{q \in \partial_Z(K)} \norm{\tau_q}_{\mathrm{net}}$) and $\tau_Z := \tau|_K$, the intermediate syndrome $\tau$ satisfies $\norm{\tau_Q}_{\mathrm{net}} + \abs{\tau_Z} \le \abs{\sigma_Q} + \abs{\sigma_Z} + \frac{4}{3}\abs{\sigma_X}$. Next, by Lemma~\ref{lem:stage2-star-cleaner}(i), Stage~2 cleaning on each $Y_q^{(3)}$ ($q \in \partial_Z(K)$) exports weight $\le \frac{4}{3}\norm{\tau_Q}_{\mathrm{net}}$ to the coarse checks $Y_z^{(3)}$, so the final three-layer syndrome $\tau_{\mathrm{final}}|_{\bigsqcup_{z \in K}Y_z^{(3)}}$ satisfies
\begin{equation}
    \label{eq:two-stage-composition-bound}
    \bigl\lvert\tau_{\mathrm{final}}\big|_{\bigsqcup_{z \in K}Y_z^{(3)}}\bigr\rvert \;\le\; \abs{\tau_Z} + \frac{4}{3}\norm{\tau_Q}_{\mathrm{net}} \;\le\; \frac{4}{3}\bigl(\abs{\tau_Z} + \norm{\tau_Q}_{\mathrm{net}}\bigr) \;\le\; \frac{4}{3}\Bigl(\abs{\sigma_Z} + \abs{\sigma_Q} + \frac{4}{3}\abs{\sigma_X}\Bigr) \;\le\; \frac{16}{9}\,\bigl\lvert\sigma\big|_{\cP_2(K)}\bigr\rvert.
\end{equation}
Dividing by $3$ via the exact three-layer replication identity $\abs{\cR_X^{\syn}(\sigma)|_K} = \frac{1}{3}\abs{\tau_{\mathrm{final}}|_{\bigsqcup_{z \in K}Y_z^{(3)}}}$ (Lemma~\ref{lem:stage2-star-cleaner}(ii)) yields $27\abs{\cR_X^{\syn}(\sigma)|_K} \le 16\abs{\sigma|_{\cP_2(K)}}$, and since $\abs{\sigma|_{\cP_2(K)}} \in \Z_{\ge 0}$ this implies the integer ceiling bound in~\eqref{eq:local-16-27}. Taking $K = M_Z$ gives~\eqref{eq:global-16-27}.
\end{proof}

\section{Positive-Temperature Self-Correction}
\label{sec:cluster-decoder-peierls}

\subsection{The Apex Walk Obstruction}
\label{subsec:apex-walk}

We next examine how syndromes propagate across the full multiscale hierarchy $\cC_m = (\cR_Z^{(3)} \circ \cR_X^{(3)})^{\circ(k+1)}(\cC_0)$ of alternating primal ($\cR_X^{(3)}$) and dual ($\cR_Z^{(3)}$) replacements. Following Section~4.2 (Theorem~4.6) of~\cite{BDL24}, when $\cR_Z^{(3)} = (\cR_X^{(3)}((\cdot)^\vee))^\vee$ is applied to $\cC_{2j-1}$, it replaces each $Z$-check $z' \in M_{Z, 2j-1}$ by a dual cylinder gadget $Y_{z'}^{(3),\vee}$ whose dual $0$-cohomology is one-dimensional ($H_2(Y_{z'}^{(3),\vee}) \cong H^0(Y_{z'}^{(3)}) \cong \F$). Summing the active fine $Z$-checks inside each dual gadget $Y_{z'}^{(3),\vee}$ modulo~$2$ onto its owner coarse check $z'$ defines a linear coarse-syndrome map $\cR_Z^{\syn}$ with zero export between distinct gadgets and exact coboundary commutation $\cR_Z^{\syn}(\delta_{\cC_{2j}}^1 e) = \delta_{\cC_{2j-1}}^1(\pi_{\mathrm{rad}}^1 e)$ under $\pi_{\mathrm{rad}}^1 := (\widetilde{\iota}_{\cC_{2j-1}^\vee}^1)^\top$, preserving weight via $\abs{\cR_Z^{\syn}(\sigma_{2j})(z')} \le \abs{\sigma_{2j}|_{Y_{z'}^{(3),\vee}}}$~\cite[Section~4.2, Theorem~4.6]{BDL24}.

\begin{figure}[t]
\centering
\includegraphics[width=0.88\textwidth]{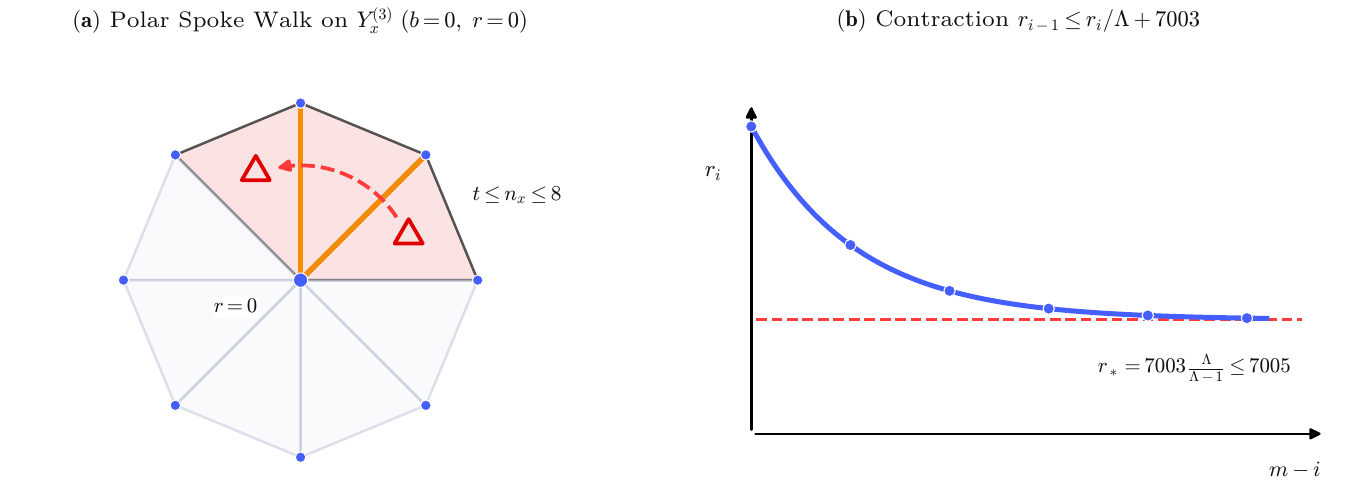}
\caption{The layer-$0$ apex walk on $Y_x^{(3)}$ and coarse-grained radius contraction (Proposition~\ref{prop:apex-walk-discontinuity} and Proposition~\ref{prop:cluster-decoder-witness}(ii)). \textbf{(a)} Polar spoke walk on $Y_x^{(3)}$ ($b = 0, r = 0$) showing the $n_x$-sector wheel ($n_x \le 8$) around the blue apex $r = 0$ with blue rim nodes, three pink-shaded walk sectors, two orange active spokes, and two open red syndrome triangles ($\triangle$) connected by a dashed red arc arrow ($t \le n_x \le 8$). \textbf{(b)} Coarse-grained protection radius $r_i$ versus scale depth $m - i$ (solid blue curve with blue dots approaching the red dashed asymptote $r_* = r_{\mathrm{fix}} := 7003\frac{\Lambda}{\Lambda - 1} \le 7005$, previewing the ball-witness radius recurrence $r_{i-1} = r_i/\Lambda + 7003$ of Proposition~\ref{prop:cluster-decoder-witness}(ii)).}
\label{fig:apex-walk}
\end{figure}

Why, then, do we depart from a bottom-up \emph{stepwise RG decoder} that reconstructs physical recovery chains across scales (cf.\ the lifted recovery map of~\cite[Section~4.1]{BDL24} and the existential coarse error chain of~\cite[Section~4.3.2, Lemma~4.10]{BDL24}) rather than outputting only a logical parity bit? Any minimum-export local cleaner $\cK_X$ on $Y_x^{(3)}$ exhibits an unavoidable single-qubit discontinuity along the \emph{layer-$0$ apex walk} (Figure~\ref{fig:apex-walk}), which motivates our syndrome-only architecture.

Flipping the horizontal radial spokes one by one around the central hub of layer $0$ sweeps the zero error continuously into a full check stabilizer while exciting only the two bordering plaquettes at every intermediate step. Because a symmetric minimum-spoke cleaner sees only those two defects, partway around the hub a single spoke flip forces its coarse residual chain to jump by the full coarse check stabilizer.

\begin{proposition}[The Layer-0 Apex Walk and Stepwise Cleaner Discontinuity]
\label{prop:apex-walk-discontinuity}
Fix a cylinder block $Y_x^{(3)}$ whose link graph $G_x$ is a single cycle on $n_x := \abs{\partial_X(x)} \ge 4$ qubits (e.g.\ an interior plaquette check), and let $e_1, \dots, e_{n_x} \in C^1(Y_x^{\mathrm{disk}}) \times \{0\}$ be the $n_x$ horizontal edges on layer $0$ incident to the apex $(v_x, 0)$, ordered cyclically around $(v_x, 0)$. For $0 \le t \le n_x$, define the partial apex walk $w_t := \sum_{j=1}^t e_j$. Then:
\begin{enumerate}[label=\textbf{(\roman*)}, leftmargin=2.2em, itemsep=2.5pt]
    \item For every $1 \le t < n_x$, the internal horizontal syndrome $\delta_{\mathrm{disk}}^1(w_t)$ consists of only the two radial plaquettes bordering $e_1$ and $e_t$ on layer $0$ (with zero vertical bridge syndrome since the interior apex $(v_x, 0)$ has no vertical bridges), while $w_0 = 0$ and $w_{n_x} = \delta_{\widetilde{\cC}}^0(v_x, 0) \in \im(\delta_{\widetilde{\cC}}^0)$ has vanishing syndrome $\delta_{\widetilde{\cC}}^1(w_{n_x}) = 0$.
    \item Any local cleaner $\cK_{X,x}$ satisfying $\cK_{X,x}(0) = 0$ leaves $w_0$ and $w_{n_x}$ unchanged, whereas the coarse homology pushforward $\widetilde{\pi}^1(w_{n_x})$ equals the full coarse stabilizer $\delta_{\cC}^0(\mathbf{1}_x)$ of Hamming weight $n_x$. Consequently, along the single-qubit walk $w_0, w_1, \dots, w_{n_x}$ there exists a single-qubit transition $w_{t-1} \to w_t$ across which the coarse residual chain changes by Hamming weight at least $1$ (and for symmetric minimum-spoke cleaners, by weight $n_x$ across $w_{t^*} \to w_{t^*+1}$ when $n_x = 2t^* + 1$ is odd, or by total weight $n_x$ across $w_{t^*-1} \to w_{t^*} \to w_{t^*+1}$ giving a step $\ge n_x/2$ when $n_x = 2t^*$ is even).
\end{enumerate}
\end{proposition}

\begin{proof}
Since consecutive radial spokes $e_j, e_{j+1}$ share the intervening sector plaquette around $(v_x, 0)$, for $1 \le t < n_x$ their horizontal coboundaries cancel pairwise on the $t - 1$ shared sector plaquettes between $e_j$ and $e_{j+1}$ ($1 \le j < t$), leaving only the two outer sector plaquettes bordering $e_1$ and $e_t$. At $t = n_x$, summing all $n_x$ horizontal spokes completes the exact coboundary $w_{n_x} = \delta_{\widetilde{\cC}}^0(v_x, 0)$, so $d_x^1(w_{n_x}) = 0$ and $\cK_{X,x}(d_x^1(w_{n_x})) = 0$. However, the difference between the actual chain $w_t$ and its local cleaner reconstruction $c_t = \cK_{X,x}(d_x^1(w_t))$ projects to a coarse chain $r_t = \widetilde{\pi}^1(w_t \oplus c_t)$ that starts at $r_0 = 0$ and ends at $r_{n_x} = \widetilde{\pi}^1(\delta_{\widetilde{\cC}}^0(v_x, 0)) = \delta_{\cC}^0(\mathbf{1}_x)$ of weight $n_x$. By the triangle inequality, $\max_{1 \le t \le n_x} \abs{r_t \oplus r_{t-1}} \ge 1$, and for any symmetric minimum-spoke cleaner ($t^* := \floor{n_x/2}$), switching from the short spoke arc $[1, t^*]$ at $w_{t^*}$ to the complementary spoke arc $[t^*+2, n_x]$ at $w_{t^*+1}$ when $n_x = 2t^* + 1$ is odd (or from $[1, t^*-1]$ to $[t^*+2, n_x]$ across $w_{t^*-1} \to w_{t^*} \to w_{t^*+1}$ when $n_x = 2t^*$ is even) shifts $r_t$ by $\delta_{\cC}^0(\mathbf{1}_x)$ of weight $n_x$.
\end{proof}

In a stepwise RG decoder that lifts physical recovery chains across levels, a collection of disconnected $O(1)$-weight thermal fluctuations sitting near their apex-walk midpoints can jump under a single physical qubit flip, destroying single-qubit continuity of physical recovery chains across scales (Figure~\ref{fig:stepwise-vs-multiscale}(a)).

\subsection{The Multiscale Cluster Decoder}
\label{subsec:cluster-decoder}

\begin{figure}[t]
\centering
\includegraphics[width=0.88\textwidth]{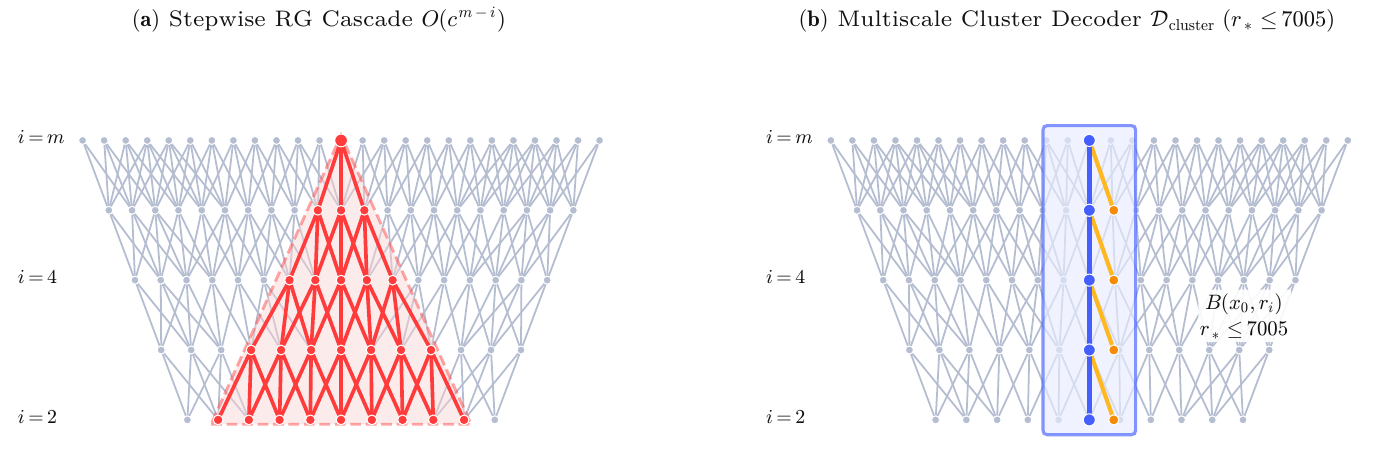}
\caption{Comparison between stepwise error-pushing RG decoding and multiscale syndrome-only parity decoding. \textbf{(a)} Stepwise RG cascade $O(c^{m-i})$ ($c > 1$) where a single red defect node at $i = m$ fans out downward into a red dashed cascade cone of active red nodes and edges across scales $i = m \to 2$. \textbf{(b)} Recursive syndrome-only parity decoding (Definition~\ref{def:multiscale-cluster-decoder}), in which any localized error perturbation at scale $m$ tracks a coarse-grained syndrome spine (blue) with non-cascading local cleaner parity readouts (amber/orange) inside a narrow blue tube $B(x_0, r_i)$ ($r_i \le r_{i+1}/\Lambda + 7003$, asymptote $r_* \le 7005$) down to $\cC_2$.}
\label{fig:stepwise-vs-multiscale}
\end{figure}

To avoid stepwise error cascades, we pass only syndromes and a single logical parity bit across renormalization scales (Figure~\ref{fig:stepwise-vs-multiscale}(b)). Given a syndrome $\sigma \in \im(\delta_{\cC_m}^1)$ at level $m = 2k + 2$, its downward \emph{syndrome trajectory} $(\sigma_i)_{i=2}^m$ is defined solely through the syndrome maps $\cR_X^{\syn}$ and $\cR_Z^{\syn}$:
\begin{equation}
    \label{eq:syndrome-trajectory}
    \sigma_m := \sigma, \qquad
    \sigma_{i-1} \;:=\; \begin{cases}
        \cR_X^{\syn}(\sigma_i) & \text{if } i \text{ is odd } (\cC_i = \cR_X^{(3)}(\cC_{i-1})),\\
        \cR_Z^{\syn}(\sigma_i) & \text{if } i \text{ is even } (\cC_i = \cR_Z^{(3)}(\cC_{i-1})),
    \end{cases} \qquad (i = m, m-1, \dots, 3).
\end{equation}
Because adding a coarse stabilizer $\delta_{\cC}^0(\mathbf{1}_x)$ leaves the coarse syndrome unchanged, $\sigma_{i-1}$ is insensitive to the midpoint jump of Proposition~\ref{prop:apex-walk-discontinuity}. Moreover, since $\dim_{\F} H^1(\cC_m) = 1$, a syndrome decoder needs only output a single parity bit $\psi_m(\sigma) \in \F$ predicting the $\F$-bilinear pairing $\ip{L_{Z,m}}{e} := \sum_{q \in Q_m} (L_{Z,m})_q e_q \in \F$ of the underlying error $e$ with a dual logical cycle $L_{Z,m} \in \ker \partial_1^{\cC_m}$.

In this formulation, the decoder updates its predicted logical parity bit recursively from the coarse-grained syndromes without ever modifying physical qubits at intermediate scales. To certify fault tolerance, we pair the parity rule with a finite pool of minimum-weight check patterns (\emph{witnesses}) such that no error perturbation confined to a ball can flip the decoded logical parity unless at least one witness pattern is fully excited before or after the perturbation.

\leanlink{def:multiscale-cluster-decoder}
\begin{definition}[{Recursive Parity Decoder $\cD_{\mathrm{cluster}}$ and $(R, w, N)$-Ball Parity Witness Families; cf.~\cite[Section~5.2]{BDL24}}]
\label{def:multiscale-cluster-decoder}
Let $(\cC_m, \Phi_m)$ ($m = 2k + 2$, $k \ge 0$) be the three-layer welded hierarchy with dual logical representatives $(L_{X,i}, L_{Z,i}) \in \ker \delta_{\cC_i}^1 \times \ker \partial_1^{\cC_i}$ ($\ip{L_{Z,i}}{L_{X,i}} = 1$) lifted inductively from $(L_{X,0}, L_{Z,0})$ via $L_{X,i} := \widetilde{\iota}^1(L_{X,i-1}),\, L_{Z,i} := (\widetilde{\pi}^1)^\top(L_{Z,i-1})$ on odd $\cR_X^{(3)}$ steps and $L_{X,i} := (\widetilde{\pi}_{\cC_{i-1}^\vee}^1)^\top(L_{X,i-1}),\, L_{Z,i} := \widetilde{\iota}_{\cC_{i-1}^\vee}^1(L_{Z,i-1})$ on even $\cR_Z^{(3)}$ steps (writing ${}^\vee$ for the CSS dual maps on $\cC^\vee$).
\begin{enumerate}[label=\textbf{(\alph*)}, leftmargin=2.2em, itemsep=2.5pt]
    \item \textbf{Recursive parity functionals and homology-projected decoder $\cD_{\mathrm{cluster}}$}: At the base welded complex $\cC_2 = (\cR_Z^{(3)} \circ \cR_X^{(3)})(\cC_0)$ ($k = 0$), set the primal and dual parity rules to zero: $\psi_2 \equiv 0$ on $C^2(\cC_2)$ and $\psi_2^\vee \equiv 0$ on $C^0(\cC_2)$. Across each step $i \ge 3$, lift $(\psi_{i-1}, \psi_{i-1}^\vee)$ inductively on realizable syndromes $\sigma \in \im(\delta_{\cC_i}^1),\, \sigma_X \in \im(\partial_1^{\cC_i})$ (extended arbitrarily outside, e.g.\ setting local cleaners to $0$; non-realizable sectors have vanishing spectral projectors in Definition~\ref{def:safe-decoded-observables}) by:
    \begin{itemize}[leftmargin=1.5em, itemsep=1.5pt]
        \item On an odd primal step $\cC_i = \cR_X^{(3)}(\cC_{i-1})$, define $\psi_i(\sigma) := \psi_{i-1}(\cR_X^{\syn}(\sigma)) \oplus \ip{L_{Z,i}}{c_{\mathrm{clean}}^{(i)}(\sigma)}$ (where $c_{\mathrm{clean}}^{(i)}(\sigma)$ is the two-stage cleaner chain of Lemma~\ref{lem:stage2-star-cleaner}(ii)) and $\psi_i^\vee(\sigma_X) := \psi_{i-1}^\vee(\cR_Z^{\syn,\vee}(\sigma_X))$ (coarse $X$-gadget parity projection).
        \item On an even dual step $\cC_i = \cR_Z^{(3)}(\cC_{i-1}) = (\cR_X^{(3)}(\cC_{i-1}^\vee))^\vee$, swap the two rules by CSS duality: $\psi_i(\sigma) := \psi_{i-1}(\cR_Z^{\syn}(\sigma))$ and $\psi_i^\vee(\sigma_X) := \psi_{i-1}^\vee(\cR_X^{\syn,\vee}(\sigma_X)) \oplus \ip{L_{X,i}}{c_{\mathrm{clean}}^{(i),\vee}(\sigma_X)}$.
    \end{itemize}
    Computing for each $\sigma \in \im(\delta_{\cC_m}^1)$ a canonical baseline preimage $e_{\mathrm{base}}(\sigma) \in C^1(\cC_m)$ solving the $\F$-linear system $\delta_{\cC_m}^1(e_{\mathrm{base}}(\sigma)) = \sigma$ and $\ip{L_{Z,m}}{e_{\mathrm{base}}(\sigma)} = 0$ via Gaussian elimination (with $e_{\mathrm{base}}(0) = 0$) defines the chain-valued decoder $\cD_{\mathrm{cluster}} : \im(\delta_{\cC_m}^1) \to C^1(\cC_m)$ (and dually $\cD_{\mathrm{cluster}}^\vee$ on $\cC_m^\vee$):
    \begin{equation}
        \label{eq:parity-decoder-recursion}
        \cD_{\mathrm{cluster}}(\sigma) \;:=\; e_{\mathrm{base}}(\sigma) \oplus \psi_m(\sigma)\,L_{X,m}, \qquad
        \psi_m(0) = 0, \qquad
        \ip{L_{Z,m}}{\cD_{\mathrm{cluster}}(\sigma)} = \psi_m(\sigma).
    \end{equation}
    \item \textbf{$(R, w, N)$-Ball Parity Witness Family}: On a 3D CSS complex $(\cC, \Phi)$ with logical pair $(L_X, L_Z)$, a pair $(\psi, \cW)$ consisting of a parity map $\psi : C^2(\cC) \to \F$ with $\psi(0) = 0$ and a finite pool of nonzero check patterns $\cW \subseteq C^2(\cC)$ is an \emph{$(R, w, N)$-ball parity witness family} ($R \ge 0$, integer $w \ge 1$, $N \ge 1$) if $\abs{\cW} \le N$, every witness $\tau \in \cW$ has Hamming weight $\abs{\tau} \ge w$, and for every pair of error chains $e, e' \in C^1(\cC)$ whose difference $e \oplus e'$ is supported inside a Euclidean ball $B(x_0, R) \subset \R^3$,
    \[
    \begin{aligned}
        &\psi(\delta_{\cC}^1 e') \oplus \ip{L_Z}{e'} \;\neq\; \psi(\delta_{\cC}^1 e) \oplus \ip{L_Z}{e} \\
        &\qquad\Longrightarrow\quad \exists\,\tau \in \cW \;\text{with}\; \supp(\tau) \subseteq \supp(\delta_{\cC}^1 e) \;\text{or}\; \supp(\tau) \subseteq \supp(\delta_{\cC}^1 e').
    \end{aligned}
    \]
\end{enumerate}
\end{definition}

It is worth pausing to interpret Definition~\ref{def:multiscale-cluster-decoder}(b): rather than asking the decoder to reconstruct the microscopic error chain, it asks only that any ball-supported error shift that fools the parity rule leave behind a check pattern from $\cW$ of weight at least $w$ either before or after the shift. At the base welded complex $\cC_2$, local cone acyclicity (Section~\ref{sec:linear-tqo2-proof}) rules out any syndrome-free error loop inside a bounded ball that could change the encoded logical qubit, so the individual checks themselves serve as weight-$1$ witnesses under the zero parity rule. Across each subsequent replacement step, the subset-local $16/27$ contraction of Theorem~\ref{thm:16-27-contraction}(ii) amplifies the minimum witness weight by a factor of $27/16$ in the cleaned sector while preserving witness weight in the coarse-projected dual sector.

\leanlink{prop:cluster-decoder-witness}
\begin{proposition}[Local Homological Immunity and Inductive Witness Pool Lifting]
\label{prop:cluster-decoder-witness}
Let $\mu_{\mathrm{step}} = O(\Lambda^2)$ be the single-step gadget check-fiber upper bound of Theorem~\ref{thm:16-27-contraction}(ii).
\begin{enumerate}[label=\textbf{(\roman*)}, leftmargin=2.2em, itemsep=2.5pt]
    \item For every radius $0 \le R_0 \le 7015$, the trivial parity rule $\psi_2 \equiv 0$ and the singleton check pool $\cW_2 := \{\mathbf{1}_z : z \in M_{Z,2}\}$ form an $(R_0, 1, \abs{M_{Z,2}})$-ball parity witness family for $\cC_2$ (and dually $(\psi_2^\vee \equiv 0, \cW_2^\vee := \{\mathbf{1}_x : x \in M_{X,2}\})$ for $\cC_2^\vee$).
    \item Suppose $(\cC, \Phi)$ has $(8, r_0, \rho_0)$-bounded geometry ($1 \le d_{QX}, d_{QZ} \le 4$, $r_0 \le 8$, $\rho_0 \le 6.6 \times 10^{10}$) at Step~1 scale $\Lambda \ge \Lambda_0$ with logical pair $(L_X, L_Z)$, and admits an $(R_{\mathrm{coarse}}, w, N)$-ball witness family $(\psi, \cW)$ on $Z$-syndromes and an $(R_{\mathrm{coarse}}, w^\vee, N^\vee)$-ball witness family $(\psi^\vee, \cW^\vee)$ on $X$-syndromes at coarse radius $R_{\mathrm{coarse}} := R/\Lambda + 7003$ ($R \ge 0$). Setting $L_X^{\mathrm{lift}} := \widetilde{\iota}^1(L_X)$ and $L_Z^{\mathrm{lift}} := (\widetilde{\pi}^1)^\top(L_Z)$ on $\widetilde{\cC} = \cR_X^{(3)}(\cC)$:
    \begin{itemize}[leftmargin=1.5em, itemsep=1.5pt]
        \item \emph{Cleaned $Z$-sector amplification}: $\widetilde{\psi}(\sigma) = \psi(\cR_X^{\syn}(\sigma)) \oplus \ip{L_Z^{\mathrm{lift}}}{c_{\mathrm{clean}}(\sigma)}$ and the lifted two-hop subset pool $\widetilde{\cW}$ form an $\bigl(R,\, \ceil{27 w / 16},\, N \cdot (2^{41\mu_{\mathrm{step}}})^w\bigr)$-ball parity witness family for $\widetilde{\cC}$.
        \item \emph{Projected $X$-sector preservation}: $\widetilde{\psi}^\vee(\sigma_X) = \psi^\vee(\cR_Z^{\syn,\vee}(\sigma_X))$ and the fiber-lifted pool $\widetilde{\cW}^\vee$ form an $\bigl(R,\, w^\vee,\, N^\vee \cdot \mu_{\mathrm{step}}^{w^\vee}\bigr)$-ball parity witness family for $\widetilde{\cC}^\vee$.
    \end{itemize}
\end{enumerate}
\end{proposition}

\begin{proof}
For (i), suppose $e, e' \in C^1(\cC_2)$ satisfy $\supp(e \oplus e') \subseteq B(x_0, R_0)$ with $R_0 \le 7015$, and assume neither $\delta_{\cC_2}^1 e$ nor $\delta_{\cC_2}^1 e'$ contains any singleton in $\cW_2$. Then $\delta_{\cC_2}^1 e = \delta_{\cC_2}^1 e' = 0$, so $e \oplus e' \in \ker \delta_{\cC_2}^1$ is a cocycle supported inside $B(x_0, 7015)$. By subcylinder Relative Cone Acyclicity on $\cC_2$ (Lemma~\ref{lem:clean-shell-subcylinder} and Theorem~\ref{thm:rca-welded-hierarchy}), $e \oplus e' = \delta_{\cC_2}^0 \alpha$ for some $\alpha \in C^0(\cC_2)$. Since $L_{Z,2} \in \ker \partial_1^{\cC_2}$ is a closed $1$-cycle, its $\F$-pairing with any coboundary vanishes:
\begin{equation}
    \label{eq:local-pairing-invariance}
    \ip{L_{Z,2}}{e' \oplus e} \;=\; \ip{L_{Z,2}}{\delta_{\cC_2}^0 \alpha} \;=\; \ip{\partial_1^{\cC_2} L_{Z,2}}{\alpha} \;=\; 0
    \qquad\Longrightarrow\qquad
    \ip{L_{Z,2}}{e'} \;=\; \ip{L_{Z,2}}{e},
\end{equation}
so $\psi_2(\delta_{\cC_2}^1 e') \oplus \ip{L_{Z,2}}{e'} = \psi_2(\delta_{\cC_2}^1 e) \oplus \ip{L_{Z,2}}{e}$, proving (i).

For (ii), consider first the cleaned $Z$-sector on $\widetilde{\cC} = \cR_X^{(3)}(\cC)$. For any fine chain $e \in C^1(\widetilde{\cC})$ with $\sigma = \delta_{\widetilde{\cC}}^1 e$, Lemma~\ref{lem:stage2-star-cleaner}(ii) shows that $e_{\mathrm{coarse}} := \widetilde{\pi}^1(e \oplus c_{\mathrm{clean}}(\sigma))$ satisfies $\delta_{\cC}^1 e_{\mathrm{coarse}} = \cR_X^{\syn}(\sigma)$, while $\ip{L_Z}{e_{\mathrm{coarse}}} = \ip{L_Z^{\mathrm{lift}}}{e \oplus c_{\mathrm{clean}}(\sigma)} = \ip{L_Z^{\mathrm{lift}}}{e} \oplus \ip{L_Z^{\mathrm{lift}}}{c_{\mathrm{clean}}(\sigma)}$. Adding $\psi(\cR_X^{\syn}(\sigma))$ to both sides gives exact discrepancy preservation:
\[
    \widetilde{\psi}(\delta_{\widetilde{\cC}}^1 e) \oplus \ip{L_Z^{\mathrm{lift}}}{e} \;=\; \psi(\delta_{\cC}^1 e_{\mathrm{coarse}}) \oplus \ip{L_Z}{e_{\mathrm{coarse}}}.
\]
If $\supp(e \oplus e') \subseteq B(x_0, R)$, then $\sigma$ and $\sigma' = \delta_{\widetilde{\cC}}^1 e'$ agree on all gadgets outside $B(x_0, R + r_0)$. Because Stage~1 and Stage~2 cleaning plus $\widetilde{\pi}^1$ cross at most three coarse-edge hops of length $\le 2001\Lambda$ plus the $\alpha\Lambda = 1000\Lambda$ spherical-cap shift, $e_{\mathrm{coarse}}$ and $e'_{\mathrm{coarse}}$ agree outside $B\bigl(x_0/\Lambda, (R + 3 \cdot 2001\Lambda)/\Lambda + 1000\bigr) = B(x_0/\Lambda, R/\Lambda + 7003)$. Hence if $\widetilde{\psi}(\sigma') \oplus \ip{L_Z^{\mathrm{lift}}}{e'} \neq \widetilde{\psi}(\sigma) \oplus \ip{L_Z^{\mathrm{lift}}}{e}$, the coarse witness property forces either $\cR_X^{\syn}(\sigma)$ or $\cR_X^{\syn}(\sigma')$ to contain $\supp(\tau)$ for some $\tau \in \cW$. Fix a $w$-element subset $K_\tau \subseteq \supp(\tau)$ ($\abs{K_\tau} = w$). Whenever $K_\tau \subseteq \supp(\cR_X^{\syn}(\sigma))$, Theorem~\ref{thm:16-27-contraction}(ii) gives $\abs{\sigma|_{\cP_2(K_\tau)}} \ge \ceil{27 w / 16} =: w'$, so $\supp(\sigma)$ contains a $w'$-element subset of $\cP_2(K_\tau)$. Since $\abs{\cP_2(K_\tau)} \le 41\mu_{\mathrm{step}} w$ has at most $2^{41\mu_{\mathrm{step}} w}$ subsets, taking $\widetilde{\cW}$ to be all $w'$-element subsets of $\cP_2(K_\tau)$ as $\tau$ ranges over $\cW$ yields $\abs{\widetilde{\cW}} \le N \cdot (2^{41\mu_{\mathrm{step}}})^w$. Finally, in the coarse-projected $X$-sector, $(\widetilde{\iota}^1)^\top$ maps $B(x_0, R)$ into $B(x_0/\Lambda, R/\Lambda + 7003)$ with $\ip{L_X^{\mathrm{lift}}}{e_Z} = \ip{L_X}{(\widetilde{\iota}^1)^\top e_Z}$, and if $\cR_Z^{\syn,\vee}(\sigma_X)$ contains a $w^\vee$-element subset $K' \subseteq \supp(\tau^\vee)$, then for each $x \in K'$ at least one of the $\le \mu_{\mathrm{step}}$ fine checks in $Y_x^{(3)}$ is active in $\sigma_X$; choosing one fine check per $x \in K'$ gives at most $\mu_{\mathrm{step}}^{w^\vee}$ lifted witnesses per $\tau^\vee \in \cW^\vee$.
\end{proof}

\subsection{The Free-Energy Barrier}
\label{subsec:walsh-fourier-peierls}

\begin{figure}[t]
\centering
\includegraphics[width=0.88\textwidth]{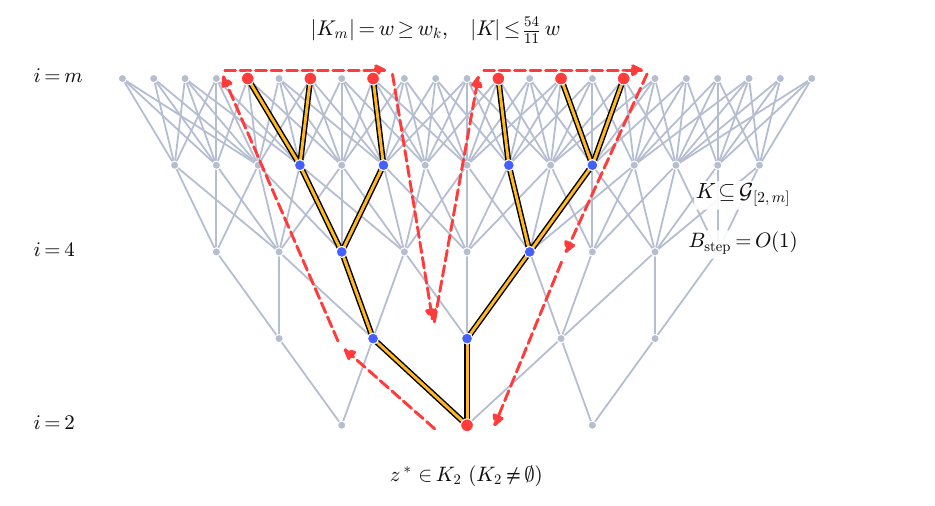}
\caption{Multiscale check-hierarchy graph $\cG_{[2,m]}$ on $\bigsqcup_{i=2}^m M_{Z,i}$ (slate-gray background lattice, witness-pool bound $B_{\mathrm{step}} = O(1)$) and a percolating multiscale witness tree $K = \bigsqcup_{i=2}^m K_i \subseteq \cG_{[2,m]}$ (thick black and amber edges lifting a coarse seed check $z^* \in K_2$ ($K_2 \neq \emptyset$, red node at $i = 2$) upward across blue intermediate nodes at scales $2 < i < m$ to $\abs{K_m} = w \ge w_k$ red physical checks at $i = m$, with total cross-scale size $\abs{K} \le \frac{54}{11} w$ and a dashed red traversal contour; Lemma~\ref{lem:percolating-cluster-bounds} and Theorem~\ref{thm:walsh-fourier-peierls}).}
\label{fig:percolating-cluster-peierls}
\end{figure}

Iterating Proposition~\ref{prop:cluster-decoder-witness}(ii) across $k$ full primal/dual cycles $(\cR_Z^{(3)} \circ \cR_X^{(3)})^{\circ k}$ starting from the weight-$1$ base witness family on $\cC_2$ amplifies the minimum witness weight exponentially in $k$ while maintaining ball protection up to the macroscopic Euclidean radius $74\Lambda^k$ (Figure~\ref{fig:percolating-cluster-peierls}).

Because integer ceiling rounding sends $1 \mapsto 2 \mapsto 4 \mapsto 7 \mapsto \cdots$, the minimum witness weight grows by at least a factor of $3/2$ at every two-step cycle, while geometric summation across scales keeps both the total multiscale witness volume and the logarithm of the witness pool cardinality linear in the top-level witness weight. Whereas~\cite[Section~5.3, Lemmas~5.9--5.12]{BDL24} counts connected witness subgraphs of the full decoding graph via depth-first traversals and distance-$4$ domination (with geometric ratio $1/2$), Lemma~\ref{lem:percolating-cluster-bounds} lifts two-hop subset witnesses inductively from $\cC_2$ with contraction ratio $16/27$ while simultaneously tracking the macroscopic $74\Lambda^k$ Euclidean protection radius.

\leanlink{lem:percolating-cluster-bounds}
\begin{lemma}[{Multiscale Witness Amplification and $74\Lambda^k$ Protection Radius, adapted from~\cite[Section~5.3, Lemmas~5.9--5.12]{BDL24}}]
\label{lem:percolating-cluster-bounds}
Define the integer witness-weight sequence $(w_j)_{j \ge 0}$ recursively by $w_0 := 1$ and $w_{j+1} := \ceil{27 w_j / 16} = \floor{(27 w_j + 15)/16}$, and set $B_{\mathrm{step}} := \mu_{\mathrm{step}}\,2^{41\mu_{\mathrm{step}}} = O(1)$. For every $k \ge 0$ ($m = 2k + 2$):
\begin{enumerate}[label=\textbf{(\roman*)}, leftmargin=2.2em, itemsep=2.5pt]
    \item Since $w_0 = 1 \mapsto w_1 = 2 \mapsto w_2 = 4 \mapsto w_3 = 7 \mapsto \cdots$, the sequence satisfies $2 w_j \le \frac{4}{3} w_{j+1}$ (that is, $w_{j+1} \ge \frac{3}{2} w_j$) for every $j \ge 0$, as well as $w_{j+1} \ge \frac{27}{16} w_j$, yielding
    \begin{equation}
        \label{eq:percolating-energy-lower-bound}
        w_k \;\ge\; \max\!\Bigl\{\Bigl(\frac{3}{2}\Bigr)^k,\, \Bigl(\frac{27}{16}\Bigr)^k\Bigr\} \;=\; \Omega\bigl(\abs{Q_m}^\eta\bigr), \qquad
        \eta \;:=\; \frac{\ln(3/2)}{6\ln \Lambda} \;>\; 0,
    \end{equation}
    where $\abs{Q_m} \le \abs{Q_0}\mu_{\mathrm{cell}}^{2k+2} = \exp(O(k))$ for the single-step cell multiplier $\mu_{\mathrm{cell}} \le 600 m_{\mathrm{sub}}^2 < \Lambda^3$. Moreover, every nontrivial logical cocycle $e \in \ker \delta_{\cC_m}^1 \setminus \im \delta_{\cC_m}^0$ (or cycle in $\ker \partial_1^{\cC_m} \setminus \im \partial_2^{\cC_m}$) has weight $\abs{e} \ge \frac{1}{4} w_k$, so $d_{\mathrm{log}}(\cC_m) \ge \frac{1}{4} w_k = \Omega((27/16)^k)$.
    \item Both $(\cC_m, \Phi_m)$ and $(\cC_m^\vee, \Phi_m^\vee)$ admit $(74\Lambda^k,\, w_k,\, B_{\mathrm{step}}^{8 w_k})$-ball parity witness families $(\psi_m, \cW_m)$ and $(\psi_m^\vee, \cW_m^\vee)$. Furthermore, across any sequence of coarse syndrome subsets $(K_i)_{i=2}^m$ obeying $\abs{K_{2j}} \le \frac{16}{27}\abs{K_{2j+1}}$ and $\abs{K_{2j+1}} \le \abs{K_{2j+2}}$, the total multiscale size is bounded linearly by the top-level weight:
    \begin{equation}
        \label{eq:linear-multiscale-volume}
        \sum_{i = 2}^m \abs{K_i} \;\le\; 2 \sum_{j=0}^{k} \Bigl(\frac{16}{27}\Bigr)^j \abs{K_m} \;\le\; \frac{2}{1 - 16/27}\,\abs{K_m} \;=\; \frac{54}{11}\,\abs{K_m}.
    \end{equation}
\end{enumerate}
\end{lemma}

\begin{proof}
For (i), since $w_j \ge 1$ is an integer, $16 w_{j+1} \ge 27 w_j$; moreover, for $w_j = 1$ we have $2 w_1 = 4 \ge 3 = 3 w_0$, and for $w_j \ge 2$ we have $2 \cdot 16 w_{j+1} \ge 54 w_j = 48 w_j + 6 w_j \ge 48 w_j$, so $2 w_{j+1} \ge 3 w_j$ holds for every $j \ge 0$, giving $w_k \ge (27/16)^k \ge (3/2)^k$. To bound the minimum distance, write any nontrivial logical cocycle $e \in \ker \delta_{\cC_m}^1 \setminus \im \delta_{\cC_m}^0$ (which satisfies $\ip{L_{Z,m}}{e} = 1$ since $\dim_{\F} H^1(\cC_m) = 1$ and $\ip{L_{Z,m}}{L_{X,m}} = 1$) as $e = \sum_{s=1}^D \mathbf{1}_{q_s}$ ($D = \abs{e}$) and set $e_{\le s} := \sum_{j=1}^s \mathbf{1}_{q_j}$. Because $\psi_m(\delta_{\cC_m}^1 e_{\le 0}) \oplus \ip{L_{Z,m}}{e_{\le 0}} = 0$ while $\psi_m(\delta_{\cC_m}^1 e_{\le D}) \oplus \ip{L_{Z,m}}{e_{\le D}} = \psi_m(0) \oplus 1 = 1$, there is an index $1 \le s \le D$ at which this parity flips across the single-qubit step $e_{\le s-1} \to e_{\le s}$ (supported on $\{q_s\} \subset B(\Phi_m(q_s), 74\Lambda^k)$). Part (ii) then forces $\max(\abs{\delta_{\cC_m}^1 e_{\le s-1}}, \abs{\delta_{\cC_m}^1 e_{\le s}}) \ge w_k$; since each qubit excites at most $d_{QZ} \le 4$ checks, $4 D \ge \max(\abs{\delta_{\cC_m}^1 e_{\le s-1}}, \abs{\delta_{\cC_m}^1 e_{\le s}}) \ge w_k$.

For (ii), at $k = 0$ ($m = 2$, zero half-steps above $\cC_2$) we have $R = 74\Lambda^0 = 74 \le 7015$, so Proposition~\ref{prop:cluster-decoder-witness}(i) directly furnishes $(74, 1, \abs{M_{Z,2}})$- and $(74, 1, \abs{M_{X,2}})$-ball parity witness families on $\cC_2$ and $\cC_2^\vee$; for $k \ge 1$, iterating the radius map $r \mapsto r/\Lambda + 7003$ of Proposition~\ref{prop:cluster-decoder-witness}(ii) over $2k \ge 2$ half-steps from $R = 74\Lambda^k$ (with $\Lambda \ge \Lambda_0 = 5.4 \times 10^{24}$) gives final radius $r_{2k} = 74\Lambda^k / \Lambda^{2k} + 7003\sum_{s=0}^{2k-1}\Lambda^{-s} < 74/\Lambda + 7004 < 7005 \le 7015$ at $\cC_2$. Over each full two-step cycle $\cR_Z^{(3)} \circ \cR_X^{(3)}$ from $2j + 2$ to $2j + 4$, Proposition~\ref{prop:cluster-decoder-witness}(ii) advances both primal and dual weights from $w_j$ to $w_{j+1} = \ceil{27 w_j / 16}$ and multiplies each pool cardinality by at most $\mu_{\mathrm{step}}^{w_{j+1}} \cdot (2^{41\mu_{\mathrm{step}}})^{w_j} \le B_{\mathrm{step}}^{w_{j+1}}$. Since $w_j \le (2/3)^{k-j} w_k$ by (i), the exponent sum telescopes via $\sum_{j=1}^k w_j \le \sum_{r=0}^{k-1}(2/3)^r w_k \le 3 w_k$; absorbing the $O(1)$ base check count $\abs{M_{Z,2}} + \abs{M_{X,2}} \le B_{\mathrm{step}}^{5}$ at $k = 0$ yields $\abs{\cW_m}, \abs{\cW_m^\vee} \le B_{\mathrm{step}}^{8 w_k}$. Summing $\abs{K_{m-2j}} + \abs{K_{m-2j-1}} \le 2(16/27)^j \abs{K_m}$ gives~\eqref{eq:linear-multiscale-volume}.
\end{proof}

Unrolling the inductive subset choice $K_{i-1} \mapsto \sigma_i|_{\cP_2(K_{i-1})}$ across scales $i = 2, \dots, m$ recovers the geometric picture of Figure~\ref{fig:percolating-cluster-peierls}: each minimal failing witness in $\cW_m$ traces out the top-level footprint of a multiscale tree $K = \bigsqcup_{i=2}^m K_i$ of active checks in the check-hierarchy graph $\cG_{[2,m]}$ on $\bigsqcup_{i=2}^m M_{Z,i}$ spanning from scale $m$ down to $\cC_2$, whose total cross-scale volume $\abs{K} := \sum_{i=2}^m \abs{K_i}$ is bounded linearly by $\frac{54}{11}\abs{K_m}$ via~\eqref{eq:linear-multiscale-volume} (cf.\ lattice-animal spanning-tree bounds~\cite{klarner1967cell} and~\cite[Section~5.3]{BDL24}). Working directly with the inductively lifted top-level pool $\cW_m$, however, keeps the Peierls proof below self-contained without global connected-component clustering or base-scale cutoffs.

To convert this witness pool into a canonical equilibrium free-energy barrier (following~\cite{peierls1936ising, griffiths1964peierls, alicki2010thermal, chesi2010thermodynamic} and~\cite[Section~5.4, Theorem~5.13]{BDL24}), we bound the Boltzmann weight of all realizable syndromes containing a given witness $\tau \in \cW_m$. In~\cite[Section~5.3, Lemma~5.8, and Section~5.4, Theorem~5.13]{BDL24}, each witness arises from a connected component of the full decoding graph and is first shown to be a $2$-coboundary in the realizable $Z$-syndrome subspace $\cS_m := \im(\delta_{\cC_m}^1) \subset \F^{M_{Z,m}}$, so adding it translates $\cS_m$ into itself. Because our localized ball-parity witnesses $\tau \in \cW_m \subset \F^{M_{Z,m}}$ are extracted on two-hop check pools without global connected-component clustering, an individual $\tau$ need not itself lie in $\cS_m$: subtracting it from a realizable syndrome shifts the remaining configuration into an affine coset $\tau \oplus \cS_m$. Expanding the Boltzmann factor in the $\F$-Fourier character basis~\cite{macwilliams1963theorem} shows directly that every affine coset carries total Boltzmann weight at most that of $\cS_m$ itself, allowing a union bound over $\cW_m$ whenever $\beta > \beta_c$. Recall also~\cite{alicki2009thermalization, alicki2010thermal, chesi2010thermodynamic} that a syndrome decoder $\cD : \im(\delta_{\cC}^1) \to C^1(\cC)$ satisfying $\delta_{\cC}^1(\cD(\sigma)) = \sigma$ suffers a \emph{logical transition failure} under a single-qubit flip at $q \in Q$ from $\sigma \in \im(\delta_{\cC}^1)$ if the residual cocycle $\cD(\sigma) \oplus \mathbf{1}_q \oplus \cD(\sigma \oplus \delta_{\cC}^1 \mathbf{1}_q) \in \ker \delta_{\cC}^1$ does not lie in $\im \delta_{\cC}^0$; we write $\cT_{\mathrm{fail}}(\cC, \cD) \subseteq \im(\delta_{\cC}^1) \times Q$ for the set of all failing single-qubit transition pairs $(\sigma, q)$.

\leanlink{thm:walsh-fourier-peierls}
\begin{theorem}[{Coset Peierls Inequality and Classical Free-Energy Barrier; cf.~\cite[Section~5.4, Theorem~5.13]{BDL24}}]
\label{thm:walsh-fourier-peierls}
For the realizable $Z$-syndrome subspace $\cS_m = \im(\delta_{\cC_m}^1) \subset \F^{M_{Z,m}}$, let $Z_{\cC_m}(\beta) := \sum_{\sigma \in \cS_m} e^{-\beta\abs{\sigma}}$ be the canonical syndrome partition function at inverse temperature $\beta > 0$.
\begin{enumerate}[label=\textbf{(\roman*)}, leftmargin=2.2em, itemsep=2.5pt]
    \item For every check pattern $\tau \in \F^{M_{Z,m}}$, let $\cE(\tau) := \{\sigma \in \cS_m : \supp(\tau) \subseteq \supp(\sigma)\}$. Then
    \begin{equation}
        \label{eq:walsh-fourier-coset-bound}
        \frac{1}{Z_{\cC_m}(\beta)}\sum_{\sigma \in \cE(\tau)} e^{-\beta\abs{\sigma}} \;\le\; e^{-\beta\abs{\tau}}.
    \end{equation}
    \item Fix a constant margin $c_\beta \ge 5$ (the formalization sets a conservative threshold $\ge \beta_c$ using linear upper bounds on $\ln B_{\mathrm{step}}$) and define the critical inverse temperature
    \begin{equation}
        \label{eq:beta-c-def}
        \beta_c \;:=\; 8\ln B_{\mathrm{step}} + c_\beta \;=\; O(1).
    \end{equation}
    Let $\cS_m^{\mathrm{unsafe}} := \bigcup_{\tau \in \cW_m} \cE(\tau)$ be the set of realizable syndromes containing some witness in $\cW_m$, and set $c_F' := \Lambda^{-12\eta} > 0$. Then for all $\beta > \beta_c$, both $\cS_m^{\mathrm{unsafe}}$ and the single-qubit transition failure set $\cT_{\mathrm{fail}}(\cC_m, \cD_{\mathrm{cluster}})$ satisfy
    \begin{equation}
        \label{eq:peierls-sum-final}
        \frac{1}{Z_{\cC_m}(\beta)}\sum_{\sigma \in \cS_m^{\mathrm{unsafe}}} e^{-\beta\abs{\sigma}} \;\le\; B_{\mathrm{step}}^{8 w_k}\,e^{-\beta w_k} \;\le\; \frac{1}{2}\exp\bigl(-(\beta - \beta_c)w_k\bigr),
    \end{equation}
    and
    \begin{equation}
        \label{eq:intro-gibbs-barrier}
        \frac{1}{\abs{Q_m}\,Z_{\cC_m}(\beta)}\sum_{(\sigma, q) \in \cT_{\mathrm{fail}}(\cC_m, \cD_{\mathrm{cluster}})} e^{-\beta\max(\abs{\sigma},\, \abs{\sigma \oplus \delta_{\cC_m}^1 \mathbf{1}_q})} \;\le\; \exp\bigl(-c_F'(\beta - \beta_c)\abs{Q_m}^\eta\bigr)
    \end{equation}
    (and identically for $\cC_m^\vee$ under $\cD_{\mathrm{cluster}}^\vee$).
\end{enumerate}
\end{theorem}

\begin{proof}
For (i), every $\sigma \in \cE(\tau)$ satisfies $\supp(\tau) \subseteq \supp(\sigma)$, so $u := \sigma \oplus \tau$ has exact Hamming-weight decomposition $\abs{\sigma} = \abs{\tau} + \abs{u}$ and $\sigma \mapsto \sigma' := \sigma \in \cS_m$ injects $\cE(\tau)$ into $\cS_m$ with $u = \tau \oplus \sigma'$. Expanding the single-bit Boltzmann factor $e^{-\beta\abs{v_z}} = e^{-\beta/2}\cosh(\beta/2)\bigl(1 + \tanh(\beta/2)\,(-1)^{v_z}\bigr)$ across $z \in M_{Z,m}$ in the characters $\chi_T(v) := (-1)^{\sum_{z \in T} v_z}$ ($T \subseteq M_{Z,m}$) of $\F^{M_{Z,m}}$~\cite{macwilliams1963theorem}, linearity of $\cS_m \subseteq \F^{M_{Z,m}}$ ensures character-sum nonnegativity:
\begin{align}
    e^{-\beta\abs{v}} &\;=\; \bigl(e^{-\beta/2}\cosh(\beta/2)\bigr)^{\abs{M_{Z,m}}}\sum_{T \subseteq M_{Z,m}}\bigl(\tanh(\beta/2)\bigr)^{\abs{T}}\chi_T(v), \nonumber\\
    \sum_{\sigma' \in \cS_m}\chi_T(\sigma') &\;=\; \abs{\cS_m}\,\mathbf{1}_{\{T \in \cS_m^\perp\}} \;\ge\; 0.
    \label{eq:walsh-fourier-expansion}
\end{align}
Since $\chi_T(\tau \oplus \sigma') = \chi_T(\tau)\chi_T(\sigma')$ with $\chi_T(\tau) \in \{\pm 1\} \le 1$ and $\tanh(\beta/2) > 0$, summing over $\sigma' \in \cS_m$ term-by-term gives
\begin{equation}
    \label{eq:peeling-injection}
    \sum_{\sigma \in \cE(\tau)} e^{-\beta\abs{\sigma}} \;\le\; e^{-\beta\abs{\tau}} \sum_{\sigma' \in \cS_m} e^{-\beta\abs{\tau \oplus \sigma'}} \;\le\; e^{-\beta\abs{\tau}} \sum_{\sigma' \in \cS_m} e^{-\beta\abs{\sigma'}} \;=\; e^{-\beta\abs{\tau}} Z_{\cC_m}(\beta).
\end{equation}

For (ii), by Lemma~\ref{lem:percolating-cluster-bounds}(ii), $\abs{\cW_m} \le B_{\mathrm{step}}^{8 w_k}$ and every $\tau \in \cW_m$ satisfies $\abs{\tau} \ge w_k \ge 1$. Taking a union bound of~\eqref{eq:walsh-fourier-coset-bound} over $\tau \in \cW_m$ yields
\[
    \frac{1}{Z_{\cC_m}(\beta)}\sum_{\sigma \in \cS_m^{\mathrm{unsafe}}} e^{-\beta\abs{\sigma}} \;\le\; \sum_{\tau \in \cW_m} e^{-\beta\abs{\tau}} \;\le\; \exp\bigl((8\ln B_{\mathrm{step}} - \beta) w_k\bigr) \;=\; e^{-c_\beta w_k}\exp\bigl(-(\beta - \beta_c) w_k\bigr),
\]
proving~\eqref{eq:peierls-sum-final} since $e^{-c_\beta w_k} \le e^{-5} < \frac{1}{2}$. Finally, if $(\sigma, q) \in \cT_{\mathrm{fail}}(\cC_m, \cD_{\mathrm{cluster}})$, setting $e := \cD_{\mathrm{cluster}}(\sigma)$ and $e' := e \oplus \mathbf{1}_q$ ($\supp(\mathbf{1}_q) \subset B(\Phi_m(q), 74\Lambda^k)$) flips $\psi_m(\delta_{\cC_m}^1 e) \oplus \ip{L_{Z,m}}{e}$, so Lemma~\ref{lem:percolating-cluster-bounds}(ii) forces $\sigma \in \cS_m^{\mathrm{unsafe}}$ or $\sigma \oplus \delta_{\cC_m}^1 \mathbf{1}_q \in \cS_m^{\mathrm{unsafe}}$. Since each $\tau_{\mathrm{end}} \in \cS_m^{\mathrm{unsafe}}$ arises as an endpoint of at most $2\abs{Q_m}$ pairs $(\sigma, q)$ and $\abs{Q_m} \le 5\mu_{\mathrm{cell}}^{2k+2} < \Lambda^{6k+12}$ gives $c_F'\abs{Q_m}^\eta \le (3/2)^k \le w_k$, applying~\eqref{eq:peierls-sum-final} yields~\eqref{eq:intro-gibbs-barrier}.
\end{proof}

\subsection{Safe Decoded Observables}
\label{subsec:safe-decoded-observables}

To pass from classical $\F$-chain syndromes to open-system quantum statistical mechanics on a regular three-dimensional spin lattice, we snap the cells of the welded code onto an open cubic grid with uniform finite site dimension (pinning unused qubit slots in $\ket{0}$) and lift the classical decoder into quantum operators. The witness pools single out a \emph{safe subspace} of joint check eigenstates containing no macroscopic witness pattern, while weighting the bare logical operators by the decoder's parity sign on each syndrome sector produces anticommuting quantum observables (commuting with $H_{0,k}$) that read out the decoded qubit.

\leanlink{def:safe-decoded-observables}
\begin{definition}[Cubic Lattice Realization, Safe Syndrome Subspace, and Decoded Logical Observables]
\label{def:safe-decoded-observables}
Fix hierarchy level $k \ge 0$ ($m = 2k + 2$) and define the open cubic lattice $\Gamma_k := \{0, \dots, L_k - 1\}^3$ of linear side length $L_k := \Lambda^{2k+4}$:
\begin{enumerate}[label=\textbf{(\alph*)}, leftmargin=2.2em, itemsep=2.5pt]
    \item \textbf{Composite-site CSS Hamiltonian $H_{0,k}$ with inactive-slot pins}: Centering $\Phi_m$ in $[0, L_k]^3$ (with boundary margin $\ge 1000$) and snapping each cell coordinate via the integer floor into $\Gamma_k$, uniform density $\rho_0 = O(1)$ and Euclidean range $r_0 = O(1)$ guarantee that every lattice site $v \in \Gamma_k$ hosts at most $q_{\mathrm{code}} = O(1)$ physical qubits of $Q_m$ and anchors at most $m_{\max} = O(1)$ CSS checks of each type ($X, Z$), each supported inside the $\ell_\infty$ ball $B_\infty(v, r_{\mathrm{chk}}) := \{u \in \Z^3 : \norm{u - v}_\infty \le r_{\mathrm{chk}}\}$ of radius $r_{\mathrm{chk}} := r_0 + 1 = 9$ around $v$. Assign every site $v \in \Gamma_k$ the uniform local Hilbert space $\cH_v \cong (\C^2)^{\otimes q_{\mathrm{code}}} \cong \C^{d_{\loc}}$ ($d_{\loc} := 2^{q_{\mathrm{code}}}$), forming $\cH_k := \bigotimes_{v \in \Gamma_k} \cH_v$ (with complex inner product $\ip{\phi}{\psi}$, state norm $\norm{\psi}$, and spectral norm $\norm{A}$ on the bounded operators $\cB(\cH_k)$). Writing $\cI_{\mathrm{pin}}$ for the set of unoccupied qubit slots $(v, s) \notin Q_m$, on each $(v, s) \in \cI_{\mathrm{pin}}$ impose the single-slot pinning check $Q_{v,s}^{\mathrm{pin}} := \frac{1}{2}(I - Z_{v,s})$ (fixing unused slots to $\ket{0}$ with unit energy penalty), so the total commuting-projector Hamiltonian is
    \begin{equation}
        \label{eq:composite-H0k-def}
        H_{0,k} \;:=\; \sum_{(v,s) \notin Q_m} \frac{I - Z_{v,s}}{2} \;+\; \sum_{x \in M_{X,m}} \frac{I - A_x}{2} \;+\; \sum_{z \in M_{Z,m}} \frac{I - B_z}{2} \;=\; \sum_{i \in \cI_k} Q_i,
    \end{equation}
    enumerating $\cI_k = \{1, \dots, N_{\mathrm{chk}}(k)\}$ ($N_{\mathrm{chk}}(k) := \abs{\cI_k}$), with code-space projector $P_{0,k} := \prod_{i \in \cI_k}(I - Q_i)$ of exact ground degeneracy $\Tr(P_{0,k}) = 2$.
    \item \textbf{Safe syndrome set $S_{\mathrm{safe},k}$ and projector $\Pi_{\mathrm{safe},k}$}: Each check bitstring $\sigma \in \F^{\cI_k}$ decomposes as $\sigma = (\sigma_{\mathrm{pin}}, s_X(\sigma), s_Z(\sigma)) \in \F^{\cI_{\mathrm{pin}}} \times C^0(\cC_m) \times C^2(\cC_m)$ and defines a joint spectral projector $\Pi_\sigma := \prod_{i \in \cI_k}(\sigma_i Q_i + (1 - \sigma_i)(I - Q_i))$, forming an orthogonal resolution of the identity $\sum_{\sigma \in \F^{\cI_k}} \Pi_\sigma = I$ with $\Pi_0 = P_{0,k}$ and $H_{0,k}\Pi_\sigma = \abs{\sigma}\Pi_\sigma$. Using the $(74\Lambda^k, w_k, B_{\mathrm{step}}^{8 w_k})$-ball parity witness families $(\psi_m, \cW_m)$ and $(\psi_m^\vee, \cW_m^\vee)$ of Lemma~\ref{lem:percolating-cluster-bounds}(ii), define
    \begin{equation}
        \label{eq:safe-syndrome-proj-def}
        \begin{aligned}
            S_{\mathrm{safe},k} \;:=\; \bigl\{\sigma \in \F^{\cI_k} \;:\;
            &\forall\,\tau^\vee \in \cW_m^\vee,\; \supp(\tau^\vee) \not\subseteq \supp(s_X(\sigma)) \\
            &\text{and}\;\; \forall\,\tau \in \cW_m,\; \supp(\tau) \not\subseteq \supp(s_Z(\sigma))\bigr\},
        \end{aligned}
    \end{equation}
    and set $\Pi_{\mathrm{safe},k} := \sum_{\sigma \in S_{\mathrm{safe},k}} \Pi_\sigma$.
    \item \textbf{Decoded logical observables $O_{\mathrm{dec},k}^Z, O_{\mathrm{dec},k}^X$}: Let $\overline{Z}_k := \prod_{q \in \supp(L_{Z,m})} Z_q$ and $\overline{X}_k := \prod_{q \in \supp(L_{X,m})} X_q$ denote the bare Pauli logical operators supported on active code slots. Weighting each syndrome sector by its decoded parity sign defines
    \begin{equation}
        \label{eq:decoded-observables-def}
        O_{\mathrm{dec},k}^Z \;:=\; \overline{Z}_k \sum_{\sigma \in \F^{\cI_k}} (-1)^{\psi_m(s_Z(\sigma))}\,\Pi_\sigma, \qquad
        O_{\mathrm{dec},k}^X \;:=\; \overline{X}_k \sum_{\sigma \in \F^{\cI_k}} (-1)^{\psi_m^\vee(s_X(\sigma))}\,\Pi_\sigma.
    \end{equation}
\end{enumerate}
\end{definition}

Because these safe-subspace projectors and decoded observables inherit the macroscopic protection radius of the witness pools, no operator supported inside a cube of side at most $37\Lambda^k$ can induce transitions between opposite decoded logical eigenvalues within the safe subspace. At the same time, every realizable syndrome sector has the same Hilbert-space multiplicity as the code space, so the quantum Gibbs weight outside the safe subspace factorizes into the classical unsafe-syndrome sums bounded in Theorem~\ref{thm:walsh-fourier-peierls}.

\leanlink{prop:safe-subspace-immunity}
\begin{proposition}[Safe-Subspace Immunity and Quantum Gibbs Suppression]
\label{prop:safe-subspace-immunity}
For every hierarchy level $k \ge 0$ on $\Gamma_k = \{0, \dots, L_k - 1\}^3$ ($L_k = \Lambda^{2k+4}$):
\begin{enumerate}[label=\textbf{(\roman*)}, leftmargin=2.2em, itemsep=2.5pt]
    \item $\Pi_{\mathrm{safe},k}$ is an orthogonal projection ($0 \neq \Pi_{\mathrm{safe},k} \neq I$) commuting with every check $Q_i$ and with $H_{0,k}$, satisfying $P_{0,k}\Pi_{\mathrm{safe},k} = P_{0,k}$. Both $O_{\mathrm{dec},k}^Z$ and $O_{\mathrm{dec},k}^X$ are self-adjoint unitaries ($(O_{\mathrm{dec},k}^{Z/X})^\dagger = O_{\mathrm{dec},k}^{Z/X}$, $(O_{\mathrm{dec},k}^{Z/X})^2 = I$, $\norm{O_{\mathrm{dec},k}^{Z/X}} = 1$), commute with every $Q_i$, $H_{0,k}$, $P_{0,k}$, and $\Pi_{\mathrm{safe},k}$, agree with $\overline{Z}_k, \overline{X}_k$ on $P_{0,k}$, and anticommute globally ($\{O_{\mathrm{dec},k}^Z, O_{\mathrm{dec},k}^X\} = 0$, with $O_{\mathrm{dec},k}^{Z/X}\Pi_{\mathrm{safe},k} \neq 0$).
    \item For every site $v \in \Gamma_k$ and single-site operator $A_v$ supported on $\{v\}$ (and more generally, for every operator $A_W$ supported on an $\ell_\infty$ cube $W := (v_0 + \{0, \dots, r-1\}^3) \cap \Gamma_k$ with $v_0 \in \Gamma_k$ and side length $1 \le r \le 37\Lambda^k$), the safe-subspace commutators vanish identically:
    \begin{equation}
        \label{eq:safe-subspace-immunity}
        \Pi_{\mathrm{safe},k}\,[A_W,\, O_{\mathrm{dec},k}^Z]\,\Pi_{\mathrm{safe},k} \;=\; 0, \qquad
        \Pi_{\mathrm{safe},k}\,[A_W,\, O_{\mathrm{dec},k}^X]\,\Pi_{\mathrm{safe},k} \;=\; 0.
    \end{equation}
    \item With $\beta_c > 0$ as in~\eqref{eq:beta-c-def}, $\eta = \frac{\ln(3/2)}{6\ln\Lambda} > 0$, and $c_F := \Lambda^{-12\eta} > 0$, for every $\beta > \beta_c$ the canonical Gibbs state $\rho_\beta(H_{0,k}) := e^{-\beta H_{0,k}}/\Tr(e^{-\beta H_{0,k}})$ satisfies
    \begin{equation}
        \label{eq:quantum-gibbs-failure-bound}
        p_{\mathrm{fail},k}(\beta) \;:=\; \Tr\!\bigl(\rho_\beta(H_{0,k})(I - \Pi_{\mathrm{safe},k})\bigr) \;\le\; \exp\bigl(-\Delta F_k(\beta)\bigr), \qquad
        \Delta F_k(\beta) \;:=\; c_F(\beta - \beta_c)\bigl(L_k^3\bigr)^\eta.
    \end{equation}
\end{enumerate}
\end{proposition}

\begin{proof}
For (i), since every witness has weight $\ge w_k \ge 1$, the zero syndrome $0 \in S_{\mathrm{safe},k}$, so $P_{0,k}\Pi_{\mathrm{safe},k} = P_{0,k} \neq 0$ (and $\Pi_{\mathrm{safe},k} \neq I$ since single-qubit errors on $\overline{X}_k$ cumulatively reach an unsafe syndrome). Because $\overline{Z}_k$ and $\overline{X}_k$ act only on active code slots and correspond to $\ker \partial_1^{\cC_m}$ and $\ker \delta_{\cC_m}^1$, they commute with every check $Q_i$ and hence with every $\Pi_\sigma$. Each sign factor $\Sigma_Z := \sum_\sigma (-1)^{\psi_m(s_Z(\sigma))}\Pi_\sigma$ is a self-adjoint involution commuting with $\overline{Z}_k, \overline{X}_k$, so $O_{\mathrm{dec},k}^Z = \overline{Z}_k \Sigma_Z$ and $O_{\mathrm{dec},k}^X = \overline{X}_k \Sigma_X$ are self-adjoint unitaries commuting with all $Q_i, H_{0,k}, P_{0,k}, \Pi_{\mathrm{safe},k}$, satisfy $O_{\mathrm{dec},k}^Z P_{0,k} = (-1)^{\psi_m(0)}\overline{Z}_k P_{0,k} = \overline{Z}_k P_{0,k}$, and anticommute since $\overline{Z}_k \overline{X}_k = (-1)^{\ip{L_{Z,m}}{L_{X,m}}}\overline{X}_k \overline{Z}_k = -\overline{X}_k \overline{Z}_k$.

For (ii), expand $A_W$ in the multi-qubit Pauli basis $W_{a,b} = Z^a X^b$ supported on the qubit slots of $W$ ($1 \le r \le 37\Lambda^k$). Conjugating the check projectors by $W_{a,b}$ shifts syndromes by $\varsigma(a,b) := (b_{\mathrm{pin}}, \partial_1^{\cC_m} a_{\mathrm{code}}, \delta_{\cC_m}^1 b_{\mathrm{code}}) \in \F^{\cI_k}$ via $W_{a,b}\Pi_{\sigma_2} = \Pi_{\sigma_2 \oplus \varsigma(a,b)} W_{a,b}$, while $W_{a,b}\overline{Z}_k = (-1)^{\ip{L_{Z,m}}{b_{\mathrm{code}}}}\overline{Z}_k W_{a,b}$ (for active-slot restrictions $a_{\mathrm{code}}, b_{\mathrm{code}} \in C^1(\cC_m)$). Sandwiching $[W_{a,b}, O_{\mathrm{dec},k}^Z]$ between $\Pi_{\sigma_1}$ and $\Pi_{\sigma_2}$ ($\sigma_1, \sigma_2 \in S_{\mathrm{safe},k}$) gives
\[
    \Pi_{\sigma_1}\,[W_{a,b},\, O_{\mathrm{dec},k}^Z]\,\Pi_{\sigma_2} \;=\; \Bigl((-1)^{\psi_m(s_Z(\sigma_2))} - (-1)^{\psi_m(s_Z(\sigma_1)) + \ip{L_{Z,m}}{b_{\mathrm{code}}}}\Bigr)\Pi_{\sigma_1} W_{a,b}\overline{Z}_k \Pi_{\sigma_2},
\]
which vanishes by orthogonality $\Pi_{\sigma_1}\Pi_{\sigma_2 \oplus \varsigma(a,b)} = 0$ unless $\sigma_1 = \sigma_2 \oplus \varsigma(a,b)$ and $\Pi_{\sigma_2} \neq 0$. Whenever $\Pi_{\sigma_2} \neq 0$, stabilizing $\ket{0}^{\otimes \cI_{\mathrm{pin}}}$ and CSS code-space surjectivity imply that $\sigma_2 = \varsigma(a_0, b_0)$ is excited from $P_{0,k}$ by some global Pauli operator $W_{a_0, b_0}$, so $s_Z(\sigma_2) = \delta_{\cC_m}^1 e$ and $s_Z(\sigma_1) = \delta_{\cC_m}^1 e'$ for $e := b_{0,\mathrm{code}}$ and $e' := e \oplus b_{\mathrm{code}}$. Every physical qubit snapped into $W$ ($r \le 37\Lambda^k$) has $\R^3$ coordinate within $\ell_\infty$ distance $\le r/2 + 1/2 \le r$ of the cube midpoint $p_W \in \R^3$, hence Euclidean distance $\le \sqrt{3}\,r < 2 r \le 74\Lambda^k$, so $\supp(e \oplus e') \subseteq B(p_W, 74\Lambda^k)$. Because $\sigma_1, \sigma_2 \in S_{\mathrm{safe},k}$, neither $s_Z(\sigma_2)$ nor $s_Z(\sigma_1)$ contains any witness in $\cW_m$; the $(74\Lambda^k, w_k, B_{\mathrm{step}}^{8 w_k})$-ball witness property of Lemma~\ref{lem:percolating-cluster-bounds}(ii) therefore forces $\psi_m(\delta_{\cC_m}^1 e') \oplus \ip{L_{Z,m}}{e'} = \psi_m(\delta_{\cC_m}^1 e) \oplus \ip{L_{Z,m}}{e}$, i.e., $\psi_m(s_Z(\sigma_1)) \equiv \psi_m(s_Z(\sigma_2)) + \ip{L_{Z,m}}{b_{\mathrm{code}}} \pmod 2$, making the sign difference vanish identically. Summing over $\sigma_1, \sigma_2 \in S_{\mathrm{safe},k}$ (and arguing identically for $O_{\mathrm{dec},k}^X$ via $\cW_m^\vee$) proves~\eqref{eq:safe-subspace-immunity}.

For (iii), a syndrome $\sigma = (\sigma_{\mathrm{pin}}, s_X, s_Z) \in \F^{\cI_k}$ has $\Pi_\sigma \neq 0$ if and only if $(s_X, s_Z) \in \im(\partial_1^{\cC_m}) \times \im(\delta_{\cC_m}^1)$, in which case choosing $W_\sigma = X_{\mathrm{pin}}^{\sigma_{\mathrm{pin}}} Z^{e_Z} X^{e_X}$ conjugates $P_{0,k} = \Pi_0$ unitarily onto $\Pi_\sigma = W_\sigma P_{0,k} W_\sigma^\dagger$. Consequently, every realizable syndrome subspace has the same dimension $\Tr(\Pi_\sigma) = \Tr(P_{0,k}) = 2$. Since $H_{0,k}\Pi_\sigma = (\abs{\sigma_{\mathrm{pin}}} + \abs{s_X} + \abs{s_Z})\Pi_\sigma$, the quantum partition function factorizes across the three independent sectors:
\[
    \Tr\bigl(e^{-\beta H_{0,k}}\bigr) \;=\; \sum_{\sigma :\, \Pi_\sigma \neq 0} \Tr(\Pi_\sigma)\,e^{-\beta\abs{\sigma}} \;=\; \Tr(P_{0,k})\,(1 + e^{-\beta})^{\abs{\cI_{\mathrm{pin}}}}\,Z_{\cC_m^\vee}(\beta)\,Z_{\cC_m}(\beta).
\]
Any realizable $\sigma \notin S_{\mathrm{safe},k}$ has either $s_Z \in \cS_m^{\mathrm{unsafe}}$ or $s_X \in \cS_m^{\vee,\mathrm{unsafe}}$. Cancelling the common prefactor $\Tr(P_{0,k})(1 + e^{-\beta})^{\abs{\cI_{\mathrm{pin}}}}$ in the ratio $\Tr(e^{-\beta H_{0,k}}(I - \Pi_{\mathrm{safe},k})) / \Tr(e^{-\beta H_{0,k}})$ and applying~\eqref{eq:peierls-sum-final} of Theorem~\ref{thm:walsh-fourier-peierls}(ii) to both sectors yields
\[
    p_{\mathrm{fail},k}(\beta) \;\le\; \frac{1}{Z_{\cC_m}(\beta)}\sum_{s_Z \in \cS_m^{\mathrm{unsafe}}} e^{-\beta\abs{s_Z}} \;+\; \frac{1}{Z_{\cC_m^\vee}(\beta)}\sum_{s_X \in \cS_m^{\vee,\mathrm{unsafe}}} e^{-\beta\abs{s_X}} \;\le\; \exp\bigl(-(\beta - \beta_c)w_k\bigr).
\]
Finally, since $L_k^3 = \Lambda^{6k + 12}$ and $\Lambda^{6\eta} = 3/2$, we have $c_F(L_k^3)^\eta = \Lambda^{-12\eta}\Lambda^{(6k+12)\eta} = (3/2)^k \le w_k$ by~\eqref{eq:percolating-energy-lower-bound}, giving~\eqref{eq:quantum-gibbs-failure-bound} (the formalization uses the conservative constants $\frac{\ln(3/2)}{6\ln\Lambda + 1} < \eta$ and $\frac{1}{6}\Lambda^{-12} < c_F$).
\end{proof}

\subsection{Dissipative Memory Lifetime}
\label{subsec:unperturbed-kms-lifetime}

We now show that the static Gibbs bound~\eqref{eq:quantum-gibbs-failure-bound} and cube safe-subspace logical immunity~\eqref{eq:safe-subspace-immunity} directly imply a stretched-exponential dynamical memory lifetime under Markovian thermalization at $V = 0$~\cite{davies1974markovian, alicki2009thermalization, alicki2010thermal, chesi2010thermodynamic}. For any self-adjoint Hamiltonian $H$ on $\cH_k$ with canonical Gibbs state $\rho_\beta = e^{-\beta H}/\Tr(e^{-\beta H})$, let $\Gamma_\beta(X) := \rho_\beta^{1/2} X \rho_\beta^{1/2}$ be the symmetric Kubo--Martin--Schwinger (KMS) weighting map, and define the real KMS inner product and KMS norm on bounded operators by
\begin{equation}
    \label{eq:kms-inner-norm-def}
    \ip{A}{B}_{\mathrm{KMS},\beta} \;:=\; \mathrm{Re}\,\Tr\bigl(\rho_\beta^{1/2} A^\dagger \rho_\beta^{1/2} B\bigr), \qquad
    \norm{A}_{\mathrm{KMS},\beta}^2 \;:=\; \ip{A}{A}_{\mathrm{KMS},\beta} \;=\; \Tr\bigl(\rho_\beta^{1/2} A^\dagger \rho_\beta^{1/2} A\bigr).
\end{equation}
A \emph{jump channel family of cube range $R_{\mathrm{bath}} \ge 1$} on $\Gamma_k$ is a finite indexed family $\cB = \{(\gamma_{\mathrm{ch}}, L_{\mathrm{ch}})\}_{\mathrm{ch} \in \cB}$ with rates $\gamma_{\mathrm{ch}} \ge 0$ and bounded jump operators $L_{\mathrm{ch}} \in \cB(\cH_k)$ each supported on an axis-aligned $\ell_\infty$ cube $W_{\mathrm{ch}} = (v_{\mathrm{ch}} + \{0, \dots, r_{\mathrm{ch}} - 1\}^3) \cap \Gamma_k$ ($v_{\mathrm{ch}} \in \Gamma_k$) of side $1 \le r_{\mathrm{ch}} \le R_{\mathrm{bath}}$; we say $\cB$ is \emph{adjoint-symmetric} if there is an involution $\mathrm{ch} \mapsto \overline{\mathrm{ch}}$ on $\cB$ with $\gamma_{\overline{\mathrm{ch}}} = \gamma_{\mathrm{ch}}$ and $L_{\overline{\mathrm{ch}}} = L_{\mathrm{ch}}^\dagger$. At $H = H_{0,k}$, define the associated \emph{KMS-symmetric double-commutator (pre-Dirichlet) generator} on $\cB(\cH_k)$ in the Heisenberg picture by
\begin{equation}
    \label{eq:kms-lindblad-generator-def}
    \cL_\beta^*(X) \;:=\; i[H_{0,k}, X] \;-\; \frac{1}{2}\sum_{\mathrm{ch} \in \cB} \gamma_{\mathrm{ch}}\,\Gamma_\beta^{-1}\!\Bigl(\bigl[L_{\mathrm{ch}}^\dagger,\, \Gamma_\beta([L_{\mathrm{ch}}, X])\bigr]\Bigr).
\end{equation}
In the real KMS inner product~\eqref{eq:kms-inner-norm-def}, $X \mapsto i[H_{0,k}, X]$ is skew-symmetric while $\cL_{\beta,\mathrm{diss}}^* := \cL_\beta^* - i[H_{0,k}, \cdot]$ satisfies $\ip{A}{\cL_{\beta,\mathrm{diss}}^*(B)}_{\mathrm{KMS},\beta} = -\frac{1}{2}\sum_{\mathrm{ch} \in \cB}\gamma_{\mathrm{ch}}\ip{[L_{\mathrm{ch}}, A]}{[L_{\mathrm{ch}}, B]}_{\mathrm{KMS},\beta}$ with $\cL_\beta^*(I) = 0$ and $\Tr(\rho_\beta(H_{0,k})\cL_\beta^*(X)) = 0$ (and $(\cL_\beta^*(X))^\dagger = \cL_\beta^*(X^\dagger)$ whenever $\cB$ is adjoint-symmetric). For arbitrary unfiltered operators $L_{\mathrm{ch}}$ mixing distinct Bohr frequencies of $H_{0,k}$, however, $e^{t\cL_\beta^*}$ need not preserve operator positivity---whereas the KMS-symmetric Markov semigroups and quantum Dirichlet forms of Goldstein and Lindsay~\cite{goldstein1995kms} and Majewski and Zegarli\'nski~\cite{majewski1995quantum} require positivity preservation. Physically valid Markovian thermal baths arise via the Gorini--Kossakowski--Sudarshan--Lindblad (GKLS) weak-coupling construction of Davies~\cite{davies1974markovian, lindblad1976generators, GKS1976, alicki2010thermal, chesi2010thermodynamic}: given a finite family of single-site Hermitian system--bath coupling operators $\{S_{v,\alpha}\}_{(v,\alpha) \in \cA}$ ($\supp(S_{v,\alpha}) \subseteq \{v\}$, $S_{v,\alpha}^\dagger = S_{v,\alpha}$) and bath spectral densities $h_{v,\alpha} : \Z \to \R_{\ge 0}$ obeying KMS detailed balance $h_{v,\alpha}(-\omega) = e^{-\beta\omega} h_{v,\alpha}(\omega)$ ($\omega \in \Z$), decomposing $S_{v,\alpha} = \sum_{\omega \in \Z} A_{v,\alpha,\omega}$ into Bohr-frequency eigenoperators $A_{v,\alpha,\omega} := \sum_{E \in \Z_{\ge 0}} P_{E-\omega}(H_{0,k}) S_{v,\alpha} P_E(H_{0,k})$ ($P_{E'}(H_{0,k}) := 0$ for $E' < 0$, so $[H_{0,k}, A_{v,\alpha,\omega}] = -\omega A_{v,\alpha,\omega}$ and $A_{v,\alpha,-\omega} = A_{v,\alpha,\omega}^\dagger$) defines the completely positive, unital Heisenberg-picture \emph{GKLS Davies thermal generator} (which satisfies Gelfand--Naimark--Segal detailed balance~\cite{alicki1976detailed, kossakowski1977quantum, carlen2017gradient})
\[
    \cL_{\beta,\mathrm{Davies}}^*(X) \;:=\; i[H_{0,k}, X] \;+\; \sum_{(v,\alpha) \in \cA}\sum_{\omega \in \Z} h_{v,\alpha}(\omega)\Bigl(A_{v,\alpha,\omega}^\dagger X A_{v,\alpha,\omega} - \frac{1}{2}\bigl\{A_{v,\alpha,\omega}^\dagger A_{v,\alpha,\omega},\, X\bigr\}\Bigr).
\]
As shown in Theorem~\ref{thm:unperturbed-kms-lifetime}(ii) below, each $A_{v,\alpha,\omega}$ vanishes for $\abs{\omega} > w_{\mathrm{site}} = O(1)$ and is supported on the check-star neighborhood of $v$ inside an $\ell_\infty$ cube of side $37 \le 37\Lambda^k$ for every $k \ge 0$, and $\cL_{\beta,\mathrm{Davies}}^*$ coincides identically with~\eqref{eq:kms-lindblad-generator-def} on the adjoint-symmetric channel family $\cB_{\mathrm{Davies}} := \{(\gamma_{v,\alpha}(\omega), A_{v,\alpha,\omega})\}_{(v,\alpha) \in \cA,\, \abs{\omega} \le w_{\mathrm{site}}}$ with symmetric KMS rates $\gamma_{v,\alpha}(\omega) := e^{-\beta\omega/2} h_{v,\alpha}(\omega) = \sqrt{h_{v,\alpha}(\omega) h_{v,\alpha}(-\omega)} \ge 0$.

Physically, because each decoded observable commutes with the bare code Hamiltonian and with every cube-supported jump operator ($r_{\mathrm{ch}} \le 37\Lambda^k$) inside the safe subspace, dissipative drift can arise only from transitions touching the rare unsafe-syndrome subspace. Following the safe-subspace and Dirichlet-form bottleneck methods of~\cite{alicki2009thermalization, alicki2010thermal, chesi2010thermodynamic} (invoked via~\cite[Theorem~1]{chesi2010thermodynamic} in~\cite[Section~2.5, Theorem~2.6 and Section~5.4, Corollary~5.14]{BDL24}; see also~\cite{baspin2025free}), we bound that boundary contribution directly in the symmetric KMS inner product for arbitrary cube-supported jump channels and specialize to completely positive trace-preserving (CPTP) GKLS Davies thermal baths across all $k \ge 0$, isolating a commutator-free fourth-root projector estimate that carries over without assuming $[H, \Pi_{\mathrm{safe},k}] = 0$ when coherent perturbations are turned on in Part~II~\cite{HaydenPartII}.

\leanlink{thm:unperturbed-kms-lifetime}
\begin{theorem}[{Unperturbed Memory Lifetime under KMS-Symmetric and GKLS Davies Dynamics, adapted from~\cite{alicki2010thermal} and~\cite[Theorem~1]{chesi2010thermodynamic}}]
\label{thm:unperturbed-kms-lifetime}
Fix level $k \ge 0$, inverse temperature $\beta > \beta_c$, and initial decoded observable $O_0 \in \{O_{\mathrm{dec},k}^Z, O_{\mathrm{dec},k}^X\}$, and set $\Delta F_k(\beta) = c_F(\beta - \beta_c)(L_k^3)^\eta$:
\begin{enumerate}[label=\textbf{(\roman*)}, leftmargin=2.2em, itemsep=2.5pt]
    \item \textbf{Bounded-range KMS-symmetric double-commutator bound}: For any jump channel family $\cB = \{(\gamma_{\mathrm{ch}}, L_{\mathrm{ch}})\}_{\mathrm{ch} \in \cB}$ on $\Gamma_k$ of cube range $R_{\mathrm{bath}} \le 37\Lambda^k$, every self-adjoint Heisenberg-picture trajectory $(O_t)_{t \ge 0}$ initialized at $O_{t=0} = O_0$ (in particular, $O_t := e^{t\cL_\beta^*}(O_0)$ whenever $\cB$ is adjoint-symmetric) satisfying $\frac{1}{2}\frac{d}{dt}\norm{O_t - O_0}_{\mathrm{KMS},\beta}^2 = \ip{O_t - O_0}{\cL_\beta^*(O_t)}_{\mathrm{KMS},\beta}$ obeys
    \begin{equation}
        \label{eq:unperturbed-kms-lifetime-bound}
        \norm{O_t - O_0}_{\mathrm{KMS},\beta}^2 \;\le\; 8 \Biggl(\sum_{\mathrm{ch} \in \cB} \gamma_{\mathrm{ch}}\,\norm{L_{\mathrm{ch}}}^2\Biggr) \exp\!\Bigl(-\frac{1}{2}\Delta F_k(\beta)\Bigr)\, t \qquad (\forall\, t \ge 0).
    \end{equation}
    \item \textbf{Completely positive GKLS Davies thermal baths}: Let $\{S_{v,\alpha}\}_{(v,\alpha) \in \cA}$ be any finite family of single-site Hermitian coupling operators on $\Gamma_k$ with KMS detailed-balance bath spectral densities $h_{v,\alpha}(-\omega) = e^{-\beta\omega} h_{v,\alpha}(\omega) \ge 0$ ($\omega \in \Z$), and set $\gamma_{v,\alpha}(\omega) := e^{-\beta\omega/2} h_{v,\alpha}(\omega) = \gamma_{v,\alpha}(-\omega) \ge 0$. Then every Bohr-frequency eigenoperator $A_{v,\alpha,\omega}$ satisfies $\norm{A_{v,\alpha,\omega}} \le \norm{S_{v,\alpha}}$, vanishes for $\abs{\omega} > w_{\mathrm{site}} := q_{\mathrm{code}} + 2 m_{\max}(2 r_{\mathrm{chk}} + 1)^3 = O(1)$, and is supported inside an axis-aligned $\ell_\infty$ cube $W_v \subseteq \Gamma_k$ of side $37 \le 37\Lambda^k$; moreover, $\cL_{\beta,\mathrm{Davies}}^*$ generates a completely positive, unital, $\dagger$-preserving semigroup $(e^{t\cL_{\beta,\mathrm{Davies}}^*})_{t \ge 0}$ (dual to a CPTP quantum Markov semigroup fixing $\rho_\beta(H_{0,k})$) and coincides identically on $\cB(\cH_k)$ with~\eqref{eq:kms-lindblad-generator-def} evaluated on $\cB_{\mathrm{Davies}} := \{(\gamma_{v,\alpha}(\omega), A_{v,\alpha,\omega})\}_{(v,\alpha) \in \cA,\, \abs{\omega} \le w_{\mathrm{site}}}$. Consequently, $O_t := e^{t\cL_{\beta,\mathrm{Davies}}^*}(O_0)$ obeys
    \[
    \begin{aligned}
        \norm{O_t - O_0}_{\mathrm{KMS},\beta}^2
        &\;\le\; 8\Biggl(\sum_{(v,\alpha) \in \cA}\sum_{\abs{\omega} \le w_{\mathrm{site}}} \gamma_{v,\alpha}(\omega)\,\norm{A_{v,\alpha,\omega}}^2\Biggr)\exp\!\Bigl(-\frac{1}{2}\Delta F_k(\beta)\Bigr)\,t \\
        &\;\le\; 8\Biggl(\sum_{(v,\alpha) \in \cA}\Bigl(\sum_{\abs{\omega} \le w_{\mathrm{site}}}\gamma_{v,\alpha}(\omega)\Bigr)\norm{S_{v,\alpha}}^2\Biggr)\exp\!\Bigl(-\frac{1}{2}\Delta F_k(\beta)\Bigr)\,t
    \end{aligned}
    \]
    for all $t \ge 0$. In particular, for any extensive local bath with $\sum_{(v,\alpha) \in \cA}(\sum_{\abs{\omega} \le w_{\mathrm{site}}}\gamma_{v,\alpha}(\omega))\norm{S_{v,\alpha}}^2 \le \gamma_{\max} L_k^3$ (or $\sum_{\mathrm{ch} \in \cB}\gamma_{\mathrm{ch}}\norm{L_{\mathrm{ch}}}^2 \le \gamma_{\max} L_k^3$ in~(i), $\gamma_{\max} > 0$) and any $\varepsilon > 0$, both decoded logical observables satisfy $\norm{O_t - O_0}_{\mathrm{KMS},\beta} \le \varepsilon$ (and KMS autocorrelation $\ip{O_0}{O_t}_{\mathrm{KMS},\beta} \ge 1 - \varepsilon$) for all times up to $\tau_{\mathrm{mem},k}(\beta, \varepsilon) := \frac{\varepsilon^2}{8\gamma_{\max} L_k^3}\exp(\frac{1}{2}\Delta F_k(\beta)) = \exp(\Omega((L_k^3)^\eta))$.
\end{enumerate}
\end{theorem}

\begin{proof}
\emph{Part~(i).} First note that if $\cB$ is adjoint-symmetric via $\mathrm{ch} \mapsto \overline{\mathrm{ch}}$, then $([L_{\mathrm{ch}}^\dagger, \Gamma_\beta([L_{\mathrm{ch}}, X])])^\dagger = [L_{\mathrm{ch}}, \Gamma_\beta([L_{\mathrm{ch}}^\dagger, X^\dagger])] = [L_{\overline{\mathrm{ch}}}^\dagger, \Gamma_\beta([L_{\overline{\mathrm{ch}}}, X^\dagger])]$ and $(i[H_{0,k}, X])^\dagger = i[H_{0,k}, X^\dagger]$, so $(\cL_\beta^*(X))^\dagger = \cL_\beta^*(X^\dagger)$ for all $X \in \cB(\cH_k)$ and hence $O_s = e^{s\cL_\beta^*}(O_0)$ remains self-adjoint for all $s \ge 0$. We bound the time derivative of $g(s) := \norm{O_s - O_0}_{\mathrm{KMS},\beta}^2$ ($s \in [0, t]$) in four short steps:
\begin{enumerate}[label=\textbf{Step \arabic*:}, leftmargin=4.2em, itemsep=2.5pt]
    \item \textbf{Vanishing coherent drift}: Because $[H_{0,k}, O_0] = 0$ (Proposition~\ref{prop:safe-subspace-immunity}(i)), we have $[H_{0,k}, O_s] = [H_{0,k}, \Delta_s]$ where $\Delta_s := O_s - O_0$. Setting $Y_s := \rho_\beta^{1/4}\Delta_s\rho_\beta^{1/4}$ and using $[H_{0,k}, \rho_\beta^{1/4}] = 0$ with $H_{0,k} = H_{0,k}^\dagger$, both $\Tr(\rho_\beta^{1/2}\Delta_s^\dagger\rho_\beta^{1/2} H_{0,k}\Delta_s) = \Tr(Y_s^\dagger H_{0,k} Y_s)$ and $\Tr(\rho_\beta^{1/2}\Delta_s^\dagger\rho_\beta^{1/2}\Delta_s H_{0,k}) = \Tr(Y_s H_{0,k} Y_s^\dagger)$ are real, so $\ip{\Delta_s}{i[H_{0,k}, O_s]}_{\mathrm{KMS},\beta} = \mathrm{Re}\bigl(i\,\Tr(\rho_\beta^{1/2}\Delta_s^\dagger\rho_\beta^{1/2}[H_{0,k}, \Delta_s])\bigr) = 0$.
    \item \textbf{Double-commutator KMS identity and Cauchy--Schwarz}: For each channel $\mathrm{ch} \in \cB$, cancelling $\rho_\beta^{1/2}\rho_\beta^{-1/2} = I$ inside~\eqref{eq:kms-inner-norm-def} and cycling $L_{\mathrm{ch}}^\dagger$ yields the symmetric KMS identity
    \[
        \Bigl\langle \Delta_s,\, \Gamma_\beta^{-1}\!\bigl([L_{\mathrm{ch}}^\dagger, \Gamma_\beta([L_{\mathrm{ch}}, O_s])]\bigr)\Bigr\rangle_{\mathrm{KMS},\beta} \;=\; \bigl\langle [L_{\mathrm{ch}}, O_s] - [L_{\mathrm{ch}}, O_0],\, [L_{\mathrm{ch}}, O_s]\bigr\rangle_{\mathrm{KMS},\beta}.
    \]
    Setting $E_s := \norm{[L_{\mathrm{ch}}, O_s]}_{\mathrm{KMS},\beta}^2 \ge 0$ and $E_0 := \norm{[L_{\mathrm{ch}}, O_0]}_{\mathrm{KMS},\beta}^2 \ge 0$, the Cauchy--Schwarz inequality for $\ip{\cdot}{\cdot}_{\mathrm{KMS},\beta}$ and AM--GM ($\sqrt{E_0 E_s} \le \frac{1}{2}E_s + \frac{1}{2}E_0$) give
    \[
    \begin{aligned}
        -\frac{\gamma_{\mathrm{ch}}}{2}\bigl\langle [L_{\mathrm{ch}}, O_s] - [L_{\mathrm{ch}}, O_0],\, [L_{\mathrm{ch}}, O_s]\bigr\rangle_{\mathrm{KMS},\beta}
        &\;=\; \frac{\gamma_{\mathrm{ch}}}{2}\Bigl(-E_s + \bigl\langle [L_{\mathrm{ch}}, O_0],\, [L_{\mathrm{ch}}, O_s]\bigr\rangle_{\mathrm{KMS},\beta}\Bigr) \\
        &\;\le\; \frac{\gamma_{\mathrm{ch}}}{2}\bigl(-E_s + \sqrt{E_0 E_s}\bigr) \;\le\; \frac{\gamma_{\mathrm{ch}}}{4}\,E_0.
    \end{aligned}
    \]
    \item \textbf{Safe-subspace commutator splitting and fourth-root projector bound}: Set $P_{\mathrm{fail}} := I - \Pi_{\mathrm{safe},k}$ and $C_{\mathrm{ch}} := [L_{\mathrm{ch}}, O_0]$, so $\norm{C_{\mathrm{ch}}} \le 2\norm{L_{\mathrm{ch}}}\norm{O_0} = 2\norm{L_{\mathrm{ch}}}$. Since $L_{\mathrm{ch}}$ is supported on an $\ell_\infty$ cube $W_{\mathrm{ch}} \subseteq \Gamma_k$ of side $1 \le r_{\mathrm{ch}} \le R_{\mathrm{bath}} \le 37\Lambda^k$, Proposition~\ref{prop:safe-subspace-immunity}(ii) gives $\Pi_{\mathrm{safe},k} C_{\mathrm{ch}} \Pi_{\mathrm{safe},k} = 0$, yielding the exact two-term splitting
    \[
        C_{\mathrm{ch}} \;=\; (P_{\mathrm{fail}} + \Pi_{\mathrm{safe},k})\,C_{\mathrm{ch}} \;=\; P_{\mathrm{fail}} C_{\mathrm{ch}} \;+\; \Pi_{\mathrm{safe},k} C_{\mathrm{ch}} P_{\mathrm{fail}}.
    \]
    For any orthogonal projector $P$ and bounded operator $K$ (even without assuming $[H, P] = 0$, a universality we reuse for perturbed Gibbs states in Part~II~\cite{HaydenPartII}), the Cauchy--Schwarz inequality for the Hilbert--Schmidt (HS) inner product ($\norm{X}_{\mathrm{HS}} := \sqrt{\Tr(X^\dagger X)}$) gives
    \[
    \begin{aligned}
        \norm{P K}_{\mathrm{KMS},\beta}^2 \;=\; \Tr\bigl((\rho_\beta^{1/2} K^\dagger P)(\rho_\beta^{1/2} P K)\bigr)
        &\;\le\; \norm{P K \rho_\beta^{1/2}}_{\mathrm{HS}}\,\norm{\rho_\beta^{1/2} P K}_{\mathrm{HS}} \\
        &\;\le\; \norm{K}^2\,\bigl(\Tr(\rho_\beta P)\bigr)^{1/2},
    \end{aligned}
    \]
    since $\norm{P K \rho_\beta^{1/2}}_{\mathrm{HS}}^2 = \Tr(\rho_\beta^{1/2} K^\dagger P K \rho_\beta^{1/2}) \le \norm{K}^2 \Tr(\rho_\beta) = \norm{K}^2$ and $\norm{\rho_\beta^{1/2} P K}_{\mathrm{HS}}^2 = \Tr(K^\dagger P \rho_\beta P K) \le \norm{K}^2 \Tr(\rho_\beta P)$; taking adjoints ($\norm{X^\dagger}_{\mathrm{KMS},\beta} = \norm{X}_{\mathrm{KMS},\beta}$) likewise gives $\norm{K P}_{\mathrm{KMS},\beta} \le \norm{K}(\Tr(\rho_\beta P))^{1/4}$. Applying the triangle inequality for $\norm{\cdot}_{\mathrm{KMS},\beta}$ with $P = P_{\mathrm{fail}}$ and $p_{\mathrm{fail}} := p_{\mathrm{fail},k}(\beta) \le \exp(-\Delta F_k(\beta))$ (Proposition~\ref{prop:safe-subspace-immunity}(iii)) yields
    \[
    \begin{aligned}
        \norm{C_{\mathrm{ch}}}_{\mathrm{KMS},\beta} &\;\le\; \norm{P_{\mathrm{fail}} C_{\mathrm{ch}}}_{\mathrm{KMS},\beta} + \norm{\Pi_{\mathrm{safe},k} C_{\mathrm{ch}} P_{\mathrm{fail}}}_{\mathrm{KMS},\beta} \;\le\; 2\norm{C_{\mathrm{ch}}}\,p_{\mathrm{fail}}^{1/4}, \\
        E_0 &\;\le\; 4\norm{C_{\mathrm{ch}}}^2 p_{\mathrm{fail}}^{1/2} \;\le\; 16\norm{L_{\mathrm{ch}}}^2 p_{\mathrm{fail}}^{1/2}.
    \end{aligned}
    \]
    \item \textbf{Time integration}: Combining Steps~1--3 gives $\frac{1}{2} g'(s) \le \sum_{\mathrm{ch} \in \cB} 4\gamma_{\mathrm{ch}}\norm{L_{\mathrm{ch}}}^2 p_{\mathrm{fail}}^{1/2}$, so $g'(s) \le 8(\sum_{\mathrm{ch} \in \cB}\gamma_{\mathrm{ch}}\norm{L_{\mathrm{ch}}}^2)p_{\mathrm{fail}}^{1/2} \le 8(\sum_{\mathrm{ch} \in \cB}\gamma_{\mathrm{ch}}\norm{L_{\mathrm{ch}}}^2)\exp(-\frac{1}{2}\Delta F_k(\beta))$. Integrating over $s \in [0, t]$ from $g(0) = 0$ establishes~\eqref{eq:unperturbed-kms-lifetime-bound}. Finally, since $O_0 = O_0^\dagger$, $O_0^2 = I$, and $[O_0, \rho_\beta] = 0$, we have $\norm{O_0}_{\mathrm{KMS},\beta} = 1$ and $\ip{O_0}{O_t}_{\mathrm{KMS},\beta} = 1 + \ip{O_0}{O_t - O_0}_{\mathrm{KMS},\beta} \ge 1 - \norm{O_t - O_0}_{\mathrm{KMS},\beta}$.
\end{enumerate}

\smallskip
\emph{Part~(ii).} For each $v \in \Gamma_k$, partition $\cI_k = \cI_{\mathrm{loc}}(v) \sqcup \cI_{\mathrm{ext}}(v)$, where $\cI_{\mathrm{loc}}(v)$ consists of the pinning checks at $v$ ($\le q_{\mathrm{code}}$) together with all CSS $X$- and $Z$-checks anchored at sites $u \in B_\infty(v, r_{\mathrm{chk}}) \cap \Gamma_k$ ($\le 2 m_{\max}(2 r_{\mathrm{chk}} + 1)^3$, so $\abs{\cI_{\mathrm{loc}}(v)} \le w_{\mathrm{site}}$), and set $H_{\mathrm{loc}}(v) := \sum_{i \in \cI_{\mathrm{loc}}(v)} Q_i$ and $H_{\mathrm{ext}}(v) := \sum_{i \in \cI_{\mathrm{ext}}(v)} Q_i$. By Definition~\ref{def:safe-decoded-observables}(a), every $Q_i$ with $i \in \cI_{\mathrm{ext}}(v)$ is supported in $\Gamma_k \setminus \{v\}$ and hence commutes with $S_{v,\alpha}$, while $[H_{\mathrm{loc}}(v), H_{\mathrm{ext}}(v)] = 0$. Writing $P_m(M) = \int_0^{2\pi} e^{i\theta(M - m I)}\frac{d\theta}{2\pi}$ for integer-spectrum $M$ cancels $e^{\pm i\theta H_{\mathrm{ext}}(v)}$ in
\[
    A_{v,\alpha,\omega} \;=\; \int_0^{2\pi} e^{i\theta(H_{0,k} + \omega I)}\,S_{v,\alpha}\,e^{-i\theta H_{0,k}}\,\frac{d\theta}{2\pi}
    \;=\; \sum_{E_{\mathrm{loc}} = 0}^{w_{\mathrm{site}}} P_{E_{\mathrm{loc}} - \omega}\bigl(H_{\mathrm{loc}}(v)\bigr)\,S_{v,\alpha}\,P_{E_{\mathrm{loc}}}\bigl(H_{\mathrm{loc}}(v)\bigr).
\]
Hence $A_{v,\alpha,\omega} = 0$ for $\abs{\omega} > w_{\mathrm{site}}$; writing $P_E := P_E(H_{\mathrm{loc}}(v))$, orthogonality of $\{P_E\}_{E=0}^{w_{\mathrm{site}}}$ gives $\norm{A_{v,\alpha,\omega}\psi}^2 = \sum_{E=0}^{w_{\mathrm{site}}}\norm{P_{E-\omega} S_{v,\alpha} P_E\psi}^2 \le \norm{S_{v,\alpha}}^2\sum_{E=0}^{w_{\mathrm{site}}}\norm{P_E\psi}^2 = \norm{S_{v,\alpha}}^2\norm{\psi}^2$; and since every $i \in \cI_{\mathrm{loc}}(v)$ has anchor $\norm{u - v}_\infty \le r_{\mathrm{chk}} = 9$ and check support inside $B_\infty(u, 9) \cap \Gamma_k \subseteq B_\infty(v, 18) \cap \Gamma_k$, $A_{v,\alpha,\omega}$ is supported inside the axis-aligned cube $W_v := (v_0 + \{0, \dots, 36\}^3) \cap \Gamma_k$ of side $37 \le L_k$ (setting $(v_0)_\mu := \max(0, v_\mu - 18) \in \{0, \dots, L_k - 1\}$; note $L_k = \Lambda^{2k+4} > 37$ and $37 \le 37\Lambda^k$ for all $k \ge 0$).

Because $h_{v,\alpha}(\omega) \ge 0$, $\cL_{\beta,\mathrm{Davies}}^*$ has standard GKLS form~\cite{lindblad1976generators, GKS1976}, so $(e^{t\cL_{\beta,\mathrm{Davies}}^*})_{t \ge 0}$ is completely positive and unital. Moreover, $[H_{0,k}, A_\omega] = -\omega A_\omega$ ($A_\omega := A_{v,\alpha,\omega}$) implies $\rho_\beta^{-1/2} A_\omega^\dagger \rho_\beta^{1/2} = e^{+\beta\omega/2} A_\omega^\dagger$ and $\rho_\beta^{1/2} A_\omega^\dagger \rho_\beta^{-1/2} = e^{-\beta\omega/2} A_\omega^\dagger$. Expanding~\eqref{eq:kms-lindblad-generator-def} on a single channel $A_\omega$ cancels the outer $\rho_\beta^{\pm 1/2}$ factors in pairs to give
\[
    -\frac{1}{2}\,\Gamma_\beta^{-1}\!\Bigl(\bigl[A_\omega^\dagger,\, \Gamma_\beta([A_\omega, X])\bigr]\Bigr)
    \;=\; \frac{1}{2}\,e^{+\beta\omega/2}\bigl(A_\omega^\dagger X A_\omega - A_\omega^\dagger A_\omega X\bigr)
    \;+\; \frac{1}{2}\,e^{-\beta\omega/2}\bigl(A_\omega X A_\omega^\dagger - X A_\omega A_\omega^\dagger\bigr).
\]
Multiplying by $\gamma_{v,\alpha}(\omega) = \gamma_{v,\alpha}(-\omega)$, summing over $\abs{\omega} \le w_{\mathrm{site}}$ (as $A_\omega = 0$ for $\abs{\omega} > w_{\mathrm{site}}$), and reindexing the second group via $\omega \mapsto -\omega$ (using $A_{-\omega} = A_\omega^\dagger$ and $\gamma_{v,\alpha}(\omega) e^{+\beta\omega/2} = h_{v,\alpha}(\omega)$) yields $\sum_{\omega \in \Z} h_{v,\alpha}(\omega)(A_\omega^\dagger X A_\omega - \frac{1}{2}\{A_\omega^\dagger A_\omega, X\})$, proving $\cL_{\beta,\mathrm{Davies}}^* = \cL_\beta^*$ on $\cB_{\mathrm{Davies}}$ (which also confirms stationarity of $\rho_\beta(H_{0,k})$). Since $\cB_{\mathrm{Davies}}$ is adjoint-symmetric under $(v, \alpha, \omega) \mapsto (v, \alpha, -\omega)$ and has cube range $37 \le 37\Lambda^k$ for every $k \ge 0$, applying Part~(i) and $\norm{A_{v,\alpha,\omega}} \le \norm{S_{v,\alpha}}$ completes the proof (and the identical argument reduces any Davies bath built from Hermitian couplings $S_\alpha$ on cubes of side $r_{\mathrm{coup}} = O(1)$ to Part~(i) once $r_{\mathrm{coup}} + 36 \le 37\Lambda^k$, with $\abs{\omega} \le r_{\mathrm{coup}}^3 w_{\mathrm{site}}$).
\end{proof}

It is instructive to compare Theorem~\ref{thm:unperturbed-kms-lifetime} with the thermodynamic stability criterion of Chesi, Loss, Bravyi, and Terhal~\cite[Section~6, Theorem~1]{chesi2010thermodynamic} (invoked in~\cite[Section~2.5, Theorem~2.6]{BDL24}). On microscopic weak-coupling baths, the two models coincide: $\cL_{\beta,\mathrm{Davies}}^*$ in Theorem~\ref{thm:unperturbed-kms-lifetime}(ii) is the exact Heisenberg-picture dual of the Davies master equation in~\cite[Section~6]{chesi2010thermodynamic}, sharing the same Bohr-frequency jump operators, KMS detailed-balance relation, and omitted Lamb shift, while any bath with bounded local spectral densities $h_{\max} = O(1)$ (and $O(1)$ couplings of norm $O(1)$ per site) satisfies our extensive rate hypothesis with $\gamma_{\max} = O(h_{\max})$. At the level of abstract lifetime bounds,~\cite[Theorem~1]{chesi2010thermodynamic} uses trace-norm contractivity to cover arbitrary Gibbs-stationary CPTP Lindbladians assuming \emph{one-sided} subspace protection $[E, O_0] P = 0$, giving decoded trace error $O(t\norm{\cL}_{1\to 1}\Tr(\rho_\beta(I - P)))$. Theorem~\ref{thm:unperturbed-kms-lifetime}(i) instead works in the symmetric KMS inner product, requiring KMS symmetry~\eqref{eq:kms-lindblad-generator-def} (automatic for all Davies thermal baths) while weakening protection to the \emph{two-sided} condition $\Pi_{\mathrm{safe},k}[L_{\mathrm{ch}}, O_0]\Pi_{\mathrm{safe},k} = 0$; the resulting $p_{\mathrm{fail}}^{1/2}$ factor yields the same stretched-exponential scaling $\tau_{\mathrm{mem}} = \exp(\Omega((L_k^3)^\eta))$, with the crucial advantage for Part~II~\cite{HaydenPartII} that Step~3's fourth-root estimate never requires the Hamiltonian to commute with $\Pi_{\mathrm{safe},k}$.

\section{Quasi-Adiabatic Continuation}
\label{sec:qac-banach}

\subsection{Quasi-Local Operator Spaces}
\label{subsec:dsuper-banach}

We now turn to the zero-temperature ($T = 0$) perturbative and dynamical memory stability branch of the proof (the right branch of Figure~\ref{fig:proof-roadmap}). Together with the unperturbed positive-temperature Gibbs and KMS-dissipative analysis of Sections~\ref{sec:two-stage-cleaner}--\ref{sec:cluster-decoder-peierls}, Sections~\ref{sec:qac-banach}--\ref{sec:telescoping-main} resolve the zero-temperature perturbative stability question raised as an open direction in~\cite[Section~1.1]{BDL24}: we prove that at every macroscopic welded hierarchy scale, the bulk spectral gap, the decoded logical observables, their safe-subspace error immunity, and the closed-system ground-space memory lifetime remain stable under arbitrary quasi-local perturbations of coupling $J \le J_0$ with superpolynomial spatial decay. Establishing simultaneous positive-temperature and perturbative passive self-correction ($T > 0, V \neq 0$) requires combining quasi-local spectral filtering with finite-temperature worldline cluster expansions and multiscale syndrome peeling, and is treated in Part~II of this series~\cite{HaydenPartII} (previewed in Section~\ref{subsec:part-ii-preview}).

Recall from Definition~\ref{def:safe-decoded-observables}(a) that at even hierarchy depth $m = 2k + 2$ ($k \ge 4$), snapping the embedded welded complex $(\cC_m, \Phi_m)$ into unit cells places the code on the open three-dimensional cubic grid
\begin{equation}
    \label{eq:open-cubic-grid}
    \Gamma_k \;:=\; \{0, 1, \dots, L_k - 1\}^3, \qquad L_k \;=\; \Lambda^{2k+4},
\end{equation}
endowed with the $\ell_\infty$ lattice metric $\dist_\infty(u, v) := \max_{\alpha \in \{1,2,3\}} \abs{u_\alpha - v_\alpha}$, centered $\ell_\infty$-balls $B(u, j) := \{v \in \Gamma_k : \dist_\infty(u, v) \le j\}$ and $j$-collars $B(X, j) := \{v \in \Gamma_k : \dist_\infty(v, X) \le j\}$ ($X \subseteq \Gamma_k$, reusing $B(\cdot, \cdot)$ on $\Gamma_k$ for $\ell_\infty$-balls), and the family $\cS(r)$ of axis-aligned (boundary-clamped) lattice cubes $Q_r(u) := (u + \{0, \dots, r-1\}^3) \cap \Gamma_k$ ($u \in \Gamma_k$) of side $r \ge 1$ (each having $\ell_\infty$-diameter at most $r - 1$), with uniform site Hilbert space $\cH_u \cong \C^{d_{\loc}}$ ($d_{\loc} = 2^{q_{\mathrm{code}}} = O(1)$), total space $\cH := \cH_k = \bigotimes_{u \in \Gamma_k} \cH_u$, and frustration-free commuting-projector code Hamiltonian
\begin{equation}
    \label{eq:coarse-grained-H0}
    H_{0,k} \;=\; \sum_{i = 1}^{N_{\mathrm{chk}}(k)} Q_i \;=\; \sum_{u \in \Gamma_k} Q_{\mathrm{site}}(u), \qquad
    P_{0,k} \;:=\; \prod_{i = 1}^{N_{\mathrm{chk}}(k)} (I - Q_i).
\end{equation}
Here each check $Q_i$ is supported within $\ell_\infty$ radius $r_{\mathrm{chk}} = 9$ of its anchor site (inside a $(2 r_{\mathrm{chk}} + 1)^3$ cube in $\cS(2 r_{\mathrm{chk}} + 1)$); for notational simplicity across Sections~\ref{sec:qac-banach}--\ref{sec:telescoping-main} (following~\cite{BH11}, either by blocking constant-size subcubes or replacing $2j + 3$ with $2j + 4 r_{\mathrm{chk}} + 1$ in~\eqref{eq:Cj-ball-def}, which preserves $\cD_{\mathrm{super}}$ identically by Lemma~\ref{lem:dsuper-closure}(iii); in our formalization, $4 r_{\mathrm{chk}} + 3 = 39 \le C$ is absorbed directly inside clamped cubes of side $\ceil{C(2j + 3)}$), we write check supports as elementary $2 \times 2 \times 2$ cubes in $\cS(2)$ ($\ell_\infty$-diameter at most $1$), with $Q_{\mathrm{site}}(u)$ summing the checks anchored at $u$ so $\supp(Q_{\mathrm{site}}(u)) \subseteq B(u, 1)$ and $\norm{Q_{\mathrm{site}}(u)} \le w_{\mathrm{chk}} = O(1)$. On $\Gamma_k$, Theorem~\ref{thm:rca-welded-hierarchy} furnishes Relative Cone Acyclicity $\RCA(L_k^*, C)$ up to the macroscopic cutoff scale $L_k^* = \Lambda^{2k}$ with universal linear collar constant $C \ge 1$.

Even when a bare perturbation acts only at finite range, rotating the Hamiltonian into its perturbed ground-state frame via quasi-adiabatic continuation (QAC)~\cite{kato1950adiabatic, hastings2005quasi, BHM10, BMNS12, MZ13} and applying a band-limited spectral filter spreads each local interaction across space. Exact annihilation of cross-gap transitions comes at a price, however: a smooth, compactly supported frequency filter cannot decay exponentially in time, so the dressed operators decay faster than any inverse power of distance rather than strictly exponentially. Moreover, when the topological stability step in Section~\ref{sec:telescoping-main} relates local checks inside a ball to a linearly dilated collar around it, spatial radii are rescaled by a constant factor. Formulating perturbations from the outset within the class $\cD_{\mathrm{super}}$ of superpolynomially decaying profiles accommodates both band-limited spectral filtering and linear collar rescaling within a single algebraically closed operator space.

\leanlink{def:banach-quasilocal}
\begin{definition}[Superpolynomial Decay Class $\cD_{\mathrm{super}}$ and Site-Anchored Quasi-Local Perturbations]
\label{def:banach-quasilocal}
Let $\Gamma_k = \{0, \dots, L_k - 1\}^3$ be the open cubic lattice~\eqref{eq:open-cubic-grid}, and for any subset $S \subseteq \Gamma_k$ let $\bbE_S : \cB(\cH) \to \cB(\cH_S) \otimes I_{S^c}$ on the bounded-operator algebra $\cB(\cH)$ denote the normalized partial-trace conditional expectation onto operators supported on $S$ (obtained by Haar-averaging $U^\dagger(\cdot)U$ over unitaries $U$ supported on $S^c$~\cite{BHV06}).
\begin{enumerate}[label=\textbf{(\alph*)}, leftmargin=2.2em, itemsep=2.5pt]
    \item \textbf{Superpolynomial decay class $\cD_{\mathrm{super}}$}: A nonnegative profile $f : \N \to [0, \infty)$ belongs to $\cD_{\mathrm{super}}$ if $\lim_{r \to \infty} r^p f(r) = 0$ for every integer $p \ge 1$, or equivalently if every polynomial moment is finite:
    \begin{equation}
        \label{eq:dsuper-def}
        \norm{f}_{(p)} \;:=\; \sup_{r \ge 0} (1 + r)^p f(r) \;<\; \infty \qquad \forall\, p \in \N.
    \end{equation}
    \item \textbf{Cube-indexed and site-anchored quasi-local perturbations $\cB_f(\Gamma_k)$}: Given $f \in \cD_{\mathrm{super}}$ and coupling $J \ge 0$, a Hermitian perturbation
    \begin{equation}
        \label{eq:cube-perturbation-decomp}
        V \;=\; \sum_{r = 1}^{L_k} \sum_{A \in \cS(r)} V_{r,A}, \qquad V_{r,A}^\dagger \;=\; V_{r,A}, \qquad \supp(V_{r,A}) \;\subseteq\; A,
    \end{equation}
    belongs to $\cB_f(\Gamma_k)$ with $\norm{V}_f \le J$ if it admits a decomposition~\eqref{eq:cube-perturbation-decomp} satisfying $\norm{V_{r,A}} \le J f(r)$ for all $1 \le r \le L_k$ and $A \in \cS(r)$ (equivalently, $\norm{V}_f := \inf\sup_{r,A}\norm{V_{r,A}}/f(r) \le J$, taking $0/0 := 0$ when $V_{r,A} = 0$). Assigning each cube $A \in \cS(r)$ to its lexicographically minimal corner $u(A) \in \Gamma_k$ (at most one cube in $\cS(r)$ has $u(A) = u$, and $A \subseteq B(u(A), r - 1) \subseteq B(u(A), r)$) yields the exact site-anchored decomposition
    \begin{equation}
        \label{eq:site-anchored-decomp}
        V \;=\; \sum_{u \in \Gamma_k} V_{\mathrm{site}}(u), \qquad
        V_{\mathrm{site}}(u) \;:=\; \sum_{r = 1}^{L_k} V_{u,r}, \qquad
        V_{u,r} \;:=\; \sum_{\substack{A \in \cS(r)\\ u(A) = u}} V_{r,A},
    \end{equation}
    satisfying $V_{\mathrm{site}}(u)^\dagger = V_{\mathrm{site}}(u)$, $\supp(V_{u,r}) \subseteq B(u, r)$, $\norm{V_{u,r}} \le J f(r)$, and $\norm{V_{\mathrm{site}}(u)} \le J \sum_{r=1}^\infty f(r)$. More generally, a dressed site-operator family $\{X_u\}_{u \in \Gamma_k}$ has $J$-proportional $\cD_{\mathrm{super}}$ conditional-expectation tails with profile $g \in \cD_{\mathrm{super}}$ if $\norm{X_u - \bbE_{B(u,j)}(X_u)} \le J g(j)$ for all $u \in \Gamma_k$ and $j \ge 1$.
\end{enumerate}
\end{definition}

Characterizing $\cD_{\mathrm{super}}$ through finite polynomial moments rather than monotone envelopes makes it straightforward to verify the standard rapid-decay closure properties used across quasi-local lattice perturbation theory~\cite{BH11, BMNS12, MZ13, NSY19}: superpolynomial decay is preserved when multiplying by polynomials, summing over spatial tails, rescaling ball radii linearly or by square roots, or optimizing across polynomial Lieb--Robinson bounds of arbitrary order.

\leanlink{lem:dsuper-closure}
\begin{lemma}[{Closure Properties of $\cD_{\mathrm{super}}$; cf.~\cite{BH11, BMNS12, MZ13, NSY19}}]
\label{lem:dsuper-closure}
The superpolynomial decay class $\cD_{\mathrm{super}}$ is closed under the following four operations:
\begin{enumerate}[label=\textbf{(\roman*)}, leftmargin=2.2em, itemsep=2.5pt]
    \item For any $f \in \cD_{\mathrm{super}}$ and $d \in \N$, $g(r) := (1 + r)^d f(r)$ belongs to $\cD_{\mathrm{super}}$ with $\norm{g}_{(p)} = \norm{f}_{(p+d)}$.
    \item For any $f \in \cD_{\mathrm{super}}$, the upper tail $F(R) := \sum_{r = R}^\infty f(r)$ belongs to $\cD_{\mathrm{super}}$ with $\norm{F}_{(p)} \le 2\norm{f}_{(p+2)}$.
    \item For any $f \in \cD_{\mathrm{super}}$ and integer constants $C \ge 1, a \ge 0$, both the linearly rescaled profile $f_{C, a}(r) := f\bigl(\floor{\max(0, r - a)/C}\bigr)$ and the square-root profile $f_{\mathrm{sqrt}}(r) := f\bigl(\floor{\sqrt{r}}\bigr)$ belong to $\cD_{\mathrm{super}}$.
    \item If a nonnegative sequence $b : \N \to [0, \infty)$ satisfies a family of polynomial bounds $b(r) \le C_m (1 + c r)^{-m}$ for some $c > 0$ and constants $C_m \ge 0$ for every integer $m \ge 1$, then $b \in \cD_{\mathrm{super}}$.
\end{enumerate}
\end{lemma}

\ifstandardproofs
\begin{proof}
Property~(i) follows immediately from $(1+r)^p(1+r)^d f(r) = (1+r)^{p+d}f(r) \le \norm{f}_{(p+d)}$. For~(ii),
\[
    (1 + R)^p \sum_{r = R}^\infty f(r) \;\le\; \sum_{r = R}^\infty (1 + r)^p f(r) \;\le\; \norm{f}_{(p+2)} \sum_{r = R}^\infty (1 + r)^{-2} \;\le\; 2\norm{f}_{(p+2)}.
\]
For~(iii), for $r \ge a + C$ we have $1 + r \le (2C + a)\bigl(1 + \floor{(r - a)/C}\bigr)$, so $(1+r)^p f_{C,a}(r) \le (2C + a)^p \norm{f}_{(p+1)}(1 + \floor{(r-a)/C})^{-1} \to 0$ as $r \to \infty$; similarly, for $r \ge 1$, $r \le 4\floor{\sqrt{r}}^2$ gives $r^p f(\floor{\sqrt{r}}) \le 4^p \floor{\sqrt{r}}^{2p} f(\floor{\sqrt{r}}) \to 0$. Finally, for~(iv), setting $m = p + 1$ and noting $1 + r \le \max(1, c^{-1})(1 + cr)$ gives $(1+r)^p b(r) \le \max(1, c^{-(p+1)}) C_{p+1}(1 + r)^{-1}$, so $\norm{b}_{(p)} < \infty$ for every $p \ge 1$.
\end{proof}
\fi

\subsection{Band-Limited Spectral Filters}
\label{subsec:bochner-lr}

To track the perturbed ground space across $H(s) = H_{0,k} + s V$ ($s \in [0, 1]$) and filter out off-diagonal transitions across a bulk spectral gap $\Delta E(s) \ge 1/2$, we construct a compact-frequency low-pass filter kernel $w_{\mathrm{BP}}$ and its associated upper-tail continuation kernel $W_{\mathrm{BP}}$ (retaining subscript $\mathrm{BP}$ for the band-limited low-pass profile, following~\cite{BH11, BMNS12, MZ13}). Fix an even, compactly supported smooth bump function $\widehat{w}_{\mathrm{BP}} \in C_c^\infty(\R)$ satisfying $0 \le \widehat{w}_{\mathrm{BP}}(\xi) \le 1$ and
\begin{equation}
    \label{eq:bump-freq-def}
    \widehat{w}_{\mathrm{BP}}(\xi) \;=\; 1 \quad \text{for } \abs{\xi} \le \frac{1}{8\pi}, \qquad
    \widehat{w}_{\mathrm{BP}}(\xi) \;=\; 0 \quad \text{for } \abs{\xi} \ge \frac{1}{4\pi}.
\end{equation}
Define its inverse Fourier transform $w_{\mathrm{BP}} : \R \to \R$ and odd upper-tail continuation weight $W_{\mathrm{BP}} : \R \to \R$ by
\begin{equation}
    \label{eq:wBP-WBP-def}
    w_{\mathrm{BP}}(t) \;:=\; \int_{-\infty}^\infty \widehat{w}_{\mathrm{BP}}(\xi)\,e^{-2\pi i \xi t}\,d\xi, \qquad
    W_{\mathrm{BP}}(\tau) \;:=\; \operatorname{sgn}(\tau)\int_{\abs{\tau}}^\infty w_{\mathrm{BP}}(t)\,dt.
\end{equation}
Because $\widehat{w}_{\mathrm{BP}}$ is real, even, and smooth with compact support, $w_{\mathrm{BP}}$ is real-valued, even, and lies in the Schwartz space $\mathscr{S}(\R)$, with finite polynomial moments $\int_{-\infty}^\infty \abs{t}^p \abs{w_{\mathrm{BP}}(t)}\,dt < \infty$ of all orders $p \ge 0$ and $\cD_{\mathrm{super}}$ upper tails $R \mapsto \int_{\abs{t} \ge c R} \abs{w_{\mathrm{BP}}(t)}\,dt$ for every $c > 0$. By Fourier inversion, $\hat{w}(\omega) := \int_{-\infty}^\infty w_{\mathrm{BP}}(t) e^{i\omega t}\,dt = \widehat{w}_{\mathrm{BP}}\bigl(\omega / (2\pi)\bigr)$, which yields unit normalization and exact spectral annihilation above the gap threshold $\Delta = 1/2$:
\begin{equation}
    \label{eq:wBP-fourier-props}
    \int_{-\infty}^\infty w_{\mathrm{BP}}(t)\,dt \;=\; 1, \qquad
    \int_{-\infty}^\infty w_{\mathrm{BP}}(t)\,e^{i\omega t}\,dt \;=\; 0 \quad \forall\,\abs{\omega} \ge \frac{1}{2}.
\end{equation}
Correspondingly, $W_{\mathrm{BP}}$ is real-valued, odd ($\int_{-\infty}^\infty W_{\mathrm{BP}}(\tau)\,d\tau = 0$), has finite polynomial moments and $\cD_{\mathrm{super}}$ upper tails of all orders, and by Fubini's theorem satisfies the tail-integration identity
\begin{equation}
    \label{eq:wBP-WBP-ibp}
    \int_{-\infty}^\infty w_{\mathrm{BP}}(t)\Bigl(\int_0^t G(\tau)\,d\tau\Bigr)dt \;=\; \int_{-\infty}^\infty W_{\mathrm{BP}}(\tau)\,G(\tau)\,d\tau
\end{equation}
for every bounded continuous map $G : \R \to \cB(\cH)$, since $w_{\mathrm{BP}}(-t) = w_{\mathrm{BP}}(t)$ turns the $\tau > 0$ and $\tau < 0$ regions of $\int_0^t G(\tau)\,d\tau$ into $\int_{\abs{\tau}}^\infty w_{\mathrm{BP}}(t)\,dt \cdot \operatorname{sgn}(\tau)$. For any Hermitian Hamiltonian $H'$ and any integrable weight $K \in L^1(\R)$, we denote the associated Heisenberg time evolution and filtered time-convolution operator of $A \in \cB(\cH)$ by
\begin{equation}
    \label{eq:two-bochner-transforms}
    \tau_t^{H'}(A) \;:=\; e^{i t H'} A e^{-i t H'}, \qquad
    \cF_K^{H'}(A) \;:=\; \int_{-\infty}^\infty K(t)\,\tau_t^{H'}(A)\,dt,
\end{equation}
and write $\tau_t^s := \tau_t^{H(s)}$ for the interpolated Hamiltonian $H(s) = H_{0,k} + s V$.

\begin{figure}[t]
\centering
\includegraphics[width=0.88\textwidth]{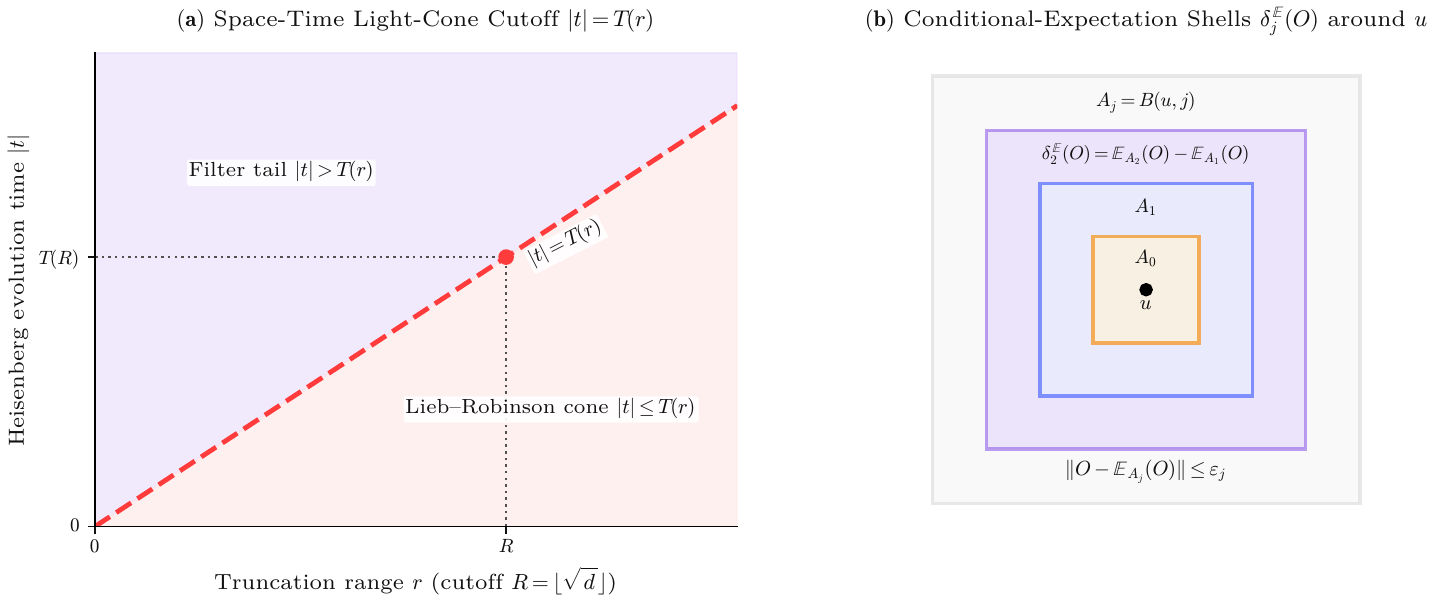}
\caption{Space-time lightcone truncation and conditional-expectation shell decomposition for the filtered time-convolution operator $\cF_K^{H(s)}(W) = \int_{-\infty}^\infty K(t)\,\tau_t^s(W)\,dt$ of a local operator $W$ (Proposition~\ref{prop:bochner-time-convolution}). \textbf{(a)} Space-time lightcone cutoff $\abs{t} = T(r)$ in the $(r, \abs{t})$ plane for truncation range $r$ (evaluated at $R = \floor{\sqrt{d}}$ for separation $d = \dist_\infty(X, Y)$), split along the red dashed line $\abs{t} = T(r)$ into the lower pink Lieb--Robinson cone $\abs{t} \le T(r)$ and upper violet filter tail $\abs{t} > T(r)$. \textbf{(b)} Conditional-expectation shells $\delta_j^{\bbE}(O) := \bbE_{A_j}(O) - \bbE_{A_{j-1}}(O)$ for a quasi-local operator $O$ around site $u$ (black dot), showing concentric nested $\ell_\infty$-balls $A_0 = \{u\} \subset A_1 \subset \dots \subset A_j = B(u, j)$ ($j \ge 1$, $\delta_0^{\bbE} := \bbE_{\{u\}}$, in amber, blue, violet, gray) with $\delta_2^{\bbE}(O) = \bbE_{A_2}(O) - \bbE_{A_1}(O)$ and superpolynomial tail bound $\norm{O - \bbE_{A_j}(O)} \le \varepsilon_j$ (e.g.\ $\varepsilon_j = J g_{\mathrm{tail}}(j)$, $g_{\mathrm{tail}} \in \cD_{\mathrm{super}}$; Section~\ref{subsec:single-cube-telescoping}).}
\label{fig:lieb-robinson-shells}
\end{figure}

To control spatial spreading under $\tau_t^s$, $\cF_{w_{\mathrm{BP}}}^{H(s)}$, $\cF_{W_{\mathrm{BP}}}^{H(s)}$, and the non-autonomous unitary flow $U(s)$ ($U(0) = I$) generated by $\cD(s) := \cF_{W_{\mathrm{BP}}}^{H(s)}(V)$ on $s \in [0, 1]$ (Figure~\ref{fig:lieb-robinson-shells}), we combine a polynomial-decay Lieb--Robinson (LR) bound (adapting~\cite[Appendix~A]{hastings2006spectral} and~\cite{lieb1972finite, nachtergaele2006propagation, NSY19}) with a short-range/long-range scale split. Specifically, for any algebraic order $p \ge 0$, writing $\diam_\infty(Z) := \max_{x, y \in Z}\dist_\infty(x, y)$, the local interaction terms $\{h_s(Z)\}_{Z \subseteq \Gamma_k}$ of $H(s) = \sum_{Z \subseteq \Gamma_k} h_s(Z)$ ($s \in [0, 1]$, $\norm{V}_f \le J \le 1$) have a finite polynomial interaction norm $s_1(p) := \max_{x \in \Gamma_k} \sum_{Z \ni x} \norm{h_s(Z)}\,\abs{Z}\,\bigl(1 + \diam_\infty(Z)\bigr)^p < \infty$ independent of $L_k$, yielding for disjoint subsets $X, Y \subseteq \Gamma_k$ at $\ell_\infty$-distance $d = \dist_\infty(X, Y) \ge 1$ and operators $A$ supported on $X$, $B$ supported on $Y$:
\begin{equation}
    \label{eq:hastings-koma-lr}
    \norm{[\tau_t^s(A), B]} \;\le\; 2\,\abs{X}\,\norm{A}\,\norm{B}\,\frac{e^{2 s_1(p)\abs{t}} - 1}{(1 + d/2)^p} \qquad (\forall\, t \in \R,\; p \ge 0).
\end{equation}
Care is required when integrating~\eqref{eq:hastings-koma-lr} over all $t \in \R$ against $K \in \{w_{\mathrm{BP}}, W_{\mathrm{BP}}\}$: because $\widehat{w}_{\mathrm{BP}}$ is compactly supported, $K(t)$ decays superpolynomially rather than exponentially, while $e^{2 s_1(p)\abs{t}}$ grows exponentially at a $p$-dependent rate. We handle this by restricting $H_{\le R}(s)$ to the range-$R$ interaction terms supported inside the local collar $B(X, d - 1)$ at $R := \floor{\sqrt{d}}$: on $\abs{t} \le T(R)$ where $T(r) := c_T r$ and $c_T := \bigl(4 e (s_1(0) + 1)\bigr)^{-1}$, the collar evolution $\tau_t^{\le R}(A)$ stays disjoint from $Y$, boundary range-$R$ terms contribute a $p$-independent LR error $\propto \abs{X} e^{-(d-R)/R + 2 e s_1(0)\abs{t}} \le e\abs{X} e^{-R/2}$, and long-range ($> R$) terms intersecting $B(X, d - 1)$ ($O(\abs{X} R^6)$ sites) contribute a Duhamel error linear in $\abs{t}$ of order $O(\abs{t}\,\abs{X}\,s_1(m+6)(1+R)^{-m})$ for every $m \in \N$, while $\abs{t} > T(R)$ is bounded directly by the rapid-decay filter tail $2\norm{A}\norm{B}\int_{\abs{t} \ge c_T R} \abs{K(t)}\,dt$; all of these terms belong to $\cD_{\mathrm{super}}$ in $d$ by Lemma~\ref{lem:dsuper-closure}(iii)--(iv). Meanwhile, for the non-autonomous QAC generator flow $\frac{d}{d\sigma}U(\sigma) = i\cD(\sigma)U(\sigma)$ on $\sigma \in [0, s] \subseteq [0, 1]$, the flow duration $\sigma \le 1$ is bounded, so~\eqref{eq:hastings-koma-lr} applies directly at $\abs{\sigma} \le 1$ for every $p = m \in \N$ and optimizes into $\cD_{\mathrm{super}}$ by Lemma~\ref{lem:dsuper-closure}(iv).

Combining this range-and-time decomposition with telescoping conditional-expectation shells $\delta_j^{\bbE} := \bbE_{B(u,j)} - \bbE_{B(u,j-1)}$ yields the standard quasi-locality estimates of~\cite[Lemmas~1 and~2]{BH11}, \cite[Section~4]{BMNS12}, \cite[Lemmas~1 and~2]{MZ13}, and~\cite{NSY19} specialized to $\cD_{\mathrm{super}}$: both band-limited spectral filtering and the quasi-adiabatic unitary flow spread local operators only up to superpolynomially decaying tails carrying an explicit factor of the perturbation strength $J$.

\leanlink{prop:bochner-time-convolution}
\begin{proposition}[{Quasi-Locality Bounds for Spectral Filters and QAC Flows~\cite[Lemmas~1 and~2]{BH11}, \cite[Section~4]{BMNS12}, \cite[Lemmas~1 and~2]{MZ13}; see also~\cite{NSY19}}]
\label{prop:bochner-time-convolution}
Let $f \in \cD_{\mathrm{super}}$, let $V \in \cB_f(\Gamma_k)$ satisfy $\norm{V}_f \le J \le 1$, and let $K \in \{w_{\mathrm{BP}}, W_{\mathrm{BP}}\}$. There exist $L_k$-independent decay profiles $g_{\mathrm{bare},V}, g_{\mathrm{bare},Q}, f_{\mathrm{gen}}, g_{\mathrm{LR}} \in \cD_{\mathrm{super}}$ (depending only on $f$ and the uniform geometric constants of $(\cC_m, \Phi_m)$) such that for all $s \in [0, 1]$, all $u \in \Gamma_k$, and all radii $j \ge 1$ with $j \le 4 L_k$:
\begin{enumerate}[label=\textbf{(\roman*)}, leftmargin=2.2em, itemsep=2.5pt]
    \item Both the filtered site perturbation $\cF_K^{H(s)}(V_{\mathrm{site}}(u))$ and the filtered check difference $\cF_{w_{\mathrm{BP}}}^{H(s)}(Q_{\mathrm{site}}(u)) - Q_{\mathrm{site}}(u)$ satisfy $J$-proportional conditional-expectation bounds on the half-radius ball $B(u, \floor{j/2})$:
    \begin{align}
        \norm{\cF_K^{H(s)}\bigl(V_{\mathrm{site}}(u)\bigr) - \bbE_{B(u, \floor{j/2})}\!\Bigl(\cF_K^{H(s)}\bigl(V_{\mathrm{site}}(u)\bigr)\Bigr)} &\;\le\; J\,g_{\mathrm{bare},V}(j), \label{eq:bochner-bare-V-tail} \\
        \norm{\Bigl(\cF_{w_{\mathrm{BP}}}^{H(s)}(Q_{\mathrm{site}}(u)) - Q_{\mathrm{site}}(u)\Bigr) - \bbE_{B(u, \floor{j/2})}\!\Bigl(\cF_{w_{\mathrm{BP}}}^{H(s)}(Q_{\mathrm{site}}(u)) - Q_{\mathrm{site}}(u)\Bigr)} &\;\le\; J\,g_{\mathrm{bare},Q}(j). \label{eq:bochner-bare-Q-tail}
    \end{align}
    \item The self-adjoint generator $\cD(\sigma) := \cF_{W_{\mathrm{BP}}}^{H(\sigma)}(V) = \sum_{u \in \Gamma_k} \cF_{W_{\mathrm{BP}}}^{H(\sigma)}(V_{\mathrm{site}}(u))$ decomposes into quasi-local terms of decay profile $f_{\mathrm{gen}} \in \cD_{\mathrm{super}}$ scaled by $J$, and the unique unitary solution $\frac{d}{d\sigma}U(\sigma) = i\cD(\sigma)U(\sigma)$ ($U(0) = I$) satisfies, for every single-site operator $A_v$ supported on $\{v\} \subseteq \Gamma_k$ and every radius $r \ge 0$,
    \begin{equation}
        \label{eq:unitary-one-site-lr}
        \norm{U(s)^\dagger A_v U(s) - \bbE_{B(v, r)}\bigl(U(s)^\dagger A_v U(s)\bigr)} \;\le\; J\,g_{\mathrm{LR}}(r)\,\norm{A_v},
    \end{equation}
    as well as analogous $J$-proportional light-cone bounds from $B(u, \floor{j/2})$ to $B(u, j)$ and for $U(s)^\dagger Q_{\mathrm{site}}(u) U(s) - Q_{\mathrm{site}}(u)$.
\end{enumerate}
\end{proposition}

\ifstandardproofs
\begin{proof}
For~(i), split $V_{\mathrm{site}}(u) = \sum_{q = 1}^{\floor{j/4}} V_{u,q} + \sum_{q > \floor{j/4}} V_{u,q}$. The outer sum over $q > \floor{j/4}$ has norm at most $J \sum_{q > \floor{j/4}} f(q)$, which contributes at most $2\norm{K}_{L^1} J \sum_{q > \floor{j/4}} f(q)$ after applying $\cF_K^{H(s)}$ and $\id - \bbE_{B(u, \floor{j/2})}$. Each inner term $V_{u,q}$ ($q \le \floor{j/4}$) is supported inside $B(u, \floor{j/4})$, which lies at $\ell_\infty$-distance $d(j) \ge \floor{j/2} + 1 - \floor{j/4} \ge j/4$ from $B(u, \floor{j/2})^c$. Applying the range-$R$ collar truncation at $R = \floor{\sqrt{d(j)}}$ and time cutoff $T(R) = c_T R$ described above bounds the conditional-expectation error of $\cF_K^{H(s)}(V_{u,q})$ outside $B(u, \floor{j/2})$ by $J f(q)$ times a profile in $\cD_{\mathrm{super}}(j)$ by Lemma~\ref{lem:dsuper-closure} (formalized in full in \leanfile[SchwartzCollarLightCone.lean]{Proofs/KatoODE/SchwartzCollarLightCone.lean}), establishing~\eqref{eq:bochner-bare-V-tail}. For~\eqref{eq:bochner-bare-Q-tail}, note that $[H_{0,k}, Q_{\mathrm{site}}(u)] = 0$, so integrating Heisenberg's equation of motion $\tau_t^s(Q_{\mathrm{site}}(u)) - Q_{\mathrm{site}}(u) = \int_0^t i\,\tau_{t'}^s\bigl([s V, Q_{\mathrm{site}}(u)]\bigr)\,dt'$ and applying $\int_{-\infty}^\infty w_{\mathrm{BP}}(t)\,dt = 1$ together with~\eqref{eq:wBP-WBP-ibp} rewrites
\begin{equation}
    \label{eq:check-duhamel-WBP}
    \cF_{w_{\mathrm{BP}}}^{H(s)}(Q_{\mathrm{site}}(u)) - Q_{\mathrm{site}}(u) \;=\; i\,\cF_{W_{\mathrm{BP}}}^{H(s)}\bigl([s V, Q_{\mathrm{site}}(u)]\bigr).
\end{equation}
Because $\supp(Q_{\mathrm{site}}(u)) \subseteq B(u, 1)$, only cubes $A \in \cS(q)$ intersecting $B(u, 1)$ contribute to $[V, Q_{\mathrm{site}}(u)]$; there are at most $(q + 2)^3$ such cubes at each scale $q$, each supported in $B(u, q + 1)$ with commutator norm at most $2 w_{\mathrm{chk}} J f(q)$. Applying the same scale split to $[s V, Q_{\mathrm{site}}(u)]$ with $K = W_{\mathrm{BP}}$ yields~\eqref{eq:bochner-bare-Q-tail} with the $J$ prefactor intact.

For~(ii), expanding $\cF_{W_{\mathrm{BP}}}^{H(\sigma)}(V_{\mathrm{site}}(u))$ into telescoping conditional-expectation shells $\delta_r^{\bbE} := \bbE_{B(u,r)} - \bbE_{B(u,r-1)}$ ($\delta_0^{\bbE} := \bbE_{\{u\}}$) and applying~\eqref{eq:bochner-bare-V-tail} places $\cD(\sigma)$ into a non-autonomous interaction family of decay profile $J f_{\mathrm{gen}}$ ($f_{\mathrm{gen}} \in \cD_{\mathrm{super}}$). For any single-site operator $A_v$, integrating $\frac{d}{d\sigma}\bigl(U(\sigma)^\dagger A_v U(\sigma)\bigr) = i\,U(\sigma)^\dagger [A_v, \cD(\sigma)] U(\sigma)$ over $\sigma \in [0, s] \subseteq [0, 1]$ shows first that $\norm{U(s)^\dagger A_v U(s) - A_v} \le \int_0^s \norm{[A_v, \cD(\sigma)]}\,d\sigma = O(J\norm{A_v})$ (yielding~\eqref{eq:unitary-one-site-lr} at small $r$), and combining this first-step commutator extraction with the non-autonomous LR bound~\eqref{eq:hastings-koma-lr} on $\sigma \in [0, s] \subseteq [0, 1]$ at arbitrary order $p = m \in \N$ yields~\eqref{eq:unitary-one-site-lr} with $g_{\mathrm{LR}} \in \cD_{\mathrm{super}}$ by Lemma~\ref{lem:dsuper-closure}(iv).
\end{proof}
\fi

\subsection{Block Diagonalization}
\label{subsec:qac-block-diag}

\begin{figure}[t]
\centering
\includegraphics[width=0.90\textwidth]{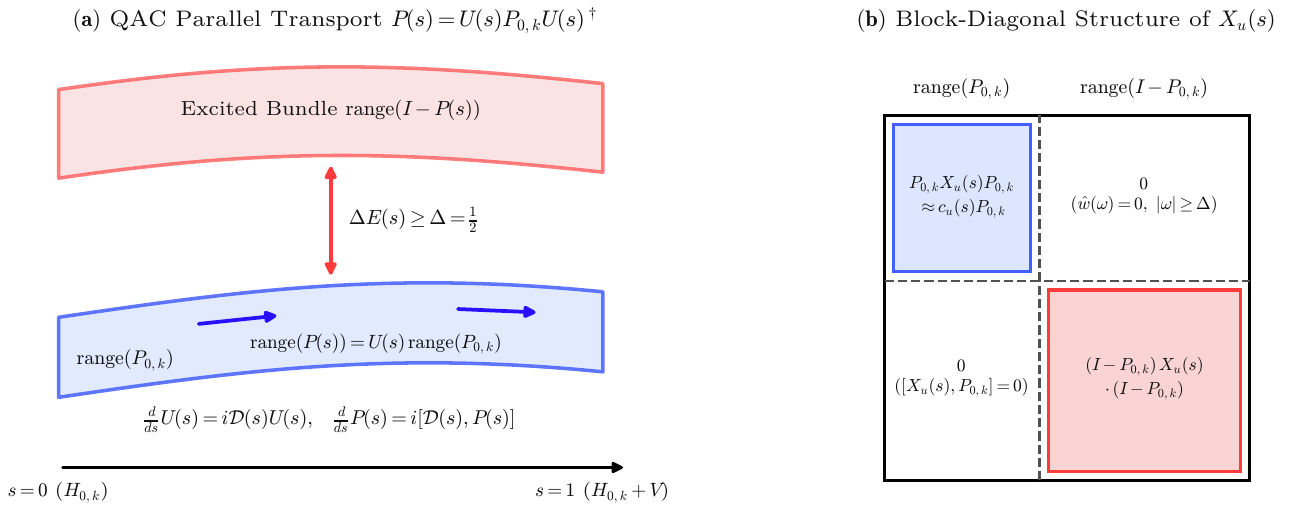}
\caption{Quasi-adiabatic parallel transport and exact block diagonalization (Proposition~\ref{prop:qac-block-diagonalization}). \textbf{(a)} QAC parallel transport $P(s) = U(s)P_{0,k} U(s)^\dagger$ along $s \in [0, 1]$, showing the blue ground-state bundle $\mathrm{range}(P(s))$ (with blue parallel-transport arrows $\frac{d}{ds}U(s) = i\cD(s)U(s)$) separated by a red double-headed arrow $\Delta E(s) \ge \Delta = 1/2$ from the pink excited bundle $\mathrm{range}(I - P(s))$. \textbf{(b)} Exact $2 \times 2$ block matrix structure of the filtered site operator $X_u(s)$~\eqref{eq:Xu-direct-def} partitioned by $\mathrm{range}(P_{0,k})$ and $\mathrm{range}(I - P_{0,k})$ into the blue ground block $P_{0,k} X_u(s) P_{0,k} \approx c_u(s) P_{0,k}$ (ground expectation $c_u(s)$ in Subsection~\ref{subsec:single-cube-telescoping}), the pink excited block $(I - P_{0,k})X_u(s)(I - P_{0,k})$, and vanishing off-diagonal blocks $0$.}
\label{fig:qac-dressed-groundspace}
\end{figure}

Let $\Delta E(\sigma)$ denote the spectral gap of $H(\sigma) = H_{0,k} + \sigma V$ separating its perturbed ground-state space $\mathrm{range}(P(\sigma))$ (of rank $\Tr(P(\sigma)) = \Tr(P_{0,k})$) from the excited spectrum $\mathrm{range}(I - P(\sigma))$. Whenever the open-gap bootstrap condition $\Delta E(\sigma) \ge 1/2$ holds for all $\sigma \in [0, s]$, the spectral projection $P(\sigma)$ is smooth in $\sigma$, and the quasi-adiabatic continuation (spectral flow) generator~\cite{kato1950adiabatic, hastings2005quasi, BH11, BMNS12}
\begin{equation}
    \label{eq:qac-generator-def}
    \cD(\sigma) \;:=\; \cF_{W_{\mathrm{BP}}}^{H(\sigma)}(V) \;=\; \int_{-\infty}^\infty W_{\mathrm{BP}}(\tau)\,e^{i\tau H(\sigma)} V e^{-i\tau H(\sigma)}\,d\tau
\end{equation}
parallel-transports $P(\sigma)$ exactly along the unitary flow $\frac{d}{d\sigma}U(\sigma) = i\cD(\sigma)U(\sigma)$ ($U(0) = I$), so that $U(s)^\dagger P(s) U(s) = P_{0,k}$ (Figure~\ref{fig:qac-dressed-groundspace}).

Once $U(s)$ rotates $P(s)$ back to the static unperturbed code projector $P_{0,k}$, the dressed Hamiltonian in the bare frame,
\[
    \widetilde{H}(s) \;:=\; U(s)^\dagger H(s) U(s),
\]
commutes globally with $P_{0,k}$ ($[\widetilde{H}(s), P_{0,k}] = 0$). However, its individual rotated site terms $U(s)^\dagger(Q_{\mathrm{site}}(u) + s V_{\mathrm{site}}(u))U(s) - Q_{\mathrm{site}}(u)$ do not yet commute with $P_{0,k}$ term by term. Following Bravyi and Hastings~\cite[Lemmas~6 and~7]{BH11} (and~\cite{BHM10, MZ13}), we apply the low-pass spectral filter $\cF_{w_{\mathrm{BP}}}^{\widetilde{H}(s)}$ directly to each rotated site term and subtract the bare check $Q_{\mathrm{site}}(u)$:
\begin{equation}
    \label{eq:Xu-direct-def}
    X_u(s) \;:=\; \cF_{w_{\mathrm{BP}}}^{\widetilde{H}(s)}\!\Bigl(U(s)^\dagger\bigl(Q_{\mathrm{site}}(u) + s V_{\mathrm{site}}(u)\bigr)U(s)\Bigr) - Q_{\mathrm{site}}(u) \qquad (u \in \Gamma_k).
\end{equation}

\begin{figure}[t]
\centering
\includegraphics[width=0.88\textwidth]{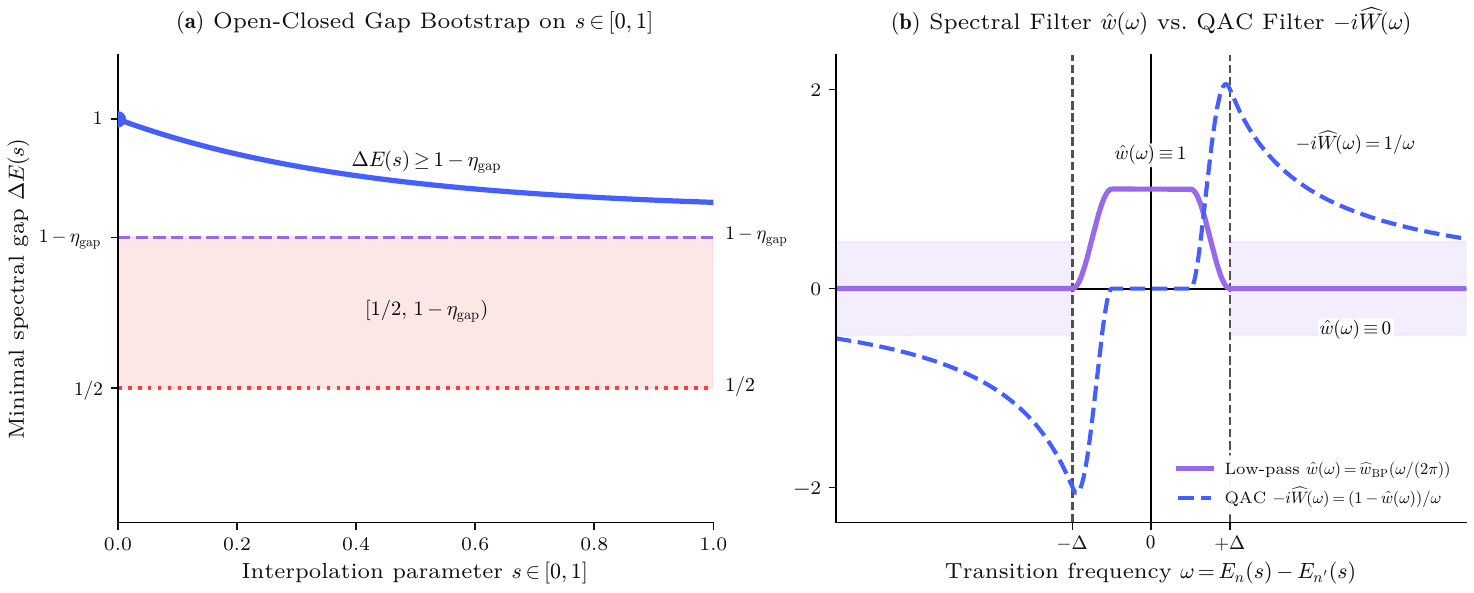}
\caption{Open-closed spectral gap bootstrap and Fourier-space filter profiles (Proposition~\ref{prop:qac-block-diagonalization} and Theorem~\ref{thm:gap-bootstrap}). \textbf{(a)} Open-closed spectral gap bootstrap on $s \in [0, 1]$, showing the minimal spectral gap $\Delta E(s) \ge 1 - \eta_{\mathrm{gap}}$ (solid blue curve) staying above the dashed violet threshold $1 - \eta_{\mathrm{gap}}$ and the pink forbidden window $[1/2, 1 - \eta_{\mathrm{gap}})$ above the dotted red line $1/2$. Here $\eta_{\mathrm{gap}} := \frac{1}{4} + \delta_{\splitE}(L_k)$ for the splitting envelope $\delta_{\splitE} \in \cD_{\mathrm{super}}$ of Theorem~\ref{thm:gap-bootstrap} ($\delta_{\splitE}(L_k) < 1/4$ above a scale threshold $k \ge k_0$), so $1 - \eta_{\mathrm{gap}} = \frac{3}{4} - \delta_{\splitE}(L_k) > \frac{1}{2}$. \textbf{(b)} Fourier-space comparison of the low-pass filter $\hat{w}(\omega) = \widehat{w}_{\mathrm{BP}}(\omega/(2\pi))$ (solid violet curve, $\hat{w}(0) = 1$, vanishing in the violet shaded bands $\abs{\omega} \ge \Delta$) and the QAC filter $-i\widehat{W}(\omega) = (1 - \hat{w}(\omega))/\omega$ (dashed blue curve, equal to $1/\omega$ for $\abs{\omega} \ge \Delta$ and vanishing on $\abs{\omega} \le \Delta/2$, where $\widehat{W}(\omega) := \int_{\R} W_{\mathrm{BP}}(\tau)e^{i\omega\tau}\,d\tau$) versus transition frequency $\omega = E_n(s) - E_{n'}(s)$.}
\label{fig:qac-bootstrap-filter}
\end{figure}

This direct site formula simultaneously achieves three properties without any reference to ground-band width (Figure~\ref{fig:qac-bootstrap-filter}):
\begin{enumerate}[label=(\arabic*), leftmargin=2.2em, itemsep=2pt]
    \item Summing over $u \in \Gamma_k$ collapses $\sum_{u \in \Gamma_k} U(s)^\dagger(Q_{\mathrm{site}}(u) + s V_{\mathrm{site}}(u))U(s) = \widetilde{H}(s)$, which commutes with $e^{it\widetilde{H}(s)}$, so $\cF_{w_{\mathrm{BP}}}^{\widetilde{H}(s)}(\widetilde{H}(s)) = \widetilde{H}(s)\int_{-\infty}^\infty w_{\mathrm{BP}}(t)\,dt = \widetilde{H}(s)$ unconditionally;
    \item Because every eigenvalue pair $(E_n(s), E_{n'}(s))$ between $\mathrm{range}(P_{0,k})$ and $\mathrm{range}(I - P_{0,k})$ satisfies $\abs{E_n(s) - E_{n'}(s)} \ge \Delta E(s) \ge 1/2$,~\eqref{eq:wBP-fourier-props} annihilates all off-diagonal blocks of $\cF_{w_{\mathrm{BP}}}^{\widetilde{H}(s)}(\cdot)$ across $P_{0,k}$, while frustration-freeness gives $Q_{\mathrm{site}}(u)P_{0,k} = P_{0,k}Q_{\mathrm{site}}(u) = 0$, so $[X_u(s), P_{0,k}] = 0$ holds identically for every site $u \in \Gamma_k$;
    \item Rewriting $X_u(s)$ in the bare frame isolates $V = 0$ cancellation term by term, ensuring that $\norm{X_u(s)}$ and all conditional-expectation tails $\norm{X_u(s) - \bbE_{B(u,j)}(X_u(s))}$ carry an explicit factor of $J$ (note that while the full site operator $X_u(s)$ commutes with $P_{0,k}$, its spatial truncations $\bbE_{B(u,j)}(X_u(s))$ only commute with $P_{0,k}$ up to an $O(J)$ superpolynomial tail; Proposition~\ref{prop:qac-block-diagonalization}(iii)).
\end{enumerate}

Putting parallel transport together with site-by-site low-pass filtering gives the exact quasi-adiabatic block diagonalization of~\cite{BHM10, BMNS12}, \cite[Lemmas~6 and~7]{BH11}, and~\cite[Lemma~1 and Proposition~1]{MZ13}: the dressed Hamiltonian difference decomposes into local site terms that each commute with the unperturbed code projector and carry superpolynomially decaying spatial tails scaled by the coupling strength $J$.

\leanlink{prop:qac-block-diagonalization}
\begin{proposition}[{Quasi-Adiabatic Continuation and Site Block Diagonalization~\cite{BHM10, BMNS12}, \cite[Lemmas~6 and~7]{BH11}, \cite[Lemma~1 and Proposition~1]{MZ13}}]
\label{prop:qac-block-diagonalization}
Let $f \in \cD_{\mathrm{super}}$. There exist an $L_k$-independent constant $m_0 \ge 0$ and a decay profile $g_{\mathrm{tail}} \in \cD_{\mathrm{super}}$ (depending only on $f$ and the uniform geometric constants of $(\cC_m, \Phi_m)$) such that for every $k \ge 4$, every $V \in \cB_f(\Gamma_k)$ with $\norm{V}_f \le J \le 1$, and every $s \in [0, 1]$ satisfying the open-gap hypothesis $\Delta E(\sigma) \ge 1/2$ for all $\sigma \in [0, s]$:
\begin{enumerate}[label=\textbf{(\roman*)}, leftmargin=2.2em, itemsep=2.5pt]
    \item Along $\sigma \in [0, s]$, $\frac{d}{d\sigma}P(\sigma) = i[\cD(\sigma), P(\sigma)]$, and the QAC unitary $U(s)$ satisfies $U(s)^\dagger P(s) U(s) = P_{0,k}$ and $[\widetilde{H}(s), P_{0,k}] = 0$.
    \item The filtered site operators $X_u(s)$ defined in~\eqref{eq:Xu-direct-def} are Hermitian ($X_u(s)^\dagger = X_u(s)$) and satisfy
    \begin{equation}
        \label{eq:rotated-H-block-decomp}
        \widetilde{H}(s) - H_{0,k} \;=\; \sum_{u \in \Gamma_k} X_u(s), \qquad
        [X_u(s), P_{0,k}] \;=\; 0, \qquad
        P_{0,k} X_u(s) (I - P_{0,k}) \;=\; 0 \qquad (\forall\, u \in \Gamma_k).
    \end{equation}
    \item For every site $u \in \Gamma_k$ and every ball radius $1 \le j \le 4 L_k$,
    \begin{equation}
        \label{eq:Xu-norm-and-tail}
        \norm{X_u(s)} \;\le\; \frac{1}{2}J m_0, \qquad
        \norm{X_u(s) - \bbE_{B(u, j)}\bigl(X_u(s)\bigr)} \;\le\; J\,g_{\mathrm{tail}}(j).
    \end{equation}
\end{enumerate}
\end{proposition}

\ifstandardproofs
\begin{proof}
For~(i), differentiating $P(\sigma)^2 = P(\sigma)$ gives $P(\sigma)P'(\sigma)P(\sigma) = (I - P(\sigma))P'(\sigma)(I - P(\sigma)) = 0$, so $P'(\sigma)$ only has matrix elements between eigenstates of $H(\sigma)$ separated in energy by at least $\Delta E(\sigma) \ge 1/2$. By~\eqref{eq:wBP-fourier-props}, $\cF_{w_{\mathrm{BP}}}^{H(\sigma)}(P'(\sigma)) = 0$. On the other hand, differentiating $[H(\sigma), P(\sigma)] = 0$ yields $[H(\sigma), P'(\sigma)] = [P(\sigma), V]$, so $\frac{d}{dt'}\tau_{t'}^\sigma(P'(\sigma)) = i\tau_{t'}^\sigma([H(\sigma), P'(\sigma)]) = i[P(\sigma), \tau_{t'}^\sigma(V)]$ since $\tau_{t'}^\sigma(P(\sigma)) = P(\sigma)$. Applying $\int_{-\infty}^\infty w_{\mathrm{BP}}(t)\,dt = 1$ and~\eqref{eq:wBP-WBP-ibp} to $\tau_t^\sigma(P'(\sigma)) - P'(\sigma) = \int_0^t i[P(\sigma), \tau_{t'}^\sigma(V)]\,dt'$ gives
\[
    -P'(\sigma) \;=\; \cF_{w_{\mathrm{BP}}}^{H(\sigma)}(P'(\sigma)) - P'(\sigma) \;=\; \int_{-\infty}^\infty W_{\mathrm{BP}}(t')\,i\bigl[P(\sigma), \tau_{t'}^\sigma(V)\bigr]\,dt' \;=\; i\bigl[P(\sigma), \cD(\sigma)\bigr],
\]
hence $P'(\sigma) = i[\cD(\sigma), P(\sigma)]$. Differentiating $U(\sigma)^\dagger P(\sigma) U(\sigma)$ therefore gives $0$ on $[0, s]$, proving $U(s)^\dagger P(s) U(s) = P(0) = P_{0,k}$ and $[\widetilde{H}(s), P_{0,k}] = U(s)^\dagger [H(s), P(s)] U(s) = 0$.

For~(ii), $X_u(s)^\dagger = X_u(s)$ follows because $w_{\mathrm{BP}}$ is real-valued and $Q_{\mathrm{site}}(u), V_{\mathrm{site}}(u), \widetilde{H}(s)$ are Hermitian. Summing~\eqref{eq:Xu-direct-def} over $u \in \Gamma_k$ using $\sum_u Q_{\mathrm{site}}(u) = H_{0,k}$ and $\sum_u (Q_{\mathrm{site}}(u) + s V_{\mathrm{site}}(u)) = H(s)$ gives
\[
    \sum_{u \in \Gamma_k} X_u(s) \;=\; \cF_{w_{\mathrm{BP}}}^{\widetilde{H}(s)}\bigl(\widetilde{H}(s)\bigr) - H_{0,k} \;=\; \Bigl(\int_{-\infty}^\infty w_{\mathrm{BP}}(t)\,dt\Bigr)\widetilde{H}(s) - H_{0,k} \;=\; \widetilde{H}(s) - H_{0,k}.
\]
Because $[\widetilde{H}(s), P_{0,k}] = 0$ and $\widetilde{H}(s)$ is unitarily equivalent to $H(s)$, $\widetilde{H}(s)$ admits a joint eigenbasis respecting $\mathrm{range}(P_{0,k}) \oplus \mathrm{range}(I - P_{0,k})$ in which every cross-gap frequency satisfies $\abs{E_n(s) - E_{n'}(s)} \ge \Delta E(s) \ge 1/2$. Thus~\eqref{eq:wBP-fourier-props} gives $P_{0,k} \cF_{w_{\mathrm{BP}}}^{\widetilde{H}(s)}(Y)(I - P_{0,k}) = 0$ and $(I - P_{0,k})\cF_{w_{\mathrm{BP}}}^{\widetilde{H}(s)}(Y) P_{0,k} = 0$ for every operator $Y$. Since $Q_{\mathrm{site}}(u)P_{0,k} = P_{0,k}Q_{\mathrm{site}}(u) = 0$ by frustration-freeness, $P_{0,k} X_u(s) (I - P_{0,k}) = 0$ and $(I - P_{0,k}) X_u(s) P_{0,k} = 0$, which is equivalent to $[X_u(s), P_{0,k}] = 0$.

For~(iii), using $e^{it\widetilde{H}(s)} U(s)^\dagger Y U(s) e^{-it\widetilde{H}(s)} = U(s)^\dagger e^{it H(s)} Y e^{-it H(s)} U(s)$ yields the bare-frame three-term identity
\begin{equation}
    \label{eq:Xu-bare-three-term}
    \begin{split}
        X_u(s) &\;=\; U(s)^\dagger \cF_{w_{\mathrm{BP}}}^{H(s)}\bigl(s V_{\mathrm{site}}(u)\bigr) U(s) \;+\; U(s)^\dagger \Bigl(\cF_{w_{\mathrm{BP}}}^{H(s)}(Q_{\mathrm{site}}(u)) - Q_{\mathrm{site}}(u)\Bigr) U(s) \\
        &\qquad +\; \Bigl(U(s)^\dagger Q_{\mathrm{site}}(u) U(s) - Q_{\mathrm{site}}(u)\Bigr).
    \end{split}
\end{equation}
Each of the three summands vanishes at $V = 0$ and has operator norm $O(J)$: the first has norm at most $\norm{w_{\mathrm{BP}}}_{L^1} \norm{V_{\mathrm{site}}(u)} \le J \norm{w_{\mathrm{BP}}}_{L^1}\sum_{r \ge 1} f(r)$; the second has norm at most $\norm{W_{\mathrm{BP}}}_{L^1}\norm{[s V, Q_{\mathrm{site}}(u)]} = O(J)$ by~\eqref{eq:check-duhamel-WBP}; and the third has norm at most $\int_0^s \norm{[Q_{\mathrm{site}}(u), \cD(\sigma)]}\,d\sigma = O(J)$ by Proposition~\ref{prop:bochner-time-convolution}(ii). This proves $\norm{X_u(s)} \le \frac{1}{2}J m_0$. Finally, for any $B_{\mathrm{in}} := B(u, \floor{j/2}) \subseteq B_{\mathrm{out}} := B(u, j)$ and any bare operator $Y$ (such as $\cF_{w_{\mathrm{BP}}}^{H(s)}(s V_{\mathrm{site}}(u))$ or $\cF_{w_{\mathrm{BP}}}^{H(s)}(Q_{\mathrm{site}}(u)) - Q_{\mathrm{site}}(u)$), splitting $Y = Y_{\loc} + (Y - Y_{\loc})$ with $Y_{\loc} := \bbE_{B_{\mathrm{in}}}(Y)$ and using contractivity of $\id - \bbE_{B_{\mathrm{out}}}$ ($\norm{\id - \bbE_{B_{\mathrm{out}}}} \le 2$) gives
\[
    \norm{U(s)^\dagger Y U(s) - \bbE_{B_{\mathrm{out}}}\bigl(U(s)^\dagger Y U(s)\bigr)} \;\le\; 2\,\norm{Y - Y_{\loc}} \;+\; \norm{U(s)^\dagger Y_{\loc} U(s) - \bbE_{B_{\mathrm{out}}}\bigl(U(s)^\dagger Y_{\loc} U(s)\bigr)}.
\]
Combining this inequality with Proposition~\ref{prop:bochner-time-convolution}(i)--(ii) bounds all three terms of~\eqref{eq:Xu-bare-three-term} outside $B(u, j)$ by $J g_{\mathrm{tail}}(j)$ with $g_{\mathrm{tail}} \in \cD_{\mathrm{super}}$.
\end{proof}
\fi

\section{Zero-Temperature Perturbative Stability}
\label{sec:telescoping-main}

\subsection{Single-Cube Telescoping}
\label{subsec:single-cube-telescoping}

We now show how Relative Cone Acyclicity $\RCA(L_k^*, C)$, and hence $\mathrm{TQO\text{-}1}(L_k^*) \wedge \mathrm{Linear\ TQO\text{-}2}(C)$ by Proposition~\ref{prop:trinity-equivalence}, decomposes the block-diagonal perturbation $\sum_{u \in \Gamma_k} X_u(s)$ of Proposition~\ref{prop:qac-block-diagonalization} across the full Hilbert space $\cH$ into a sum of local excited-band terms plus a superpolynomially small global remainder.

First, for each site $u \in \Gamma_k$ and integer ball radius $j \ge 1$ (whose centered $\ell_\infty$-ball $B(u, j)$ has side length at most $2j + 1$), define the $1$-site check collar and the linearly dilated $\RCA$ collar by
\begin{equation}
    \label{eq:Cj-ball-def}
    B_0(u, j) \;:=\; B(u, j + 1), \qquad
    C_j(u) \;:=\; B\bigl(u, \ceil{C(2j + 3)}\bigr), \qquad
    P_S \;:=\; \prod_{\supp(Q_i) \subseteq S} (I - Q_i) \quad (S \subseteq \Gamma_k).
\end{equation}
Before telescoping across spatial scales, we need a way to compare local ground-state projections around a site with the global code projection. In~\cite[Corollary~1]{BH11}, sandwiching an operator inside its $1$-site check collar produces an observable that commutes with all global checks, and constant-collar $\mathrm{TQO\text{-}2}$ transfers its ground-space norm to a $2$-site collar. Lemma~\ref{lem:norm-transfer} applies the same operator-compression argument under $\RCA(L_k^*, C)$, replacing the fixed $2$-site collar of~\cite[Corollary~1]{BH11} with the linearly dilated collar $C_j(u) = B(u, \ceil{C(2j + 3)})$.

\leanlink{lem:norm-transfer}
\begin{lemma}[Local-to-Global Norm Transfer under $\RCA(L_k^*, C)$, extending~{\cite[Corollary~1]{BH11}}]
\label{lem:norm-transfer}
Assume $(\cC_m, \Phi_m)$ satisfies $\RCA(L_k^*, C)$ on $\Gamma_k$. For every site $u \in \Gamma_k$, every ball radius $j \ge 1$ with $2j + 3 \le L_k^*$, and every operator $S \in \cB(\cH)$ supported on $B(u, j)$, the local ground projector $P_{C_j(u)}$ on $C_j(u)$ satisfies
\begin{equation}
    \label{eq:exact-norm-transfer}
    \norm{P_{C_j(u)} S} \;\le\; \norm{P_{0,k} S}, \qquad
    \norm{S P_{C_j(u)}} \;\le\; \norm{S P_{0,k}}.
\end{equation}
\end{lemma}

\begin{proof}
Because every check projector $Q_i$ has $\ell_\infty$-diameter at most $1$, any check intersecting $B(u, j)$ is supported entirely inside the $1$-collar $B_0(u, j) = B(u, j + 1)$ (whose side length is at most $2j + 3 \le L_k^*$), whereas every check outside $B_0(u, j)$ is disjoint from $B(u, j)$ and therefore commutes with $S$ and $S^\dagger$. Consequently, the positive operator
\[
    M \;:=\; P_{B_0(u, j)}\, S S^\dagger\, P_{B_0(u, j)}
\]
is supported inside $B_0(u, j)$, annihilates every check inside $B_0(u, j)$, and commutes with every check outside $B_0(u, j)$, so $M$ commutes with all global checks $\{Q_i\}$ of $H_{0,k}$. Since $C_j(u) = B(u, \ceil{C(2j + 3)})$ contains the $\RCA(L_k^*, C)$ collar cube of $B_0(u, j)$, Proposition~\ref{prop:trinity-equivalence} ($\mathrm{Linear\ TQO\text{-}2}(C)$ and $\mathrm{TQO\text{-}1}(L_k^*)$) implies that $M$ compresses to the same nonnegative scalar $c(M) \ge 0$ on $P_{C_j(u)}$ and on $P_{0,k}$:
\[
    P_{C_j(u)} M P_{C_j(u)} \;=\; c(M)\,P_{C_j(u)}, \qquad
    P_{0,k} M P_{0,k} \;=\; c(M)\,P_{0,k}.
\]
Because $B_0(u, j) \subseteq C_j(u) \subseteq \Gamma_k$, the commuting check projectors satisfy $P_{C_j(u)} P_{B_0(u,j)} = P_{C_j(u)}$ and $P_{0,k} P_{B_0(u,j)} = P_{0,k}$, whence $P_{C_j(u)} S S^\dagger P_{C_j(u)} = c(M)\,P_{C_j(u)}$ and $P_{0,k} S S^\dagger P_{0,k} = c(M)\,P_{0,k}$. Since $P_{0,k} \neq 0$, taking operator norms yields $\norm{P_{C_j(u)} S}^2 = \norm{P_{C_j(u)} S S^\dagger P_{C_j(u)}} \le c(M) = \norm{P_{0,k} S S^\dagger P_{0,k}} = \norm{P_{0,k} S}^2$ (with equality whenever $P_{C_j(u)} \neq 0$). Applying the first bound to $S^\dagger$ gives $\norm{S P_{C_j(u)}} = \norm{P_{C_j(u)} S^\dagger} \le \norm{P_{0,k} S^\dagger} = \norm{S P_{0,k}}$.
\end{proof}

Next, we calibrate the finite telescoping cutoff radius $K$ at hierarchy scale $k \ge 4$ to be the intermediate geometric scale
\begin{equation}
    \label{eq:cutoff-K-calibration}
    K \;:=\; \Lambda^{k - 4}.
\end{equation}
By~\eqref{eq:linear-collar-radius-final} ($R_{\mathrm{base}} = \Lambda(10005\Lambda + 13006)$), the lattice collar constant satisfies $C \le 2 \times 10^4 \Lambda^3 \ll \Lambda^4$ at $\Lambda = 5.4 \times 10^{24}$, placing the outermost dilated collar $C_K(u) = B(u, \ceil{C(2K + 3)})$ simultaneously inside both fundamental radii of the welded code at scale $m = 2k + 2$:
\begin{equation}
    \label{eq:double-scale-inclusion}
    2K + 3 \;\le\; 2\ceil{C(2K + 3)} + 1 \;\le\; \Lambda^k \;\ll\; L_k^* \;=\; \Lambda^{2k}.
\end{equation}
The upper bound $2K + 3 \le L_k^*$ places every telescoping ball $B(u, j)$ ($1 \le j \le K$) well inside the $\RCA(L_k^*, C)$ topological order radius, while the tighter bound $2\ceil{C(2K + 3)} + 1 \le \Lambda^k$ ensures that every dilated collar $C_j(u)$ ($1 \le j \le K$) is supported inside an axis-aligned cube of side at most $\Lambda^k$, lying well inside the macroscopic $37\Lambda^k$-cube protection radius of Proposition~\ref{prop:safe-subspace-immunity} and yielding dressed safe-subspace logical immunity in Subsection~\ref{subsec:zero-temp-dressed-memory}.

\begin{figure}[t]
\centering
\includegraphics[width=0.90\textwidth]{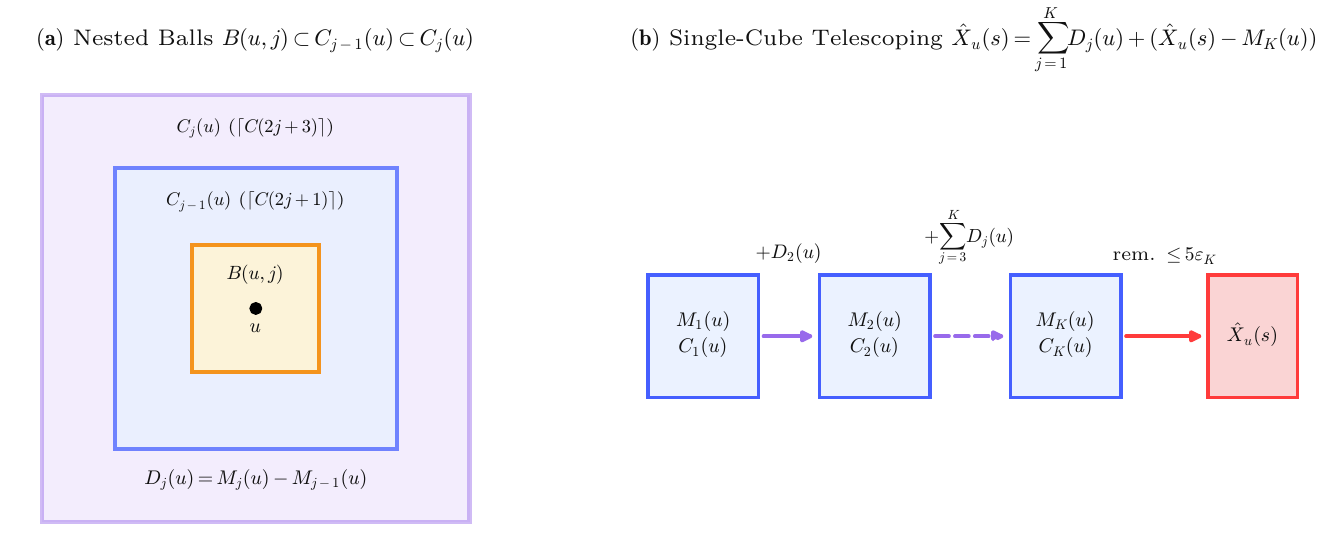}
\caption{Single-cube telescoping of the centered site operator $\hat{X}_u(s) = X_u(s) - c_u(s)I$ along concentric linearly dilated balls $C_j(u) = B(u, \ceil{C(2j+3)})$ up to cutoff $K = \Lambda^{k-4}$ (Proposition~\ref{prop:single-cube-telescoping}). \textbf{(a)} Nested spatial balls centered at site $u$ (black dot): the inner conditional-expectation support $B(u, j)$ (amber square) is enclosed within the stage-$(j-1)$ collar $C_{j-1}(u)$ (blue square) and the stage-$j$ collar $C_j(u)$ (violet outer square). \textbf{(b)} Single-index telescoping chain $M_1(u) \to M_2(u) \to \dots \to M_K(u) \to \hat{X}_u(s)$: using local-to-global norm transfer $\norm{P_{C_j(u)}\bbE_{B(u,j)}(\hat{X}_u)} \le \norm{P_{0,k}\bbE_{B(u,j)}(\hat{X}_u)} \le \varepsilon_j + \varepsilon_K$ (Lemma~\ref{lem:norm-transfer}) and the two-term projection identity $S_j - (I-P)S_j(I-P) = P S_j + (I-P)S_j P$, each stage satisfies $\norm{\hat{X}_u(s) - M_j(u)} \le 3\varepsilon_j + 2\varepsilon_K$ (residual arrow $\le 5\varepsilon_K$ at the cutoff $K$) and each consecutive difference $D_j(u) = M_j(u) - M_{j-1}(u)$ (violet arrows) is supported inside $C_j(u)$ and annihilates $P_{C_j(u)}$.}
\label{fig:single-cube-telescoping}
\end{figure}

For each site $u \in \Gamma_k$, instead of centering infinitely many individual shells, we center $X_u(s)$ once at the cutoff ball $B(u, K)$: since $\bbE_{B(u, K)}(X_u(s))$ is Hermitian and supported on $B(u, K)$ of side length $\le 2K + 1 \le L_k^*$, $\mathrm{TQO\text{-}1}(L_k^*)$ furnishes a unique real scalar $c_u(s) \in \R$ ($\abs{c_u(s)} \le \norm{X_u(s)}$) such that $P_{0,k}\,\bbE_{B(u, K)}(X_u(s))\,P_{0,k} = c_u(s)\,P_{0,k}$. Setting
\begin{equation}
    \label{eq:centered-Xu-and-Mk}
    \hat{X}_u(s) := X_u(s) - c_u(s)I, \quad
    E_0(s) := \sum_{u \in \Gamma_k} c_u(s), \quad
    M_j(u) := \bigl(I - P_{C_j(u)}\bigr)\bbE_{B(u, j)}\bigl(\hat{X}_u(s)\bigr)\bigl(I - P_{C_j(u)}\bigr)
\end{equation}
for $1 \le j \le K$ (with $M_0(u) := 0$ and tail bound $\varepsilon_j := J g_{\mathrm{tail}}(j)$), we telescope $\hat{X}_u(s) = \sum_{j = 1}^K D_j(u) + (\hat{X}_u(s) - M_K(u))$ via $D_j(u) := M_j(u) - M_{j-1}(u)$ directly on the full Hilbert space $\cH$ (Figure~\ref{fig:single-cube-telescoping}), using the two-term projection identity
\begin{equation}
    \label{eq:noncomm-projection-identity}
    S_j - (I - P)S_j(I - P) \;=\; P S_j \;+\; (I - P)S_j P
\end{equation}
with $S_j := \bbE_{B(u,j)}(\hat{X}_u(s))$ and $P := P_{C_j(u)}$.

Subtracting the ground-space expectation value once at the outer cutoff ball and taking consecutive differences along the nested collars expresses each site perturbation on the full Hilbert space as a finite sum of local operators supported entirely in the excited subspace, up to a superpolynomially small global remainder. In~\cite[Lemma~3]{BH11} and~\cite[Section~5, Lemma~4]{MZ13}, local block diagonalization centers every shell individually up to $L^*$, expands the identity into orthogonal projector differences $E_m = P_{B_{m-1}} - P_{B_m}$ all the way up to the system size $L$, and collects double-index cross sums $\sum_{p+r=j} E_p(\cdots)E_r$. Building on conditional-expectation telescoping in~\cite{NSY22, NSY24}, Proposition~\ref{prop:single-cube-telescoping} instead centers $X_u(s)$ once at $B(u, K)$ and telescopes the single-index sequence $M_j(u)$ only up to the intermediate cutoff $K = \Lambda^{k-4} \ll L_k^*$, avoiding double-sum cross terms while keeping every local difference $D_j(u)$ inside a $\Lambda^k$-cube for dressed safe-subspace immunity.

\leanlink{prop:single-cube-telescoping}
\begin{proposition}[{Centered-Ball Single-Cube Telescoping on the Full Hilbert Space, refining~\cite[Lemma~3]{BH11} and~\cite[Section~5, Lemma~4]{MZ13}; cf.~\cite{NSY22, NSY24}}]
\label{prop:single-cube-telescoping}
Under the hypotheses of Proposition~\ref{prop:qac-block-diagonalization} on $\Gamma_k$ ($k \ge 4$), set $\varepsilon_j := J g_{\mathrm{tail}}(j)$ for $1 \le j \le K = \Lambda^{k-4}$. Then for every site $u \in \Gamma_k$:
\begin{enumerate}[label=\textbf{(\roman*)}, leftmargin=2.2em, itemsep=2.5pt]
    \item The centered Hermitian operator $\hat{X}_u(s) = X_u(s) - c_u(s)I$ satisfies $\norm{\hat{X}_u(s)} \le J m_0$, $\norm{P_{0,k}\hat{X}_u(s)} = \norm{\hat{X}_u(s)P_{0,k}} \le \varepsilon_K$, and for every $1 \le j \le K$,
    \begin{equation}
        \label{eq:3-eps-j-bound}
        \norm{\hat{X}_u(s) - M_j(u)} \;\le\; 3\varepsilon_j + 2\varepsilon_K, \qquad
        \norm{\hat{X}_u(s) - M_K(u)} \;\le\; 5\varepsilon_K.
    \end{equation}
    \item Defining $D_j(u) := M_j(u) - M_{j-1}(u)$ for $1 \le j \le K$ (with $M_0(u) := 0$), each $D_j(u)$ is Hermitian, supported inside $C_j(u) = B(u, \ceil{C(2j+3)})$, annihilates the local ground projector from both sides ($P_{C_j(u)} D_j(u) = D_j(u) P_{C_j(u)} = 0$, hence $[D_j(u), P_{0,k}] = 0$), and satisfies
    \begin{equation}
        \label{eq:Dj-norm-bound}
        \norm{D_1(u)} \;\le\; J m_0 + 3\varepsilon_1 + 2\varepsilon_K, \qquad
        \norm{D_j(u)} \;\le\; 3\varepsilon_j + 3\varepsilon_{j-1} + 4\varepsilon_K \quad (2 \le j \le K).
    \end{equation}
    Summing $\hat{X}_u(s) = M_K(u) + (\hat{X}_u(s) - M_K(u))$ over $u \in \Gamma_k$ yields the exact full-Hilbert-space decomposition
    \begin{equation}
        \label{eq:full-space-telescoping}
        \widetilde{H}(s) - H_{0,k} - E_0(s)I = W_{\mathrm{tel}}(s) + R_{\mathrm{rem}}(s), \quad
        W_{\mathrm{tel}}(s) := \sum_{u \in \Gamma_k}\sum_{j = 1}^K D_j(u), \quad
        \norm{R_{\mathrm{rem}}(s)} \le 5 L_k^3 \varepsilon_K,
    \end{equation}
    where $[W_{\mathrm{tel}}(s), P_{0,k}] = 0$ and $[R_{\mathrm{rem}}(s), P_{0,k}] = 0$.
\end{enumerate}
\end{proposition}

\begin{proof}
For~(i), $\norm{\hat{X}_u(s)} \le \norm{X_u(s)} + \abs{c_u(s)} \le 2\norm{X_u(s)} \le J m_0$. Because $[\hat{X}_u(s), P_{0,k}] = [X_u(s), P_{0,k}] = 0$ by Proposition~\ref{prop:qac-block-diagonalization}(ii) and $P_{0,k}\bbE_{B(u,K)}(\hat{X}_u(s))P_{0,k} = P_{0,k}(\bbE_{B(u,K)}(X_u(s)) - c_u(s)I)P_{0,k} = 0$ by definition of $c_u(s)$, we have
\[
    P_{0,k}\hat{X}_u(s) \;=\; P_{0,k}\hat{X}_u(s)P_{0,k} \;=\; P_{0,k}\Bigl(X_u(s) - \bbE_{B(u,K)}\bigl(X_u(s)\bigr)\Bigr)P_{0,k},
\]
so $\norm{P_{0,k}\hat{X}_u(s)} \le \norm{X_u(s) - \bbE_{B(u,K)}(X_u(s))} \le \varepsilon_K$, and taking adjoints gives $\norm{\hat{X}_u(s)P_{0,k}} \le \varepsilon_K$. For any $1 \le j \le K$, write $S_j := \bbE_{B(u,j)}(\hat{X}_u(s)) = \bbE_{B(u,j)}(X_u(s)) - c_u(s)I$; since $\bbE_{B(u,j)}(I) = I$, Proposition~\ref{prop:qac-block-diagonalization}(iii) gives $\norm{\hat{X}_u(s) - S_j} = \norm{X_u(s) - \bbE_{B(u,j)}(X_u(s))} \le \varepsilon_j$. Therefore $\norm{P_{0,k} S_j} \le \norm{P_{0,k}(S_j - \hat{X}_u(s))} + \norm{P_{0,k}\hat{X}_u(s)} \le \varepsilon_j + \varepsilon_K$, and similarly $\norm{S_j P_{0,k}} \le \varepsilon_j + \varepsilon_K$. Since $S_j$ is supported on $B(u, j)$ with $2j + 3 \le 2K + 3 \le L_k^*$, Lemma~\ref{lem:norm-transfer} transfers both bounds to $P_{C_j(u)}$:
\[
    \norm{P_{C_j(u)} S_j} \;\le\; \varepsilon_j + \varepsilon_K, \qquad
    \norm{S_j P_{C_j(u)}} \;\le\; \varepsilon_j + \varepsilon_K.
\]
Applying the projection identity~\eqref{eq:noncomm-projection-identity} with $P = P_{C_j(u)}$ gives $\norm{S_j - M_j(u)} \le \norm{P_{C_j(u)} S_j} + \norm{I - P_{C_j(u)}}\norm{S_j P_{C_j(u)}} \le 2(\varepsilon_j + \varepsilon_K)$, and hence $\norm{\hat{X}_u(s) - M_j(u)} \le \norm{\hat{X}_u(s) - S_j} + \norm{S_j - M_j(u)} \le 3\varepsilon_j + 2\varepsilon_K$, proving~\eqref{eq:3-eps-j-bound}.

For~(ii), at $j = 1$ we have $\norm{D_1(u)} = \norm{M_1(u)} \le \norm{\hat{X}_u(s)} + \norm{\hat{X}_u(s) - M_1(u)} \le J m_0 + 3\varepsilon_1 + 2\varepsilon_K$, while for $2 \le j \le K$, $\norm{D_j(u)} \le \norm{M_j(u) - \hat{X}_u(s)} + \norm{\hat{X}_u(s) - M_{j-1}(u)} \le 3\varepsilon_j + 3\varepsilon_{j-1} + 4\varepsilon_K$. Because $C_{j-1}(u) \subseteq C_j(u)$, the local projectors satisfy $P_{C_j(u)} \le P_{C_{j-1}(u)}$, so $(I - P_{C_{j-1}(u)})P_{C_j(u)} = 0$ and both $M_j(u)$ and $M_{j-1}(u)$ annihilate $P_{C_j(u)}$ from left and right; since $P_{0,k} \le P_{C_j(u)}$, this also implies $D_j(u)P_{0,k} = P_{0,k}D_j(u) = 0$. Finally, summing $\sum_{j=1}^K D_j(u) = M_K(u)$ and $R_{\mathrm{rem}}(s) := \sum_{u \in \Gamma_k}(\hat{X}_u(s) - M_K(u))$ over the $\abs{\Gamma_k} = L_k^3$ sites yields~\eqref{eq:full-space-telescoping} with $\norm{R_{\mathrm{rem}}(s)} \le L_k^3 \cdot 5\varepsilon_K$.
\end{proof}

\subsection{Spectral Gap Bootstrap}
\label{subsec:gap-bootstrap}

Because each local telescoping term acts non-trivially only on states that violate at least one stabilizer check inside its collar, its expectation value can be bounded against the local check energy of the unperturbed Hamiltonian. Theorem~\ref{thm:gap-bootstrap} adapts the relative-boundedness and gap-bootstrap framework of~\cite[Lemmas~4--5, Proposition~1, and Section~7]{BH11} and~\cite[Proposition~2 and Section~7]{MZ13} to linear $\RCA$ collars: rather than squaring $(H_{0,k} - \lambda I)\psi = -W\psi$ across box-syndrome sectors as in~\cite[Lemma~4 and Proposition~1]{BH11}, we bound the quadratic form $\abs{\ip{\psi}{D_j(u)\psi}}$ directly at first order by Cauchy--Schwarz as in~\cite[Proposition~2]{MZ13}, absorb the polynomial $O(C^3 j^3)$ volume of the linearly dilated collars $C_j(u)$ into the superpolynomial decay of $g_{\mathrm{tail}} \in \cD_{\mathrm{super}}$, and close the open--closed continuity bootstrap along $s \in [0, 1]$ (Figure~\ref{fig:qac-bootstrap-filter}(a)).

\leanlink{thm:gap-bootstrap}
\begin{theorem}[{Full-Space Relative Form Bound and Open--Closed Gap Bootstrap, adapted from~\cite[Lemmas~4--5, Proposition~1, and Section~7]{BH11} and~\cite[Proposition~2 and Section~7]{MZ13}}]
\label{thm:gap-bootstrap}
Let $(\cC_m, \Phi_m)$ ($m = 2k + 2$) be the coarse-grained multiscale welded code hierarchy on $\Gamma_k$, and let $f \in \cD_{\mathrm{super}}$. There exist a scale threshold $k_0 \ge 4$, a universal coupling threshold $J_0 \in (0, 1]$ (depending only on $f, C$, and the uniform geometric constants, and independent of $k$ and $L_k$), and a splitting envelope $\delta_{\splitE} \in \cD_{\mathrm{super}}$ with $\delta_{\splitE}(L_k) < 1/4$ for all $k \ge k_0$ such that for every $k \ge k_0$, every $V \in \cB_f(\Gamma_k)$ with $\norm{V}_f \le J \le J_0$, and every $s \in [0, 1]$:
\begin{enumerate}[label=\textbf{(\roman*)}, leftmargin=2.2em, itemsep=2.5pt]
    \item On the entire Hilbert space $\cH$, the rotated Hamiltonian $\widetilde{H}(s) = U(s)^\dagger H(s) U(s)$ satisfies
    \begin{equation}
        \label{eq:relative-form-bound}
        \Bigl\lvert\ip{\psi}{\bigl(\widetilde{H}(s) - H_{0,k} - E_0(s)I\bigr)\psi}\Bigr\rvert \;\le\; \frac{1}{4}\ip{\psi}{H_{0,k}\psi} \;+\; \frac{1}{2}\delta_{\splitE}(L_k)\norm{\psi}^2 \qquad \forall\, \psi \in \cH.
    \end{equation}
    \item The ground-band splitting and bulk spectral gap of $H(s) = H_{0,k} + s V$ satisfy, for all $s \in [0, 1]$,
    \begin{equation}
        \label{eq:gap-and-splitting-bootstrap}
        \norm{P(s)\bigl(H(s) - E_0(s)I\bigr)P(s)} \;\le\; \frac{1}{2}\delta_{\splitE}(L_k), \qquad
        \Delta E(s) \;\ge\; \frac{3}{4} - \delta_{\splitE}(L_k) \;>\; \frac{1}{2}.
    \end{equation}
\end{enumerate}
\end{theorem}

\begin{proof}
Assume first that $s \in [0, 1]$ satisfies $\Delta E(\sigma) \ge 1/2$ for all $\sigma \in [0, s]$, so Proposition~\ref{prop:qac-block-diagonalization} and Proposition~\ref{prop:single-cube-telescoping} apply on $[0, s]$. For each $u \in \Gamma_k$, $1 \le j \le K$, and $\psi \in \cH$, Cauchy--Schwarz and the commuting-projector union bound $P_{C_j(u)}^\perp := I - P_{C_j(u)} = I - \prod_{\supp(Q_i) \subseteq C_j(u)} (I - Q_i) \le \sum_{\supp(Q_i) \subseteq C_j(u)} Q_i$ give
\begin{equation}
    \label{eq:cauchy-schwarz-Dj}
    \abs{\ip{\psi}{D_j(u)\psi}} = \abs{\ip{P_{C_j(u)}^\perp\psi}{D_j(u)P_{C_j(u)}^\perp\psi}} \le \norm{D_j(u)}\ip{\psi}{P_{C_j(u)}^\perp\psi} \le \norm{D_j(u)} \sum_{\supp(Q_i) \subseteq C_j(u)} \ip{\psi}{Q_i\psi}.
\end{equation}
For any fixed check $Q_i$, if $\supp(Q_i) \subseteq C_j(u) = B(u, R_j)$ with $R_j := \ceil{C(2j + 3)}$, then picking a site $u_0 \in \supp(Q_i)$ restricts $u \in B(u_0, R_j)$, so at most $(2 R_j + 1)^3$ sites $u \in \Gamma_k$ satisfy $\supp(Q_i) \subseteq C_j(u)$. (Equivalently, as in~\cite[Proposition~1]{BH11} and~\cite[Proposition~2]{MZ13}, partitioning $\Gamma_k$ into $(2 R_j + 1)^3$ subgrids of spacing $2 R_j + 1$ renders the balls $\{C_j(u)\}$ within each subgrid pairwise disjoint, so each subgrid sum is bounded by $\max_u \norm{D_j(u)} \sum_i Q_i = \max_u \norm{D_j(u)} H_{0,k}$.) Summing~\eqref{eq:cauchy-schwarz-Dj} over $u \in \Gamma_k$ and $1 \le j \le K$ via~\eqref{eq:Dj-norm-bound} therefore yields
\begin{equation}
    \label{eq:sum-Dj-form-bound}
    \begin{split}
        \abs{\ip{\psi}{W_{\mathrm{tel}}(s)\psi}} \;\le\; J\Biggl[&\bigl(m_0 + 3 g_{\mathrm{tail}}(1)\bigr)(2 R_1 + 1)^3 + \sum_{j = 2}^\infty 3\bigl(g_{\mathrm{tail}}(j) + g_{\mathrm{tail}}(j-1)\bigr)(2 R_j + 1)^3 \\
        &\quad +\; 4 g_{\mathrm{tail}}(K)\sum_{j = 1}^K (2 R_j + 1)^3\Biggr]\ip{\psi}{H_{0,k}\psi}.
    \end{split}
\end{equation}
Since $R_j = O(C j)$ and $g_{\mathrm{tail}} \in \cD_{\mathrm{super}}$, Lemma~\ref{lem:dsuper-closure}(i) guarantees that the series in~\eqref{eq:sum-Dj-form-bound} (including $\sup_{K \ge 1} 4 g_{\mathrm{tail}}(K)\sum_{j=1}^K (2 R_j + 1)^3 < \infty$) is bounded by a finite constant $C_{\mathrm{form}} > 0$ independent of $k$ and $L_k$. That is essentially the end of the form-bound argument; all that remains is to make appropriate choices for the threshold parameters $J_0$ and $k_0$. Setting $J_0 := \min(1, (4 C_{\mathrm{form}})^{-1})$ gives $\abs{\ip{\psi}{W_{\mathrm{tel}}(s)\psi}} \le \frac{1}{4}\ip{\psi}{H_{0,k}\psi}$ for all $J \le J_0$. Meanwhile, since $L_k = \Lambda^{2k+4}$ and $K = \Lambda^{k-4} = \Lambda^{-6}\sqrt{L_k}$ for $k \ge 4$ (with $k_0 \ge 7$ in the formalization so $L_k = \Lambda^{2k+4} \le K^6$), Lemma~\ref{lem:dsuper-closure}(i),~(iii) ensures that $\delta_{\splitE}(L) := 10 L^3 J_0 g_{\mathrm{tail}}\bigl(\floor{\Lambda^{-6}\sqrt{L}}\bigr)$ (satisfying $10 L_k^3 J_0 g_{\mathrm{tail}}(K) = \delta_{\splitE}(L_k)$) belongs to $\cD_{\mathrm{super}}$, so choosing $k_0 \ge 4$ large enough gives $\delta_{\splitE}(L_k) < 1/4$ for all $k \ge k_0$. Combining~\eqref{eq:full-space-telescoping} with $\norm{R_{\mathrm{rem}}(s)} \le 5 L_k^3 \varepsilon_K \le \frac{1}{2}\delta_{\splitE}(L_k)$ yields~\eqref{eq:relative-form-bound}.

For~(ii), under the same hypothesis on $[0, s]$, evaluating~\eqref{eq:relative-form-bound} on $\psi \in \mathrm{range}(P_{0,k})$ (where $H_{0,k}\psi = 0$) and conjugating by $U(s)$ gives $\norm{P(s)(H(s) - E_0(s)I)P(s)} = \norm{P_{0,k}(\widetilde{H}(s) - E_0(s)I)P_{0,k}} \le \frac{1}{2}\delta_{\splitE}(L_k)$, so every ground-band eigenvalue lies in $[E_0(s) - \frac{1}{2}\delta_{\splitE}(L_k), E_0(s) + \frac{1}{2}\delta_{\splitE}(L_k)]$ (hence ground-space splitting $\Delta E_{\splitE}(s) \le \delta_{\splitE}(L_k)$). On $\psi_\perp \in \mathrm{range}(I - P_{0,k})$ (where $\ip{\psi_\perp}{H_{0,k}\psi_\perp} \ge \norm{\psi_\perp}^2$),~\eqref{eq:relative-form-bound} gives
\[
    \ip{\psi_\perp}{\bigl(\widetilde{H}(s) - E_0(s)I\bigr)\psi_\perp} \;\ge\; \frac{3}{4}\ip{\psi_\perp}{H_{0,k}\psi_\perp} - \frac{1}{2}\delta_{\splitE}(L_k)\norm{\psi_\perp}^2 \;\ge\; \Bigl(\frac{3}{4} - \frac{1}{2}\delta_{\splitE}(L_k)\Bigr)\norm{\psi_\perp}^2,
\]
so every excited eigenvalue is at least $E_0(s) + \frac{3}{4} - \frac{1}{2}\delta_{\splitE}(L_k)$, giving the sharpened gap $\Delta E(s) \ge \frac{3}{4} - \delta_{\splitE}(L_k) > \frac{1}{2}$ (since $\delta_{\splitE}(L_k) < 1/4$). Now let $I_{\mathrm{gap}} := \{s \in [0, 1] : \Delta E(\sigma) \ge 1/2 \text{ for all } \sigma \in [0, s]\}$. At $s = 0$, $H(0) = H_{0,k}$ has gap $\Delta E(0) \ge 1$, so $0 \in I_{\mathrm{gap}}$. By Weyl's eigenvalue perturbation inequality ($\abs{\lambda_n(\sigma') - \lambda_n(\sigma)} \le \abs{\sigma' - \sigma}\norm{V}$), $\Delta E(\sigma)$ is continuous in $\sigma$: thus $I_{\mathrm{gap}}$ is closed in $[0, 1]$, and for any $s \in I_{\mathrm{gap}}$ the sharpened bound $\Delta E(s) \ge \frac{3}{4} - \delta_{\splitE}(L_k) > \frac{1}{2}$ ensures $\Delta E(\sigma) \ge 1/2$ on a neighborhood of $s$ in $[0, 1]$, so $I_{\mathrm{gap}}$ is also open in $[0, 1]$. By connectedness of $[0, 1]$, $I_{\mathrm{gap}} = [0, 1]$, completing the proof.
\end{proof}

\subsection{Dressed Logical Memory}
\label{subsec:zero-temp-dressed-memory}

Evaluating the bootstrapped QAC unitary flow at $s = 1$ defines the quasi-local dressing unitary $U_{\mathrm{dress}} := U(1)$ and reference energy $E_0 := E_0(1)$ for the full perturbed Hamiltonian $H_k := H(1) = H_{0,k} + V$. Conjugating the bare code projector $P_{0,k}$, the bare low-weight-clean safe-subspace projector $\Pi_{\mathrm{safe}} \equiv \Pi_{\mathrm{safe},k}$ of Proposition~\ref{prop:safe-subspace-immunity} ($P_{0,k} \le \Pi_{\mathrm{safe}}$, suppressing the subscript $k$ throughout this subsection), and the bare cluster-decoded logical observables $O_{\mathrm{dec}}^{Z/X} \equiv O_{\mathrm{dec},k}^{Z/X}$ ($\norm{O_{\mathrm{dec}}^{Z/X}} = 1$, $(O_{\mathrm{dec}}^{Z/X})^2 = I$, $[O_{\mathrm{dec}}^{Z/X}, P_{0,k}] = 0$, and $\{O_{\mathrm{dec}}^Z, O_{\mathrm{dec}}^X\} P_{0,k} = 0$) into the dressed quasiparticle frame yields
\begin{equation}
    \label{eq:dressed-memory-objects}
    P_{0,\mathrm{dress}} \;:=\; U_{\mathrm{dress}} P_{0,k} U_{\mathrm{dress}}^\dagger, \qquad
    \Pi_{\mathrm{safe,dress}} \;:=\; U_{\mathrm{dress}} \Pi_{\mathrm{safe}} U_{\mathrm{dress}}^\dagger, \qquad
    O_{\mathrm{dress}}^{Z/X} \;:=\; U_{\mathrm{dress}} O_{\mathrm{dec}}^{Z/X} U_{\mathrm{dress}}^\dagger.
\end{equation}
By unitary conjugation, $O_{\mathrm{dress}}^Z$ and $O_{\mathrm{dress}}^X$ remain Hermitian involutions commuting with $P_{0,\mathrm{dress}}$ ($P_{0,\mathrm{dress}} \le \Pi_{\mathrm{safe,dress}}$) and anticommuting on the perturbed ground space ($\{O_{\mathrm{dress}}^Z, O_{\mathrm{dress}}^X\} P_{0,\mathrm{dress}} = 0$). Because the dressing unitary spreads single-site operators only up to superpolynomially small tails outside the cutoff ball, and because every local telescoping term of the rotated Hamiltonian fits inside the unperturbed decoder's protection radius, both single-site safe-subspace immunity and coherent ground-space memory stability carry over to the perturbed system.

\leanlink{thm:zero-temp-dressed-memory}
\begin{theorem}[Zero-Temperature Perturbative Spectral, Safe-Subspace Immunity, and Dynamical Memory Stability]
\label{thm:zero-temp-dressed-memory}
Under the hypotheses of Theorem~\ref{thm:gap-bootstrap}, after enlarging $k_0 \ge 4$ and replacing $\delta_{\splitE} \in \cD_{\mathrm{super}}$ by an $L_k$-independent superpolynomial envelope (still denoted $\delta_{\splitE} \in \cD_{\mathrm{super}}$, with $\delta_{\splitE}(L_k) < 1/4$ for $k \ge k_0$), for every $k \ge k_0$, every $J \in [0, J_0]$, and every Hermitian perturbation $V \in \cB_f(\Gamma_k)$ with $\norm{V}_f \le J$, the perturbed Hamiltonian $H_k = H_{0,k} + V$ and dressed memory package~\eqref{eq:dressed-memory-objects} satisfy:
\begin{enumerate}[label=\textbf{(\roman*)}, leftmargin=2.2em, itemsep=2.5pt]
    \item For every site $v \in \Gamma_k$, every operator $A_v \in \cB(\cH)$ supported on $\{v\}$, and every radius $r \ge 0$,
    \begin{equation}
        \label{eq:dressed-unitary-lr}
        \norm{U_{\mathrm{dress}}^\dagger A_v U_{\mathrm{dress}} - \bbE_{B(v, r)}\bigl(U_{\mathrm{dress}}^\dagger A_v U_{\mathrm{dress}}\bigr)} \;\le\; g_{\mathrm{LR}}(r)\,\norm{A_v},
    \end{equation}
    where $g_{\mathrm{LR}} \in \cD_{\mathrm{super}}$ is independent of $k$ and $L_k$.
    \item The perturbed ground projector $P_{0,\mathrm{dress}} = P(1)$ satisfies $[H_k, P_{0,\mathrm{dress}}] = 0$ (hence $\norm{[H_k, P_{0,\mathrm{dress}}]} \le \delta_{\splitE}(L_k)$), ground-band splitting $\norm{P_{0,\mathrm{dress}}(H_k - E_0 I)P_{0,\mathrm{dress}}} \le \frac{1}{2}\delta_{\splitE}(L_k)$, bulk spectral gap $\Delta E(1) \ge \frac{3}{4} - \delta_{\splitE}(L_k) > \frac{1}{2}$, and the full-space relative form bound~\eqref{eq:relative-form-bound} at $s = 1$; equivalently, in terms of the dressed stabilizer Hamiltonian $H_{0,\mathrm{dress}} := U_{\mathrm{dress}} H_{0,k} U_{\mathrm{dress}}^\dagger$,
    \begin{equation}
        \label{eq:dressed-energy-sandwich}
        \frac{3}{4}\,H_{0,\mathrm{dress}} - \frac{1}{2}\delta_{\splitE}(L_k)I \;\le\; H_k - E_0 I \;\le\; \frac{5}{4}\,H_{0,\mathrm{dress}} + \frac{1}{2}\delta_{\splitE}(L_k)I.
    \end{equation}
    \item For each dressed decoded observable $O_{\mathrm{dress}} \in \{O_{\mathrm{dress}}^Z, O_{\mathrm{dress}}^X\}$:
    \begin{enumerate}[label=\textbf{(\alph*)}, leftmargin=1.8em, itemsep=2pt]
        \item \emph{Dressed safe-subspace coherent drift}: $\norm{\Pi_{\mathrm{safe,dress}} [H_k, O_{\mathrm{dress}}] \Pi_{\mathrm{safe,dress}}} \le \delta_{\splitE}(L_k)$;
        \item \emph{Dressed single-site error immunity}: for every site $v \in \Gamma_k$ and every operator $A_v$ supported on $\{v\}$,
        \begin{equation}
            \label{eq:dressed-1site-immunity}
            \norm{\Pi_{\mathrm{safe,dress}} [A_v, O_{\mathrm{dress}}] \Pi_{\mathrm{safe,dress}}} \;\le\; \delta_{\splitE}(L_k)\,\norm{A_v};
        \end{equation}
        \item \emph{Autonomous zero-temperature unitary memory lifetime}: for all evolution times $t \ge 0$,
        \begin{equation}
            \label{eq:zero-temp-heisenberg-lifetime}
            \norm{\bigl(e^{it H_k} O_{\mathrm{dress}} e^{-it H_k} - O_{\mathrm{dress}}\bigr) P_{0,\mathrm{dress}}} \;\le\; t\,\delta_{\splitE}(L_k).
        \end{equation}
    \end{enumerate}
\end{enumerate}
\end{theorem}

\begin{proof}
Part~(i) is Proposition~\ref{prop:bochner-time-convolution}(ii) evaluated at $s = 1$ (since $J \le J_0 \le 1$). Part~(ii) follows directly from Proposition~\ref{prop:qac-block-diagonalization}(i) ($[H_k, P_{0,\mathrm{dress}}] = U_{\mathrm{dress}}[\widetilde{H}(1), P_{0,k}]U_{\mathrm{dress}}^\dagger = 0$) and Theorem~\ref{thm:gap-bootstrap}(i)--(ii) at $s = 1$, where conjugating~\eqref{eq:relative-form-bound} by $U_{\mathrm{dress}}$ gives~\eqref{eq:dressed-energy-sandwich}.

For~(iii)(a), unitary invariance of the operator norm pulls the dressed safe-subspace commutator directly back to the bare frame via~\eqref{eq:full-space-telescoping} at $s = 1$:
\begin{equation}
    \label{eq:safe-commutator-pullback}
    \begin{split}
        \norm{\Pi_{\mathrm{safe,dress}} [H_k, O_{\mathrm{dress}}] \Pi_{\mathrm{safe,dress}}} &\;=\; \norm{\Pi_{\mathrm{safe}} \bigl[\widetilde{H}(1) - E_0 I,\, O_{\mathrm{dec}}\bigr] \Pi_{\mathrm{safe}}} \\
        &\;=\; \norm{\Pi_{\mathrm{safe}} \bigl[H_{0,k} + W_{\mathrm{tel}}(1) + R_{\mathrm{rem}}(1),\, O_{\mathrm{dec}}\bigr] \Pi_{\mathrm{safe}}}.
    \end{split}
\end{equation}
By Proposition~\ref{prop:safe-subspace-immunity}(ii), the pinched commutator $\Pi_{\mathrm{safe}} [W, O_{\mathrm{dec}}] \Pi_{\mathrm{safe}}$ vanishes identically for every operator $W$ on $\Gamma_k$ supported inside an axis-aligned cube of side at most $\Lambda^k$. Because every check $Q_i$ of $H_{0,k}$ has $\ell_\infty$ cube side at most $2 r_{\mathrm{chk}} + 1 = 19 \le \Lambda^k$ (written as side $2$ under the constant-range rescaling convention of Section~\ref{sec:qac-banach}), $\Pi_{\mathrm{safe}} [H_{0,k}, O_{\mathrm{dec}}] \Pi_{\mathrm{safe}} = 0$. Moreover, by our intermediate cutoff calibration~\eqref{eq:cutoff-K-calibration}--\eqref{eq:double-scale-inclusion} ($K = \Lambda^{k-4}$), every local telescoping operator $D_j(u)$ ($1 \le j \le K$) is supported inside $C_j(u) = B(u, \ceil{C(2j+3)})$, which is contained in an axis-aligned cube of side at most $2\ceil{C(2K+3)} + 1 \le \Lambda^k$; hence $\Pi_{\mathrm{safe}} [D_j(u), O_{\mathrm{dec}}] \Pi_{\mathrm{safe}} = 0$ for every $u \in \Gamma_k$ and $1 \le j \le K$, so $\Pi_{\mathrm{safe}} [W_{\mathrm{tel}}(1), O_{\mathrm{dec}}] \Pi_{\mathrm{safe}} = 0$. Only the global tail remainder $R_{\mathrm{rem}}(1)$ survives in~\eqref{eq:safe-commutator-pullback}, giving
\[
    \norm{\Pi_{\mathrm{safe,dress}} [H_k, O_{\mathrm{dress}}] \Pi_{\mathrm{safe,dress}}} = \norm{\Pi_{\mathrm{safe}} [R_{\mathrm{rem}}(1), O_{\mathrm{dec}}] \Pi_{\mathrm{safe}}} \le 2\norm{R_{\mathrm{rem}}(1)}\norm{O_{\mathrm{dec}}} \le 10 L_k^3 J_0 g_{\mathrm{tail}}(K).
\]

Similarly, for~(iii)(b), set $\widetilde{A}_v := U_{\mathrm{dress}}^\dagger A_v U_{\mathrm{dress}}$ and truncate $\widetilde{A}_v$ to the ball $B(v, K)$ via $\bbE_{B(v,K)}(\widetilde{A}_v)$. Because $B(v, K)$ has side length at most $2K + 1 \le \Lambda^k$, Proposition~\ref{prop:safe-subspace-immunity} gives $\Pi_{\mathrm{safe}} [\bbE_{B(v,K)}(\widetilde{A}_v), O_{\mathrm{dec}}] \Pi_{\mathrm{safe}} = 0$, so unitary invariance and~\eqref{eq:dressed-unitary-lr} leave only the light-cone tail:
\begin{align*}
    \norm{\Pi_{\mathrm{safe,dress}} [A_v, O_{\mathrm{dress}}] \Pi_{\mathrm{safe,dress}}}
    &\;=\; \norm{\Pi_{\mathrm{safe}} [\widetilde{A}_v, O_{\mathrm{dec}}] \Pi_{\mathrm{safe}}} \\
    &\;\le\; 2\,\norm{\widetilde{A}_v - \bbE_{B(v, K)}(\widetilde{A}_v)}\,\norm{O_{\mathrm{dec}}} \;\le\; 2 g_{\mathrm{LR}}(K)\,\norm{A_v}.
\end{align*}
Absorbing $\max\bigl(10 L_k^3 J_0 g_{\mathrm{tail}}(K),\, 2 g_{\mathrm{LR}}(K)\bigr)$ into the superpolynomial envelope $\delta_{\splitE}(L_k)$ (via Lemma~\ref{lem:dsuper-closure}(i),~(iii) since $K = \Lambda^{k-4} = \Lambda^{-6}\sqrt{L_k}$) and enlarging $k_0$ so $\delta_{\splitE}(L_k) < 1/4$ establishes both~(iii)(a) and~\eqref{eq:dressed-1site-immunity}.

Finally, for~(iii)(c), set $H' := H_k - E_0 I$, $P := P_{0,\mathrm{dress}}$, and $O := O_{\mathrm{dress}}$. Because $e^{it H_k} O e^{-it H_k} = e^{it H'} O e^{-it H'}$ (scalar shifts commute with Heisenberg evolution) and $[H', P] = [H_k, P_{0,\mathrm{dress}}] = 0$ (so $e^{-it H'} P = P e^{-it H'}$ and $e^{it H'}(O P)e^{-it H'} = e^{it H'} O e^{-it H'} P$), differentiating $\tau_t^{H'}(O P) - O P = \int_0^t i\,\tau_{t'}^{H'}\bigl([H', O P]\bigr)\,dt'$ gives
\begin{equation}
    \label{eq:heisenberg-drift-integral}
    \norm{\bigl(e^{it H_k} O e^{-it H_k} - O\bigr) P} \;=\; \norm{\tau_t^{H'}(O P) - O P} \;\le\; t\,\norm{[H', O P]}.
\end{equation}
Using $O P = P O = P O P$ and $[H', P] = 0$ (whence $H' P = P H' P = P H'$), the commutator $[H', O P]$ simplifies directly into the ground-band splitting:
\[
    [H', O P] \;=\; H' P O P - O P H' \;=\; (P H' P) O P - O (P H' P),
\]
so $\norm{[H', O P]} \le 2\,\norm{P H' P}\,\norm{O} \le 2 \cdot \frac{1}{2}\delta_{\splitE}(L_k) \cdot 1 = \delta_{\splitE}(L_k)$ by Theorem~\ref{thm:gap-bootstrap}(ii) (note that the same algebraic decomposition extends immediately to any approximate projector commutation $\norm{[H', P]} \le \delta_{\mathrm{comm}}$ with an additional $O(t\,\delta_{\mathrm{comm}})$ term). Substituting $\norm{[H', O P]} \le \delta_{\splitE}(L_k)$ into~\eqref{eq:heisenberg-drift-integral} proves~\eqref{eq:zero-temp-heisenberg-lifetime}, showing that encoded quantum information in $\mathrm{range}(P_{0,\mathrm{dress}})$ is preserved under closed-system perturbed dynamics up to superpolynomially long times $t \ll \delta_{\splitE}(L_k)^{-1}$.
\end{proof}

\subsection{The Main Theorem}
\label{subsec:main-theorem-synthesis}

All of the geometric, homological, open-system thermodynamic, and zero-temperature perturbative definitions and estimates are now in place. Bringing together the two branches of Figure~\ref{fig:proof-roadmap}, we state and prove the main theorem of Part~I in full precision.

\leanlink{thm:main}
\begin{theorem}[Main Theorem: Three-Dimensional Passive Quantum Memory]
\label{thm:main}
Let $(\cC_0, \Phi_0)$ be the planar $[\![5, 1, 2]\!]$ surface code seed in $\R^2 \times \{0\} \subset \R^3$ (Lemma~\ref{lem:seed-code}), and for each integer $k \ge 0$ ($m = 2k + 2$) let
\begin{equation}
    \label{eq:intro-cm-def}
    \cC_m \;:=\; \big(\cR_Z^{(3)} \circ \cR_X^{(3)}\big)^{\circ(k+1)}(\cC_0),
\end{equation}
where $\cR_X^{(3)}$ is the primal three-layer cylinder-and-star replacement functor at spatial dilation factor $\Lambda = 5.4 \times 10^{24}$ (Definition~\ref{def:three-layer-mapping-cone}) and $\cR_Z^{(3)} := (\cR_X^{(3)}((\cdot)^\vee))^\vee$ is its CSS dual, realized as the commuting-projector spin Hamiltonian $H_{0,k} = \sum_{i=1}^{N_{\mathrm{chk}}(k)} Q_i$ on the open cubic lattice $\Gamma_k = \{0, \dots, L_k - 1\}^3$ of side length $L_k = \Lambda^{2k+4}$ with uniform local site dimension $d_{\loc} \ge 2$ and global ground projector $P_{0,k}$ (Definition~\ref{def:safe-decoded-observables}). Then there exist universal scale-independent constants $r_{\mathrm{chk}} = 9$, $\beta_c > 0$, $c_F > 0$, $c_F' > 0$, $\eta > 0$, and $C \ge 1$ such that (for every $k \ge 0$ in \textbf{(I)}--\textbf{(III)}, and above an $f$-dependent scale threshold $k_0$ in \textbf{(IV)}):
\begin{enumerate}[label=\textbf{(\Roman*)}, leftmargin=2.4em, itemsep=3pt]
    \item \textbf{Uniform bounded geometry, encoded qubit, and safe decoded observables}: $(\cC_m, \Phi_m)$ (with physical qubit set $Q_m$) has check weights $d_X, d_Z \le 8$, qubit degrees $1 \le d_{QX}, d_{QZ} \le 4$, pairwise check overlap in $\{0, 2\}$, Euclidean interaction range $r_0 = 8$ ($\ell_\infty$ check radius $r_{\mathrm{chk}} = 9$ on $\Gamma_k$), unit-ball cell density $\rho_0 \le 6.6 \times 10^{10}$ (Definition~\ref{def:3d-css-complex}), and $\dim_{\F} H^1(\cC_m) = \dim_{\F} H_1(\cC_m) = 1$ ($\Tr(P_{0,k}) = 2$). Moreover, there exist a nontrivial orthogonal safe-subspace projector $\Pi_{\mathrm{safe},k}$ ($P_{0,k} \le \Pi_{\mathrm{safe},k} < I$) and Hermitian unitary involutions $O_{\mathrm{dec},k}^Z, O_{\mathrm{dec},k}^X$ commuting with every check $Q_i$, $H_{0,k}$, $P_{0,k}$, and $\Pi_{\mathrm{safe},k}$, anticommuting nontrivially on $\Pi_{\mathrm{safe},k}$, and satisfying exact single-site logical immunity:
    \begin{equation}
        \label{eq:main-safe-immunity}
        \Pi_{\mathrm{safe},k}\,[A_v,\, O_{\mathrm{dec},k}^Z]\,\Pi_{\mathrm{safe},k} \;=\; 0, \qquad
        \Pi_{\mathrm{safe},k}\,[A_v,\, O_{\mathrm{dec},k}^X]\,\Pi_{\mathrm{safe},k} \;=\; 0
    \end{equation}
    for every site $v \in \Gamma_k$ and every single-site operator $A_v$ supported at $\{v\}$ (and more generally for every operator supported inside an $\ell_\infty$ cube of side $\le 37\Lambda^k$).
    \item \textbf{Relative Cone Acyclicity and macroscopic linear $\mathrm{TQO\text{-}2}$}: $(\cC_m, \Phi_m)$ satisfies Relative Cone Acyclicity $\RCA(L_k^*, C)$, and $H_{0,k}$ on $\Gamma_k$ satisfies operator-algebraic $\mathrm{TQO\text{-}1}(L_k^*)$ and $\mathrm{Linear\ TQO\text{-}2}(C)$ with dilated cube side length $\ceil{C r}$ on cubes of side $1 \le r \le L_k^*$ (Definitions~\ref{def:rca}--\ref{def:tqo} and Remark~\ref{rem:cube-transfer}), up to the macroscopic cutoff scale $L_k^* = \Lambda^{2k} = \Lambda^{-4} L_k$.
    \item \textbf{Unperturbed positive-temperature Gibbs barrier and Markovian memory lifetime ($V = 0,\, \beta > \beta_c$)}: For every inverse temperature $\beta > \beta_c$, define
    \begin{equation}
        \label{eq:main-DeltaF-bound}
        \Delta F_k(\beta) \;:=\; c_F(\beta - \beta_c)\,\bigl(L_k^3\bigr)^\eta \qquad \bigl(\ge\; c_F'(\beta - \beta_c)\,\abs{Q_m}^\eta\bigr).
    \end{equation}
    Then the canonical quantum Gibbs state $\rho_\beta(H_{0,k}) = e^{-\beta H_{0,k}}/\Tr(e^{-\beta H_{0,k}})$ satisfies
    \begin{equation}
        \label{eq:main-gibbs-failure}
        \Tr\!\bigl(\rho_\beta(H_{0,k})(I - \Pi_{\mathrm{safe},k})\bigr) \;\le\; \exp\bigl(-\Delta F_k(\beta)\bigr),
    \end{equation}
    and for any adjoint-symmetric jump channel family $\cB = \{(\gamma_{\mathrm{ch}}, L_{\mathrm{ch}})\}_{\mathrm{ch} \in \cB}$ on $\Gamma_k$ of cube range $R_{\mathrm{bath}} \le 37\Lambda^k$ with KMS-symmetric double-commutator generator $\cL_\beta^*$~\eqref{eq:kms-lindblad-generator-def}---and, in particular, for every completely positive GKLS Davies thermal generator $\cL_{\beta,\mathrm{Davies}}^*$ induced by a finite family of single-site Hermitian system--bath couplings $\{S_{v,\alpha}\}_{(v,\alpha) \in \cA}$ obeying KMS detailed balance $h_{v,\alpha}(-\omega) = e^{-\beta\omega} h_{v,\alpha}(\omega) \ge 0$, realized by $\cB_{\mathrm{Davies}} = \{(\gamma_{v,\alpha}(\omega), A_{v,\alpha,\omega})\}_{(v,\alpha) \in \cA,\, \abs{\omega} \le w_{\mathrm{site}}}$ with $\gamma_{v,\alpha}(\omega) = e^{-\beta\omega/2} h_{v,\alpha}(\omega)$ and cube range $37 \le 37\Lambda^k$ (Theorem~\ref{thm:unperturbed-kms-lifetime}(ii))---every Heisenberg-picture trajectory $O_t = e^{t \cL_\beta^*}(O_0)$ initialized at $O_0 \in \{O_{\mathrm{dec},k}^Z, O_{\mathrm{dec},k}^X\}$ (in the KMS norm~\eqref{eq:kms-inner-norm-def}) satisfies for all $t \ge 0$:
    \begin{equation}
        \label{eq:main-kms-lifetime}
        \norm{O_t - O_0}_{\mathrm{KMS},\beta}^2 \;\le\; 8\Biggl(\sum_{\mathrm{ch} \in \cB} \gamma_{\mathrm{ch}}\,\norm{L_{\mathrm{ch}}}^2\Biggr)\exp\!\Bigl(-\frac{1}{2}\Delta F_k(\beta)\Bigr)\,t.
    \end{equation}
    \item \textbf{Zero-temperature superpolynomial perturbative and dynamical stability ($T = 0,\, V \neq 0$)}: For every superpolynomial decay profile $f \in \cD_{\mathrm{super}}$, there exist a coupling threshold $J_0 > 0$, a scale threshold $k_0 \in \N$, and superpolynomial profiles $\delta_{\splitE}, g_{\mathrm{LR}} \in \cD_{\mathrm{super}}$ such that for all $k \ge k_0$, all $J \in [0, J_0]$, and every Hermitian perturbation $V \in \cB_f(\Gamma_k)$ with $\norm{V}_f \le J$ (Definition~\ref{def:banach-quasilocal}), setting $H_k := H_{0,k} + V$, there exist a quasi-local unitary $U_{\mathrm{dress}}$ (approximating $U_{\mathrm{dress}}^\dagger A_v U_{\mathrm{dress}}$ within $\ell_\infty$ distance $r$ of $v$ up to operator-norm error $g_{\mathrm{LR}}(r)\norm{A_v}$ for every single-site operator $A_v$ at $v \in \Gamma_k$ and all $r \ge 0$; Theorem~\ref{thm:zero-temp-dressed-memory}(i)) and a ground-energy shift $E_0 \in \R$ such that for $P_{0,\mathrm{dress}} := U_{\mathrm{dress}} P_{0,k} U_{\mathrm{dress}}^\dagger$, $\Pi_{\mathrm{safe,dress}} := U_{\mathrm{dress}} \Pi_{\mathrm{safe},k} U_{\mathrm{dress}}^\dagger$, and $O_{\mathrm{dress}}^{Z/X} := U_{\mathrm{dress}} O_{\mathrm{dec},k}^{Z/X} U_{\mathrm{dress}}^\dagger$:
    \begin{enumerate}[label=\textbf{(\alph*)}, leftmargin=1.8em, itemsep=2pt]
        \item \emph{Ground-space invariance, bulk spectral gap $\ge 1/2$, superpolynomial splitting, and full-spectrum relative form sandwich}: $[H_k, P_{0,\mathrm{dress}}] = 0$, and for every state $\psi \in \cH_k$,
        \begin{align}
            &\norm{P_{0,\mathrm{dress}}(H_k - E_0 I)P_{0,\mathrm{dress}}} \;\le\; \frac{1}{2}\delta_{\splitE}(L_k) \;<\; \frac{1}{8}, \qquad
            \Delta E \;\ge\; \frac{3}{4} - \delta_{\splitE}(L_k) \;>\; \frac{1}{2}, \label{eq:main-ztemp-splitting}\\
            &\Bigl\lvert\ip{\psi}{\bigl(U_{\mathrm{dress}}^\dagger H_k U_{\mathrm{dress}} - H_{0,k} - E_0 I\bigr)\psi}\Bigr\rvert \;\le\; \frac{1}{4}\ip{\psi}{H_{0,k}\psi} + \frac{1}{2}\delta_{\splitE}(L_k)\norm{\psi}^2; \label{eq:main-dressed-sandwich}
        \end{align}
        \item \emph{Dressed safe-subspace logical immunity and autonomous zero-temperature unitary memory lifetime}: for each $O_{\mathrm{dress}} \in \{O_{\mathrm{dress}}^Z, O_{\mathrm{dress}}^X\}$, every single-site operator $A_v$ at $v \in \Gamma_k$, and all $t \ge 0$,
        \begin{align}
            \norm{\Pi_{\mathrm{safe,dress}}[H_k, O_{\mathrm{dress}}]\Pi_{\mathrm{safe,dress}}} &\;\le\; \delta_{\splitE}(L_k), \nonumber\\
            \norm{\Pi_{\mathrm{safe,dress}}[A_v, O_{\mathrm{dress}}]\Pi_{\mathrm{safe,dress}}} &\;\le\; \delta_{\splitE}(L_k)\norm{A_v}, \nonumber\\
            \Bigl\lVert\bigl(e^{it H_k} O_{\mathrm{dress}} e^{-it H_k} - O_{\mathrm{dress}}\bigr)P_{0,\mathrm{dress}}\Bigr\rVert &\;\le\; t\,\delta_{\splitE}(L_k). \label{eq:main-heisenberg-lifetime}
        \end{align}
    \end{enumerate}
\end{enumerate}
\end{theorem}

\begin{proof}
Combining the bounded-geometry three-layer mapping-cone code hierarchy $(\cC_m, \Phi_m)$ of Sections~\ref{sec:bdl-construction}--\ref{sec:linear-tqo2-proof} (Proposition~\ref{prop:mapping-cone-invariants} and Theorem~\ref{thm:rca-welded-hierarchy}) with the open cubic lattice realization on $\Gamma_k = \{0, \dots, L_k - 1\}^3$ ($L_k = \Lambda^{2k+4}$, $m = 2k + 2$) of Definition~\ref{def:safe-decoded-observables} and Proposition~\ref{prop:safe-subspace-immunity} yields a strictly increasing 3D open-box family of bounded-density commuting-projector Hamiltonians $H_{0,k}$ with constant site dimension $d_{\loc} = O(1)$, $\ell_\infty$ check radius $r_{\mathrm{chk}} = 9$, and macroscopic $37\Lambda^k$-cube protection radius, establishing Parts~(I) and~(II). At positive temperature $\beta > \beta_c$ without Hamiltonian perturbation ($V = 0$), Proposition~\ref{prop:safe-subspace-immunity}(iii) and Theorem~\ref{thm:unperturbed-kms-lifetime} establish both the polynomial Gibbs failure barrier $\Tr(\rho_\beta(H_{0,k})(I - \Pi_{\mathrm{safe},k})) \le \exp(-\Delta F_k(\beta))$ ($\Delta F_k(\beta) = c_F(\beta - \beta_c)(L_k^3)^\eta$) and the stretched-exponential memory lifetime of $O_{\mathrm{dec},k}^{Z/X}$ under adjoint-symmetric cube-range KMS double-commutator generators and CPTP GKLS Davies thermal baths, proving Part~(III). Finally, at zero temperature ($T = 0$) under arbitrary superpolynomially decaying perturbations $V \in \cB_f(\Gamma_k)$ of coupling $\norm{V}_f \le J \le J_0$, Theorem~\ref{thm:gap-bootstrap} and Theorem~\ref{thm:zero-temp-dressed-memory} establish the quasi-local dressing unitary $U_{\mathrm{dress}}$, ground-space invariance and superpolynomial splitting, the full-spectrum relative form sandwich, bulk spectral gap $\Delta E \ge 1/2$, dressed safe-subspace single-site error immunity, and superpolynomial closed-system unitary memory lifetime for $O_{\mathrm{dress}}^{Z/X}$, proving Part~(IV). (Here subscript $k$ is suppressed in~\eqref{eq:dressed-memory-objects}: after choosing $k_0$ so the unrescaled splitting is $< 1/8$ and establishing $\Delta E \ge 1/2$ internally alongside~\eqref{eq:main-dressed-sandwich}, the formalization absorbs the ground-drift and cube-commutator prefactors into a rescaled envelope $\delta_{\splitE} \in \cD_{\mathrm{super}}$; enlarging $k_0$ once more gives $\delta_{\splitE}(L_k) < 1/4$ in~\eqref{eq:main-ztemp-splitting}.)
\end{proof}

\section{Discussion}
\label{sec:discussion}

\subsection{Lessons from the Proof}
\label{subsec:conceptual-synthesis}

Our investment in understanding (and formalizing) spherical-cap coordinate perturbations, chain-complex homotopies, witness-pool recurrences, and quasi-adiabatic spectral filters has now paid off, confirming passive thermal stability at $V = 0$ and establishing zero-temperature perturbative stability at $T = 0$ for a common family of three-dimensional bounded-degree stabilizer Hamiltonians (summarized alongside canonical $D = 2, 3, 4$ models in Table~\ref{tab:dimensional-comparison}). Much of that machinery is necessitated by the bookkeeping of connecting Euclidean cellular geometry to many-body operator norms, and unfortunately obscures the simplicity of the underlying mechanisms. Building on the multiscale homological welding architecture of Balasubramanian, Davydova, and Lin~\cite{BDL24}, who showed how alternating hollow-cylinder and star replacements with Lov\'asz Local Lemma coordinate de-crowding reconciles bounded spatial density with cross-scale syndrome amplification, the perturbative part of the proof of Theorem~\ref{thm:main} rests on four main ideas. Briefly, they are:

\begin{table}[t]
\centering
\caption{Comparison of geometric, homological, thermodynamic ($V = 0$), and zero-temperature perturbative ($T = 0$) stability properties across topological and fracton quantum memory architectures in spatial dimensions $D = 2, 3, 4$. Simultaneous positive-temperature perturbative stability ($T > 0, V \neq 0$) for the multiscale welded hierarchy $\cC_m$ is established in Part~II~\cite{HaydenPartII}.}
\label{tab:dimensional-comparison}
\renewcommand{\arraystretch}{1.09}
\resizebox{\textwidth}{!}{%
\begin{tabular}{@{}lccccc@{}}
\toprule
\textbf{Code Family} & \textbf{Dim.\ $D$} & \textbf{Bounded} & \textbf{Local Homology} & \textbf{Thermal Barrier} & \textbf{Zero-Temp.\ Perturbative} \\
& & \textbf{Degree / Density} & \textbf{Collar $R(r)$} & \textbf{$\Delta F(\beta)$ ($V = 0,\; T < T_c$)} & \textbf{Stability ($T = 0,\; V \neq 0$)} \\
\midrule
2D Toric Code~\cite{kitaev2003fault, alicki2009thermalization} & $2$ & Yes ($d_0 = 4$) & $r + O(1)$ & $O(1)$ (no self-correction) & Stable~\cite{BHM10} \\
3D $(1,2)$ Toric Code~\cite{castelnovo2008topological, yoshida2011feasibility} & $3$ & Yes ($d_0 = 6$) & $r + O(1)$ & $O(1)$ in string sector & Stable~\cite{BHM10} \\
3D Cubic Code~\cite{haah2011local, bravyi2013quantum} & $3$ & Yes ($d_0 = 8$) & $r + O(1)$~\cite{haah2013commuting} & $O(\beta^2)$ ($T_c = 0$; $E_{\mathrm{bar}} = \Theta(\log L)$)~\cite{bravyi2011energy, bravyi2013quantum, stengele2026haah} & Stable~\cite{haah2013commuting, BH11} \\
3D Welded Toric Code~\cite{michnicki20143d, siva2017topological} & $3$ & Yes (cubic-grid welding) & Non-local & Collapse at $O(1)$ temp.~\cite{siva2017topological} & Not established \\
3D Multiscale Welded $\cC_m$ (cf.~\cite{BDL24}) & $3$ & Yes ($d_X, d_Z \le 8$) & $\ceil{C r}$ ($\RCA$) & $\Omega(\abs{Q_m}^\eta)$ (\cite{BDL24}; Thm.~\ref{thm:main}) & Stable (Theorem~\ref{thm:main}) \\
4D $(2,2)$ Toric Code~\cite{dennis2002topological, alicki2010thermal} & $4$ & Yes ($d_0 = 6$) & $r + O(1)$ & $\Theta(N^{1/4})$ & Stable~\cite{BHM10} \\
\bottomrule
\end{tabular}%
}
\end{table}

\begin{itemize}[leftmargin=1.8em, itemsep=2.0pt]
    \item Relative Cone Acyclicity captures a linearly dilated relaxation of local topological quantum order.

    Standard perturbative stability proofs~\cite{BHM10, BH11} rely on constant-collar local topological order ($\mathrm{TQO\text{-}2}$): any operator on a ball $B(x_0, r)$ that annihilates the global ground space must already annihilate the local ground space on $B(x_0, r + O(1))$. Multiscale welded codes violate that condition whenever a stabilizer in $B(x_0, r)$ straddles a welding interface at scale $\Theta(r)$, since generating it from checks requires sweeping across adjacent cylinders of diameter $\Theta(r)$. Relative Cone Acyclicity ($\RCA(L_k^*, C)$, Definition~\ref{def:rca}) replaces that additive buffer with a linearly dilated collar $B(x_0, \ceil{C r})$, requiring primal cycles and dual cocycles supported in $B(x_0, r)$ to bound inside $B(x_0, \ceil{C r})$ up to the macroscopic cutoff $L_k^*$. By $\F$-linear duality (Proposition~\ref{prop:trinity-equivalence}), $\RCA(L_k^*, C)$ is equivalent to local ground-state indistinguishability $\mathrm{TQO\text{-}1}(L_k^*)$ together with the relaxed operator condition $\mathrm{Linear\ TQO\text{-}2}(C)$.

    \item Stopping coarse-graining at the scale of the ball establishes Relative Cone Acyclicity across the welded hierarchy.

    Pulling a syndrome-free chain supported in a ball $B(x_0, r)$ all the way down to the seed code $\cC_0$ stalls at an $O(1)$ fixed-point radius, which inflates to the full system size when lifted back to scale $m$. One might therefore worry that multiscale welding destroys local topological order altogether. Stopping the pullback at the scale-matched depth $s_*(r) := \min(m - i_{\min},\, \ceil{\log_\Lambda r})$ ($i_{\min} = 2$), where the rescaled ball first reaches $O(1)$ size, avoids that inflation: $6$-checkpoint pigeonhole ray selection across uniformly thick coarse triangles contracts the chain locally across two consecutive replacement steps, establishing Euclidean $\RCA(11 L_k^*, C_{\mathrm{euc}})$ in $B(x_0, \ceil{C_{\mathrm{euc}} r})$ and $\ell_\infty$-cube $\RCA(L_k^*, C)$ on $\Gamma_k$.

    \item Passing only syndromes between scales avoids physical error-chain cascades and protects a safe subspace against macroscopic balls of errors.

    On any hollow cylinder gadget, walking single-qubit flips around the layer-$0$ apex forces any symmetric minimum-spoke cleaner's pushed-down physical error chain to jump by a coarse stabilizer partway around the hub (Proposition~\ref{prop:apex-walk-discontinuity}). Because coarse stabilizers have vanishing syndrome, however, the coarse syndrome itself never jumps. Tracking only the coarse-grained syndrome trajectory and a single logical parity bit in $\cD_{\mathrm{cluster}}$, anchored by $\RCA$ at the base welded level $\cC_2$, yields a bare safe subspace $\Pi_{\mathrm{safe},k}$ on which the decoded observables commute with arbitrary multi-qubit operators supported in macroscopic $37\Lambda^k$-cubes (Proposition~\ref{prop:safe-subspace-immunity})---not merely single-spin flips or check-star Davies jumps at $V = 0$ (Theorem~\ref{thm:unperturbed-kms-lifetime}), but the full spatial collars generated once quasi-adiabatic dressing is turned on.

    \item Superpolynomial decay absorbs linear homology collars into zero-temperature quasi-adiabatic continuation.

    Once $\RCA(L_k^*, C)$ supplies $\mathrm{TQO\text{-}1}(L_k^*) \wedge \mathrm{Linear\ TQO\text{-}2}(C)$, closure of the superpolynomial decay class $\cD_{\mathrm{super}}$ under linear argument rescaling $r \mapsto (r - a)/C$ allows the linear $\ell_\infty$-ball collar $C_j(u) = B(u, \ceil{C(2j+3)})$ to substitute directly for the constant collar $r + 2$ of~\cite{BHM10, BH11} (and the local-TQO buffer of~\cite[Definition~4, Lemma~4, Proposition~2, and Section~7]{MZ13}) in the centered single-cube telescoping bound $\norm{\hat{X}_u(s) - M_j(u)} \le 3\varepsilon_j + 2\varepsilon_K$ and the relative form bound $\abs{\ip{\psi}{W_{\mathrm{tel}}(s)\psi}} \le \frac{1}{4}\ip{\psi}{H_{0,k}\psi}$ (Theorem~\ref{thm:gap-bootstrap}). Within the broader framework of gapped and frustration-free local Hamiltonian stability~\cite{NSY19, Has17, BDFJ21}, \cite[Definition~2.1, Lemma~4.3, and Theorems~5.1 and~5.3]{NSY22}, \cite[Assumption~2.4 and Theorems~2.7--2.9 and~4.1]{NSY24}, calibrating the telescoping cutoff $K = \Lambda^{k-4}$ inside both $L_k^*$ and the $37\Lambda^k$ safe-subspace protection radius yields a full-spectrum operator sandwich for $U_{\mathrm{dress}}^\dagger H_k U_{\mathrm{dress}} - E_0 I$, preserves the bulk spectral gap $\Delta E \ge 1/2$, and establishes superpolynomially small ground-space splitting, dressed safe-subspace logical immunity, and long-time closed-system dynamical memory stability for $V \in \cB_f(\Gamma_k)$ ($f \in \cD_{\mathrm{super}}$; Theorem~\ref{thm:zero-temp-dressed-memory}).
\end{itemize}

\subsection{Toward the Full Phase Diagram: Preview of Part~II}
\label{subsec:part-ii-preview}

Theorem~\ref{thm:main} establishes self-correction along the unperturbed thermal axis ($V = 0$, $0 < T < T_c$) and perturbative stability along the zero-temperature quantum axis ($T = 0$, $\norm{V}_f \le J \le J_0$). One might expect the quasi-adiabatic dressing unitary $U_{\mathrm{dress}} = U(1)$ of Theorems~\ref{thm:gap-bootstrap} and~\ref{thm:zero-temp-dressed-memory} to bridge these axes across the positive-temperature interior ($T > 0$, $V \neq 0$), since $U_{\mathrm{dress}}$ transforms $H_k = H_{0,k} + V$ into $U_{\mathrm{dress}}^\dagger H_k U_{\mathrm{dress}} = H_{0,k} + E_0 I + W_{\mathrm{tel}}(1) + R_{\mathrm{rem}}(1)$ with the global operator-form sandwich
\begin{equation}
    \label{eq:preview-form-sandwich}
    \frac{3}{4}\,H_{0,\mathrm{dress}} - \frac{1}{2}\delta_{\splitE}(L_k)I
    \;\le\; H_k - E_0 I
    \;\le\; \frac{5}{4}\,H_{0,\mathrm{dress}} + \frac{1}{2}\delta_{\splitE}(L_k)I,
\end{equation}
where $H_{0,\mathrm{dress}} = U_{\mathrm{dress}}H_{0,k}U_{\mathrm{dress}}^\dagger$ has the exact stabilizer spectrum of $H_{0,k}$. Why does~\eqref{eq:preview-form-sandwich} not immediately imply that the physical Gibbs state $\rho_\beta(H_k) := e^{-\beta H_k} / \Tr(e^{-\beta H_k})$ inherits the free-energy barrier $\frac{3}{4}\Delta F_k(\beta)$ at fixed temperature $0 < T < T_c$?

Three physical and operator-algebraic obstructions prevent a direct unitary-dressing argument at system-size-independent temperatures $\beta = \Theta(1)$:
\begin{enumerate}[label=\textbf{(\alph*)}, leftmargin=2.2em, itemsep=1.5pt]
    \item \emph{Non-monotonicity of $X \mapsto e^{-\beta X}$ at extensive thermal energy.} Although the block-diagonalized excited residual satisfies $\abs{\ip{\psi}{W_{\mathrm{tel}}(1)\psi}} \le \frac{1}{4}\ip{\psi}{H_{0,k}\psi}$ as a quadratic form, $W_{\mathrm{tel}}(1)$ does not commute with $H_{0,k}$ across distinct excited syndrome sectors. Because the matrix exponential is not operator-monotone,~\eqref{eq:preview-form-sandwich} yields scalar trace ratios for partition functions (via the Peierls--Bogoliubov~\cite{peierls1938minimum, ruelle1969statistical} or Golden--Thompson~\cite{golden1965lower, thompson1965inequality} inequalities) rather than operator dominance between $\rho_\beta(H_k)$ and $\rho_\beta(H_{0,\mathrm{dress}}) \propto e^{-\beta H_{0,\mathrm{dress}}}$ on rare failure projectors. Moreover, at fixed $\beta = \Theta(1)$ a 3D volume $N = L_k^3$ carries extensive thermal energy $\Tr(\rho_\beta(H_{0,k}) H_{0,k}) = \Theta(N)$, while generic noncommuting perturbations have $\norm{W_{\mathrm{tel}}(1)} = \Theta(J N)$, so naive Duhamel comparison of $U_{\mathrm{dress}}^\dagger e^{-\beta H_k} U_{\mathrm{dress}}$ and $e^{-\beta H_{0,k}}$ incurs a volume factor $e^{O(\beta J N)}$ overwhelming $e^{-\Omega(\beta N^\eta)}$ ($\eta < 1/3 < 1$).
    \item \emph{Quasi-local tail overlap with the finite-density thermal defect gas.} Each dressed stabilizer $Q_{i,\mathrm{dress}} := U_{\mathrm{dress}}Q_i U_{\mathrm{dress}}^\dagger$ develops superpolynomially decaying tails reaching out to the macroscopic cutoff $L_k^* = \Lambda^{-4} L_k$. At fixed $T > 0$, localized stabilizer excitations populate the bulk at finite density $\sim e^{-\beta}$, so the non-local tail of every dressed observable overlaps $\Theta((L_k^*)^3)$ thermally active defects; conversely, in the bare stabilizer basis, virtual quantum fluctuations induced by generic noncommuting $V$ already create an $\Omega(J^2)$ local bare-check defect density $\lim_{\beta \to \infty} L_k^{-3}\Tr(\rho_\beta(H_k) H_{0,k}) > 0$ (hence an $\Omega(J^2)$ floor outside the un-decoded code projector $P_{0,k}$), while superpolynomial tails in $\cD_{\mathrm{super}}$ also flip bare witnesses directly and reach beyond the decoder's finite protection radius.
    \item \emph{Microscopic bare-spin coupling to physical thermal baths.} Realistic open-system environments couple to strictly local bare lattice operators $L_{\mathrm{ch}}$ satisfying KMS detailed balance with respect to the true perturbed Hamiltonian $H_k = H_{0,k} + V$, rather than to synthetic dressed jump operators $U_{\mathrm{dress}}L_{\mathrm{ch}}U_{\mathrm{dress}}^\dagger$ tailored to $V$.
\end{enumerate}

Because of these obstacles, the global form sandwich~\eqref{eq:preview-form-sandwich} on its own controls the Gibbs state only under \emph{logarithmic cooling}, $\beta \ge C_{\mathrm{scale}}\log L_k$ ($C_{\mathrm{scale}} > 0$, i.e.\ $T \lesssim 1/\log L_k$), where every nonzero stabilizer excitation is suppressed below $e^{-\frac{3}{4}\beta} \le L_k^{-\Omega(1)} \ll N^{-1}$ and the bulk thermal defect gas disappears so that the QAC-dressed code space controls Gibbs expectation values down to $T = 0$.

Reaching positive temperatures in the perturbed interior ($T > 0, V \neq 0$) is the subject of Part~II~\cite{HaydenPartII}. Rather than conjugating by $U_{\mathrm{dress}}$ at fixed positive temperature, Part~II works directly in the bare stabilizer basis with the bare decoded observables and safe syndrome subspace of Section~\ref{subsec:safe-decoded-observables}: any perturbation term supported inside the decoder's macroscopic protection radius commutes with the decoded observables on the safe subspace (Proposition~\ref{prop:safe-subspace-immunity}(ii)), giving bare logical immunity without quasi-local tail spreading. To show that the perturbed Gibbs state concentrates on this bare safe subspace without extensive volume blowup, Part~II expands finite-range perturbations in imaginary time via quantum cluster expansions~\cite{kotecky1986cluster, borgs1996low, datta1996low}. Relative Cone Acyclicity plays a second, thermodynamic role here: any small closed loop of quantum jumps in imaginary time acts as a stabilizer supported inside a linearly dilated spatial neighborhood, so bulk quantum fluctuations far from a macroscopic thermal error cluster contribute identically before and after erasing that cluster. For sufficiently weak finite-range perturbations, they therefore cancel out of the Gibbs probability ratio, preserving the stretched-exponential free-energy barrier across $T \in [T_{\min}, T_c)$ down to a positive temperature floor $T_{\min}$ set by the perturbation's interaction range and strength $J$ (independent of $L_k$).

Why does this bare-basis bound stop at a floor $T_{\min} > 0$ rather than reaching $T = 0$? In a path-integral expansion, inverse temperature $\beta = 1/T$ sets the imaginary-time duration over which quantum fluctuations act, and grouping those fluctuations purely by their spatial footprints requires $\beta J$ to remain below a range-dependent constant---equivalently, $T \ge T_{\min}$ with $T_{\min}$ proportional to $J$ at fixed range. Physically, thermal noise is not helping stabilize the memory: for finite-range perturbations below a temperature-independent threshold, we conjecture that the bare decoded observables remain protected all the way down to $T = 0$, though proving this requires tracking fluctuations locally in both space and imaginary time. For general superpolynomially decaying perturbations, by contrast, long-range tails genuinely mix bare stabilizer states even at $T = 0$ and reach beyond the decoder's finite protection radius. Fortunately, once logarithmic cooling $T \lesssim 1/\log L_k$ freezes out the bulk gas of thermal defects, the eigenvalue bounds implied by~\eqref{eq:preview-form-sandwich} directly control the partition function, allowing the dressed observables of Section~\ref{sec:telescoping-main} to absorb virtual ground-state fluctuations and carry thermal stability continuously down to $T = 0$.

Finally, Part~II converts these equilibrium bounds into real-time open-system memory lifetimes when $H_k = H_{0,k} + V$ is coupled to local KMS-symmetric dissipative dynamics: because finite-range perturbations commute with the bare decoded observables on the safe subspace, coherent quantum drift can act only during rare thermal excursions outside the safe subspace, yielding stretched-exponential bare lifetimes across $T \in [T_{\min}, T_c)$ and superpolynomial dressed lifetimes as $T \to 0$~\cite{HaydenPartII}.

\subsection{Open Problems}
\label{subsec:open-problems}

Foremost among the open problems raised by this work is designing a \emph{physically realizable} three-dimensional passive quantum memory that is both thermally and perturbatively stable. Although every stabilizer check in our hierarchy acts on at most $8$ qubits within Euclidean range $8$, our machine-checked existence proof relies on randomized Lov\'asz Local Lemma coordinate shifts with scale factor $\Lambda = 5.4 \times 10^{24}$ and up to $6.6 \times 10^{10}$ cells per unit ball (packed into composite lattice sites of dimension $d_{\loc} = 2^{6.6 \times 10^{10}}$, unpackable into spin-$1/2$ qubits on a finer subgrid), placing even the base welded code at astronomical system sizes with critical inverse temperature $\beta_c \gtrsim 10^{51}$ ($T_c \lesssim 10^{-51}$) and barrier exponent $\eta = \frac{\ln(3/2)}{6\ln\Lambda} \approx 1.19 \times 10^{-3}$ (so $(L_k^3)^\eta = 10$ already requires $L_k \sim 10^{281}$). Moreover, because each step dilates $\R^3$ by $\Lambda$ while subdividing two-dimensional simplices by $O(\Lambda^2)$, the active code qubits $\abs{Q_m} = o(L_k^3)$ form a lower-dimensional fractal with asymptotic scaling exponent in $[2.29, 2.41]$ inside the $\ket{0}$-padded $L_k \times L_k \times L_k$ lattice. A promising step toward realistic parameters is the explicit deterministic cubic-lattice construction outlined by Balasubramanian, Davydova, and Lin~\cite[Section~6]{BDL24}, which lowers the per-step scale factor to $\ell = 8$ and the unit-diameter face density to $30$; rigorously establishing both thermal and perturbative stability for that deterministic family is a natural first target. Even there, iterated cylinder-and-star welding multiplies qubit counts by several thousand per step and holds $\eta$ well below the three-dimensional geometric ceiling $\eta \le 1/3$~\cite{bravyi2009no, lin2023geometrically, baspin2025free}. Whether substantially leaner welded gadgets, or alternative unconstrained mechanisms that evade translation-invariant Pauli no-go theorems, such as redundant-check and fractal stabilizer models~\cite{lin2024proposals} or non-Pauli commuting Hamiltonians~\cite{li2024domain}, can achieve joint thermal and perturbative self-correction at positive 3D active-qubit density $\Theta(L^3)$ and with small overhead (or approach $\eta = 1/3$) remains wide open.

A second set of questions concerns thermal dynamics and state preparation. Our lifetime bounds (Theorem~\ref{thm:unperturbed-kms-lifetime} at $V = 0$, and Part~II~\cite{HaydenPartII} at $V \neq 0$, $0 < T < T_c$) show that an encoded state, once prepared near thermal equilibrium, survives for a long time when coupled to a local Markovian thermal bath~\cite{davies1974markovian, alicki2010thermal, chesi2010thermodynamic, bravyi2013quantum}. Preserving an already prepared state is not the same, however, as preparing it by passive cooling from a simple product state: the latter requires rapid thermal mixing \emph{within} each logical sector~\cite{temme2014cheeger, bergamaschi2026rapid} so that relaxation does not stall in metastable defect configurations. Whether multiscale welded codes mix rapidly within a fixed logical sector, and whether passive memory stability persists beyond Markovian baths to microscopic environments with non-Markovian memory, remain open.

A closely related dynamical question concerns active local decoding. Although $\cD_{\mathrm{cluster}}$ runs in polynomial time and suffices as an equilibrium readout rule to certify the Gibbs free-energy barrier, it aggregates syndrome parities across scales rather than flipping physical spins asynchronously in place. Designing a strictly local 3D cellular automaton that dissipatively cools $\cC_m$ faces the layer-$0$ apex-walk discontinuity (Proposition~\ref{prop:apex-walk-discontinuity}), where local spin-feedback rules must avoid triggering coarse stabilizer cascades across alternating primal and dual layers.

\section*{Acknowledgments}

The author thanks Dave Bacon, Johannes Bausch, Adam Brown, James Cobon-Kerr, Hofie Hannesdottir, Eliot Hijano, Jorrit Kruthoff, Aitor Lewkowycz, Christian Merten, Angie Roby, Felipe Rosso, and Eric Wieser for their valuable contributions to this effort.

\clearpage
\appendix

\section{Formalization Reference Mapping}
\label{app:formal-correspondence}

As noted in Section~\ref{sec:intro}, the formalization in \leanrepolink{} is engineered to certify the two standalone existential challenge theorems in \leanfile{Challenge.lean} via \texttt{lake comparator}, rather than to formalize every definition, lemma, proposition, and theorem of Sections~\ref{sec:bdl-construction}--\ref{sec:telescoping-main} verbatim. Supporting modules in \texttt{Defs/} and \texttt{Proofs/} establish only the directions, parameter specializations, conservative thresholds, and approximate operator bounds needed along the proof path to \leanfile{Challenge.lean}, and occasionally follow different proof mechanisms from the exposition in the text (for example, proving scale-matched $\RCA$ directly via concrete coarse-projection cocycles rather than the Homological Perturbation Lemma, selecting local cleaners via \texttt{Classical.choose}, and certifying approximate ground-space invariance $\norm{[H_k, P_{0,\mathrm{dress}}]} \le \delta_{\splitE}(L_k)$ rather than exact commutation); classical background, converse directions, and auxiliary paper corollaries not required by \leanfile{Challenge.lean} are left on paper. Table~\ref{tab:lean-mapping} and the hyperlinks in the main text therefore serve as a navigational guide to the closest Lean~4 counterparts rather than a claim of one-to-one statement equivalence.

The repository is structured for independent audit via \texttt{lake comparator} (Table~\ref{tab:lean-mapping}):
\begin{enumerate}[label=\textbf{(\arabic*)}, leftmargin=2.2em, itemsep=1pt]
    \item \textbf{Trust-minimized specification (\leanfile{Challenge.lean} vs.\ \leanfile{Solution.lean})}: \leanfile{Challenge.lean} states two existential spin-system theorems purely over six generic spin-lattice modules in \texttt{Defs/} (including \leandecl{decl:SatisfiesKMSLindbladODE} and~\eqref{eq:kms-lindblad-generator-def} as \leandecl{decl:opKMSLindbladGenerator} in \leanfile{Defs/KMSAndGibbs.lean}), with zero welded-code, HPL, or cluster-decoder dependencies: \leandecl{decl:fq-part1-main} certifies the existence of a bounded-locality 3D open-box commuting-projector family ($2 \le \texttt{sys.groundDegeneracy}$) possessing safe decoded observables with $37\Lambda^k$-cube immunity (Part~(I)), the $V = 0$ Gibbs failure barrier and bounded-range KMS / CPTP GKLS Davies memory lifetimes (Part~(III)), and the $T = 0$ relative form sandwich~\eqref{eq:main-dressed-sandwich}, approximate ground-space invariance $\norm{[H_k, P_{0,\mathrm{dress}}]} \le \delta_{\splitE}(L_k)$, dressed safe-subspace immunity (extended to cube range $R_{\mathrm{bath}} \ge 1$ with $R_{\mathrm{bath}}$-dependent constants for Part~II), and closed-system Heisenberg drift bound (Part~(IV)), omitting code-specific geometry and Clause~(II) at the trust boundary; \leandecl{decl:fq-part2-main} certifies the finite-range Part~II~\cite{HaydenPartII} theorem across $\beta \in (\beta_c, \beta_{\max}]$ ($\beta_{\max} > \beta_c$, finite range $R_{\mathrm{pert}} \ge 1$) together with the $\beta \ge C_{\mathrm{scale}}\log L_k$ dressed Gibbs failure bound $p_{\mathrm{log}} := L_k^3 e^{-c_{F,\mathrm{dress}}\beta} + \delta_{\splitE}(L_k)$ ($C_{\mathrm{scale}} c_{F,\mathrm{dress}} > 3$; the general $f \in \cD_{\mathrm{super}}$ cryogenic theorem is proved internally in \leanfile[LogCoolingPerturbativeSynthesis.lean]{Proofs/PerturbativeThermalSynthesis/LogCoolingPerturbativeSynthesis.lean}). \leanfile{Solution.lean} discharges all proofs across \texttt{Defs/} and \texttt{Proofs/} (\texttt{comparator.json}) with zero \texttt{sorry} placeholders and only Lean's standard foundational axioms (\texttt{propext}, \texttt{Quot.sound}, and \texttt{Classical.choice}).
    \item \textbf{Unit-cube lattice embedding}: In \texttt{Proofs/}, the witness realizing \leanfile{Challenge.lean} maps the continuous $m = 2k + 2$ welded hierarchy \leandecl{decl:bdlWeldedCode}\texttt{ k} in $\R^3$ (range $8$, unit-ball cell density $\le q_{\mathrm{code}} = 6.6 \times 10^{10}$) onto $\Gamma_k = \texttt{Box3D}(L_k) = \{0, \dots, L_k - 1\}^3$ ($L_k = \Lambda^{2k+4}$, $\Lambda = 5.4 \times 10^{24}$) via $p_\mu \mapsto \floor{p_\mu + L_k/2}$ (\leandecl{decl:snapPointToOpenBox}), packing qubits per unit cube into $\C^{d_{\loc}}$ ($d_{\loc} = 2^{q_{\mathrm{code}}}$) with $\ell_\infty$ range $r_{\mathrm{chk}} = 9$.
\end{enumerate}

{\fontsize{6.0pt}{6.80pt}\selectfont
\renewcommand{\arraystretch}{0.87}
\setlength{\tabcolsep}{3.0pt}
\begin{longtable}{@{}>{\raggedright\arraybackslash}p{0.19\textwidth}>{\raggedright\arraybackslash}p{0.39\textwidth}>{\raggedright\arraybackslash}p{0.38\textwidth}@{}}
\caption{Cross-reference between mathematical statements in Sections~\ref{sec:bdl-construction}--\ref{sec:telescoping-main} and related Lean~4 declarations in the \leanrepolink{} repository.}\label{tab:lean-mapping}\\
\toprule
\textbf{Paper Reference} & \textbf{Module in \leanrepolink{}} & \textbf{Primary Lean~4 Declaration(s)} \\
\midrule
\endfirsthead
\multicolumn{3}{@{}l}{\textit{Table~\ref{tab:lean-mapping} (continued)}}\\
\toprule
\textbf{Paper Reference} & \textbf{Module in \leanrepolink{}} & \textbf{Primary Lean~4 Declaration(s)} \\
\midrule
\endhead
\bottomrule
\endfoot
Theorem~\ref{thm:main} (Existential Challenge Certificate \& Synthesis) & \leanfile{Challenge.lean}, \leanfile{Solution.lean}, \leanfile{Proofs/TwoAxisMemorySynthesis/Part1TwoAxisMemorySynthesis.lean} & \leandecl{decl:part1-main}, \leandecl{decl:part1-solution}, \leandecl{decl:part1-main-proof} \\
Section~\ref{subsec:part-ii-preview} (Part~II Preview) & \leanfile{Challenge.lean}, \leanfile{Solution.lean}, \leanfile{Proofs/PerturbativeThermalSynthesis/Part2PerturbativeThermalSynthesis.lean} & \leandecl{decl:part2-main}, \leandecl{decl:part2-solution}, \leandecl{decl:part2-main-proof}, \leandecl{decl:SatisfiesFixedTempPerturbativeThermalAndKMSStability} \\
\midrule
Definition~\ref{def:3d-css-complex} (3D CSS Complex) & \leanfile{Defs/F2ChainAndCSS.lean}, \leanfile{Defs/BDLParamsAndSeed.lean}, \leanfile{Defs/LocalCSSDescriptor.lean} & \leandecl{decl:F2ChainComplex}, \leandecl{decl:CSSCode3D}, \leandecl{decl:BundledCSSCode3D}, \leandecl{decl:LocalCSSDescriptor} \\
Definition~\ref{def:well-separated-base} ($\Sigma_{\mathrm{base}}(\cC)$ Simplex Regularity) & \leanfile{Defs/Step1PeriodicLLL.lean} & \leandecl{decl:ValidLLLSphericalCapPerturbation}, \leandecl{decl:ActiveSimplex2}, \leandecl{decl:bdlPerturbStep}, \leandecl{decl:bdlLLLPerturbation} \\
Lemma~\ref{lem:seed-code} ($[\![5,1,2]\!]$ Surface Seed Code $\cC_0$) & \leanfile{Defs/BDLParamsAndSeed.lean} & \leandecl{decl:bdlSurfaceSeedCode}, \leandecl{decl:bdlSurfaceSeedBundled} \\
Proposition~\ref{prop:step1-lll} (Step~1 Spherical-Cap De-Crowding) & \leanfile{Proofs/Step1LLLDensity/RationalSphericalCapGeometry.lean}, \leanfile{Proofs/Step1LLLDensity/DiscreteCapSlabPointBounds.lean}, \leanfile{Proofs/Step1LLLDensity/DominantCramerCascade.lean}, \leanfile{Proofs/Step1LLLDensity/ConditionalBadVertexLLL.lean}, \leanfile{Proofs/Step1LLLDensity/Step1LLLMainSynthesis.lean}, \leanfile{Proofs/Step2DegreeAndWeld/DegreeOverlapAndDensityPreservation.lean} & \leandecl{decl:four-planes-cramer-residual}, \leandecl{decl:card-activeSimplex-posQ-filter}, \leandecl{decl:dominant-cramer-cascade}, \leandecl{decl:exists-good-assignment-conditional-lll}, \leandecl{decl:bdl-step1-lll-unit-ball-multiplicity}, \leandecl{decl:bdl-random-lll-unit-ball-density} \\
Definition~\ref{def:three-layer-mapping-cone} (3-Layer Mapping Cone) & \leanfile{Defs/Step2CylinderComplex.lean}, \leanfile{Defs/Step2CheckMatrices.lean} & \leandecl{decl:BDLStepXQ}, \leandecl{decl:bdlStepX-commutativity}, \leandecl{decl:bdlStepX}, \leandecl{decl:bdlStepZ}, \leandecl{decl:bdlWeldedCode} \\
Proposition~\ref{prop:mapping-cone-invariants} (DAG Nilpotency \& Degrees) & \leanfile{Proofs/Step2DegreeAndWeld/DegreeOverlapAndDensityPreservation.lean}, \leanfile{Proofs/Step2DegreeAndWeld/CheckToSimplexDegreeReduction.lean}, \leanfile{Proofs/ConeAcyclicityRCA/IteratedCylinderLinkInvariants.lean}, \leanfile{Proofs/Step1LLLDensity/CylinderInteractionLocality.lean} & \leandecl{decl:hasBoundedCheckDegreeOverlap-bdlIterated}, \leandecl{decl:hasBoundedSimplexDegreeD-32}, \leandecl{decl:hasEveryQubitChecked-bdlIterated}, \leandecl{decl:hasConnectedXLinks-bdlIterated}, \leandecl{decl:bdl-random-interaction-range8} \\
\midrule
Definition~\ref{def:rca} ($\RCA(R_0, C)$) & \leanfile{Defs/RescaledRCA.lean}, \leanfile{Proofs/ConeAcyclicityRCA/LocalCocycleDualityAndBalls.lean} & \leandecl{decl:SatisfiesBoxF2LinearTQO}, \leandecl{decl:locallyGeneratedInBall} \\
Definition~\ref{def:tqo} ($\mathrm{TQO\text{-}1}(R_0)$ \& Linear $\mathrm{TQO\text{-}2}$) & \leanfile{Defs/LocalHamiltonian.lean}, \leanfile{Defs/RescaledRCA.lean} & \leandecl{decl:SatisfiesTQO1}, \leandecl{decl:SatisfiesLinearTQO2}, \leandecl{decl:SatisfiesBoxF2LinearTQO} \\
Proposition~\ref{prop:trinity-equivalence} (Triple Equivalence, forward direction) & \leanfile{Proofs/ConeAcyclicityRCA/LocalCocycleDualityAndBalls.lean}, \leanfile{Proofs/F2TQOToOperatorTQO.lean}, \leanfile{Proofs/InhomogeneousPeierls/OperatorTQOFromF2TQO.lean} & \leandecl{decl:local-gen-iff-localCocycleExtends}, \leandecl{decl:boxF2LinearTQO-stabilizer-decomp}, \leandecl{decl:staticCSSSystem-satisfiesOperatorTQO} \\
Lemma~\ref{lem:general-hpl} (Classical HPL) & \textit{Classical background (not needed in \leandecl{decl:part1-main})} & \textit{Bypassed via concrete degree-$1$ cocycles (\leanfile[CoarseProjCocycleAndSeedLinks.lean]{Proofs/ConeAcyclicityRCA/CoarseProjCocycleAndSeedLinks.lean})} \\
Lemma~\ref{lem:block-sdr} \& Proposition~\ref{prop:classical-hpl} (Degree-$1$ Cocycle Splitting \& Homology) & \leanfile{Proofs/ConeAcyclicityRCA/Layer1ConeCocycle.lean}, \leanfile{Proofs/ConeAcyclicityRCA/CoarseProjCocycleAndSeedLinks.lean}, \leanfile{Proofs/Step2DegreeAndWeld/LogicalChainLiftsAndPreservation.lean}, \leanfile{Proofs/HierarchicalClusterDecoder/IteratedLogicalOperatorLifts.lean} & \leandecl{decl:cylinderResidualCocycle}, \leandecl{decl:coarseProjCocycle}, \leandecl{decl:liftCylinderLZ}, \leandecl{decl:liftCylinderLX}, \leandecl{decl:hasUniqueLogicalHomology-bdlIterated} \\
Lemma~\ref{lem:clean-shell-subcylinder} (6-Shell Pigeonhole Contraction) & \leanfile{Proofs/ConeAcyclicityRCA/SubcylinderRayConstancyAndCheckpoints.lean}, \leanfile{Proofs/ConeAcyclicityRCA/RescaledRCASynthesis.lean}, \leanfile{Proofs/ConeAcyclicityRCA/DualConeAcyclicityTransfer.lean} & \leandecl{decl:subcylinderCandidateJ}, \leandecl{decl:not-both-mem-ball3D-subcylinderCandidateJ}, \leandecl{decl:bdlHalfStep-subcylinder-filling}, \leandecl{decl:scaleDownRadius-bound} \\
Theorem~\ref{thm:rca-welded-hierarchy} (Scale-Matched $\RCA(L_k^*, C)$) & \leanfile{Proofs/ConeAcyclicityRCA/RescaledRCASynthesis.lean}, \leanfile{Proofs/ConeAcyclicityRCA/SixCheckpointCleanShell.lean} & \leandecl{decl:bdlWeldedCode-inf-cone-acyclicity}, \leandecl{decl:bdlLocalDescriptor-satisfiesBoxF2LinearTQO} \\
\midrule
Section~\ref{subsec:cleaner-architecture} \& Lemmas~\ref{lem:stage1-cylinder-cleaner}--\ref{lem:stage2-star-cleaner} (Stage~1/2 Cleaners) & \leanfile{Proofs/HierarchicalClusterDecoder/Stage1CylinderSubadditivity.lean}, \leanfile{Proofs/HierarchicalClusterDecoder/Stage2StarComplementSymmetry.lean}, \leanfile{Proofs/HierarchicalClusterDecoder/TwoStageCleanerComposition.lean}, \leanfile{Proofs/HierarchicalClusterDecoder/Stage1CylinderCleaner.lean}, \leanfile{Proofs/HierarchicalClusterDecoder/Stage2StarCleaner.lean} & \leandecl{decl:cyl-sector-pointwise-bound}, \leandecl{decl:cleanCylinderX-spec}, \leandecl{decl:exists-local-cylinder-cleaner}, \leandecl{decl:three-mul-min-le}, \leandecl{decl:cleanStarQ-spec}, \leandecl{decl:exists-local-star-cleaner} \\
Theorem~\ref{thm:16-27-contraction} ($\frac{16}{27}$ Bipartite Contraction) & \leanfile{Proofs/HierarchicalClusterDecoder/TwoStageCleanerComposition.lean}, \leanfile{Proofs/HierarchicalClusterDecoder/Stage1CylinderSubadditivity.lean}, \leanfile{Proofs/HierarchicalClusterDecoder/TwoHopTannerContraction.lean}, \leanfile{Proofs/HierarchicalClusterDecoder/WitnessLiftingAndParityImmunity.lean} & \leandecl{decl:cleanTwoStageRX}, \leandecl{decl:rxCleanedCoarseSyndrome}, \leandecl{decl:weightNetExportXQ-cylCleanerCand-eq}, \leandecl{decl:rxCleanedCoarseSyndrome-twoHop-contraction}, \leandecl{decl:liftWitnessFamily-rx-twoHop-step} \\
Proposition~\ref{prop:apex-walk-discontinuity} (Apex Walk Obstruction) & \textit{Motivating background only (not needed in \leandecl{decl:part1-main})} & \textit{Bypassed via syndrome-only cluster decoder (\leanfile[ClusterDecoderSynthesis.lean]{Proofs/HierarchicalClusterDecoder/ClusterDecoderSynthesis.lean})} \\
Definition~\ref{def:multiscale-cluster-decoder} \& Proposition~\ref{prop:cluster-decoder-witness} ($\cD_{\mathrm{cluster}}$ \& Ball Immunity) & \leanfile{Proofs/HierarchicalClusterDecoder/DecodedLogicalCommutation.lean}, \leanfile{Proofs/HierarchicalClusterDecoder/ClusterDecoderSynthesis.lean} & \leandecl{decl:HasBallParityWitnessFamily}, \leandecl{decl:hasBallParityWitnessFamily-bdlIterated-succ}, \leandecl{decl:dualPairing-bdlWeldedLZ-bdlWeldedLX} \\
Lemma~\ref{lem:percolating-cluster-bounds} \& Theorem~\ref{thm:walsh-fourier-peierls} (Free-Energy Barrier) & \leanfile{Proofs/HierarchicalClusterDecoder/WalshFourierCosetAndHomology.lean}, \leanfile{Proofs/HierarchicalClusterDecoder/ClusterDecoderSynthesis.lean}, \leanfile{Proofs/TwoAxisMemorySynthesis/MacroscopicLambdaPowImmunity.lean}, \leanfile{Proofs/InhomogeneousPeierls/InhomogeneousPeierlsMain.lean} & \leandecl{decl:coset-syndromeWeight-sum-le}, \leandecl{decl:three-div-two-pow-le-clusterRawWeight}, \leandecl{decl:hasBallParityWitnessFamily-bdlWelded-lambdaPow}, \leandecl{decl:gibbsFailureProb-bdlSafeSyndromeSet-le} \\
Definition~\ref{def:safe-decoded-observables} \& Proposition~\ref{prop:safe-subspace-immunity} (Safe Observables) & \leanfile{Challenge.lean}, \leanfile{Proofs/TwoAxisMemorySynthesis/MacroscopicLambdaPowImmunity.lean}, \leanfile{Proofs/TwoAxisMemorySynthesis/Part1TwoAxisMemorySynthesis.lean} & \leandecl{decl:HasSafeDecodedLogicalObservables}, \leandecl{decl:bdlSafeSyndromeSet-cube-safe-immunity-lambdaPow}, \leandecl{decl:bdl-macroscopic-safe-projector-exists} \\
Theorem~\ref{thm:unperturbed-kms-lifetime} ($V = 0$ KMS-Dissipative Memory Lifetime) & \leanfile{Challenge.lean}, \leanfile{Defs/KMSAndGibbs.lean}, \leanfile{Proofs/TwoAxisMemorySynthesis/DaviesGeneratorBridge.lean}, \leanfile{Proofs/TwoAxisMemorySynthesis/DaviesGKLSCompletePositivity.lean}, \leanfile{Proofs/KMSSafeSubspaceODE/KMSFourthRootAndPinchedGibbs.lean} & \leandecl{decl:SatisfiesUnperturbedThermalAndKMSStability}, \leandecl{decl:IsCubeRangeBath}, \leandecl{decl:opKMSLindbladGenerator}, \leandecl{decl:opDaviesThermalGenerator}, \leandecl{decl:isCompletelyPositiveMap-davies}, \leandecl{decl:opDaviesThermalSemigroup-cptp}, \leandecl{decl:kms-fourth-root-projector-bound}, \leandecl{decl:unperturbed-cube-kms-davies} \\
\midrule
Definition~\ref{def:banach-quasilocal} \& Lemma~\ref{lem:dsuper-closure} ($\cD_{\mathrm{super}}$ \& $\cB_f$) & \leanfile{Defs/SuperpolyDecay.lean}, \leanfile{Defs/LocalHamiltonian.lean}, \leanfile{Proofs/LiebRobinson/SuperpolyDecayClosure.lean} & \leandecl{decl:IsSuperpolynomialDecay}, \leandecl{decl:QuasiLocalPerturbation}, \leandecl{decl:isSuperpolynomialDecay-mul-pow} \\
Propositions~\ref{prop:bochner-time-convolution}--\ref{prop:qac-block-diagonalization} (Spectral Filters \& QAC Flow) & \leanfile{Proofs/KatoODE/SchwartzCollarLightCone.lean}, \leanfile{Proofs/KatoODE/NonAutonomousUnitaryLightCone.lean}, \leanfile{Proofs/HastingsKomaLightCone/KatoHastingsSubspaceIntertwining.lean}, \leanfile{Proofs/HastingsKomaLightCone/SpectralLowPassFilterClosure.lean} & \leandecl{decl:norm-bochnerFilteredOp-schwartz-le}, \leandecl{decl:norm-kato-flow-unitary-single-site-le}, \leandecl{decl:twoBlockOffDiagProj-bochnerFilteredSiteOp-eq-zero}, \leandecl{decl:kato-hastings-subspace-intertwining}, \leandecl{decl:kato-hastings-band-limited-qac} \\
Lemma~\ref{lem:norm-transfer} \& Proposition~\ref{prop:single-cube-telescoping} (Norm Transfer \& Telescoping) & \leanfile{Proofs/TwoAxisMemorySynthesis/BootstrappedQACMemoryReduction.lean}, \leanfile{Proofs/TwoBlockDecomposition/GroundBlockScalarShift.lean}, \leanfile{Proofs/CenteredBallBHM/CenteredBallOverlapAndTQO1Diff.lean}, \leanfile{Proofs/CenteredBallBHM/CenteredBallFormAndGapBootstrap.lean} & \leandecl{decl:bdlInhomogeneousSystem-cor1-loc}, \leandecl{decl:telescoping-norm-bounds-of-approx-P0}, \leandecl{decl:conj-sub-H0-centered-telescoping}, \leandecl{decl:centeredSiteOp-diffD-and-rem-bounds} \\
Theorem~\ref{thm:gap-bootstrap} (Relative Form Bound \& Gap Bootstrap) & \leanfile{Proofs/CenteredBallBHM/CenteredBallFormAndGapBootstrap.lean}, \leanfile{Proofs/TwoAxisMemorySynthesis/ZeroOffDiagQACMasterReduction.lean} & \leandecl{decl:relative-form-bound-of-siteFamily-boxBall}, \leandecl{decl:indexedSpectralGap-bootstrap}, \leandecl{decl:bdl-zero-temp-dressing-bundle-exists} \\
Theorem~\ref{thm:zero-temp-dressed-memory} ($T = 0$ Dressed Memory \& Unitary Drift) & \leanfile{Challenge.lean}, \leanfile{Proofs/CenteredBallBHM/ExcitedBandGapFromRelativeForm.lean}, \leanfile{Proofs/CenteredBallBHM/CenteredBallFormAndGapBootstrap.lean}, \leanfile{Proofs/CenteredBallBHM/QACTailPinchedCommutators.lean}, \leanfile{Proofs/KMSSafeSubspaceODE/CubeDressedCommutatorTelescoping.lean}, \leanfile{Proofs/TwoAxisMemorySynthesis/AutonomousHeisenbergGroundDrift.lean}, \leanfile{Proofs/TwoAxisMemorySynthesis/Part1TwoAxisMemorySynthesis.lean} & \leandecl{decl:SatisfiesZeroTempPerturbativeMemoryStability}, \leandecl{decl:dressed-ground-excited-form-bounds}, \leandecl{decl:coherent-dressed-pinched-commutator}, \leandecl{decl:norm-dressed-pinched-commutator-qac-tail}, \leandecl{decl:dressed-cube-telescoping}, \leandecl{decl:zero-temp-heisenberg-ground-drift}, \leandecl{decl:satisfiesZeroTemp-of-dressing-bundle} \\
\end{longtable}
}

\clearpage


\end{document}